\documentclass[prc,twocolumn,showpacs,showkeys,superscriptaddress,nofootinbib,floatfix]
{revtex4-2}
\usepackage{graphicx,amsmath,amssymb}
\def\be{\begin{eqnarray}}
\def\ee{\end{eqnarray}}
\def\bc{\begin{center}}
\def\ec{\end{center}}

\def\rmd{{\rm d}}
\def\om{\omega}

\def\half{{\textstyle \frac12}}

\def\rmd{{\rm d}}
\def\Tr{{\rm Tr}}
\def\tr{{\rm tr}}

\newcommand{\lsim}{\stackrel{\scriptstyle <}{\phantom{}_{\sim}}}
\newcommand{\gsim}{\stackrel{\scriptstyle >}{\phantom{}_{\sim}}}
\def\braket#1{\langle #1 \rangle}
\def\bra#1{\langle #1 |}
\def\ket#1{| #1 \rangle}

\def\YYint#1#2#3{{\setbox0=\hbox{$#1{#2#3}{\int}$}   \vcenter{\hbox{$#2#3$}}\kern-.52\wd0}}

\def\Ucor{U}
\def\Umom{\mathcal{U}}

\makeatletter
\let\old@ssect\@ssect % Store how ifacconf defines \@ssect
\makeatother
\usepackage[plainpages=false,pdfpagelabels=false]{hyperref}
\makeatletter
\def\@ssect#1#2#3#4#5#6{%
  \NR@gettitle{#6}% Insert key \nameref title grab
  \old@ssect{#1}{#2}{#3}{#4}{#5}{#6}% Restore ifacconf's \@ssect
}
\makeatother
\begin{document}
\title{Conserving approximations to dilute equilibrium  systems.\\ Pair interaction potential.}
\author{E.E. Kolomeitsev}
\affiliation{Matej Bel University, SK-97401 Banska Bystrica, Slovakia}
\affiliation{BLTP, Joint Institute for Nuclear Research, RU-141980 Dubna, Russia}
\author{P.D. Lukianov}
\affiliation{National Research Nuclear University (MEPhI), 115409 Moscow, Russia}
%\affiliation{BLTP, Joint Institute for Nuclear Research, RU-141980 Dubna, Russia}
\author{D.N. Voskresensky}
\affiliation{BLTP, Joint Institute for Nuclear Research, RU-141980 Dubna, Russia}
\affiliation{National Research Nuclear University (MEPhI), 115409 Moscow, Russia}
%=================================================================%
\date{\today}
\begin{abstract} We study the conserving $\Phi$-derivable and virial approaches for dilute strongly-interacting systems in equilibrium. We consider a system of non-relativistic fermions of one kind (e.g., neutron matter) interacting via a pair potential in the limit $n\lambda^3\ll 1$, where $n$ is the fermion number density and $\lambda$ is the thermal wavelength. First, thermodynamic quantities are expressed in terms of various spectral functions in the $\Phi$ derivable approximation scheme. Explicit expressions are derived in the case of the tadpole and sandwich diagrams in $\Phi$. The variety of potentials of the quantum and classical origin are considered and then used for the quantitative comparison of the results obtained within both approaches. We demonstrate that, in all cases relevant to the nucleon-nucleon interaction, the attraction volume of the interaction is larger than the repulsion one. Expressions for the second and third virial coefficients are derived and analyzed for the $\Phi$ functional including the tadpole and sandwich diagrams, and the relations between the Hartree and Fock contributions are found. It is shown that the inclusion of these two diagrams in $\Phi$ is sufficient to obtain explicit expressions for all virial coefficients within the Born approximation. We show that the classical high-temperature limit of the $\Phi$ functional is reproduced by the tadpole diagram. Next, classical, semiclassical (up to order $\hbar^3$), and purely quantum (Beth-Uhlenbeck) virial expansions of the equation of state are studied in detail. Different extrapolations of the virial equation of state are considered including the van~der~Waals form and the excluded volume models. We derive the expression for the second virial coefficient using the effective range approximation for the scattering amplitude and compare the result with the purely quantum Beth-Uhlenbeck result using the experimental phase shifts, showing their approximate agreement for low temperatures, $T\lsim 20 $ MeV. Then, we focus attention on the problem of the anomalously large empirical value of the nucleon-nucleon scattering length, appearing due to the presence of the quasi-bound state in nucleon-nucleon scattering. The latter can be destroyed in the matter because of the Pauli blocking. Subtracting the quasi-bound state contribution, we show that the second virial coefficient has the same low-temperature dependence as one obtained within the $\Phi$ derivable approach in the Born approximation. We discuss the validity of such a procedure to describe the equation of state of the nuclear matter in the virial limit. The $\Phi$-derivable approach with the tadpole and sandwich diagrams allows to recover the classical part of the third virial coefficient evaluated with the scattering amplitude taken in the Born approximation. Also, we discuss the origins of the differences appearing beyond the region of the common validity of the $\Phi$ derivable and virial approaches.
\end{abstract}

\maketitle
\tableofcontents
%=======================================================================
\section{Introduction}
Any actual calculation requires the usage of certain approximations. The simplest approximation scheme is the perturbation theory. However, in many cases, perturbation theory is not applicable due to the absence of a small parameter, as for systems with strong coupling.

The appropriate frame for the description of many-particle systems is Green's function formalism. A quasiparticle Green's function approach to the many-body theory developed by Landau, Migdal, and others, cf.~\cite{Land56,Galitskii,Mig,LL9,Pines-Nozieres}, proved to be very successful. For example, it resulted in the construction of a phenomenological Fermi liquid theory. However, the approach requires a renormalization procedure, since exact sum rules for the Green's functions are not fulfilled within the standard quasiparticle approximation (QPA). Therefore, in many cases, the quasiparticle description is not sufficient due to a lack of the effects of particles' finite widths. For example, without a proper treatment of these effects it is impossible to describe the dynamics of resonances, cf.~\cite{Knoll:1995gs,KV1996,IKV-2}. Even stable particles acquire finite mass-widths in the strongly interacting matter and their properties become similar to those of resonances. In some cases, mass-width effects in matter prove to be so strong that one deals with completely blurred particles~\cite{Dunn,Shklovskii,Dyugaev:1993mn,Voskresensky:2004ux}.

Thus, the appropriate frame to construct an approximation scheme for the description of many-particle systems is Green's function formalism developed beyond the perturbation theory and the QPA. In the case of thermal equilibrium, one may start with an exact expression for the grand-canonical thermodynamic potential presented as infinite series of diagrams expressed in terms of the full Green's functions obeying the exact Dyson equations. These diagrams can be treated within the Matsubara diagrammatic technique~\cite{Abrikos,Kohn-Luttinger,Luttinger-Ward,Schwinger}. For non-equilibrium systems, an appropriate real-time formalism was developed
by Schwinger, Kadanoff, Baym and Keldysh~\cite{Schw,KadanoffBaym,Keld64}.

Within the perturbation theory and the QPA, the conservations of the charge, energy, and momentum can be satisfied only approximately.
However, in the numerical codes describing collisions and production and absorption of particles even a tiny non-conservation of the number of particles or energy-momentum in each individual act of interaction may result in an accumulation of a serious error after processing many acts of particle interactions. Namely, a necessity to provide exact conservation laws forces one to employ the simple kinetic Boltzmann equation and its Boltzmann-Uhling-Uhlenbeck modification in actual simulations of heavy-ion collisions, although it is clear that these equations are derived within the perturbation theory, which is not justified in application to the strong interactions.

Instead of the usage of the perturbation theory and the quasiparticle approach one may sum up certain sub-series of diagrams in order to obtain a reasonable approximation. However, quite not any approximation to the Green's functions provides exact charge, energy and momentum conservations. For the first time, this problem of conservation laws was considered by Baym and Kadanoff~\cite{KadanoffBaym,Baym-Phi} in the equilibrium formalism within the theory of the linear response to an external perturbation. There, a so-called $\Phi$-functional was introduced. A similar formalism was introduced by Luttinger and Ward in~\cite{Luttinger-Ward} in relation to the thermodynamic potential (see also~\cite{Abrikos}). The $\Phi$-functional is determined in terms of the full Green's functions and free vertices, the vertex parameter can be considered as an expansion parameter of the given approximation theory. In Ref.~\cite{Baym-Phi}, Baym showed that any approximation, in order to keep the exact conservation laws of the 4-current and energy-momentum, i.e. to be conservable and thermodynamically consistent, must be so-called $\Phi$-derivable. The problem of conservable approximations in equilibrium formalism was further considered in many works, see~\cite{Riedel,Carneiro-Pethick,Weinhold,Weinhold1,Rios} and references therein. Even if one keeps one diagram of $\Phi$-functional and calculates the pressure with its help, the approximation proves to be conserving and thermodynamically consistent. The thermodynamic consistency is provided by the fact that all thermodynamic quantities are related to the pressure by the exact thermodynamic relations. The price paid for the consistency of the model is that one deals not with the initial exact theory of interacting particles but with a model described by the given sub-series of the diagrams.

The problem of conserving approximations in non-equilibrium formalism is even more severe, since in this case, the system may exercise a violent time evolution. Conservations of the 4-current and energy-momentum in the non-equilibrium formalism were studied in~\cite{KadanoffBaym,Dan84,BotMalf,IKV-1,IKV-2,Leupold2000,KIV,IKV-YaF,Ivanov:2000ma,vanHees:2001ik,VanHees:2001pf,vanHees:2002bv,Voskresensky:2008ur,Ivanov:2009wc,KVJPG40} and other works, and a number of examples of conserving approximations were presented. The quantum kinetic Kadanoff-Baym equations were derived within the $\Phi$-derivable scheme~\cite{IKV-2} up to the first order in gradients. The system of equations contains the proper kinetic Kadanoff-Baym equation and the mass-shell equation for the 4-momentum and 4-coordinate distribution and the equation for the spectral function. It proved to be that the collision term of the kinetic Kadanoff-Baym equation contains not only a local (Markovian) contribution but also a first-order-gradient (memory) term. The latter term comes from the gradient contributions of the diagrams with three and more vertices in $\Phi$. Reference~\cite{BotMalf} showed that the Kadanoff-Baym first-gradient-order kinetic equation can be further simplified if the system is very close to the equilibrium. Such a kinetic equation is called in the literature the Kadanoff-Baym equation in the Botermans-Malfliet (BM) form. It proves to be~\cite{IKV-2} that the Kadanoff-Baym equation in the BM form and the mass-shell equation coincide. Also in~\cite{IKV-2} expression for a flow of the kinetic entropy, $S_{\rm loc}^\mu$, was derived, $\mu =0,1,2,3$. It includes the purely Markovian contribution and the non-Markovian memory term, $S_{\rm mem}^\mu$, appearing from the gradient contributions of the diagrams with three and more vertices in the $\Phi$ yielding the collision term. In special cases, it was demonstrated that the derived expressions of the entropy satisfy the $H$-theorem. Memory effects in the kinetic terms provided contributions to the kinetic entropy flow that in the equilibrium limit recovered the famous boson type $T^3\ln T$ correction to the specific heat as was derived in Ref.~\cite{Riedel,Carneiro-Pethick} for the Fermi liquids like Helium-3.

It was demonstrated in~\cite{KIV,IKV-YaF} that the Noether 4-current, and the energy-momentum tensor related to the Kadanoff-Baym first-order-gradient kinetic equation are conserved exactly if the $\Phi$-derivable approximation is used, and the memory effects are consistently taken into account in the collision term up to the first gradients. Reference~\cite{Leupold2000} showed that the BM equation with the local collision term conserves exactly not the Noether current but another one -- the so-called BM 4-current. Reference~\cite{IKV-YaF} demonstrated that for the pair-potential interaction ($U$), the Noether and the BM 4-currents derived from the $\Phi$ functional with both the Hartree-Fock tadpole and sandwich (TS) diagrams (linear in $U$) and the so-called Born diagrams (quadratic in $U$) coincide approximately up to terms containing the first gradients. Actual calculations of heavy-ion collisions are performed using the BM form of the Kadanoff-Baym kinetic equation in the first-order gradients and exploiting the local collision term neglecting memory effects. As a result, these calculations conserve the BM 4-current rather than the original Noether 4-current. Therefore, there appears a question about the relation between the Noether and the BM 4-currents. These problems will be considered below in our paper on examples of 0-components of the 4-currents (particle densities) in equilibrium systems. An example of the nucleon-nucleon pair potential for the first, TS diagrams of $\Phi$ will be considered explicitly.

The other so-called virial approach to the equation of state was developed by many authors for the description of dilute finite temperature classical gases, cf. in textbooks~\cite{LLV,Huang,Reichl}, and by Beth and Uhlenbeck~\cite{Uhlenbeck-Beth,Beth} for the description of dilute high-temperature quantum gases. The parameter of the expansion of the free energy $F(n,T)$ is $n\lambda^3\ll 1$, where $n$ is the density of the gas, $T$ is the temperature, and $\lambda =\hbar\sqrt{2\pi /(mT)}$ is the thermal wavelength, here $\hbar$ is the Planck constant and $m$ is the particle mass. One usually stops the expansion already on the second-order term in the parameter $n\lambda^3$, since expressions for the expansion parameters become very cumbersome already at the third order. Pressure and other thermodynamic characteristics are then found with the help of exact thermodynamic relations. As the $\Phi$-derivable approach, the method of the virial equation of state in the above-mentioned interpretation is the self-consistent approximation scheme. This means that all thermodynamic quantities are found with the help of the exact thermodynamic relations starting from the expression for the free energy $F(n,T)$ written as an expansion in the $n\lambda^3$ parameter up to a certain order. One also can expand the pressure $P(z(\mu), T)=b_1 z+b_2 z^2 +...$ in powers of the fugacity $z=e^{\mu/T}$ up to certain order and find the density following the exact thermodynamic relation $n(z,T)=\frac{\partial P}{\partial \mu}$, where $\mu$ is the chemical potential, $b_i$ are the virial coefficients.
In the approximation scheme one may drop $O(z^k)$ terms (e.g. for $k\geq 3$) in both $P(z,T)$ and $n(z,T)$ expansions. The approximation scheme becomes fully thermodynamically consistent if, constructing $P(n,T)$, one resolves the equation $n=n(z,T)$ and replaces the function $z(n,T)$ in the expression for the pressure $P(z(n,T),T)$ considering the so-obtained $P(n,T)$ not as an expansion in $n$ but as an exact algebraic expression.

It is remarkable that  the second-order virial correction to the pressure of the ideal gas can be expressed in terms of such convenient  observable quantities as the scattering phase shifts, cf.~\cite{Uhlenbeck-Beth,Beth,Dashen}. Reference~\cite{RMS-NPA379} first payed attention to the application of the Beth-Uhlenbeck approach to dilute nuclear systems.
Further applications to the equation of state of the dilute nuclear systems can be found in~\cite{Horowitz2006,Horowitz2006_2,OConnor,Du2019}. Since phase shifts for the particle scattering in vacuum are experimentally measured, one could expect that the virial equation of state at $n\lambda^3\ll 1$, i.e., for $T\gg 100\,(n/n_0)^{2/3}$\,MeV, where $n_0=0.16\,{\rm fm}^{-3}$ is the nuclear saturation density,  can be treated as the model-independent one.
However a problem appears due to the fact that the $nn$ scattering length has a very large negative value, $a_{nn}\simeq -19$\,fm, cf.~\cite{Gardestig2009} owing to the presence of a quasi-bound state with a very small energy. Therefore, when at increasing density the interparticle distances become $l\sim 1/n^{1/3}\lsim |a_{nn}|$, i.e., for $n\gsim 10^{-3}\,n_0$, this quasi-bound state can be destroyed by the Pauli blocking effects. At these conditions $a_{nn}$ becomes not informative quantity. For the description of the in-medium scattering Refs.~\cite{ZS,SRS1990} suggested to employ the so-called generalized Beth-Uhlenbeck approach including the Pauli blocking effects within a quasiparticle treatment, cf. discussion in~\cite{Ropke:2012qv}. These effects also modify the relativistic mean-field description of the equation of state of nuclear matter at low densities~\cite{Voskresenskaya}. It was demonstrated, cf. Fig.~2 in~\cite{RopkeSchnell}, that, e.g., for $T=10$\,MeV the $^1$S$_0$ phase shifts in the medium change their vacuum low-energy behavior already for $n\gsim 0.01\,n_0$, spoiling model independence of the consideration.

The chiral effective field theory based on the symmetries of QCD is the ab-initio approach to the  dilute nuclear matter. It works very well for the description of hadron interactions~\cite{LK02} employing a small parameter of the chiral expansion (a quark mass $\sim$ few MeV divided by the chiral symmetry breaking scale $\sim$ GeV). It provides a powerful tool for  construction of the low-density expansion for the nuclear equation of state at zero temperature ~\cite{DHW21}. Although with increasing density accuracy of expansion deminishes, nevertheless the saturation phenomenon of the energy density of the isospin symmetric nuclear matter can be  reproduced  by tuning one parameter of the contact nucleon-nucleon interaction  and by a variation of the momentum cut-off in diagrams including only pion-nucleon interactions, cf.  ~\cite{LFA00,KFW02}.  Another approach is to apply the chiral perturbation theory to establish the hierarchy for the two-body, three-body, etc. nucleon-nucleon forces and to use these forces in the non-perturbative many body calculations, see, e.g., \cite{GIESHS13,DHS19}. In ~\cite{FKW02} the  approach was extended to finite temperatures. However perturbative expansions have  limitations, e.g., not reproducing the virial limit of the equation of state and  losing the predictive power when extended to densities above the saturation one. Operating with the bare Green's functions these approaches are not of the $\Phi$-derivable type. Our aim in this paper is different, namely to compare the self-consistent $\Phi$ derivable and virial approximation schemes.

After presenting the general formalism applicable for multi-species system, we use the example of a one-component system of non-relativistic fermions, e.g., neutrons, interacting via a spin-independent pair potential. We will show the coincidence of the  results obtained in both approaches within the Born approximation\footnote{
One should not confuse the first Born approximation (or simply the Born approximation), in which the particle-particle scattering amplitude and the contributions to the thermodynamic characteristics are linear in $U$, and the Born diagrams in $\Phi$ yielding contribution proportional to $U^2$.} for $n\lambda^3\ll 1$ or $z\ll 1$ and then we discuss origins of the differences appearing beyond the region of the common validity of the mentioned approaches. At low temperatures in nuclear systems nucleons are paired \cite{LL9,Schuck}.  Considering $\Phi$ derivable approach we assume that the Cooper pairing phenomenon does not occur in our system at considered temperatures and densities. It does not occur also for $n\lambda^3\ll 1$.

The presentation is organized as follows. The first part of the manuscript, see Sections~\ref{Phi}, \ref{ssec:test-pot}, and \ref{sec:Phi-U},
is devoted to the consideration of the $\Phi$ derivable approximations to strongly interacting systems. In Section~\ref{Phi} we discuss general properties of the $\Phi$ derivable approach. In Subsection~\ref{nonideal-Phi} we consider variational properties of the $\Phi$ functional. Then in Subsection~\ref{nonideal-appr} we introduce the $\Phi$ derivable approximation scheme. Subsection~\ref{NoetherSpec} introduces the spectral function and the Noether density. Then Subsection~\ref{QPA} discusses the QPA and so called improved QPA (IQPA) and Subsection~\ref{Other} introduces two other relevant spectral functions. In Subsection~\ref{nonideal-number} we first introduce the BM number density and then derive a relation between the Noether and BM number densities. Subsection~\ref{Thermod} derives expressions for the entropy and the energy within a general $\Phi$ derivable scheme. In Section~\ref{ssec:test-pot} we apply the $\Phi$ derivable approach to the system of one $1/2$ spin fermion species interacting via a pair potential. We give several examples of test pair potentials. First, in Subsection~\ref{app:NNpot} we study Yukawa potentials between nucleons. Then in Subsection~\ref{LJpotentials} we consider classical potentials employed to describe inter-molecular interaction and then in Subsection~\ref{squarewell} we consider model square-barrier-well potentials. In Section~\ref{sec:Phi-U} we focus on the example of $\Phi$ presented by the first two TS diagrams. First, in Subsection~\ref{TermodTS} by direct calculation we show the coincidence of the Noether and BM number densities in this case. Then we derive expressions for thermodynamic quantities. In Subsection~\ref{ssec:fugacity-Phi} we focus on the low-density limit developing a fugacity expansion within the $\Phi$ derivable TS approximation and we derive expressions for virial coefficients of the fugacity expansion in this TS approximation. Then in Subsections~\ref{testTSnuclear} and~\ref{testTSbarrier} we demonstrate results on examples of various test potentials.

The second part of the manuscript is devoted to the properties of the virial expansion. In Section~\ref{virial} we study  virial expansion for equation of state. In Subsection~\ref{Densfug} we introduce the density and fugacity expansions. Then in Subsection~\ref{ssec:selfcons-virial} we study thermodynamically consistent approximations within the virial expansion for the equation of state. Next, in Section~\ref{ssec:class-virial} we discuss the virial coefficients within the classical  approach. In Subsection~\ref{MayerFunc} we introduce Mayer's function and find general expressions for classical virial coefficients. Subsection~\ref{clashightemp} studies high temperature limit. In Section~\ref{classicalspec} we find classical second-virial coefficients for specific potentials. In Subsection \ref{ssec:extrapol-vEoS} we consider extrapolated virial equations of state, which allow to discuss possibilities of the phase transitions. In Section~\ref{sssec:cluster} we introduce the virial coefficients within the fully quantum approach. Subsection~\ref{Generalap} presents general expressions. Then in Section~\ref{ssec:fullquant-b2} we study the Beth-Uhlenbeck approach. In Section~\ref{Virsemicl} we introduce semiclassical approximation scheme deriving expressions to the first virial coefficients. First, in Section~\ref{semiclsecvircoef} we find expressions for the semiclassical second virial coefficients. In Section~\ref{semicltest} we perform their estimations using test potentials. In Section~\ref{sec:comp} we compare the fugacity expansion performed within the $\Phi$ derivable approach with taking into account of two TS diagrams of $\Phi$ and virial expansion of equation of state.

The results derived in this paper and our conclusions are formulated in Section~\ref{Conclusion}. Details of calculations are given in Appendices.

\section{$\Phi$ functional approach to strongly interacting fermions}\label{Phi}

\subsection{Thermodynamic potential.  $\Phi$ functional}\label{nonideal-Phi}
\subsubsection{General relations}

We consider  equilibrium system of strongly interacting nonrelativistic fermions  characterized by the temperature $T$, volume $V$ and chemical potentials $\{\mu_f\}$ associated with some conserving quantities, like electric charge, baryon number, strangeness, etc. In general case the system may contain various species of fermions.

The thermodynamic potential can be written as \cite{LL9,Abrikos,FetterWalecka,Ropkebook},
\begin{align}
\Omega(T,\mu_f)=\Omega_{f,0}(T,\mu_f)- \intop_0^{{1}} \frac{\rmd \lambda}{\lambda} \Big\langle \int \rmd^3x \hat{\mathcal{L}}_{{\rm int},\lambda} \Big\rangle\,.
\label{Om-gen-int}
\end{align}
The thermodynamic potential of the ideal fermion  gas is given by
\begin{align}
\Omega_{f,0}(T,\mu_f) &= - V\Tr_p[\tr\ln(-\hat{G}^{-1}_0(\epsilon_n,\vec{p}\,)) ]\,.
\label{Om-0}
\end{align}
Symbol $\Tr_p$ here stays for the integration over the momentum and summation over the fermion Matsubara frequencies $\epsilon_n=(2n+1) i\pi T+\mu_f$, integer number $n$ runs from $-\infty$ to $\infty$,
\begin{align}
\Tr_{p}=\Tr_{\epsilon_n,\vec{p}} = \lim_{\eta\to 0}
T\sum_{n}e^{i\epsilon_n \eta}\intop \frac{\rmd^3 p}{(2\pi)^3} \,,
\label{Tr-def}
\end{align}
where $e^{i\epsilon_n \eta}$ ensures convergency of the integrals.

To sum over Mansubara frequencies, $\epsilon_n$,  a function $F(z)$  analytic in a complex plane $z$ with the property $F(z^*)=F^*(z)$, we may use the following rule, cf.~\cite{IKV-1},
\begin{align}
&\lim_{\eta\to 0}T\!\!\!\!\sum_{n=-\infty}^{+\infty} \!\! F(\epsilon_n) e^{i\epsilon_n\eta}
%\nonumber\\
%&\qquad \qquad
= -\intop_{-\infty}^{+\infty}\!\!
\frac{\rmd \epsilon}{2\pi} f(\epsilon-\mu_f) 2\Im F(\epsilon +i 0)\,,
\label{Msum-f}
\end{align}
where
\begin{align}
&\quad f(\epsilon)=\frac{1}{e^{\epsilon/T}+1}
\,,
\label{f-n-def}
\end{align}
is the equilibrium Fermi-Dirac distribution function, and on the right hand side of Eq.~(\ref{Msum-f}) there appeared functions analytically continued to the upper complex semi-plane. The regulator $e^{i\epsilon_n\eta}$  in Eq.~(\ref{Msum-f}) makes it applicable to the function of the type $1/(\epsilon_n-a)$, whereas in the sum over two and more Green's functions this regulator can be dropped. Applying Eq.~(\ref{Msum-f}) we obtain, for instance, the  following relation
\begin{align}
&\lim_{\eta\to 0}T \sum_{n=-\infty}^{\infty} \frac{e^{i\epsilon_n\eta}}{\epsilon_n - \epsilon } = f(\epsilon-\mu_f)\,.
\label{bose-sum}
\end{align}

The Green's function $\hat{G}_0$ is in general case a matrix in the spin and isospin spaces, and traces in these spaces are denoted as $\tr$.  In such a way we may consider the system consisted of various species. We will use $G_0$  notation for the structure-less Green's functions, i.e. $G_0=\tr(\hat{G}_0)/d_f$, where $d_f$ are degeneracy factors of fermions.

Applying Eq.~(\ref{Msum-f}) we present the first term in Eq.~(\ref{Om-0}) as
\begin{align}
\Omega_{f,0}(T,\mu_f) &= d_f\intop
\frac{V\rmd^3 p}{(2\pi)^3}\intop_{-\infty}^{+\infty}\frac{\rmd \epsilon}{2\pi}
2 \Im \ln(-G^{-1}_0(\epsilon,\vec{p}\,))
\nonumber\\
&\times f(\epsilon-\mu_f)\,,
\label{free-Omega-f}
\end{align}
where $f(\epsilon)=\frac{1}{e^{\epsilon/T}+1}$ is ordinary equilibrium Fermi-Dirac distribution function. Explicit expression for the integral in (\ref{free-Omega-f}) is given by Eq.~(\ref{free-Omega}) of Appendix~\ref{ideal}.

The term $\langle\hat{\mathcal{L}}_{{\rm int},\lambda} \rangle$ stands for the thermal average of the interaction Lagrangian density  constructed from the field operators in the interaction representation. Sub-index $\lambda$ indicates that $\hat{\mathcal{L}}_{\rm int}$ is scaled by a common scaling factor $\lambda$, i.e. $\hat{\mathcal{L}}_{{\rm int},\lambda}=\lambda\hat{\mathcal{L}}_{\rm int}$, and the thermal averaging is performed with account for this scaling. Summation over the couplings as summation over the species is implied.

It was shown in~\cite{Luttinger-Ward} for fermions that general expression (\ref{Om-gen-int}) can be written as a functional of the full Matsubara Green's functions $G$,
\begin{align}
&{\Omega[\hat{G} ]} = - V\Tr_p \big[\tr(
\ln(-\hat{G}^{-1}_{p}) + \hat{\Sigma}_{p} \hat{G}_{p})\big]
+ {\Phi[\hat{G}]}\,.
\label{Om-gen}
\end{align}
We use   shorten notations for the arguments $\hat{G}_{p} = \hat{G} (\epsilon_n,\vec{p}\,)$ and the similar ones for the  fermion  self-energies, $\hat{\Sigma}$,
\begin{align}
\hat{\Sigma}_p = \hat{G}_{0,p}^{-1} - \hat{G}_p^{-1}\,.
\label{Dyson-eq}
\end{align}
The functional $\Phi [\hat{G}]$ is defined as the infinite series of all closed skeleton two-particle irreducible graphs, where all lines are full Green's functions, $\hat{G}$, which are matrices in the spin-isospin space. The diagrammatic rules may vary slightly for various interaction terms. The general rules can be found in Refs.~\cite{LL9,FetterWalecka,Kapusta,KadanoffBaym,IZ,IKV-1}. Each diagram contains a symmetry factor depending on the specific form of the interaction.

\subsubsection{Variational properties}

Functional (\ref{Om-gen}) has stationary properties with respect to variation of the Green's functions, which are assured by the relation between $\Phi$ functional and the self-energies,
\begin{align}
\delta\Phi= V\Tr_p[\tr(\hat{\Sigma}_p \delta\hat{G}_p)] \,.
\label{dPhi-variation}
\end{align}
This relation implies the  rule for determining the self-energies from the $\Phi$ functional written in terms of full Green's functions,
\begin{align}
\hat{\Sigma}_p=\frac{\delta \Phi}{\delta \hat{G}_p}\Big|_{\hat{G}_0}\,.
\label{Sigma-Phi}
\end{align}
Diagrammatically, this means that the expressions for the self-energies $\hat\Sigma$ can be obtained from $\Phi$ by removing one of the Green's functions.

The definition of the self-energies as a variation of the $\Phi$ functional imposes the stationarity properties of the $\Omega$ functional. Indeed, using that $\Tr_p \tr \delta\ln(-\hat{G}^{-1}_{p})=\Tr_p\big[\tr(\hat{G}_p\delta \hat{G}_p^{-1})\big]$ and Dyson Eqs.~(\ref{Dyson-eq}) we find that the variations of $\Omega$ with respect to ${G}$  vanish, i.e.,
\begin{align}
\frac{\delta \Omega}{\delta \hat{G}_p}\Big|_{\hat{G}_0}=\frac{\delta \Phi}{\delta \hat{G}_p}-\hat{\Sigma}_p=0 \,.\label{Om-G-var}
\end{align}
%If now one formally rewrites the functional $\Omega[\hat{G},\hat{D}]$ with the help of relations (\ref{Dyson-eq}) through the self-energies as $\Omega[\hat{\Sigma},\hat{\Pi}]$ then this functional
%has the property
%From this relation we can also find
%\begin{align}
%\frac{\delta \Omega}{\delta \hat{\Sigma}}\Big|_{\hat{\Pi}} =
%\frac{\delta\hat{G}}{\delta\hat{\Sigma}}\frac{\delta \Omega}{\delta \hat{G}}\Big|_{\hat{\Pi}}
%=0\,,
%\quad
%\frac{\delta \Omega}{\delta \hat{\Pi}}\Big|_{\hat{\Sigma}}=
%\frac{\delta\hat{D}}{\delta\hat{\Pi}}\frac{\delta \Omega}{\delta \hat{D}}\Big|_{\hat{G}}
%=0\,.
%\label{Om-Sigma-var}
%\end{align}
%These relations simplify largely  calculations of the derivatives of $\Omega$ with respect to $T$ and $\mu$, as the only explicit dependence on these quantities has to be taken into account, whereas the dependence entering through the self-energies can be formally ignored, cf. \cite{Carneiro-Pethick}.

\subsubsection{Matsubara and retarded Green's functions}

The full Green's functions satisfy the Kramers-Kronig dispersion relations
\begin{align}
\hat{G}_p=\hat{G}(\epsilon_n,\vec{p}\,) &= \intop_{-\infty}^\infty\frac{\rmd {\epsilon}}{2\pi} \frac{\hat{A}_f({\epsilon},\vec{p}\,)}{\epsilon_n -{\epsilon}}\,.
\label{GD-def}
\end{align}
As the free Green's functions (\ref{G0D0-R}), the full Green's functions can be continued analytically to the upper complex semi-plane as the retarded Green's functions~\cite{Baym-Mermin},
\begin{align}
&\hat{G}^R(\epsilon,\vec{p}\,)=\hat{G}(\epsilon+i0,\vec{p}\,)\,,
\label{GD-R-def}
\end{align}
having poles only in the lower complex semi-plane.
Then the spectral functions are expressed as
\begin{align}
&\hat{A}_f(\epsilon,\vec{p}\,)= -2\Im \hat{G}^R(\epsilon,\vec{p}\,)\,.
\label{Spec-dens-def}
\end{align}

Analogously, using the Dyson equations we can present  the retarded self-energies as functions of the retarded Green's functions,
\begin{align}
\hat{\Sigma}^R(\epsilon,\vec{p}\,) &= [\hat{G}^R_0(\epsilon,\vec{p}\,)]^{-1} - [\hat{G}^R(\epsilon,\vec{p}\,)]^{-1}\,.
\label{SE-ret-def}
\end{align}
Thus the complete information about in-medium properties of fermion  excitations is encoded in the corresponding spectral functions.

To avoid a complicated spin-isospin algebra we will further limit ourselves by consideration of system when all Green's functions are proportional to the unit matrix of the corresponding dimensionality. This is so, e.g., for unpolarized  neutron matter and isospin-symmetric nuclear matter. Then, as for the Green's functions of ideal gas, we will use the notation $G$ for the structure-less Green's functions, i.e. $G=\tr(\hat{G})/d_f$. In this case summation over Matsubara frequencies in Eq.~(\ref{Om-gen}) can be performed explicitly with the help of relation (\ref{Msum-f}) of Appendix A resulting in
\begin{align}
&\Omega[G] =  d_f\intop\frac{V\rmd^3 p}{(2\pi)^3}
\intop_{-\infty}^{+\infty}\frac{\rmd \epsilon}{2\pi}f(\epsilon-\mu_f)
\nonumber\\
&\qquad\times 2\Im\big[\ln(-[G^R(p)]^{-1}) +\Sigma^R(p)G^R(p)\big]+\Phi[G]
  \,,
 \label{Omega-int-1}
\end{align}
where
%the retarded  Green's functions $G^R$ and $D^R$ and self-energies are understood as analytical %continuations of the Matrsubara quantities $G$ and $D$, see Eq.~(\ref{GD-R-def}) and (\ref{SE-ret-def}).
$\Phi$ is still written in terms of the Matsubara Green's functions.

\subsection{$\Phi$ derivable approximations}\label{nonideal-appr}

\begin{figure*}
\centering
$$\parbox{17cm}{\includegraphics[width=17cm]{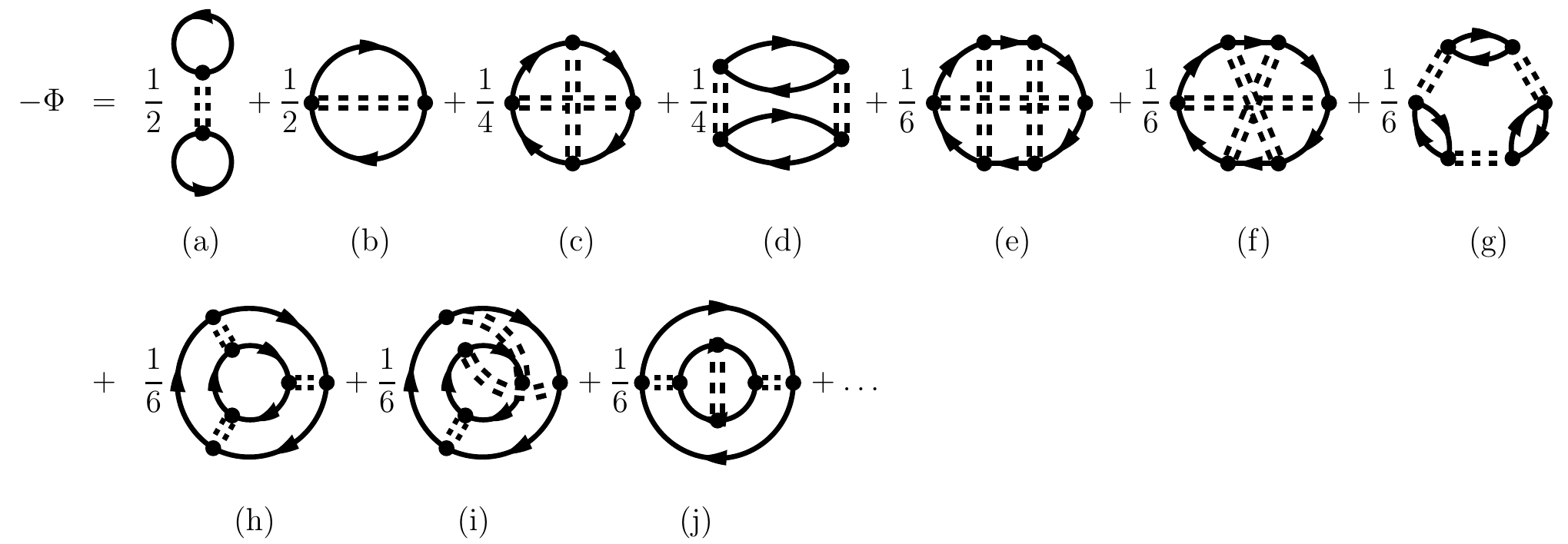}}$$
\caption{Diagrams contributing to the $\Phi$ functional for the theory of nonrelativistic fermions interacting via a pair potential. Fermion Green's functions are shown by the solid lines, the pair potential $\Ucor$, by double-dashed lines. All diagrams up to the third order in the potential are  presented explicitly. }
\label{fig:Phi-U-ser}
\end{figure*}

In derivation of Eq.~(\ref{Om-gen}) from Eq.~(\ref{Om-gen-int}) no approximations are done and the $\Phi$ functional contains infinite set of diagrams expressed in terms of full Green's functions. For a specific choice of the interaction Lagrangian, the selection of the skeleton diagrams contributing to the $\Phi$ functional can be quite different.

For a  case of a model describing fermions interacting via a pair-potential, we show in Fig.~\ref{fig:Phi-U-ser} the diagrams contributing to $\Phi$ up to order $O(\Umom^3)$. The vertex is given by the Fourier transform of the potential   $\Umom(|\vec{q}\,|)$, where $\vec{q}$ is the exchange momentum between fermions. The symmetry factors $1/I_f$, where $I_f$ is the number of internal fermion lines, differ diagrammatic rules for the $\Phi$ from the rules for self-energies and interaction amplitudes. $I_f$ is related to the number of potential double-dashed lines in the diagram, $v$, as $I_f=2\,v$\,, therefore the additional factor is $1/2v$ in this case.
For the case of the non-relativistic fermions interacting via a pair potential the diagrammatic rules are formulated in Appendix~\ref{DiagramRules}.
The first two tadpole-sandwich diagrams (a,b) in Fig.~\ref{fig:Phi-U-ser} expressed in terms of the full Green's functions constitute the Hartree-Fock contribution ($\propto U$). The next two so-called Born diagrams (c,d) produce the Boltzmann or Landau collision terms ($\propto U^2$) in the kinetic equation, cf. Appendix C in \cite{IKV-YaF}. Other terms are of higher orders in the pair potential.

A remarkable (so-called $\Phi$-derivable) approximation scheme was suggested by Baym in Ref.~\cite{Baym-Phi}, who showed that one may cut
the series of diagrams in $\Phi$ on any term preserving nevertheless the thermodynamic consistency of the theory.
Indeed, as was shown in~\cite{Baym-Phi}, this approximation for the `full' Green's functions $G$ defined in (\ref{Dyson-eq}), (\ref{Sigma-Phi}), leads to the 4-current densities and the energy-momentum tensor obeying the continuity equations. Note that within the $\Phi$-derivable approximation scheme with even one diagram retained in the $\Phi$ functional one deals with an infinite sub-series of diagrams in the perturbation theory. Therefore, the term ``full'' takes the sense of the sum of this sub-series.

Further we will consider derivatives of $\Omega$ with respect to $T$ and $\mu$ and derive other thermodynamic quantities within the $\Phi$ derivable approximation scheme. We shall show that expressions for thermodynamic quantities like, the particle number, entropy and energy can be expressed in the closed form through the $\Phi$ functional and the self-consistently calculated Green's functions.

\subsection{The Noether's number density and the spectral function}\label{NoetherSpec}

The conserved number of non-relativistic fermions is expressed via the chemical potential as
\begin{align}
N_f=-\frac{\partial \Omega}{\partial \mu_f}\Big|_{T} \,.
\label{nf-dOmdmu}
\end{align}
In view of the stationarity properties (\ref{Om-G-var})
%, (\ref{Om-Sigma-var})
we have to differentiate $\Omega$ with respect to explicit dependence on $\mu_f$ only, which enters through $\epsilon_n$. Therefore, the external derivatives $\partial/\partial\mu_f$ can be translated in the derivative $\partial/\partial\epsilon_n$ under the trace operation. From (\ref{Om-gen}) we find, cf. derivation in \cite{Rios},
\begin{align}
&\frac{\partial \Omega}{\partial \mu_f}\Big|_T
= -V\Tr_p \big[\frac{\partial}{\partial\epsilon_n}\tr(\ln(-\hat{G}^{-1}_{p}) + \hat{\Sigma}_{p} \hat{G}_{p})\big]
+\frac{\partial \Phi}{\partial\mu_f}\Big|_{\hat{A}_f}
\nonumber\\
&=-V\Tr_p \big[\tr\big(\hat{G}_p + \hat{\Sigma}_p\frac{\partial \hat{G}_p}{\partial\epsilon_n}\big)\big]+\frac{\partial \Phi}{\partial\mu_f}\Big|_{\hat{A}_f} \,,
\label{dOmdmu-1}
\end{align}
where we used Eq.~(\ref{Dyson-eq}). The derivative of $\Phi$ is taken at the fixed spectral density $\hat{A}_f$. The Green's function $\hat{G}$ is assumed to be expressed  through $\hat{A}_f$ via the Kramers-Kronig dispersion relations (\ref{GD-def}), and the self-energy is expressed through $\hat{G}$.

Because of the variational properties (\ref{Sigma-Phi}) we have
\begin{align}
\frac{\partial \Phi}{\partial\mu_f}\Big|_{\hat{A}_f} &=
V\Tr_p\big[\tr\big(\hat{\Sigma}_p \frac{\partial \hat{G}_p}{\partial \mu_f} \big)\big]\Big|_{\hat{A}_f}
\nonumber\\
&=
V\Tr_p\big[\tr\big(\hat{\Sigma}_p \frac{\partial \hat{G}_p}{\partial \epsilon_n} \big)\big]\Big|_{\hat{A}_f} \,.
\label{dPhidmu-1}
\end{align}
Combining Eqs.~(\ref{dOmdmu-1}) and (\ref{dPhidmu-1}) we obtain
\begin{align}
n_f =-\frac{1}{V}\frac{\partial \Omega}{\partial \mu_f}\Big|_T=n_f^{\rm (Noeth)}=\Tr_p[\tr\hat{G}_p]\,.
\label{dOmdmu-TrG}
\end{align}
Thus, we obtain that the fermion particle density, $n_f=N_f/V$, determined from the exact thermodynamic relation, coincides with the particle density found from definition of the Green's function, which is the zeroth component of the Noether's current.

With the help of
%the retarded Green's functions, following
Eq.~(\ref{Msum-f}) we rewrite the particle density $n_f^{\rm (Noeth)}$ in (\ref{dOmdmu-TrG}) as
\begin{align}
n_f^{\rm (Noeth)}&=d_f\intop\frac{\rmd^3 p}{(2\pi)^3}
\intop_{-\infty}^{+\infty}\frac{\rmd \epsilon}{2\pi} f(\epsilon-\mu_f) A_f(\epsilon,\vec{p}\,) \,.
\label{nf-Noether}
\end{align}
%Each fermionic mode with energy $\epsilon$ contributes with the weight of the spectral function  $A_f(\epsilon,\vec{p}\,)$.

The normalization of the spectral functions is determined with the help of the commutation relations for the field operators \cite{BotMalf}. Thereby, the spectral functions (spectral densities) satisfy the exact sum-rules
\begin{align}
&\intop_{-\infty}^\infty \frac{\rmd \epsilon}{2\pi} {A}_f(\epsilon,\vec{p}\,)=1\,,
\label{A-norm}
\end{align}
where ${A}_{f}$  are defined in Eq.~(\ref{Spec-dens-def}).  As follows from the Dyson equations (\ref{SE-ret-def}),
\begin{align}
A_{f}=\frac{\Gamma_{f}}{M^2_{f} +\frac14\Gamma_{f}^2} \,.
\label{A-GM}
\end{align}
Here we introduced the mass function of non-relativistic fermions
\begin{align}
M_f(\epsilon,\vec{p}\,)=\epsilon-\epsilon_{\vec{p}} -\Re\Sigma^R(\epsilon,\vec{p}\,)\,,\quad
\epsilon_{\vec{p}}=\frac{\vec{p\,}^2}{2m_f}
\,,
\label{M-f-def}
\end{align}
and
 the mass width
\begin{align}
\Gamma_f(\epsilon,\vec{p}\,)=-2\Im\Sigma^R(\epsilon,\vec{p}\,)\,,
\label{Gamma-f-def}
\end{align}
cf. \cite{IKV-2}.

\subsection{Quasiparticle and improved quasiparticle approximations}\label{QPA}

One often employs the  QPA, which initially has been introduced for description of Fermi liquids at low temperatures, see \cite{Mig,LL9,Pines-Nozieres}, where it
constitutes a consistent approximation scheme.    For the validity of the QPA for fermions one formally needs to put  $\Gamma_{f}\to 0$ in the expression for the fermion Green's function.  Within the QPA  the equation
\begin{align}
M^{\rm (QPA)}_f(\widetilde{\epsilon}_{\vec{p}} ,\vec{p}\,)= 0\,
\label{MfG}
\end{align}
defines the in-medium mass shell $\epsilon=\widetilde{\epsilon}_{\vec{p}}$ for fermion excitations.
The quasiparticle self-energy is then calculated with such a determined fermion quasiparticle Green's function.
In practice, $\Gamma_{f}(\epsilon)$  should be less than all the relevant energy scales in the problem under consideration.

The quasiparticle picture allows for a transparent physical interpretation of sub-processes included in diagrams. In the QPA the fermion spectral function reads
\begin{align}
A_f^{\rm (QPA)}(\epsilon,\vec{p}\,) &= 2\pi \delta \big(M_f(\epsilon,\vec{p}\,)\big) = {2\pi Z_f(\widetilde{\epsilon}_{\vec{p}}, \vec{p}\,)} \delta (\epsilon -\widetilde{\epsilon}_{\vec{p}})\,,
\nonumber\\
Z_f^{-1}(\widetilde{\epsilon}_{\vec{p}}, \vec{p}\,) &= 1-{\partial_\epsilon {\Re}\Sigma^R}
\big|_{\epsilon = \tilde{\epsilon}_{\vec{p}}}\,.
\label{Af-QPA}
\end{align}

Since the derivative of $\Re\Sigma^R$ with respect to the energy variable can be expressed with the help of the Kramers-Kronig relation as
\begin{align}
\frac{\partial {\Re}\Sigma^R(\epsilon,\vec{p}\,)}{\partial \epsilon}\Big|_{\epsilon=\widetilde{\epsilon}_{\vec{p}}}
=&-\lim_{\eta\to 0} \int_{-\infty}^{\infty}
\frac{\rmd \epsilon'}{2\pi} \Gamma_f(\epsilon',\vec{p}\,)
\nonumber\\
&\times
\frac{(\epsilon' -\widetilde{\epsilon}_{\vec{p}})^2 - \eta^2}{[(\epsilon' -\widetilde{\epsilon}_{\vec{p}})^2+\eta^2]^2}
\label{KramersKroning}
\end{align}
we conclude that the residue $Z_f$ is positively-defined quantity.
The Green's function in the QPA is equal to
\begin{align}
G^{\rm (QPA)}_R(\epsilon,\vec{p}\,)=\frac{Z_f(\widetilde{\epsilon}_{\vec{p}}, \vec{p}\,)}{\epsilon-\widetilde{\epsilon}_{\vec{p}}+i\eta}\,.
\label{Gf-QPA}
\end{align}
Since $Z_f \neq 1$,  within the QPA the sum-rule renders
\begin{align}
\intop_{-\infty}^{\infty} A_f^{\rm (QPA)}(\epsilon,\vec{p}\,)\frac{\rmd\epsilon}{2\pi}
= Z_f(\widetilde{\epsilon}_{\vec{p}}, \vec{p}\,)\,,
\label{sumrule-A-QPA}
\end{align}
that differs from the exact sum-rule (\ref{A-norm}). To avoid the problem one performs renormalization of the interaction. The price paid for that is that some effects of the continuum are not treated within a quasiparticle framework. The renormalization is not needed, if one uses the so-called ``improved'' or ``extended'' quasiparticle approximation (IQPA) proposed in Ref.~\cite{Kremp-Kraeft-Lambert-84}, see also book~\cite{Kraeft-Kremp-Ebeling-Ropke-book86}. This approximation is frequently used in the literature \cite{SRS1990,Kohler-Malfliet-93,Dukelsky-Ropke-Schuck-98}. It suggests to retain an additional term in the expansion  of Eq.~(\ref{A-GM}) in $\Gamma_f$,
\begin{align}
&A_f^{\rm (IQPA)}(\epsilon,\vec{p}\,)
\simeq {2\pi Z_f(\widetilde{\epsilon}_{\vec{p}}, \vec{p}\,)} \delta \big(\epsilon -\widetilde{\epsilon}_{\vec{p}}\big )
\nonumber\\
&\quad+{\Gamma_f (\epsilon,\vec{p}\,)
\lim_{\eta\to 0}\frac{(\epsilon -\widetilde{\epsilon}_{\vec{p}})^2 - \eta^2}{[(\epsilon -\widetilde{\epsilon}_{\vec{p}})^2+\eta^2]^2}
}+O(\Gamma_f^2) \,,
\label{IQPA1}
\end{align}
For   $|\partial_\epsilon \Re\Sigma^R|\ll 1$ one uses
\begin{align}
Z_f(\widetilde{\epsilon}_{\vec{p}}, \vec{p}\,)\simeq 1 + \partial_\epsilon \Re\Sigma^R(\epsilon,\vec{p}\,)
\Big|_{\widetilde{\epsilon}_{\vec{p}}}
+O(\Gamma_f^2)\,.
\nonumber
\end{align}
All the retained terms are formally of the linear order in $\Gamma_f$, since $\Re\Sigma^R$ is related to $\Gamma_f$ via the Kramers-Kronig dispersion relation (\ref{KramersKroning}).
Using now Eq.~(\ref{KramersKroning}), we immediately verify that $A^{\rm (IQPA)}(\epsilon,\vec{p}\,)$  obeys the sum-rule (\ref{A-norm}), namely
\begin{align}
\intop_{-\infty}^{\infty} A_f^{\rm (IQPA)}(\epsilon,\vec{p}\,)\frac{\rmd\epsilon}{2\pi}
= 1+ O(\Gamma_f^2)\,.
\end{align}

The particle Noether  density (\ref{nf-Noether}) takes different forms in QPA and IQPA. So, using (\ref{Af-QPA}) we find
\begin{align}
n_f^{\rm (Noeth,QPA)} =d_f\int\frac{\rmd^3 p}{(2\pi)^3} Z_f(\widetilde{\epsilon}_{\vec{p}}, \vec{p}\,) f(\widetilde{\epsilon}_{\vec{p}} -\mu_f)\,,
\label{QPAdistr}
\end{align}
whereas,
\begin{align}
n_f^{\rm (Noeth,IQPA)} =d_f\int\frac{\rmd^3 p}{(2\pi)^3}
f(\widetilde{\epsilon}_{\vec{p}} -\mu_f)+{O}(\Gamma_f^2)\,.
\label{IQPAdistr}
\end{align}

\subsection{{Other spectral functions}}\label{Other}

%\subsubsection{General relations}
We will show that the thermodynamic potential (\ref{Omega-int-1}) can be rewritten in different forms through the various spectral functions.

%To facilitate comparison with the pressure of ideal fermion gas given by  Eq.~(\ref{free-Omega}),  l
Let us convert occupation numbers $f(\epsilon)$ in Eq.~(\ref{Omega-int-1}) in logarithms following Eq.~(\ref{nf-to-log}). Then we rewrite Eq.~(\ref{Omega-int-1}) as
\begin{align}
\Omega &= -d_f\intop\frac{V\rmd^3 p}{(2\pi)^3}
\intop_{-\infty}^{+\infty}\frac{\rmd \epsilon}{2\pi}
B_f(\epsilon,\vec{p}\,) T\ln\big(1+e^{-(\epsilon-\mu_f)/T}\big)\nonumber\\
& +\Delta \Omega \,.
 \label{Omega-int-B}
\end{align}
The first term generalizes that for ideal fermion gas given by  Eq.~(\ref{free-Omega}).
The second term   given by
\begin{align}
&\Delta \Omega = d_f\!\!\intop\!\!\frac{V\rmd^3 p}{(2\pi)^3}\!\!
\intop_{-\infty}^{+\infty}\!\!\frac{\rmd \epsilon}{2\pi} f(\epsilon-\mu_f)2\Im\big[\Sigma^R(\epsilon,\vec{p}\,)G^R(\epsilon,\vec{p}\,)\big]\nonumber\\
&+\Phi
\label{DOm1}
\end{align}
depends through a specific form of $\Phi$ on the particular model of the particle interaction.

The standard logarithm in (\ref{Omega-int-B}) is weighted not with the particle spectral density $A_{f}$, as it was in case of the Noether fermion particle number (\ref{nf-Noether}),  rather with the $B$-spectral function:
\begin{align}
&B_f(\epsilon,\vec{p}\,) =
-2 \Im{\partial}_\epsilon
%{\partial \epsilon}
\ln(-[G^R(\epsilon,\vec{p}\,)]^{-1})
\nonumber\\
&\,\,=-2\Im G^R(\epsilon,\vec{p}\,)
+2\Im \big(G^R(\epsilon,\vec{p}\,)
{\partial_\epsilon \Sigma^R(\epsilon,\vec{p}\,)}
%{\partial \epsilon}
\big)
\nonumber\\
%&\,\,=A_f(\epsilon,\vec{p}\,) +
%2\Im\big(G^R(\epsilon,\vec{p}\,)\frac{\partial \Sigma^R(\epsilon,\vec{p}\,)}{\partial \epsilon}\big)
%\nonumber\\
&\,\,=A_f(\epsilon,\vec{p}\,) - A_f(\epsilon,\vec{p}\,)
 {\partial_\epsilon \Re\Sigma^R(\epsilon,\vec{p}\,)}
 %{\partial \epsilon}
 \nonumber\\
&\,\,\,\,-\Re G_f^R(\epsilon,\vec{p}\,)
{\partial_\epsilon \Gamma_f (\epsilon,\vec{p}\,)}
%{\partial \epsilon}
\label{Bf-def}
\end{align}
for non-relativistic fermions.
The $B_f$ spectral function has been employed in various works, e.g., cf. ~\cite{BotMalf,Weinhold,IKV-2,Ivanov:2000ma,Ivanov:2009wc,KVJPG40}.
Using (\ref{M-f-def}) we may present the spectral function $B_{f}$ in another form, cf.~\cite{IKV-2},
\begin{align}
B_{f} &=   A_{f}\, \Big[\partial_\epsilon M_{f} - \frac{M_{f}}{\Gamma_{f}}\partial_\epsilon\Gamma_{f}
\Big]\,.
\label{B-GM}
%\arctan\frac{2M_{f}(\epsilon,\vec{p}\,)}{\Gamma_{f}(\epsilon,\vec{p}\,)}
\end{align}
Owing to the analytical  properties of the retarded quantities one can close the integration contour in the upper half-plane and thus one has
\begin{align}
\intop_{-\infty}^{+\infty}\rmd \epsilon \Im\Big(G^R(\epsilon,\vec{p}\,){\partial_\epsilon \Sigma^R(\epsilon,\vec{p}\,)}
%{\partial \epsilon}
\Big)
=0\,,
\end{align}
and thereby $B_{f}$ satisfies the sum-rule
\begin{align}
\intop_{-\infty}^{+\infty}\frac{\rmd \epsilon}{2\pi} B_{f} (\epsilon,\vec{p}\,)=1 \,.
\label{B-norm}
\end{align}
Within the QPA by setting $\Gamma_{f} \to 0$ in (\ref{Bf-def})  we find  that the spectral function $B_{f}^{(\rm QPA)}$ satisfies the exact sum-rule in difference with the spectral function $A_{f}^{(\rm QPA)}$.

The spectral function $B_{f}$ can be expressed in terms of the phase of the retarded Green's function
\begin{align}
G^R(\epsilon,\vec{p}\,) &= |G^R(\epsilon,\vec{p}\,)|e^{i\delta_f^{(G)}(\epsilon,\vec{p}\,)}\,.
\end{align}
As follows from Eq. (\ref{Dyson-eq}), we can write
\begin{align}
|G^R(p)|^2 &=\frac{1}{M_f^2(p)+\frac14\Gamma_f^2(p)}\,,\,\,
\nonumber\\
\delta_{f}^{(G)}(\epsilon,\vec{p}\,) &= \frac{\pi}{2}
+ \arctan\frac{2M_{f}(\epsilon,\vec{p}\,)}{\Gamma_{f}(\epsilon,\vec{p}\,)}\,.
\label{deltaFerm}
\end{align}
The phase  $\delta_f^{(G)}$ is defined such that it remains regular at $M_f=0$ and vanishes for $\epsilon$ such that $\Gamma_f\to 0$ and $M_f<0$.

The spectral function $B_{f}$ can be written as the frequency derivative of the phase shift,
\begin{align}
B_f &= -2{\partial}_\epsilon
%{\partial\epsilon}
\Im\ln\big[|G_f^R|^{-1}\exp(-i\delta_f^{(G)})\big]=
2{\partial_\epsilon\delta_f^{(G)}}
\,,
\label{B-delt}
\end{align}
 cf. \cite{KVJPG40}.

%Presence of a finite value of the mass-width, $\Gamma_{f}$, means that the state is unstable and, being produced in some collision process, will decay after a while. Then

Note that the function $B_{f}$ has the meaning of the delay time in the scattering through the resonance state ~\cite{KVJPG40}, cf. with the resonance scattering in quantum mechanics \cite{Smith}. If such a resonant state gives the dominant contribution to some scattering process, then the phase  $\delta_{f}^{(G)}$ can be related to the scattering phase shift measurable experimentally, cf. ~\cite{Weinhold}.
%In this case function $B_{f,b}$ resembles the factor in the virial expansion of the thermodynamic potential.
%, as we will discuss below in Section~\ref{sec:virial}.

A new form of the thermodynamic potential can be obtained, if one re-groups a part of the residual term $\Delta \Omega$   in (\ref{Omega-int-B}). One can present
\begin{align}
\Omega &= -d_f\!\!\intop\frac{V\rmd^3 p}{(2\pi)^3}
\intop_{-\infty}^{+\infty}\frac{\rmd \epsilon}{2\pi}
A_{s,f}(\epsilon,\vec{p}\,) T\ln\big(1+e^{-(\epsilon-\mu_f)/T}\big)
\nonumber\\
& +\Delta\Omega_s \,.
\label{Omega-int-As}
\end{align}
Here the logarithm is weighted with new so called ``entropy spectral function'', cf.~\cite{IKV-2} and expression %(\ref{SlocMem}), (\ref{DSp-1}), (\ref{sig-def}), (\ref{DSp-0}) below,
(\ref{DSp-1}) below,
\begin{align}
A_{s,f}(\epsilon,\vec{p}\,) &= B_f(\epsilon,\vec{p}\,)
+ {\partial}_\epsilon
%{\partial \epsilon}
\big(\Gamma_f(\epsilon,\vec{p}\,)
\Re G^R(\epsilon,\vec{p}\,)\big)\nonumber\\
&= A_f (\epsilon,\vec{p}\,)-\mathcal{C}_f(\epsilon,\vec{p}\,)\,,
%-A_f \frac{\partial}{\partial \epsilon} \Re \Sigma^R (\epsilon,\vec{p}\,)+ \Gamma_f  \frac{\partial}{\partial \epsilon} \Re G^R (\epsilon,\vec{p}\,)\,,
\label{Asf-def}
\end{align}
where
\begin{align}
\mathcal{C}_f(\epsilon,\vec{p}\,)&=
A_f(\epsilon,\vec{p}\,){\partial }_\epsilon
%{\partial \epsilon}
\Re\Sigma^R(\epsilon,\vec{p}\,)
-\Gamma_f (\epsilon,\vec{p}\,){\partial}_\epsilon
%{\partial \epsilon }
\Re G^R(\epsilon,\vec{p}\,)\,.
\label{C-def}
\end{align}
To obtain these expressions we used Eq.~(\ref{Bf-def}).
In the residual contribution $\Delta \Omega_s$  we separated the terms with $\Re\Sigma^{R}$ as,
\begin{align}
\Delta \Omega_s =&  -
d_f\!\!\intop\frac{V\rmd^3 p}{(2\pi)^3}\!\!\!
\intop_{-\infty}^{+\infty}\frac{\rmd \epsilon}{2\pi} f(\epsilon-\mu_f)\Re\Sigma^R(\epsilon,\vec{p}\,)
A_f(\epsilon,\vec{p}\,)
\nonumber\\
&+\Phi\,.
\label{DOm}
\end{align}

New spectral function $A_{s,f}$ enters expression for the Markovian contribution to the entropy
\cite{IKV-2}, and  the conserved baryon number density \cite{Leupold2000}.  This spectral function has the same normalization as
$B_{f}$ function in (\ref{B-norm}), since $A_{s,f}$ differs from $B_{f}$ only by the full frequency derivative,
\begin{align}
\intop_{-\infty}^{+\infty}\frac{\rmd \epsilon}{2\pi} A_{s,f}(\epsilon,\vec{p}\,)=1\,.
\label{Asf-norm}
\end{align}

Using relations (\ref{A-GM}) and (\ref{B-GM}) we rewrite Eq. (\ref{Asf-def})  as
%useful relation among $A_{s,f}$, $B_f$ and $A_f$,
\begin{align}
&A_{s,f}=
%A_{f}\, M_{f}\Big[\frac{\partial_\epsilon M_{f}}{M_{f}} - \frac{\partial_\epsilon\Gamma_{f}}{\Gamma_{f}}
%\Big]
%+ \frac{\partial}{\partial \epsilon} \Big(
%\frac{\Gamma_f M_f}{M_f^2+\frac14\Gamma_f^2}
%\Big)
%\nonumber\\
%&=A_{f}\, \Big[\partial_\epsilon M_{f}
%- M_{f}\frac{\partial_\epsilon\Gamma_{f}}{\Gamma_{f}}
%\Big]
%    +
%A_f\big(M_f\frac{\partial_\epsilon\Gamma_f}{\Gamma_f} + \partial_\epsilon M_f\big)
%\nonumber\\
%&
%- 2 \frac{A_f^2}{\Gamma_f^2}M_f\Gamma_f
%\big(M_f\partial_\epsilon M_f +\frac14\Gamma_f\partial_\epsilon \Gamma_f\big)
%\nonumber\\
%&=2A_{f}\, \partial_\epsilon M_{f}
%- 2 \frac{A_f^2M_f^2}{\Gamma_f}\partial_\epsilon M_f
%-\frac12 A_f^2M_f\partial_\epsilon \Gamma_f
%\nonumber\\
%&=2A_{f}(1-A_f M_f^2\Gamma_f^{-1})\, \partial_\epsilon M_{f}
%- \frac12 A_f^2M_f\partial_\epsilon \Gamma_f
%\nonumber\\
%&=
\frac12 A_{f}^2\Big(\Gamma_f\, \partial_\epsilon M_{f}
- M_f\partial_\epsilon \Gamma_f\Big)= \frac12 A_f\Gamma_fB_f\,.
\label{As-AGB}
\end{align}

Within the QPA by setting $\Gamma_{f} \to 0$ in (\ref{As-AGB}) we find  that the spectral function $A_{s,f}^{(\rm QPA)}$, as $B_{f}^{(\rm QPA)}$, satisfies the exact sum-rule, in difference with the spectral function $A_{f}^{(\rm QPA)}$.
In the IQPA all three spectral functions $A_{s,f}^{(\rm IQPA)}$, $A_{f}^{(\rm IQPA)}$ and $B_{f}^{(\rm IQPA)}$ satisfy the exact sum-rules.

Now with the help of the relation $\sin^2\delta_{f}^{(G)}=\frac14\Gamma_{f} A_{f}$ we find
\begin{align}
A_{s,f}(\epsilon,\vec{p}\,) &= 4\sin^2\delta_{f}^{(G)}(\epsilon,\vec{p}\,){\partial }_\epsilon
%{\partial \epsilon}
\delta_{f}^{(G)}(\epsilon,\vec{p}\,)\,.
\label{As-sin2delta}
\end{align}
In the case when the Green's function phase $\delta_{f}^{(G)}$ can be associated with a phase shift in some scattering process, the pre-factor $\sin^2\delta_{f}^{(G)}$, by which $A_{s,f}$ in (\ref{As-sin2delta}) differs from $B_{f}$ in (\ref{B-delt}), has the meaning of the probability that the particle has entered the region of the resonance interaction in the scattering~\cite{KVJPG40}, and $A_{s,f}$  is the time delay for the scattered particles. Note that  the factor $\sin^2\delta_{f}^{\rm med}$ appears in the virial limit $n\lambda^3\ll 1$ in the so called generalized Beth-Uhlenbeck approach \cite{SRS1990,Ropke:2012qv} where however enters the $T$-matrix phase shift $\delta_{f}^{\rm med}$ rather than the phase shift of the Green function $\delta_{f}^{(G)}$. In the QPA the entropy spectral function, $A_{s,f}$, proves to be a more peaked  function than the Noether spectral function $A_{f}$, cf. \cite{IKV-2}, whereas in the IQPA they are similarly picked functions.

At the end, note that the spectral functions $A_{f}$, $B_{f}$ and $A_{s,f}$ have the same dimensionality.

\subsection{Botermans-Malfliet particle number}\label{nonideal-number}

We will derive the particle number density with the help of Eq.~(\ref{nf-dOmdmu}) using the thermodynamic potential given in Eqs.~(\ref{Omega-int-As}) and (\ref{DOm}). Taking into account the $\mu_f$ dependence of $\Omega$ entering explicitly through the fermion occupation number, we separate two terms:
\begin{align}
-\frac{1}{V}\frac{\partial\Omega}{\partial \mu_f}\Big|_T = n^{\rm (BM)}_f +\Delta n_f  \,,\label{BMdelta}
\end{align}
where the quantity
\begin{align}
n^{\rm (BM)}_f=
d_f\intop\frac{\rmd^3 p}{(2\pi)^3}
\intop_{-\infty}^{+\infty}\frac{\rmd \epsilon}{2\pi}
 f(\epsilon-\mu) A_{s,f}(\epsilon,\vec{p}\,)
 \label{nBM-def}
\end{align}
is calculated with the weight function $A_s$ given by Eq. (\ref{Asf-def}) rather than with $A$.
The fermion density $n_f^{\rm (BM)}$ proves to be the zeroth component of the so-called Botermans-Malfliet (BM) current, cf. \cite{IKV-2}. This current follows from the first-gradient-order quantum kinetic Kadanoff-Baym equation written in the BM form valid  for the system not too far from the equilibrium~\cite{BotMalf}. As demonstrated in~\cite{Leupold2000}, this current is conserved  by the Kadanoff-Baym  kinetic equation in the BM form with the local collision term (i.e. disregarding memory effects).   This form of the kinetic equation allows to construct a test-particle method, which is behind some extended numerical transport codes used for simulations of heavy-ion collisions. Exact conservation laws allow to keep control of numerical codes. As we mentioned $A_{s,f}^{(\rm QPA)}$ satisfies the exact sum-rule in difference with $A_{f}^{(\rm QPA)}$.
Thereby in cases, when QPA is appropriate approximation, as in Fermi liquid theory, one may approximate $n_f \simeq n_f^{(\rm BM, QPA)}$. However there remains  question about a relation between the BM and Noether expressions for the 4-currents.

The second term in (\ref{BMdelta}) is as follows,
\begin{align}
\Delta n_f &= -\frac{\partial \Delta\Omega_s}{\partial \mu_f}\Big|_{T,A_f} =- \frac{1}{V}\frac{\partial\Phi}{\partial \mu_f}\Big|_{A_f}
\nonumber\\
& + d_f\intop\frac{\rmd^3 p}{(2\pi)^3}
\intop_{-\infty}^{+\infty}\frac{\rmd \epsilon}{2\pi}
\frac{\partial f(\epsilon-\mu_f)}{\partial \mu_f}
\Re\Sigma^R(\epsilon,\vec{p}\,) A(\epsilon,\vec{p}\,)
\,.
\label{Dnf-def}
\end{align}
Here, the derivative of  $\Phi$ has to be calculated with the fermion  spectral functions kept fixed. Applying Eq.~(\ref{Msum-f}) to Eq.~(\ref{dPhidmu-1}) we can write
\begin{align}
\frac{\partial\Phi}{\partial \mu_f}\Big|_{A_f}& {= -}
d_f\!\!\intop\frac{V\rmd^3 p}{(2\pi)^3} \!\!\!
\intop_{-\infty}^{+\infty}\frac{\rmd \epsilon}{2\pi}
f(\epsilon-\mu_f)
\nonumber\\
&\times 2\Im \Big[\Sigma^R(\epsilon,\vec{p}\,){\partial}_\epsilon
%{\partial \epsilon }
G^R(\epsilon,\vec{p}\,)\Big]\,.
\label{dPhidmu-int}
\end{align}
Combining now Eqs.~(\ref{Dnf-def}) and (\ref{dPhidmu-int}), using that $\frac{\partial f(\epsilon-\mu_f)}{\partial \mu_f}=-\frac{\partial f(\epsilon-\mu_f)}{\partial \epsilon}$, and integrating by parts,  we obtain
\begin{align}
\Delta n_f &=d_f\intop\frac{\rmd^3 p}{(2\pi)^3}
\intop_{-\infty}^{+\infty}\frac{\rmd \epsilon}{2\pi}
f(\epsilon-\mu)\, \mathcal{C}_f(\epsilon,\vec{p}\,)\,,
\label{Dnf-Cf}
\end{align}
with the function $\mathcal{C}_f(\epsilon,\vec{p}\,)$ defined in Eq. (\ref{C-def}).

Although expressions for $n_f^{\rm(BM)}$ and $\Delta n_f$ are rather complex, their sum is rather simple. Indeed, using Eqs. (\ref{Bf-def}) and (\ref{Asf-def})  we observe cancellations of the terms with self-energies so that
\begin{align}
n^{\rm (BM)}_f +\Delta n_f  =n_f^{\rm (Noeth)}\,,
\label{nBM-nNoeth-Dn}
\end{align}
where $n_f^{\rm (Noeth)}$ is given by Eq.~(\ref{nf-Noether}) and is expressed solely through the spectral function $A_f$. The non-trivial meaning of this result is that the Noether density,
which follows from the definition of the Green's function through the creation and annihilation operators
and also the quantity that follows directly from the thermodynamic relation coincide exactly for any approximation of $\Phi$, cf.~(\ref{dOmdmu-TrG}).

Using the definition of $\mathcal{C}_f$ in Eq.~(\ref{C-def}) and Eq.~(\ref{As-AGB}) we can write
\begin{align}
\mathcal{C}_f&=\frac{\Gamma_f}{M_f^2+\frac14\Gamma_f^2} -
\frac12 \frac{\Gamma_f^3}{\big(M_f^2+\frac14\Gamma_f^2\big)^2}\Big(\, \partial_\epsilon M_{f}
- M_f\frac{\partial_\epsilon \Gamma_f}{\Gamma_f}\Big)\,.
\end{align}
In the QPA we have
\begin{align}
\lim_{\Gamma \to 0}\frac{\Gamma_f}{M_f^2+\frac14\Gamma_f^2} &=2\pi \delta( M_f)\,,
\nonumber\\
\lim_{\Gamma \to 0} \frac{\frac12\Gamma_f^3}{\big(M_f^2+\frac14\Gamma_f^2\big)^2} &=2\pi \delta( M_f)\,,
\nonumber
\end{align} and  $\mathcal{C}_f$ is simplified as
\begin{align}
\mathcal{C}_f^{\rm (QPA)}&= 2\pi\delta(M_f)\Big( 1- \partial_\epsilon M_{f} - M_f\frac{\partial_\epsilon \Gamma_f}{\Gamma_f}\Big)
\nonumber\\
& =2\pi\delta(M_f)\partial_\epsilon\Re\Sigma\,.
\label{Cf-QPA}
\end{align}
The integral over energy of this function is not equal to zero that reflects the violation of the exact sum-rule for the spectral function $A_{f}$, i.e. Eq.~(\ref{sumrule-A-QPA}) instead of Eq.~(\ref{A-norm}). On the other hand, in the IQPA the $\mathcal{C}_f$ function is expanded as
\begin{align}
\mathcal{C}_f^{\rm (IQPA)}
& \approx 2\pi\delta(M_f)\partial_\epsilon\Re\Sigma+\Gamma_f\lim_{\eta\to 0}
\frac{M_f^2 - \eta^2}{(M_f^2+\eta^2)^2}
\,.
\label{Cf-IQPA}
\end{align}
The integration of this function over epsilon gives zero because of relation (\ref{KramersKroning}) that keeps the proper normalization of functions $A_{s,f}^{\rm (IQPA)}$ and $A_f^{(\rm IQPA)}$.

In Refs.~\cite{IKV-YaF,KIV} it was shown within nonequilibrium diagram technique that both Noether and BM currents are conserved by the BM quantum kinetic equation and that these currents should coincide for near equilibrium configurations up to the first gradients. This implies that not only the integral of the spectral function (\ref{C-def}) over the energy vanishes in $\Phi$ derivable theories, but also so does the integral of the spectral function weighted with $f(\epsilon-\mu)$. We return to this question in Sect.~(\ref{sec:Phi-U}).

\subsection{Entropy and energy}\label{Thermod}
Above we derived various expressions for $\Omega$, cf. (\ref{Omega-int-B}), (\ref{DOm1}), and
(\ref{Omega-int-As}),   (\ref{DOm}), with which we obtain the pressure $P=-\Omega/V$. Now let us similarly derive expressions for the entropy and the energy.

 \subsubsection{Entropy}\label{nonideal-Entropy}

The entropy  can be obtained by differentiation of the thermodynamic potential over $T$. Thus, we begin with expression (\ref{Omega-int-1}), where due to variational  relations (\ref{Om-G-var}) we have to differentiate
{with respect to $T$ only the particle occupation functions $f(\epsilon-\mu_f)$  and the explicit dependence of $\Phi$}. Thus, we have
%%%%%%%%%%%%%%%%%%%%%%%%%%%%
\begin{align}
S&=-\frac{\partial\Omega}{\partial T}\Big|_{\mu_f} = -\frac{\partial \Phi}{\partial T}\Big|_{A_f}-
d_f\intop\frac{V\rmd^3 p}{(2\pi)^3}
\intop_{-\infty}^{+\infty}\frac{\rmd \epsilon}{2\pi}\frac{\partial f(\epsilon-\mu)}{\partial T}
\nonumber\\
&\times 2\Im\big[\ln(-[G^R(\epsilon,\vec{p}\,)]^{-1}) +\Sigma^R(\epsilon,\vec{p}\,)G^R(\epsilon,\vec{p}\,)\big]\,,
\label{dOmdT-0}
\end{align}
with the derivative $\frac{\partial \Phi}{\partial T}\Big|_{A_f}$  taken at fixed spectral functions.

Using Eqs.~(\ref{dOmdT-1}) and (\ref{DS-def-app}) derived in Appendix ~\ref{app:deriv} we rewrite expression for the entropy as the sum of two terms,
\begin{align}
&S=S^{\rm (loc)}+S^{\rm (mem)}\,,\label{SlocMem}
\end{align}
with
\begin{align}
S^{\rm (loc)} &=d_f\intop\frac{V\rmd^3 p}{(2\pi)^3} \intop_{-\infty}^{+\infty}\frac{\rmd \epsilon}{2\pi} \sigma_f(\epsilon-\mu_f) A_{s,f}(\epsilon,\vec{p}\,)\,,
\label{DSp-1}
\end{align}
where
\begin{align}
\sigma_f(\epsilon) &= -(1-f(\epsilon))\ln(1-f(\epsilon))-f(\epsilon)\ln f(\epsilon)\,,
\label{sig-def}
\end{align}
and
\begin{align}
S^{\rm (mem)} &=
d_f\!\!\!\intop\!\frac{V\rmd^3 p}{(2\pi)^3}\!\!\!
\intop_{-\infty}^{+\infty}\!\!\!\frac{\rmd \epsilon}{2\pi}
 \sigma_f(\epsilon-\mu_f)
\nonumber\\
&\times {\partial}_\epsilon
 \big[\Re\Sigma^R(\epsilon,\vec{p}\,)A_f (\epsilon,\vec{p}\,)\big]
- \frac{\partial\Phi}{\partial T}\Big|_{A_f} ,
\label{DSp-0}
\end{align}
Such a decomposition of the entropy was introduced in Ref~\cite{Carneiro-Pethick}. The interpretation of these two terms from the point of view of the quantum kinetics was clarified in Ref.~\cite{IKV-2}, cf. Eqs.~(6.7)--(6.10) there. The term $S^{\rm (loc)}$ is determined by the purely Markovian contributions, whereas  $S^{\rm (mem)}$ is non-Markovian (memory) term appearing from gradient contributions of the diagrams with three and more vertices containing more than two time-arguments.

As we show in Appendix~\ref{app:deriv}, the last term in (\ref{DSp-0}) can be written as in Eq.~(\ref{dPhidT-fin}), and then $S^{\rm (mem)}$ can be cast in the form of Eq.~(\ref{Smem-app-1}). Thus we obtain
\begin{align}
S^{\rm (mem)} &=
\frac{d_f}{T}\intop\frac{V\rmd^3 p}{(2\pi)^3}
\intop_{-\infty}^{+\infty}\frac{\rmd \epsilon}{2\pi}
 f(\epsilon-\mu_f)\Re\Sigma^R(\epsilon,\vec{p}\,)  A_f (\epsilon,\vec{p}\,)
 \nonumber\\
& - \Big(\frac{\partial \Phi}{\partial T}\Big)_{\rm vert}
 \nonumber\\
&+ d_f\intop\frac{V\rmd^3 p}{(2\pi)^3}
\intop_{-\infty}^{+\infty}\frac{\rmd \epsilon}{2\pi}
f(\epsilon-\mu_f)\frac{\epsilon-\mu_f}{T}\mathcal{C}_f(\epsilon,\vec{p}\,)
\,,
\label{Smem-2}
\end{align}
with $\mathcal{C}_f$ defined in Eq.~(\ref{C-def}).
Here $\big(\frac{\partial \Phi}{\partial T}\big)_{\rm vert}$ is the derivative of $\Phi$ with respect to $T$ standing explicitly in front of the Matsubara sums in each diagram contributing to $\Phi$. Hence it  depends on the type of the theory used for the description of particle interactions and the set of diagrams included in $\Phi$. However for a theory with only one type of interaction, as for the interaction in form of a pair potential of our interest here, one can obtain a closed expression for $\big(\frac{\partial \Phi}{\partial T}\big)_{\rm vert}$.

Performing Matsubara summations  in Eq.~(\ref{dDPhidT-vert}) of Appendix \ref{app:ddPhi} with the help of Eqs.~(\ref{Msum-f})  substituting the result in Eq.~(\ref{Smem-2}) we obtain
\begin{align}
S^{\rm (mem)} &=- \frac{\Phi}{T} + \frac{1}{2}
\frac{d_f}{T}\!\!\intop\!\!\frac{V\rmd^3 p}{(2\pi)^3}\!\!
\intop_{-\infty}^{+\infty}\!\!\frac{\rmd \epsilon}{2\pi}
f(\epsilon-\mu_f)
\nonumber\\
&\times\Big[\Re\Sigma^R(\epsilon,\vec{p}\,) A_f(\epsilon,\vec{p}\,)
-\Gamma_f(\epsilon,\vec{p}\,) \Re G^R(\epsilon,\vec{p}\,)\Big]
\nonumber\\
&+ d_f\intop\frac{V\rmd^3 p}{(2\pi)^3}
\intop_{-\infty}^{+\infty}\frac{\rmd \epsilon}{2\pi}
f(\epsilon-\mu_f)\frac{\epsilon-\mu_f}{T}\mathcal{C}_f(\epsilon,\vec{p}\,)\,.
 \label{Smem-fb}
\end{align}

In the QPA  expression for the local entropy (\ref{DSp-1}) simplifies as
\begin{align}
S^{\rm (loc,QPA)}&
= d_f\intop\frac{V\rmd^3 p}{(2\pi)^3}  \sigma_f(\widetilde{\epsilon}_{\vec{p}}-\mu_f)\,,
\end{align}
being the  contribution of the ideal gas of quasiparticles with the spectrum (\ref{MfG}). Equation~(\ref{Smem-fb}) in the QPA yields
\begin{align}
S^{\rm (mem,QPA)} &= - \frac{\Phi^{\rm (QPA)}}{T} +
\frac{d_f}{T}\!\!\intop\!\!\frac{V\rmd^3 p}{(2\pi)^3}
f(\widetilde{\epsilon}_{\vec{p}}-\mu_f)
Z_f(\vec{p}\,)
\nonumber\\
&\times \Big[\widetilde{\epsilon}_{\vec{p}}-\mu_f +
\frac{\Re\Sigma^R(\widetilde{\epsilon}_{\vec{p}},\vec{p}\,)}{2}
\Big]
\nonumber\\
&- \frac{d_f}{T}\intop\frac{V\rmd^3 p}{(2\pi)^3}
f(\widetilde{\epsilon}_{\vec{p}}-\mu_f)
(\widetilde{\epsilon}_{\vec{p}}-\mu_f)\,,
 \label{Smem-QPA}
\end{align}
where we used Eq.~(\ref{Cf-QPA}) and that
$\lim_{\Gamma_f\to 0}\Gamma_f\Re G=2\pi\delta(M_f)M_f=0$. We also used the quasiparticle renormalization factors $Z_f$  defined in Eq.~(\ref{Af-QPA}).
The term  $\Phi^{\rm (QPA)}$ in (\ref{Smem-QPA}) is to be calculated with the Green's functions taken in the QPA, see, e.g., Eq.~(\ref{Gf-QPA}).

\subsubsection{Energy}\label{nonideal-Energy}

The energy of the system can be calculated with the help of the thermodynamic identity
\begin{align}
E=\Omega+T\,S+\mu_f\, N_f
 =\Omega
 -   T\frac{\partial \Omega}{\partial T}\Big|_{\mu_f}
 - \mu_f\frac{\partial\Omega}{\partial \mu_f}\Big|_{T}\,.
 \label{thermoidentity}
\end{align}
Using the representation for the entropy in the form of Eq.~(\ref{dOmdT-2}) derived in Appendix~\ref{app:deriv} we immediately obtain from Eq.~(\ref{thermoidentity}) the following expression for the energy:
\begin{align}
E&=
d_f\intop\frac{V\rmd^3 p}{(2\pi)^3}
\intop_{-\infty}^{+\infty}\frac{\rmd \epsilon}{2\pi}
f(\epsilon-\mu_f)\epsilon
A_f(\epsilon,\vec{p}\,)
 + \Delta E\,,
\label{E-result}
\end{align}
where the latter term  is given by
\begin{align}
\Delta E= \Phi- T \Big(\frac{\partial \Phi}{\partial T}\Big)_{\rm vert}\,.
\label{DeltaE-def}
\end{align}
The form of $\Delta E$ depends on a particular choice of the particle interaction.

Performing in Eq.~(\ref{DE-general}) derived in Appendix~\ref{app:ddPhi} the Matsubara summations, with the help of Eq.~(\ref{Msum-f}) for fermions interacting via a pair potential we obtain
\begin{align}
&\Delta E =
\frac{
d_f}{2}\!\!\intop\!\!\frac{V\rmd^3 p}{(2\pi)^3}\!\!
\intop_{-\infty}^{+\infty}\!\!\frac{\rmd \epsilon}{2\pi}
f(\epsilon-\mu)2\Im\big(\Sigma^R(\epsilon,\vec{p}\,) G^R(\epsilon,\vec{p}\,)\big)
\,.
\label{DE-def}
\end{align}

From the Dyson's Eq. (\ref{Dyson-eq}) written in the form $G_0^{-1} G^R = 1 + \Sigma^R G^R$ we have
\begin{align}
2\Im(\Sigma^R G^R) =  2\Im (G_0^{-1}G^R) = -(\epsilon - \epsilon_p) A_f \,.
\label{ImSG}
\end{align}
Expressing $\epsilon A_f$ in (\ref{E-result}) with the help of this relation,
we can cast  Eq.~(\ref{DE-def}) in the form
\begin{align}
E&=-
d_f\intop\frac{V\rmd^3 p}{(2\pi)^3}
\intop_{-\infty}^{+\infty}\frac{\rmd \epsilon}{2\pi}
f(\epsilon-\mu_f)
\nonumber\\
&\quad\times 2\Im\big[(\epsilon_p-\half\Sigma^R(\epsilon,\vec{p}\,)) G^R(\epsilon,\vec{p}\,)\big]\,.
\end{align}
Thereby, we recover the relation derived by Galitsky and Migdal in Ref.~\cite{Galitskii} for $T=0$.
Alternatively, using (\ref{ImSG}) the energy can be written as
\begin{align}
E&=
d_f\intop\frac{V\rmd^3 p}{(2\pi)^3}
\intop_{-\infty}^{+\infty}\frac{\rmd \epsilon}{2\pi}
f(\epsilon-\mu_f)
 \frac{(\epsilon+\epsilon_p)}{2} A_f(\epsilon,\vec{p}\,)\,.
\label{E-short}
\end{align}
It is remarkable that the energy can be calculated using only the spectral density of the fermions, whereas to calculate the entropy, Eq.~(\ref{Smem-fb}) and (\ref{Smem-QPA}), and the pressure, Eqs.~(\ref{Omega-int-B}) and (\ref{DOm1}), using finite number of diagrams in the $\Phi$ derivable approach, in general case one needs to evaluate the $\Phi$ functional explicitly. This peculiarity of the expression (\ref{E-short}) is used to calculate the energy of the fermionic system when some specific approximation for the fermion self-energy is assumed, e.g., G-matrix and T-matrix approximations, cf.~\cite{Bozek01}. However, Eq.~(\ref{E-short}) is derived here for a pair-potential interaction among fermions, which produces, particularly, Eq.~(\ref{dDPhidT-vert}). Such an interaction does not account for retardation effects, i.e. frequency dependence of the interaction. For a more general case, e.g., for the one-meson exchange model taking into account the  retardation effects in meson propagators, i.e, the energy dependence, Eq.~(\ref{E-short}) would be inapplicable.

The obtained results for thermodynamic quantities show that even if one has some scheme to obtain a single-excitation spectrum or a spectral function, which is tuned to reproduce the phenomenological response of a system to external excitations, one cannot proceed in calculations of the pressure and entropy without the information about the set of closed diagrams, which generate the chosen spectral function.
Remarkably, in Ref.~\cite{IKV-2}, starting from the energy-momentum tensor the authors obtained a simple expression for the entropy written
solely in terms of the spectral function $A$. This is in astonishing contrast with Eqs.~(\ref{DSp-1}) and (\ref{Smem-fb}) which look complicated and explicitly dependent on the $\Phi$ functional. The energy-momentum tensor was written in~\cite{IKV-2} for interactions with non-derivative coupling. This is not suitable for the momentum-dependent pair potential employed in our paper. The momentum dependence of the interaction changes the structure of the energy-momentum tensor as shown in Ref.~\cite{IKV-YaF}. However, the expression for the entropy was not reconsidered there. In Sect.~IV for the $\Phi$ functional with two first diagrams in Fig.~\ref{fig:Phi-U-ser}, we show that Eqs.~(\ref{DSp-1}) and (\ref{Smem-fb}) can be written in a simpler form similar to Eq.~(6.14) of \cite{IKV-2} but with some corrections.

Summarizing, in this Section we discussed various general properties of the $\Phi$ derivable approximations. We demonstrated that three spectral functions $A$, $B$, and $A_s$ determine properties of various thermodynamic quantities. Two of these functions, $A$ and $A_s$, enter respectively the Noether and Botermans-Malfliet particle densities. We derived explicit expressions for the latter quantities.  We also demonstrated that the knowledge of only the spectral functions and particle occupations does not allow, in the general case without additional calculation of the corresponding $\Phi$ functional diagrams to calculate the entropy and the pressure.

\section{Test pair potentials}\label{ssec:test-pot}

For the quantitative analysis of the relations derived above and comparison with the virial approximation below, we have to specify the spin-independent pair potential. We are mainly interested in effects related to such properties of the potential as its shape, depth, and range. So, we consider several test potentials. Having in mind the nuclear interaction scales, we consider a system of one fermion species with the nucleon mass ($m_f=m_N=938$\,MeV). More specifically we consider neutrons to avoid complications with electromagnetic interactions.
As an example of the potential of purely quantum origin, we employ a boson-exchange Yukawa-type potentials, which resemble the central part of the nucleon-nucleon interaction. For comparison, we also consider potentials similar to a classical molecular potential (Lennard-Jones) and square barrier-well potentials.

\subsection{One-boson exchange nucleon-nucleon interaction  potentials}\label{app:NNpot}

The nucleon-nucleon pair potential in vacuum is motivated by the one-boson-exchange model.
As known, exchange  by  vector bosons between fermions of one-sign  charge (electric, baryon, lepton, etc) describes their repulsion, whereas exchange by scalar and pseudo-scalar  bosons describes an attraction. In the early 1990s the Nijmegen group \cite{Nijmegen} examined
carefully all NN scattering data at energies below 350 MeV published between 1955 and 1992. The constructed potentials should describe this rich experimental information on the nucleon-nucleon scattering phase shifts. Note that the inverse problem of the scattering has no unique solution. However analysis of the data demonstrates that any reliable $NN$ potential should be repulsive at short distances and attractive at longer distances. The reconstructed potentials  include the Nijmegen models \cite{Nijmegen1} Nijmegen I, II and Reid-93, the Argonne~v18 \cite{Wiringa}, and the CD Bonn \cite{Machleidt1}. All these potentials have a complicated structure, cf.~\cite{Naghli}. Since our aim now is to demonstrate the abilities of the $\Phi$-derivable approximations rather than to describe the whole set of experimental nucleon-nucleon scattering data, we start by considering a simplified Yukawa potential with parameters~\cite{Walecka:1974qa}. Then we use another set of parameters fitted to the central part of the CD Bonn potentials and finally discuss the Reid potential.

\begin{figure}
\centering
\includegraphics[width=8cm]{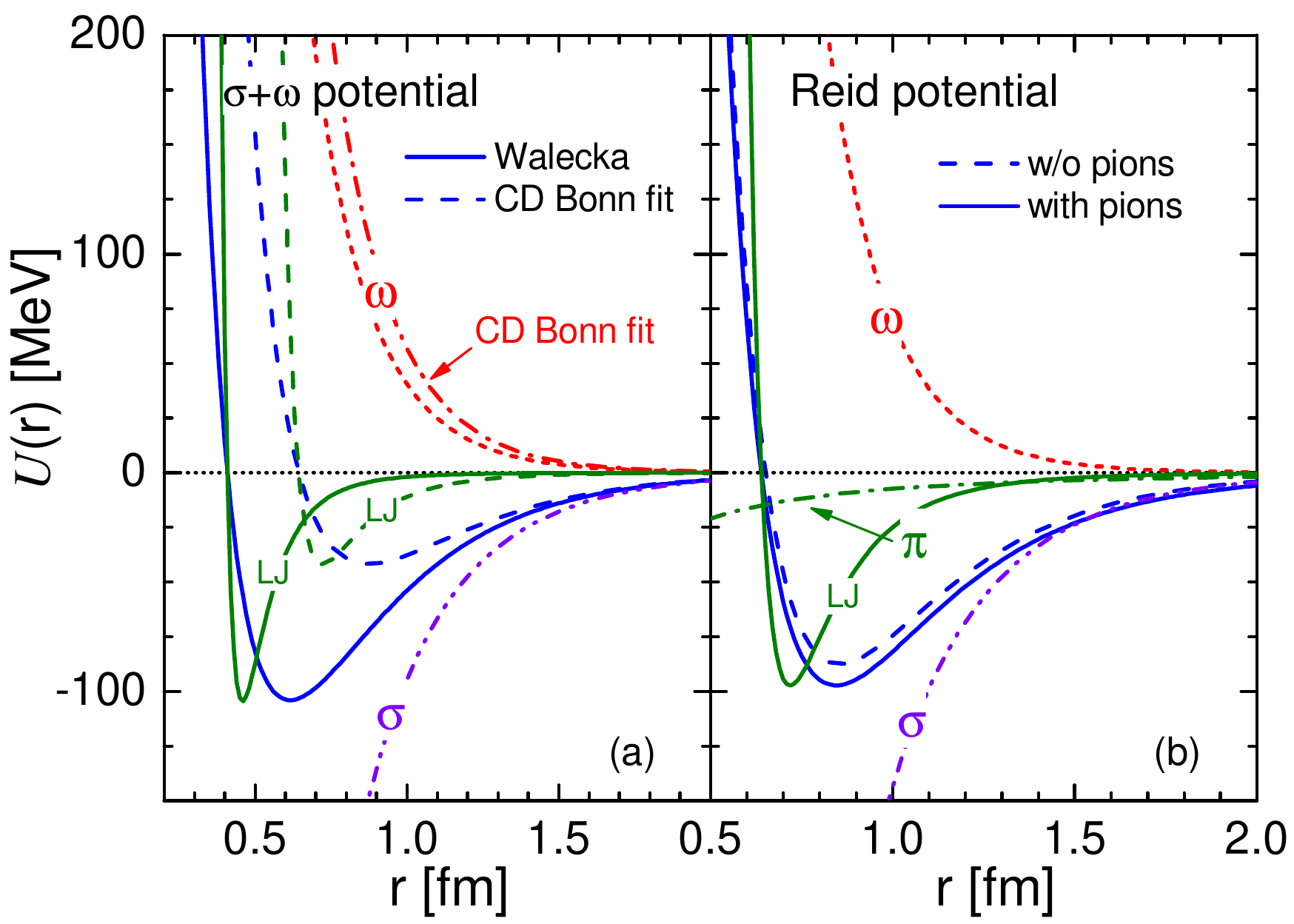}
\caption{Panel (a): Pair potential  as a function of the distance between two nucleons. Solid line (Walecka) is the Yukawa potential (\ref{U-potR}) of the Walecka relativistic mean-field model ~\cite{Walecka:1974qa} with parameters (\ref{Uomsig-par-W}). Dash-double-dotted and short dashed lines labeled as $\sigma$ and $\omega$ show separated contributions of $\sigma$ and $\omega$.
Dashed line (CD Bonn fit) characterized by parameters (\ref{Uomsig-par-Bonn})  reproduces approximately the central part of the CD Bonn potential~\cite{Machleidt89} of the nucleon-nucleon interaction, see Ref.~\cite{Shuryak-Torres19}. The corresponding $\omega$ contribution is shown  by the dash-dotted line labeled $\omega$ CD Bonn fit. On both panels solid and dashed lines labeled as LJ show the Lennard-Jones-type potentials (\ref{LJ-pot}) with parameters fitted to reproduce the same depths of potentials and the points, where the potentials cross zero line, as for the Walecka and CD Bonn fit potentials, respectively.
\\
Panel (b): Soft-core Reid potential (\ref{U-Reid}) as a function of the distance between two nucleons with and without inclusion of the pion-exchange term (solid and dashed lines, respectively). Dash-double-dotted, short dashed and dash-dotted lines show separated contributions of $\sigma$, $\omega$ and $\pi$ exchanges.
}
\label{fig:U-pot}
\end{figure}

As a simplest pair potential describing repulsion between identical nucleons  at short distances ($r< 0.3$\,fm) and attraction at intermediate distances, $0.3\,{\rm fm}< r\lsim 1\,{\rm fm}$,  first consider the model, where  it is assumed that nucleons are exchanged only by the vector $\omega$ and scalar $\sigma$ mesons:
\begin{align}
\Ucor(r) &= \Ucor_{\om}(r) + \Ucor_{\sigma}(r)
=h_{\om} Y(m_{\om} r) + h_{\sigma} Y(m_{\sigma} r) \,,
\nonumber\\
h_\om &= \frac{g_\om^2}{4\pi}m_\om\,,\,\,
h_\sigma=-\frac{g_\sigma^2}{4\pi}m_\sigma\,,\,\,
Y(x)=\frac{\exp(-x)}{x}\,.
\label{U-potR}
\end{align}
The functions $Y(x)$ describe the Yukawa potentials.
In the momentum space this potential is given by
\begin{align}
\Umom(|\vec{q}\,|) &= \intop \rmd^3 r e^{-i\vec{q}\,\vec{r}} \Ucor(r)
\nonumber\\
&=
\frac{g_\om^2}{\vec{q\,}^2+m_\om^2}- \frac{g_\sigma^2}{\vec{q\,}^2+m_\sigma^2} \,.
\label{U-pot}
\end{align}

To get a necessary repulsion at short distances, in (\ref{U-potR}) we need to employ  $g_\om >g_\sigma$.
The attraction at larger densities is realized, provided  masses of the mesons are $m_\om>m_\sigma$.
The original relativistic mean-field model proposed by Walecka~\cite{Walecka:1974qa}, describing  nucleon binding in the nuclear matter at a reasonable value of the nucleon density  used the  values
$m_\om    = 782.6\,{\rm MeV}$ and $m_\sigma = 500.0\,{\rm MeV}$  and  $g_\om =11.67$, $g_\sigma =8.71$, i.e. $g_\om >g_\sigma$, as it is necessary  to qualitatively describe behavior of the nucleon-nucleon pair  potential in vacuum. With these parameters one gets
\begin{align}
C_\om^2 = g_\om^2 \frac{m_N^2}{m_\om^2}=195.70\,,\,\,
C_\sigma^2= g_\sigma^2 \frac{m_N^2}{m_\sigma^2}=267.10
\label{Uomsig-par-W}
\end{align}
and $h_\om=8473.04$\,MeV and $h_\sigma=-3019.73$\,MeV.

The Walecka pair potential is shown in Fig.~\ref{fig:U-pot}a by the solid line labeled ``Walecka'' together with separate contributions  of $\omega$ and $\sigma$ mesons shown by the corresponding dash and dash-double-dotted lines.
The position of the minimum of the potential (\ref{U-pot}) can be evaluated as
\begin{align}
r_m \approx \hbar\frac{\ln\big(h_\om/|h_\sigma|\big)}{m_\om -m_\sigma}
\Big[1-\frac{(m_\om-m_\sigma)^2}{m_\om m_\sigma\ln^2\big(h_\om/|h_\sigma|\big)}\Big]\,,
\label{Uoms-rmin}
\end{align}
that for parameters (\ref{Uomsig-par-W}) produces $r_m\approx 0.58$\,fm, being close to the value $0.62$\,fm found numerically.
As follows from Eq.~(\ref{Uoms-rmin}),  the potential has a minimum  at a finite distance, provided $h_\om/|h_\sigma|>1$.
%that corresponds to
% $C^2_\om/C^2_\sigma> m_\sigma^2/m_\om^2\sim 0.4$ for $m_\sigma\sim 500$\,MeV.
Parameters of the Walecka model, which we employ here, satisfy this constraint.

Another parameter characterizing behavior of  the potential is the closest-approach distance of two static nucleons, which can be  defined as $U(2r_0)=0$. For the potential given by (\ref{U-potR}),
\begin{align}
2r_0=\hbar\frac{\ln(g_\om^2/g_\sigma^2)}{m_\om-m_\sigma} \,.
\label{r0-U}
\end{align}
With parameters (\ref{Uomsig-par-W}) we find $2r_0=0.41$\,fm. In Eqs.~(\ref{Uoms-rmin}) and (\ref{r0-U}) we retained $\hbar$ to emphasize that typical  scales, at which   the nuclear potential is changed, are  associated  with the quantum Compton lengthes of the $\sigma$ and $\om$ mesons, i.e. these scales are of  purely quantum  origin. In contrast, the typical sizes of intermolecular potentials are much larger, and thereby one uses   classical potentials like the Lennard-Jones (LJ) potential.

Expansion of the potential (\ref{U-potR}) for $r$ near $r_m$ yields
\begin{align}
U(r\sim r_m)\approx U_m+\alpha_U \frac{(r-r_m)^2}{2r_m^2}\,.
\label{U-exp-rm}
\end{align}
With parameters (\ref{Uomsig-par-W}) we have
\begin{align}
U_m\approx-104\,{\rm MeV},\,\, r_m\approx 0.615\,{\rm fm},\,\,
\alpha_U\approx 815\,{\rm MeV}.
\label{Uexp-par-W}
\end{align}

The potential (\ref{U-potR}) is not perturbative in the sense that being substituted in the Schr\"odinger equation it produces  a very large s-wave scattering length. Employing parameters (\ref{Uomsig-par-W}) we get $a_0^{\rm [W]}\approx -41$\,fm and the effective range of 1.8\,fm. However, the scattering length is very sensitive to tiny variations of potential parameters, so increasing the $\om$ repulsive term by only  1\% we may obtain  value of the scattering amplitude $a_0^{\rm [W]}\approx -23$\,fm which  rather appropriately fits  the nucleon-nucleon scattering amplitude~\cite{Rijken-06}. The effective range changes very little under this variation. Thus, the simple potential (\ref{U-potR})  with the original parameters (\ref{Uomsig-par-W}) fitted to reproduce nuclear saturation density and binding energy grasps the non-perturbative feature of the s-wave nucleon-nucleon scattering. Below, in Sect.~\ref{ssec:effrange} and Appendix~\ref{app:double-scatt} we discuss the experimental neutron-neutron scattering and non-perturbative feature of the nuclear potential in a more detail.

The original Walecka model has a drawback as it produces  too small effective nucleon mass, $m_N^* (n=n_0)=0.54 m_N$, at the nuclear saturation density $n_0 =0.19\,{\rm fm}^{-3}$, and too high compressibility parameter, $K=540$ MeV,  compared to experimentally motivated range of values $m_N^* (n=n_0)=(0.6\mbox{--}1)m_N$~\cite{Furnstahl,mEff-1,mEff-2} and  $K=240\pm 30$\,MeV~\cite{K-exp1,K-exp2}. Note here that in reality the effective nucleon mass is a complicated function of the frequency-momentum-dependent self-energy. However these effects are not incorporated within the mean-field approaches. References~\cite{Boguta-Bodmer,Boguta} constructed a  modified (or nonlinear) Walecka model introducing non-linear $\sigma$ self-interaction potential characterized by two extra fitting parameters.
The model proposed in~\cite{Boguta-Bodmer,Boguta} and its further modifications, like~\cite{Cubero}, allow to better satisfy  empirical constraints at $n\sim n_0$ however yielding (e.g. model 2 in~\cite{Cubero}) $g_\om =6.67$, $g_\sigma =7.21(m_\sigma/500\,{\rm MeV})$, i.e. $g_\om <g_\sigma$ for $m_\sigma =500$ MeV. Thereby,  these models are not appropriate for our needs, at least without modifications, since the pair potential in them has no repulsive core.
The  discrepancy between parameters of the various relativistic mean-field models needed  to describe saturation property of the isospin-symmetric nuclear matter and the parameters needed to describe the nucleon-nucleon interaction in vacuum are discussed, e.g.,  in~\cite{Voskresenskaya}.
To improve the description of the isospin-asymmetric matter within relativistic mean-field models one adds the contribution from the mean $\rho$ meson field, which induces the dependence on the iso-vector nucleon density. Some relativistic mean-field models also introduce attraction dependent on the iso-scalar density mediated by the putative $\delta$ meson and many other modifications. It is not our aim here to enter these complications.

To describe behavior of the pair potential at large distances one should include contribution of  the pion. To do it one should go beyond the mean-field model.
As an example of a pair nucleon-nucleon potential with included pion tuned to describe the experimental  $1{\rm S}_0$ $NN$ phase shifts in the isospin-one channel, one can take the soft core Reid potential~\cite{Reid} shown in Fig.~\ref{fig:U-pot}b by the solid line,
\begin{align}
&U_{\rm Reid}(r) = h_\pi Y(\mu r) + h_\sigma Y(4\mu r) + h_\om Y(7\mu r),
\nonumber\\
&\mu=0.7\,{\rm fm}^{-1}=138.131\,{\rm MeV}\simeq m_\pi\,,\,\,
h_\pi=-10.463\,{\rm MeV}\,,
\nonumber\\
&h_\sigma=- 6602.4\,{\rm MeV}\,,\,\, h_\om=45389.4\,{\rm MeV}\,.
%\nonumber\\
%& m_\sigma=4\mu\approx 552.5\,{\rm MeV}\,,\,\,
%m_\om=7\mu\approx 966.9\,{\rm MeV}\,.
\label{U-Reid}
\end{align}
Dashed line demonstrates central part of the Reid potential with the pion contribution switched off. Comparing solid and dashed lines we see that the pion contribution to the central part of the potential is rather small and in general the Reid potential behaves similarly to the potential (\ref{U-potR}) with the Walecka model parameters (\ref{Uomsig-par-W}). The Reid potential crosses zero at $2r_0=0.64\,{\rm fm}$. The parameters of the expansion (\ref{U-exp-rm}) for this potential are
\begin{align}
U_m\approx -97.2\,{\rm MeV},\,\, r_m\approx 0.845\,{\rm fm},\,\,
\alpha_U\approx 1523\,{\rm MeV}\,.
\label{Uexp-par-Reid}
\end{align}

Note that  the structure of the central part of the potential is not totally constrained by the data.  One of possible choices used in   Ref.~\cite{Shuryak-Torres19} demonstrated that the central part of the CD~Bonn potential~\cite{Machleidt89}  can be reproduced with the help of the Eq.  (\ref{U-potR}), if in the  parameter set (\ref{Uomsig-par-W}) the repulsion is increased by 40\%, i.e. with  modified values
\begin{align}
C_\om^2\to 1.4C_\om^2 [\rm{Walecka}] = 273.98\,,
\label{Uomsig-par-Bonn}
\end{align}
$g_\om = 13.81$ and the same values of the $\sigma$ and $\om$ meson masses and $g_\sigma$ and the coefficient $C_\sigma$ as in (\ref{Uomsig-par-W}).  The corresponding potential is shown in Fig.~\ref{fig:U-pot}a by the dashed line labeled CD Bonn fit. The potential becomes more shallow with the minimum equal to $-41.6$\,MeV in comparison with $-104$\,MeV for the Walecka parameter set (\ref{Uomsig-par-W}). The potential changes sign at $2r_0=0.64$\,fm. The minimum is now shifted to a larger distance, Eq.~(\ref{Uoms-rmin}) gives $r_m=0.85$\,fm that is close to 0.87\,fm found numerically.

Note that in the momentum representation Fourier components of  the Walecka and CD Bonn  central parts  of potentials in the long-distance limit,
$${\cal{U}} (q\to 0)=(C^2_\om -C^2_\sigma)/m_N^2\,,$$ have even different signs. For the Walecka potential ${\cal{U}} (0)<0$ and for CD Bonn fit  potential it is a small positive value. Thus, although the full CD Bonn potential (including of non-central interactions) describes well the experimental $NN$ scattering, its central part behaves differently compared to the Reid potential. Contrary, the Walecka potential is closer to the Reid potential, which describes well the experimental $1{\rm S}_0$ $NN$ phase shifts  in the isospin one channel.

The following remark is in order. In general, fermions interacting via potentials with attractive pockets may form bound states. For instance, in isospin-symmetric matter there may exist deuterons and other clusters, cf.~\cite{APR,Ropke2010,Schuck2016,Pais,Krukov} and references there in.
To analyze, if this occurs for the potentials considered above, we use the Bohr-Sommerfeld quantization condition for the radial motion with the zero angular momentum and calculate the reduced action, $A_{\rm red}$ for the negative energy of the relative motion, $-\varepsilon_{\rm b}$ ($\varepsilon_{\rm b}$ being the binding energy),
\begin{align}
A_{\rm red}(\varepsilon_{\rm b})=\intop_{r_1(\varepsilon_{\rm b})}^{r_2(\varepsilon_{\rm b})} \rmd r
\sqrt{2m_{\rm red}(-\varepsilon_{\rm b}-U(r))}\,,
\end{align}
where $r_{1,2}(\varepsilon_{\rm b})$ are the turning points depending on the energy, and $m_{\rm red}$ is the reduced mass for two-particle scattering problem, $m_{\rm red}=m_N/2$ in our example. Since potentials we consider are repulsive at $r\to 0$ there are two turning points even for the case $l=0$.
The discrete spectrum of bound states is determined then by equation $I(\varepsilon_{{\rm b},n})=A_{\rm red}(\varepsilon_{{\rm b},n})/2\pi= n+\frac12$, where $n>0$ is integer number. Since $A_{\rm red}(\varepsilon_{{\rm b}})$ is a decreasing function of $\varepsilon_{{\rm b}}$, the necessary condition for the existence of a bound state is that
$I_{\rm max}=
I(\varepsilon_{{\rm b}}\to 0)=\int_{2r_0}^{\infty}\frac{\rmd r}{2\pi}
\sqrt{2m_{\rm red}|U(r)|}>1/2\,.$
For potential (\ref{U-potR}) we find $I_{\rm max}=0.26$ with the Walecka model parameters (\ref{Uomsig-par-W}) and $I_{\rm max}=0.17$ for parameters (\ref{Uomsig-par-Bonn}). The Reid potential (\ref{U-Reid}) gives $I_{\rm max}=0.28$. Thus in the considered above nuclear potentials, bound states of two fermions with the nucleon mass $m_N$ are absent.

For the description of the lightest two-nucleon bound state, the deuteron, one has to take into account non-central (tensor) nucleon forces. Clusters with more nucleons remain under-bound, if only a two-body nucleon-nucleon  potential is used.
Besides the pair potential one needs to include three-body forces \cite{APR,Ariola}. As known  resonant two-body forces may give rise to a series of levels in three-particle systems (Efimov states). Number of these states is estimated as
$N_{\rm bound}\simeq \pi^{-1}\ln (|a|/r_0)$, cf. \cite{Efimov1970}. In the case of the neutron gas we have $N_{\rm bound}\simeq 0.67<1$.

\subsection{Classical potentials used to describe interaction between molecules}\label{LJpotentials}

Interparticle potentials used for description of monoatomic gases (like nobel gases) are taken in the form of a LJ potential, which includes an attractive part at a large distance approaching zero asymptotically $\propto -1/r^6$ and a strong repulsion at short distance $\propto +1/r^{12}$ corresponding to the mutual ``impenetrability'' of the atoms. The potential can be written as
\begin{align}
U_{\rm LJ}(r)=4U_0\Big[\Big(\frac{2r_0}{r}\Big)^{12} - \Big(\frac{2r_0}{r}\Big)^6\Big]\,,\quad U_0>0\,,
\label{LJ-pot}
\end{align}
where as before $2r_0$ is the point of the balance between attraction and repulsion, $U_{\rm LJ}(2r_0)=0$, for which reason the
distance $r_0$ can be interpreted as the radius of the atom. The LJ  potential has a minimum at $r_m=2^{1/6}(2r_0)$ and $U_{\rm LJ,min}=U_{\rm LJ}(r_m)=-U_0$. The curvature parameter $\alpha_{U}$ defined in expansion (\ref{U-exp-rm}) is $\alpha_U=72 U_0$.

In Fig.~\ref{fig:U-pot}a,b we compare the form of the LJ potential with those given by the Yukawa interactions (\ref{U-potR}) and (\ref{U-Reid}). We choose the parameters of the LJ potential to reproduce the depths of the Yukawa potentials, as given in Eqs.~(\ref{Uexp-par-W}),  (\ref{Uexp-par-Reid}) and (\ref{Uomsig-par-Bonn}), and also the parameter $r_0$ for each potential.
The corresponding lines are labeled as LJ in Fig.~\ref{fig:U-pot}a,b. We see that the repulsive part of the LJ potential is much steeper than for the nuclear potentials and the characteristic range of the potential is shorter. In Section~\ref{ssec:class-virial} we discuss how these differences in the LJ and Yukawa potentials influence the existence of \emph{the Boyle temperature}, at which the second virial coefficient vanishes and  a gas behaves almost like an ideal gas at low densities. The Boyle temperature exists for the classical gases and this fact was utilized to argue for the form of the potential acting between the gas constituents.

Note that although the LJ potential with the parameters, which we have chosen, visually looks similar to the Walecka and Reid potentials
the former requires modifications, if one would like to employ the Fourier transformation. Indeed, the Fourier transform, which we have exploited above within the $\Phi$-derivable approach, does not exist for the potential behaving as $1/r^{12}$ at $r\to 0$.

\subsection{Square barrier-well potentials}\label{squarewell}

The thermodynamic characteristics of the interacting system are expressed in terms of  the inter-particle potentials. Discussion of the second virial coefficients in the classical limit in Sections~\ref{classicalspec} and \ref{ssec:extrapol-vEoS} will show that they are determined by the volumes occupied by the attractive and repulsive parts of the potential. The simplest form of the potential, for which these quantities can be naturally defined is the square barrier-well potential that we will consider here.

Simplest potential describing a repulsive core and an attractive exterior is the square barrier-well (sw) potential
\begin{align}
\Ucor_{\rm sw}(r)=\left\{
\begin{array}{rl}
U_{\rm r},& r<2r_0\\
% 0,&r=2r_0\\
-U_{\rm a}, & 2r_0<r<r_{\rm U}\\
0, & r\ge r_{U}
\end{array}
\right. \,,\,\, U_{\rm r,a}>0\,,\,\, 2r_0<r_{U}\,,
\label{Usw}
\end{align}
where parameters $U_{\rm r}$ and $U_{\rm a}$ characterize magnitudes of the repulsive and attractive parts of the potential, $r_{\rm U}$ is the potential range, and  $2r_0$ has the sense of the repulsive core radius.
If we put $U_{\rm r}\to \infty$ and $U_{\rm a}=0$, we obtain the hard-sphere potential.

We will also employ the smoothed version of the square barrier-well potential,
\begin{align}
\Ucor_{{\rm sw},\sigma}(r)&=U_{\rm r}
- \big(U_{\rm r} + U_{\rm a}\big)\theta_\sigma(r-2r_0)
+ U_{\rm a}\theta_\sigma(r-r_{U})\,,
\nonumber\\
\theta_\sigma(x)&=\frac12[1+\tanh(x/\sigma)]\,,
\label{Usw-smooth}
\end{align}
where the sharp steps in the potential (\ref{Usw}) are blurred by the parameter $\sigma\ll 2r_0$.

In the momentum representation, the square barrier-well potential (\ref{Usw}) is given by
\begin{align}
\Umom_{\rm sw}(q) &= 4\pi(2r_0)^3 \big(U_{\rm r} + U_{\rm a}\big) f_{\rm sw}(2r_0q)
- 4\pi r_U^3 U_{\rm a} f_{\rm sw}(r_Uq)\,,
\nonumber\\
f_{\rm sw}(x) &= (\sin x-x\cos x)/x^3.
\label{Usw-mom}
\end{align}
We notice that the square barrier-well potential in the momentum representation takes the simple form in the long-wave limit
\begin{align}
\Umom_{\rm sw}(q\to 0) = U_{\rm r} v^{\rm(sw)}_{\rm r} - U_{\rm a} v^{\rm(sw)}_{\rm a}.
\label{Ufourier}
\end{align}
Two quantities introduced here, which are important for comparison of different potentials, are the effective \emph{volumes} of the repulsive core and attractive pockets. For the rectangular  potentials (\ref{Usw}) they are defined as
\begin{align}
v^{\rm(sw)}_{\rm r}&=\frac{4\pi}{3} (2r_0)^3\,,\quad
v^{\rm(sw)}_{\rm a}=
%\frac{4\pi}{3}
v^{\rm(sw)}_{\rm r} [(r_U/(2r_0))^3-1]\,,
\nonumber\\
v_U &= \frac{4\pi}{3}r_U^3 \,,
\label{vr-va-def}
\end{align}
where $v_U$ is the total potential volume, $v_U=v_{\rm r}+v_{\rm a}$. The attraction volume exceeds the repulsion volume for $r_U>2^{1/3}(2r_0)$.

We see that in the momentum representation the square barrier-well potential (\ref{Usw}) in the long-wave limit carries the information about the attractive and repulsive volumes. As we will show  in Section~\ref{semicltest}, the smoothed form of the potential (\ref{Usw-smooth}) is very handy in the discussion of the semiclassical expansion for the second virial coefficients.

The variety of potentials of the quantum and classical origin considered in this Section will be used below for the quantitative comparison of the results obtained within the $\Phi$-derivable and virial approaches, particularly, for the classical and quantum limits of the second viral coefficient.

\section{Tadpole-Sandwich approximation}\label{sec:Phi-U}

In this Section we  apply the $\Phi$ derivable approach restricted to the first two TS diagrams to a system of single fermion species of spin $1/2$ ($d_f=2$) interacting via a pair potential, which we denote as $\Umom(|\vec{p}\,|)$ in the momentum representation and as $\Ucor(|\vec{r}\,|)$ in the coordinate representation.

\subsection{Thermodynamic quantities in case of  $\Phi$  restricted to the TS diagrams }\label{TermodTS}

\begin{figure}
\centering
$$-\Sigma={\includegraphics[width=1cm]{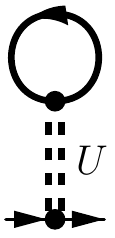}}
\,\,+\,\,
{\includegraphics[width=3cm]{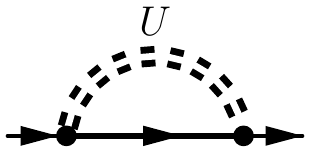}}
$$
\caption{Diagrams determining the fermion self-energy in Eq.~(\ref{Sig-U-Mats}) as it follows from  diagrams (a) and (b) for the $\Phi$ functional in Fig.~\ref{fig:Phi-U-ser}.
}
\label{fig:Sig-U}
\end{figure}

Let us consider the $\Phi$ functional restricted to the first two TS diagrams, (a) and (b) in Fig.~\ref{fig:Phi-U-ser}, which are of the first order in the pair potential $\Umom$, provided the result is expressed in terms of the full Green functions. Such an approximation is sometimes called the self-consistent Hartree-Fock approximation to contrast it from approximations including the tadpole and sandwich diagrams calculated with free Green's functions or dressed Green's functions but not consistent with the $\Phi$ functional.
Following rules listed in Appendix~\ref{DiagramRules} we open the TS diagrams, as shown in Eq.~(\ref{Phi-deriv-exampl}), and obtain
\begin{align}
\Phi^{\rm (TS)} =&
\frac12V\Tr_{p_1}[\tr \hat{G}_{p_1}]\,\Umom(|\vec{p}=0|)\,\Tr_{p_2}[\tr \hat{G}_{p_2}]
\nonumber\\
 -&
\frac12V\Tr_{p_1}\Tr_{p_2}\big[\tr\big( \hat{G}_{p_1} \Umom(|\vec{p}_1-\vec{p}_2|) \hat{G}_{p_2}\big)\big]\,.
\label{Phi-U-Mats}
\end{align}
Using Eq.~(\ref{Sigma-Phi}) from (\ref{Phi-U-Mats}) we derive expression for the fermion self-energy, given by the diagrams shown in Fig.~\ref{fig:Sig-U},
\begin{align}
\Sigma_p^{\rm (TS)}=\Umom(|\vec{p}=0|)\Tr_{p_1}[\tr \hat{G}_{p_1}] - \Tr_{p_1}[\Umom(|\vec{p}_1-\vec{p}\,|)G_{p_1}]\,.
\label{Sig-U-Mats}
\end{align}
 The first term is  the Hartree term and the second one is the Fock term. Comparing Eqs.~(\ref{Phi-U-Mats}) and  (\ref{Sig-U-Mats}) we observe  simple relation
\begin{align}
\Phi^{\rm (TS)}=\frac12 V\Tr_p[\tr (\hat{\Sigma}_p \hat{G}_p)]\,.
\label{Phi-U-Tr}
\end{align}

With the help of Eqs.~(\ref{Msum-f}) and (\ref{dOmdmu-TrG}), (\ref{nf-Noether})
we can rewrite $\Phi$ from Eq.~(\ref{Phi-U-Mats})
explicitly as
\begin{widetext}
\begin{align}
 &{\Phi}^{\rm (TS)} =\frac{V}{2} \Umom(|\vec{p}=0|) n_f^2
-\frac{V}{2}\intop\!\frac{\rmd^3 p_1}{(2\pi)^3}\intop_{-\infty}^{+\infty}\!\frac{\rmd \epsilon_1}{2\pi}\!\!
\intop\!\frac{\rmd^3 p_2}{(2\pi)^3} \intop_{-\infty}^{+\infty}\!\frac{\rmd \epsilon_2}{2\pi}
%\nonumber\\
%&\times
2A_f (\epsilon_1,\vec{p}_1\,) f(\epsilon_1-\mu_f) \Umom(|\vec{p}_1-\vec{p}_2|)
A_f (\epsilon_2,\vec{p}_2\,) f(\epsilon_2-\mu_f)\,.
\label{Phi-HF-eq}
\end{align}
\end{widetext}
Additional factor 2 in the second term is due to summation over spins. The fermion density $n_f$ is given by relations (\ref{dOmdmu-TrG}), (\ref{nf-Noether}).
From Eq.~(\ref{Sig-U-Mats}) we obtain expression for the self-energy
\begin{align}
&\Sigma^{\rm (TS)}(\vec{p}\,) = \Umom(|\vec{p}=0|)\, n_f
\nonumber\\
&\quad - \intop\frac{ \rmd^3 p'}{(2\pi)^3}
\intop_{-\infty}^{+\infty}\frac{\rmd \epsilon'}{2\pi} A_f(\epsilon',\vec{p\,}'\,)\,
\Umom(|\vec{p}-\vec{p\,}'|) f(\epsilon'-\mu_f)
\,,
\label{Sig-HF-eq}
\end{align}
 which  is a real function of the momentum and it does not depend on the frequency $\epsilon$.

Since both $\Im\Sigma^{\rm (TS)}$ and $\partial_\epsilon \Re\Sigma^{\rm (TS)}$ vanish in the TS approximation, the spectral function for fermions in this case  is a $\delta$-function,
\begin{align}
A_f(\epsilon,\vec{p} \,)=2\pi\delta(\epsilon-\epsilon_{\vec{p}}-\Sigma^{\rm (TS)}(\vec{p}\,))\,.
\label{Af-SCHF}
\end{align}

Substituting (\ref{Af-SCHF}) in (\ref{Sig-HF-eq}) and in (\ref{nf-Noether}) we find the system of equations for the self-energy $\Sigma^{\rm (TS)}$ and for the chemical potential $\mu_f$,
\begin{align}
\Sigma^{\rm (TS)}(\vec{p}\,) &=   \intop\frac{\rmd^3 p'}{(2\pi)^3}
\frac{2\Umom(|\vec{p}=0|) - \Umom(|\vec{p}-\vec{p\,}'|)}{e^{(\epsilon_{\vec{p\,}'} + \Sigma^{\rm (TS)}(\vec{p\,}')-\mu_f)/T }+1} \,,
\nonumber\\
n_f^{({\rm Noeth})} &=\intop\frac{2\rmd^3 p}{(2\pi)^3}
\frac{1}{e^{(\epsilon_{\vec{p}} + \Sigma^{\rm (TS)}(\vec{p}\,) - \mu_f)/T }+1}
\,.
\label{Sig-mu-HF}
\end{align}
%cf. Eqs. (\ref{M-f-def}), (\ref{MfG}), (\ref{IQPAdistr}).
We observe that a particular combination of potentials, $2\,\Umom(|\vec{p}=0|) - \Umom(|\vec{p}-\vec{p\,}'|)$,  enters the self-energy and then  all the other relations. The quantity $\Umom(|\vec{p}-\vec{p\,}'|)$ is the matrix element of the potential operator between the momentum states that determines the scattering of two fermions, see Appendix~\ref{app:ME-NN-pot}. Since we are interested in the connection of the results obtained in the $\Phi$ derivable approach with characteristics of the fermion scattering, it is instructive to identify  combination $\Umom(|\vec{p}=0|) - \frac12\Umom(|\vec{p}-\vec{p\,}'|)$ with the spin averaged scattering potential defined in Eq.~(\ref{U-spinaver}), which is taken here for the forward kinematics. As the result, we can present
\begin{align}
\Sigma^{\rm (TS)}(\vec{p}\,) &=   \intop\frac{2\rmd^3 p'}{(2\pi)^3}
\frac{\overline{\Umom^{(S)}}(\frac{\vec{p}-\vec{p\,}'}{2},\frac{\vec{p}-\vec{p\,}'}{2}) }{e^{(\epsilon_{\vec{p\,}'} + \Sigma^{\rm (TS)}(\vec{p\,}')-\mu_f)/T }+1} \,,
\label{SigST-U}
\end{align}
where $\overline{\Umom^{(S)}}(\frac{\vec{p}-\vec{p\,}'}{2},\frac{\vec{p}-\vec{p\,}'}{2})=
\Umom(|\vec{p}=0|) - \frac12\Umom(|\vec{p}-\vec{p\,}'|).$

Using Eqs.~(\ref{Af-SCHF}) and (\ref{SigST-U}) we verify that our spectral function satisfies the relation derived in Ref.~\cite{Kalash-Fradkin} for the general case of a system of fermions interacting via a pair potential
\begin{align}
&\intop_{-\infty}^{+\infty} \frac{\rmd \epsilon}{2\pi} (\epsilon -\epsilon_p)
A_f(\epsilon,\vec{p}\,)
= \intop\frac{2\rmd^3 p'}{(2\pi)^3}
\intop_{-\infty}^{+\infty}\frac{\rmd \epsilon}{2\pi}
\nonumber\\
&\qquad \times f(\epsilon-\mu_f) A_f(\epsilon,\vec{p\,}') \,\overline{\Umom^{(S)}}\big(\frac{\vec{p}-\vec{p\,}'}{2},\frac{\vec{p}-\vec{p\,}'}{2}\big) .
\end{align}

Since  for the chosen set of diagrams the self-energy is independent on the energy variable and $\Im\Sigma^{\rm (TS)}=0$, the spectral functions $B_f$ and $A_{s,f}$ defined in Eq.~(\ref{Bf-def}) and (\ref{Asf-def}) coincide with $A_f$ in (\ref{Af-SCHF}), i.e.,
\begin{align}
&A_{s,f}(\epsilon,\vec{p} \,)=B_f(\epsilon,\vec{p} \,)=A_f(\epsilon,\vec{p} \,)\nonumber
\\
&=2\pi\delta(\epsilon-\epsilon_{\vec{p}}-\Sigma^{\rm (TS)}(\vec{p}\,)) \,.\label{As-Bf-Af-eq}
\end{align}
From Eq.~(\ref{deltaFerm}) follows that for $\Im\Sigma^{\rm (TS)}\to 0$ the phase of the Green's function is
\begin{align}
\delta_f^{(G)}(\epsilon,\vec{p}\,)=\frac{\pi}{2}-\lim_{\eta\to + 0}\arctan\frac{2(\widetilde{\epsilon}_{\vec{p}}-\epsilon)}{\eta}=\pi\theta\big(\epsilon-\widetilde{\epsilon}_{\vec{p}}\big)\,,
\end{align}
where the Heaviside $\theta$-function is defined so that $\theta(0)=1/2$, and $\widetilde{\epsilon}_{\vec{p}}=\epsilon_{\vec{p}}+\Sigma^{\rm (TS)}(\vec{p}\,)$. Using this expression for the derivatives of the phase $\delta_f^{(G)}$ we obtain  the following relations
\begin{align}
&\partial_\epsilon \delta_f^{(G)}(\epsilon,\vec{p}\,) = \lim_{\eta\to 0}
\frac{\half\eta}{(\epsilon-\widetilde{\epsilon}_{\vec{p}})^2+\frac{\eta^2}{4}}
=\pi\delta(\epsilon-\widetilde{\epsilon}_{\vec{p}})\,,
\label{ddelta-eta}\\
&2\sin^2\delta_f^{(G)}(\epsilon,\vec{p}\,)\partial_\epsilon \delta_f^{(G)}(\epsilon,\vec{p}\,)
= \lim_{\eta\to 0}
\frac{\frac{1}{4}\eta^3}{\big[(\epsilon-\widetilde{\epsilon}_{\vec{p}})^2+\frac{\eta^2}{4}\big]^2}
\nonumber\\
&\qquad\qquad = \pi\delta(\epsilon-\widetilde{\epsilon}_{\vec{p}})\,.
\label{ddeltsin-eta}
\end{align}
Using Eqs.~(\ref{As-Bf-Af-eq}) and (\ref{ddelta-eta}) we can verify the relation (\ref{B-delt}) and with the help of Eq.~(\ref{ddeltsin-eta}) we recover Eq.~(\ref{As-sin2delta}).

Since $\Im\Sigma^{\rm TS}=0$, the correction terms to the pressure (\ref{DOm1}) and (\ref{DOm}) coincide, and we have
\begin{align}
\Delta\Omega^{\rm (TS)} &= \Delta\Omega_s^{\rm (TS)} =-\Phi^{\rm (TS)}\,,
\end{align}
where the quantities $\Delta \Omega^{\rm (TS)}$ and $\Delta \Omega_s^{\rm (TS)}$ are defined in Eqs.~(\ref{DOm1}) and (\ref{DOm}), respectively. Taking into account Eq.~(\ref{Phi-HF-eq}) and substituting   there  (\ref{Af-SCHF}) we find
\begin{align}
&\Phi^{\rm (TS)}
=\frac12\intop\!\frac{V2\rmd^3 p_1}{(2\pi)^3}f(\epsilon_{\vec{p}_1}+\Sigma^{\rm (TS)}(\vec{p}_1) - \mu_f)\Sigma^{\rm (TS)}(\vec{p}_1)
\nonumber\\
&=
\frac12\intop\!\frac{V2\rmd^3 p_1}{(2\pi)^3}\intop\!\frac{2\rmd^3 p_2}{(2\pi)^3}
%\big(2\Umom(|\vec{p}=0|)- \Umom(|\vec{p}_1-\vec{p}_2|)\big)
\, \overline{\Umom^{(S)}}\big(\frac{\vec{p}_1-\vec{p}_2}{2},\frac{\vec{p}_1-\vec{p}_2}{2}\big)
\nonumber\\
& \times f(\epsilon_{\vec{p}_1}+\Sigma^{\rm (TS)}(\vec{p}_1) - \mu_f)
f(\epsilon_{\vec{p}_2}+\Sigma^{\rm (TS)}(\vec{p}_2)- \mu)\,.\label{Phi-U-expl}
\end{align}

Since $\Im\Sigma^{\rm (TS)}=0$ and $\Sigma^{\rm (TS)}$ does not depend on the frequency and all the spectral functions are the same, cf. Eq.~(\ref{As-Bf-Af-eq}), we immediately obtain that the function $\mathcal{C}_f$ in Eq.~(\ref{C-def}) and, consequently, $\Delta n_f$ in Eq.~(\ref{Dnf-Cf}) vanish. Hence, within the self-consistent Hartree-Fock approximation we obtain
\begin{align}
n_f^{\rm (Noeth)}=n_f^{\rm (BM)}\,.\label{N-BM}
\end{align}
The non-Markovian part of the entropy (\ref{Smem-fb}) for the chosen set of diagrams, when $\Im\Sigma^{\rm (TS)}=0$ and $\Sigma^{\rm (TS)}$ does not depend on the frequency, is
\begin{align}
S^{\rm (TS)}_{\rm mem} &=- \frac{\Phi^{\rm (TS)}}{T}
+
\frac{1}{T}\!\!\intop\!\!\frac{V\rmd^3 p}{(2\pi)^3}\!\!
\intop_{-\infty}^{+\infty}\!\!\frac{\rmd \epsilon}{2\pi}
f(\epsilon-\mu_f)
\nonumber\\
&\times \Sigma^{\rm (TS)}(\vec{p}\,) A_f(\epsilon,\vec{p}\,)\,.
\label{Smem-U-ex}
\end{align}
Substituting here  $\Phi^{\rm (TS)}$ from Eq.~(\ref{Phi-U-expl}) and the spectral density  $A_f$ from Eq.~(\ref{Af-SCHF}) we conclude that $$S^{(\rm TS)}_{\rm mem} = 0,$$ as it should be for the chosen set of diagrams in $\Phi$~\cite{IKV-2}.

The correction term to the energy $\Delta E^{\rm (TS)}$
 in Eq.~(\ref{DE-def}) after substitution there $A_f$ from Eq.~(\ref{Af-SCHF}) yields
\begin{align}
\Delta E^{\rm (TS)}&=-\frac12\intop\!\frac{2V\rmd^3 p}{(2\pi)^3}
\frac{\Sigma^{\rm (TS)}(\vec{p}\,)}{e^{(\epsilon_{\vec{p}} + \Sigma^{\rm (TS)}(\vec{p}\,) -\mu_f)/T }+1}\,.
\label{DE-U}
\end{align}
Expression for the energy is obtained from Eq.~(\ref{E-result}) after substitution there Eqs.~(\ref{Af-SCHF}) and (\ref{DE-U}),
\begin{align}
E^{\rm (TS)}=V\intop\!\frac{2\rmd^3 p}{(2\pi)^3}
\frac{\epsilon_{\vec{p}} + \frac12\Sigma^{\rm (TS)}(\vec{p}\,)}{e^{(\epsilon_{\vec{p}} + \Sigma^{\rm (TS)}(\vec{p}\,) - \mu_f)/T }+1}\,.
\label{E-U-fin}
\end{align}
Taking into account that $S^{\rm (TS)}_{\rm mem}=0$, for the entropy  from Eq.~(\ref{DSp-1}) we get
\begin{align}
S^{\rm (TS)} &= V\intop\frac{2\rmd^3 p}{(2\pi)^3}
\sigma_f (\epsilon_{\vec{p}} + \Sigma^{\rm (TS)}(\vec{p}\,)-\mu_f)\,.
\label{S-U-fin}
\end{align}

For the  pressure, $P=-\Omega/V$, after substituting the spectral functions
(\ref{As-Bf-Af-eq})
and $\Phi^{\rm (TS)}$ from Eq.~(\ref{Phi-U-expl})
in Eq.~(\ref{Omega-int-As}) we obtain
\begin{align}
P^{\rm (TS)} &=\intop\frac{2\rmd^3 p}{(2\pi)^3}
T\ln\big(1+e^{-(\epsilon_{\vec{p}} + \Sigma^{\rm (TS)}(\vec{p}\,) - \mu_f)/T}\big)
\nonumber\\
& +\frac12\intop\!\frac{2\rmd^3 p}{(2\pi)^3}
\frac{\Sigma^{\rm (TS)}(\vec{p}\,)}{e^{(\epsilon_{\vec{p}} + \Sigma^{\rm (TS)}(\vec{p}\,) - \mu_f)/T }+1}
\,.
\label{P-U-fin}
\end{align}
After integration by parts in the first term we find
\begin{align}
P^{\rm (TS)} &= \intop\!\frac{2\rmd^3 p}{(2\pi)^3}
\frac{\frac{\vec{p\,}^2}{3m_f}\big(1+\frac{m_f}{|\vec{p}\,|}\frac{\partial}{\partial |\vec{p}\,|}\Sigma^{\rm (TS)}(\vec{p}\,)\big) +\frac12\Sigma^{\rm (TS)}(\vec{p}\,)}{e^{(\epsilon_{\vec{p}} + \Sigma^{\rm (TS)}(\vec{p}\,)-\mu_f)/T }+1}\,.
\label{P-U-fin-part}
\end{align}

Now it is instructive to show explicitly that  for fermions interacting via pair potential obtained thermodynamic quantities satisfy exact thermodynamic relations, i.e. they are thermodynamically consistent. First, we show that the derivative of the pressure with respect to the chemical potential gives the particle density given by the second Eq.  (\ref{Sig-mu-HF}), i.e.,
\begin{align}
\partial P^{\rm (TS)}/\partial \mu_f|_T =n_f^{({\rm Noeth})}\,.
\label{dPdmu-nf}
\end{align}
Differentiating Eq.~(\ref{P-U-fin}) with respect to $\mu_f$ we find
\begin{align}
\frac{\partial P^{\rm (TS)}}{\partial \mu_f}
%&=\intop\frac{2\rmd^3 p}{(2\pi)^3} \,\tilde{f}_{\vec{p}}
%\Big\{
%\Big(1-\frac{\partial \Sigma({\vec{p}})}{\partial\mu}\Big)
%+\frac12\frac{\partial \tilde{f}_{\vec{p}}}{\partial \mu_f} \Sigma({\vec{p}})
%+\frac12\frac{\partial \Sigma({\vec{p}})}{\partial \mu_f} \tilde{f}_{\vec{p}}
%\Big\}
%\nonumber\\
&=\intop\frac{2\rmd^3 p}{(2\pi)^3} \,\tilde{f}_{\vec{p}}\nonumber\\
&+\frac12\intop\frac{2\rmd^3 p}{(2\pi)^3}
\Big(\frac{\partial \tilde{f}_{\vec{p}}}{\partial \mu_f} \Sigma^{\rm (TS)}({\vec{p}}\,)
-\frac{\partial \Sigma^{\rm (TS)}({\vec{p}})}{\partial \mu_f} \tilde{f}_{\vec{p}}\Big)\,,
\label{dPdmu-1}
\end{align}
where we introduced the short-hand notation
$$\tilde{f}_{\vec{p}} = f(\epsilon_{\vec{p}\,} + \Sigma^{\rm (TS)}({\vec{p}})-\mu_f)\,.$$
Using  Eq. (\ref{Sig-mu-HF}) we obtain
\begin{align}
&\intop\frac{2\rmd^3 p}{(2\pi)^3}
\frac{\partial \Sigma^{\rm (TS)}({\vec{p}}\,)}{\partial \mu_f} \tilde{f}_{\vec{p}}\nonumber\\
&=\intop\frac{2\rmd^3 p2\rmd^3 p'}{(2\pi)^6}
%\big(2\Umom(\vec{p}=0)- \Umom(|\vec{p}-\vec{p\,}'|)\big)
\overline{\Umom^{(S)}}\big(\frac{\vec{p}-\vec{p\,}'}{2},\frac{\vec{p}-\vec{p\,}'}{2}\big)
\frac{\partial \tilde{f}_{\vec{p\,}'}}{\partial \mu_f} \tilde{f}_{\vec{p}}\nonumber\\
&
 = \intop\frac{2\rmd^3 p}{(2\pi)^3}
\frac{\partial \tilde{f}_{\vec{p\,}}}{\partial \mu_f}\Sigma^{\rm (TS)}(\vec{p}\,)\,.
\label{dSdmu-f-int}
\end{align}
Thus, last term in (\ref{dPdmu-1}) is cancelled and thereby we proved correctness of Eq.~(\ref{dPdmu-nf}).

{
Now we differentiate the pressure with respect to $T$. From Eq.~(\ref{P-U-fin}) we find
\begin{align}
\frac{\partial P^{\rm (TS)}}{\partial T}\Big|_{\mu_f} &=\intop\frac{2\rmd^3 p}{(2\pi)^3}
\sigma_f(\epsilon_{\vec{p}}+\Sigma^{\rm (TS)}(\vec{p}\,)-\mu_f)
\nonumber\\
&
-\intop\frac{2\rmd^3 p}{(2\pi)^3} \widetilde{f}_{\vec{p}} \frac{\partial \Sigma^{\rm (TS)}(\vec{p}\,)}{\partial T}
\nonumber\\
& +\frac12\intop\!\frac{2\rmd^3 p}{(2\pi)^3}\Big[
\frac{\partial\Sigma^{\rm (TS)}(\vec{p}\,)}{\partial T}  \widetilde{f}_{\vec{p}}
+\Sigma^{\rm (TS)}(\vec{p}\,) \frac{\partial \widetilde{f}_{\vec{p}}}{\partial T}
\Big]
\,.
\label{dPdT-1}
\end{align}
Taking into account that without loss of the validity  the derivatives in $\mu_f$ in Eq.~(\ref{dSdmu-f-int}) can be replaced by the derivatives in $T$,  we immediately observe that in (\ref{dPdT-1}) last
%three
 two
%terms
integrals with the self-energy cancel each other. Therefore, we recover  correct thermodynamic relation
\begin{align}
(\partial P^{\rm (TS)}/\partial T)|_{\mu_f}=S^{\rm (TS)}/V\,,
\end{align}
here with the pressure from (\ref{P-U-fin-part}) and entropy from Eq. (\ref{S-U-fin}), since, as we have shown, the memory term is absent in the given case. Thus, we {\em explicitly demonstrated} that  thermodynamic quantities given by expressions (\ref{S-U-fin})  -- (\ref{dPdmu-nf}), which we have derived for the $\Phi$ functional described by the TS diagrams, are thermodynamically consistent.

Expressions for the energy and the pressure can be written in a compact form if we introduce the notations for the in-medium spectrum $\tilde{\epsilon}_{\vec{p}}=\epsilon_{\vec{p}} +\Sigma^{\rm (TS)}(\vec{p}\,)$ and the slope parameter of the spectrum $\frac{1}{\widetilde{m}_f(\vec{p}\,)}=2\frac{\partial \tilde{\epsilon}_{\vec{p}}}{\partial\vec{p\,}^2}$,
\begin{align}
E^{\rm (TS)} &=V\intop\!\frac{2\rmd^3 p}{(2\pi)^3}\big( \tilde{\epsilon}_{\vec{p}} - \half\Sigma^{\rm (TS)}(\vec{p}\,)\big)f_{\vec{p}} \,,
\nonumber\\
P^{\rm (TS)} &=\intop\!\frac{2\rmd^3 p}{(2\pi)^3}\big( \frac{\vec{p\,}^2}{3\widetilde{m}_f(\vec{p}\,)} + \half\Sigma^{\rm (TS)}(\vec{p}\,)\big)f_{\vec{p}} \,.
\end{align}
These expressions have contributions in the form of an ideal gas of quasiparticles with energy $\widetilde{\epsilon}_{\vec{p}}$
%  and the momentum-dependent slop parameter $\widetilde{m}_f(\vec{p}\,)$,
and the quasiparticle interactions encoded in $\Sigma^{\rm (TS)}$. The entropy (\ref{S-U-fin}) also has the standard combinatoric form for the quasiparticle distribution. On the other hand using the thermodynamical identity (\ref{thermoidentity}) we obtain the following simple relation for the entropy
\begin{align}
TS &= E+PV-\mu_f N
\nonumber\\
&=V\intop\!\frac{2\rmd^3 p}{(2\pi)^3}\big( \tilde{\epsilon}_{\vec{p}} + \frac{\vec{p\,}^2}{3\widetilde{m}_f(p)}-\mu_f\big)\,f_{\vec{p}}
\,.
\label{TS-short}
\end{align}
In Ref.~\cite{IKV-2}, expressing $E+PV$ through the trace of the energy-momentum tensor, a similarly simple exact relation for the entropy was derived for  interactions with not-derivative coupling, see Eq.~(6.14) there. Possible corrections to the energy-momentum tensor because of derivative couplings (as in our case of the momentum dependent Fourier transform of the potential) were discussed in Ref.~\cite{IKV-YaF} but no explicit  modified expression for the entropy was written. Equation (\ref{TS-short}) derived above is a specific extension of the result~\cite{IKV-2} for the case of a momentum dependent pair potential
(the specific case of the derivative coupling)
with an account of two (TS) diagrams in the $\Phi$ functional. We see that the derivative interaction induced by the pair potential leads to the appearance of the spectrum-slope parameter $\widetilde{m}_f(p)$ instead of the bare mass $m_f$ in Eq.~(6.14) in Ref.~\cite{IKV-2}.

\subsection{Fugacity expansion for pressure $P^{\rm (TS)}$}\label{ssec:fugacity-Phi}

Consider  expansion of the pressure $P^{\rm (TS)}$, Eq.~(\ref{P-U-fin}), with respect to fugacity
\begin{align}
z=e^{\mu_f/T}\,.%= n\lambda^3(1+O(n\lambda^3))\ll 1\,, \quad \lambda =\sqrt{\frac{2\pi}{m\,T}}\label{lambda-def}
\label{z-def}
 \end{align}
The expansion in the parameter $z$ is used in the construction of the virial equation of state~\cite{Uhlenbeck-Beth}.
Our aim here is to present the pressure in the form
\begin{align}
P^{\rm (TS)} &= \frac{T}{\lambda^3}\Big(2z+b_{2,\Phi}^{\rm (TS)}z^2
+ b_{3,\Phi}^{\rm (TS)}z^3 + O(z^4)
\Big)\,,
\label{P_TS}
\\
\lambda &=\sqrt{\frac{2\pi}{m_f\,T}}\,,
\label{lambda-def}
\end{align}
where $\lambda$ is the thermal wave length of the particle.
The first term is the classical expression for the ideal Boltzmann gas. The next terms contain both the quantum-statistical corrections to the ideal Boltzmann gas term and the parts induced by the interaction.

In order to keep track to the effects of the quantum Fermi-Dirac statistics, we modify the fermion occupation factors in Eqs.~(\ref{Sig-mu-HF}) and (\ref{P-U-fin}),
\begin{align}
f(\epsilon-\mu_f)\to f^{(\alpha)}(\epsilon-\mu_f)=\frac{1}{e^{(\epsilon-\mu_f)/T}+\alpha},
\end{align}
so that for $\alpha=1$ we have Fermi-Dirac distribution and for $\alpha=0$ the Boltzmann limit is recovered.
Now, using expansion of $f^{(\alpha)}$ in $z$,
\begin{align}
f^{(\alpha)}(\epsilon-\mu_f)=\sum_{n=0}^\infty (-1)^n \alpha^n z^{n+1} e^{-(n+1)\epsilon/T}\,,
\end{align}
and taking into account  dependence  of $\Sigma^{\rm (TS)} (z)$ in both sides of Eq.  (\ref{Sig-mu-HF}),
we obtain coupled chain of equations for the expansion coefficients. As we can see in Eq.~(\ref{P-U-fin}), the self-energy is multiplied at least by one power of $z$. Hence to generate the third order term in $z$ in the expansion (\ref{P_TS}), it is enough to  retain terms up to the second order in the expansion of $\Sigma^{\rm (TS)}$,
\begin{align}
&\Sigma^{\rm (TS)}(\vec{p}\,) \approx a_1(\vec{p},T)\,z -  z^2\,\big(\alpha a_1(\vec{p},T/2)+ a_2(\vec{p},T)\big) \,,
\label{Sig-U-exp}
\end{align}
where
\begin{align}
a_1(\vec{p},T) &= \intop\frac{2\rmd^3 p'}{(2\pi)^3}
%(2\Umom(\vec{p}=0)-\Umom(|\vec{p}-\vec{p\,}'|))
\overline{\Umom^{(S)}}\big(\frac{\vec{p}-\vec{p\,}'}{2},\frac{\vec{p}-\vec{p\,}'}{2}\big)
e^{-\epsilon_{\vec{p\,}'}/T }\,,
%\nonumber\\
%&\quad= \frac{2g^2}{m_Y^2\lambda^3}
%-
%g_Y^2\intop\frac{\rmd p' %p'}{8\pi^2p}\log\frac{(p+p')^2+m_Y^2}{(p-p')^2+m_Y^2}e^{-\frac{{p'}^2}{2mT}}
\nonumber\\
a_2(\vec{p},T) &= \intop\frac{2\rmd^3 p'}{(2\pi)^3}\frac{a_1(\vec{p\,}',T)}{T}
e^{-\epsilon_{\vec{p\,}'}/T }
%\big(2\Umom(|\vec{p}=0|)
\nonumber\\
&\times %-\Umom(|\vec{p}-\vec{p\,}'|)\big)
\overline{\Umom^{(S)}}\big(\frac{\vec{p}-\vec{p\,}'}{2},\frac{\vec{p}-\vec{p\,}'}{2}\big)
 \,.
\label{as-def}
\end{align}
Note that expansion in $z$ mixes various orders of interactions $\Umom$, e.g., coefficient in the $z^2$-term in Eq.~(\ref{Sig-U-exp}) contains terms $\propto \Umom$ and $\Umom^2$, since $a_1\propto \Umom$ and thereby $a_2\propto \Umom^2$. Also we observe that the term $a_1(\vec{p},T/2)$ in the $z^2$-coefficient in expansion (\ref{Sig-U-exp}) is of the quantum  origin. Here we should recall that $\Phi^{\rm (TS)}$, being expressed in terms of the full Green's functions, contains potential only in the linear order. The non-linear terms in $\Umom$ appear from expansion of the full Green's functions in series of the Green's functions of the ideal gas.

The first term  in Eq.~(\ref{P-U-fin}) enters the quantity
 \begin{align}
&\frac{1}{\alpha}\ln\big(1+\alpha\,z\,e^{-(\epsilon_{\vec{p}}+\Sigma^{\rm (TS)}(\vec{p}\,))/T}\big)=
\frac1{\alpha}\ln\big(1+\alpha z e^{-\epsilon_{\vec{p}}/T}\big)
\nonumber\\
&+
\frac{1}{\alpha} \ln\Big[1+\alpha f^{(\alpha)}(\epsilon_{\vec{p}}-\mu_f)\left(e^{-\Sigma^{\rm (TS)}(\vec{p}\,)/T}-1\right)\Big]\,.
\label{Log-exp}
\end{align}
Thereby one can separate the terms in the pressure corresponding to the ideal gas, the first term in (\ref{Log-exp}),  and the terms induced by the interaction.
Substituting Eq.~(\ref{Sig-U-exp}) in (\ref{Log-exp}) and then in (\ref{P-U-fin})  we can cast the pressure as expansion in $z$ in the form (\ref{P_TS}) with the second term, being
\begin{align}
b_{2,\Phi}^{\rm (TS)}(T)&=-\frac{\lambda^3}{2T}\intop\frac{2\rmd^3 p}{(2\pi)^3} \big\{a_1(\vec{p},T)e^{-\epsilon_{\vec{p}}/T} +\alpha Te^{-2\epsilon_{\vec{p}}/T} \big\}
\nonumber\\
&= \alpha b_2^{(0)} +b_{2,\Phi}^{\rm (int,TS)}(T)\,,\quad b_2^{(0)}=-\frac{1}{2^{3/2}}\,,
\end{align}
where in the last line the first term is of quantum-statistic origin and the interaction-dependent part is
\begin{align}
b_{2,\Phi}^{\rm (int, TS)}(T) &= -\frac{\lambda^3}{2T} \intop\frac{2\rmd^3 p2\rmd^3 p'}{(2\pi)^6}\nonumber
%\label{b2}
\\
&\times
\overline{\Umom^{(S)}}\big(\frac{\vec{p}-\vec{p\,}'}{2},\frac{\vec{p}-\vec{p\,}'}{2}\big)
e^{-(\epsilon_{\vec{p}} + \epsilon_{\vec{p\,}'})/T} \,.
\label{b2-U}
\end{align}
The latter term  does not depend, whether classical or quantum statistics is chosen. However it contains the genuine quantum effect depending on the spin of fermions. Spin manifests itself in different coefficients and signs of the TS diagrams contributing to the $\Phi$ functional and the self-energy (see two first diagrams in Fig. \ref{fig:Phi-U-ser} and diagrams in Fig.~\ref{fig:Sig-U}), which follow from the summation over the spin indices, as illustrated in Eq.~(\ref{Phi-deriv-exampl}).

The next coefficient is
\begin{align}
b_{3,\Phi}^{\rm (TS)}(T) &=\frac{\lambda^3}{2T}\intop\frac{2\rmd^3 p}{(2\pi)^3}\Big\{(2\alpha-1)
a_1(\vec{p},T) e^{-2\epsilon_{\vec{p}}/T}
\nonumber\\
&+ \big(\alpha a_1(\vec{p},T/2)
 +  a_2(\vec{p},T)\big)e^{-\epsilon_{\vec{p}}/T}
 + \frac23 e^{-3\epsilon_{\vec{p}}/T}T
 \Big\}
 \nonumber\\
&=\alpha^2 b_{3}^{(0)} + b_{3,\Phi}^{\rm(int, TS)}(T)\,,\quad
b_3^{(0)}=\frac{2}{3^{5/2}} \,,
\label{b3}
\end{align}
where we also separated the interaction independent term $b_3^{(0)}$ of the quantum origin and the term $b_{3,\Phi}^{\rm(int, TS)}$ induced by the inter-particle potential.
Substituting Eq.~(\ref{as-def}) in Eq.~(\ref{b3}) we  cast the potential dependent term, $b_{3,\Phi}^{\rm (int, TS)}$, in the form
\begin{align}
& b_{3,\Phi}^{\rm(int, TS)}(T) =\frac{\lambda^3}{2T}\intop\frac{2\rmd^3 p}{(2\pi)^3}
\intop\frac{2\rmd^3 p'}{(2\pi)^3}\intop\frac{2\rmd^3 p''}{(2\pi)^3}
e^{-\frac{\epsilon_{\vec{p}}+\epsilon_{\vec{p\,}'}+\epsilon_{\vec{p\,}''} }{T} }
 \nonumber\\
&\quad\times \Big\{\frac{1}{T}\,
\overline{\Umom^{(S)}}\big(\frac{\vec{p\,}'-\vec{p}}{2},\frac{\vec{p\,}'-\vec{p}\,}{2}\big)
\overline{\Umom^{(S)}}\big(\frac{\vec{p\,}''-\vec{p}}{2},\frac{\vec{p\,}''-\vec{p}\,}{2}\big)
\nonumber\\
&\quad +\frac{3\alpha -1}{2}
(2\pi)^3\delta^{(3)}(\vec{p\,}''-\vec{p}\,)
\overline{\Umom^{(S)}}\big(\frac{\vec{p\,}'-\vec{p}}{2},\frac{\vec{p\,}'-\vec{p}\,}{2}\big) \Big\}\,.
\label{b3-U}
\end{align}
The structure of terms $b_{2,{\Phi}}^{\rm(int, TS)}$ and $b_{3,{\Phi}}^{\rm(int, TS)}$ shows (see exponential terms) that $b_{2,{\Phi}}^{\rm(int, TS)}$ is generated by purely two-particle processes  and  $b_{3,{\Phi}}^{\rm(int, TS)}$ involves interaction between  three particles. The first term in (\ref{b3-U}) corresponds to the process when one particle interacts sequentially with another two. In the second term the particle interacts only once and the third particle is a spectator. We note that the latter term depends  on the quantum statistics: it changes sign, if one switches from the Boltzmann statistics ($\alpha=0$) to the Fermi-Dirac statistics ($\alpha=1$).

In Eq.~(\ref{b2-U}) one can further explicitly separate the center-of-mass motion. We introduce the variables $\vec{p}=\vec{P}/2 + \vec{q}$ and $\vec{p\,}'=\vec{P}/2-\vec{q}$ in which $\epsilon_p+\epsilon_{p'}=\frac{\vec{P\,}^2}{4m_f}+\frac{\vec{q\,}^2}{m_f}$. Integrating over $\vec{P}$ we obtain
\begin{align}
&b_{2,\Phi}^{\rm (int, TS)}(T)
=-\frac{2^{3/2}}{T}\!\!\intop\!\!\frac{2\rmd^3 q}{(2\pi)^3}
e^{-\frac{\vec{q\,}^2}{m_f T}}
\overline{\Umom^{(S)}}(\vec{q},\vec{q}\,)
\,,
\label{b2-2}
\end{align}
where as before $\overline{\Umom^{(S)}}(\frac{\vec{p}-\vec{p\,}'}{2},\frac{\vec{p}-\vec{p\,}'}{2})=
\Umom(|\vec{p}=0|) - \frac12\Umom(|\vec{p}-\vec{p\,}'|).$
For the term $b_{3,{\Phi}}^{{\rm(int, TS)}}$, such a variable separation does not lead to  simplification.

Separating  contributions of the Hartree  and Fock  diagrams (the  TS  diagrams in Fig.~\ref{fig:Sig-U}) we derive
%  to the $b_{2,\Phi}^{\rm (int)}$ coefficient given in Eq.~(\ref{b2-2}) we have $b_{2,\Phi}^{\rm (int)}(T) = b_{2,\Phi}^{\rm (int,H)}(T)+b_{2,\Phi}^{\rm (int,F)}(T)$.
%\nonumber \\
\begin{align}
&b_{2,\Phi}^{\rm (int, TS)}(T)=b_{2,\Phi}^{\rm (int,H)}(T)+b_{2,\Phi}^{\rm (int,F)}(T)\nonumber \\
&=-\frac{ 2 \Umom(0)}{T\lambda^3}   +\frac{2^{3/2}}{T}\!\!\! \intop\!\!\frac{\rmd^3p}{(2\pi)^3}
e^{-\frac{p^2}{mT}}\Umom(2\vec{p}\,)\,,
\label{b2-2a}
\end{align}
where the first term in the Hartree- and the second one is the Fock term. The final expression (136) for $b_2$ is linear in the potential. It is important to note that this linearity is not because
the tadpole and sandwich diagrams included in $\Phi$ appear to be formally linear in $U$. It contains the full Green's functions, which include the interaction through the self-energies. As the result, e.g, the term $b_3$, Eq.~(\ref{b3-U}) contains higher-order terms in $U$

From the third coefficient $b_{3,\Phi}^{\rm (int)}$ given by Eq.~(\ref{b3-U}) we  derive a  transparent expression for
the Hartree contribution
\begin{align}
b_{3,\Phi}^{(\rm int,H)} = \frac{4}{\lambda^6}\frac{\Umom^2(0)}{T^2} +(3\alpha-1)\frac{\Umom(0)}{\sqrt{2}\lambda^3 T}\,.
\label{b3Phi-H}
\end{align}
We see that the second term here is of sub-leading order in the parameter $\lambda^3 T/|U(0)|$. It constitutes a correction of a purely quantum origin, since $\lambda\propto\hbar$. The correction  is small for  $T\gg  \pi^3/[\Umom(0)]^2 m_N^3$.

As will be seen  below, the term  $b_{2,\Phi}^{\rm (int, TS)}$ coincides with the second virial coefficient appearing in the virial equation of state taken at the leading order in the interaction $U$, see Eq.~(\ref{b2-exp}) below.

The above expressions are written in terms of the spin averaged potential (\ref{U-spinaver}).
It is instructive to relate this potential to  physical observables connected with the two-particle scattering. The formal scattering theory is shortly outlined in Appendix~\ref{app:scattering}. From Eq.~(\ref{f-Born-spinaver})  we see that the spin-averaged potential is directly related to the spin-averaged forward two-particle scattering amplitude calculated in the Born approximation, $\overline{F_{\rm Born}^{(S)}}$. Using representation of this amplitude in terms of the phase shifts $\delta_{{\rm Born},l}^{(S)}$ in (\ref{F-delta-Born}), we can rewrite Eq.~(\ref{b2-2}) for the coefficient $b_{2,\Phi}^{\rm (int,TS)}$ in the  form
\begin{align}
b_{2,\Phi}^{\rm (int,TS)}(T)
&=\frac{2^{3/2}}{T}\!\!\intop\!\!\frac{2\rmd^3 q}{(2\pi)^3}
e^{-\frac{\vec{q\,}^2}{m_f T}}
\nonumber\\
&\times
\frac{2\pi}{m_f}\sum_S\sum_{l[S]}(2S+1)(2l+1)\frac{\delta_{{\rm Born},l}^{(S)}}{q}
\,.
\label{b2-U-Born}
\end{align}
Here the summation goes over spin $S=0$ and 1, and even $l$ for $S=0$, and odd $l$ for $S=1$.
Integration by parts reduces this expression to
\begin{align}
b_{2,{\Phi}}^{\rm (int, TS)} &= 2^{3/2}\intop_0^\infty \frac{\rmd q}{\pi} e^{-\frac{q^2}{m_f T}}
\sum_{S,l[S]}(2S+1)(2l+1)\frac{\rmd \delta_{{\rm Born},l}^{(S)}}{\rmd q}\,,
\label{b2-U-Born-delt}
\end{align}
that resembles the full quantum-mechanical Beth-Uhlenbeck virial expansion result for the second virial coefficient~\cite{Uhlenbeck-Beth,Beth}, which we will recower in Section~\ref{ssec:fullquant-b2}. The only difference is that in Eqs.~(\ref{b2-U-Born}), (\ref{b2-U-Born-delt}) the phase shifts are taken in the Born approximation. This is not a trivial result of our study, that two different approaches ($\Phi$ approach  and the Beth-Uhlenbeck virial expansion one) produce the same second virial coefficients at the above mentioned conditions.

Comparison of the third coefficient $b_{3,\Phi}^{\rm (int,TS)}$ calculated here
with the full quantum mechanical expression for the third virial coefficient,
see discussion in Section~\ref{ssec:3dvirial}, shows that the first term in the curly brackets in Eq.~(\ref{b3-U}) coincides with the contribution to the third virial coefficient determined by two on-shell scattering amplitudes, see Eq.~(\ref{b3-tt}) below, provided the latter ones are taken in the Born approximation.

Inspecting the derivation of the higher order expansion coefficients from
Eqs.~(\ref{SigST-U}) and (\ref{P-U-fin}) we find that the $n$-th coefficient,
$b_{n,\Phi}$, contains a tower of $n$ terms with the potential in different powers,
starting from  the free term $\propto\Umom^{0}$, then  $\Umom^{1}$, till $\Umom^{n-1}$.
Each term of a specific order $\Umom^k$ receives contributions only from those diagrams of the $\Phi$ functional, which are of the order less or equal to $k$. Therefore, with  two diagrams taken into account in our TS approximation, we can guarantee  correct expressions only for the terms linear in $\Umom$ in each coefficient $b_{n,\Phi}$. Diagrams (c), (d), etc. in Fig.~\ref{fig:Phi-U-ser} contain extra interaction lines and, hence, do not contribute to the linear term  $\propto \Umom^{1}$. The contribution of  diagram (c) starts with $\propto \Umom^{2}$ terms, of diagram (d)  with $\propto \Umom^{3}$ terms, etc.

Linear Born contributions  can be separated after linearization of Eqs.~(\ref{SigST-U})
and (\ref{P-U-fin}) with respect to $\Umom$ and expansion in terms of $z$: we find for the self-energy
$\Sigma^{\rm (TS)}(\vec{p\,})=\sum_{n=0}^\infty (-1)^n z^{n+1} a_1(\vec{p}, T/(n+1))$ and for the pressure
\begin{widetext}
\begin{align}
P^{\rm (TS)} &\approx\intop\frac{2\rmd^3 p}{(2\pi)^3}
T\ln\big(1+e^{-(\epsilon_{\vec{p}} - \mu_f)/T}\big)
-\frac12\intop\!\frac{2\rmd^3 p}{(2\pi)^3} \Sigma^{\rm (TS)}(\vec{p}\,)f(\epsilon_{\vec{p}}  - \mu_f)
\nonumber\\
&=\frac{T}{\lambda^3}\sum_{n=1}^\infty\frac{2(-1)^{n-1}}{n^{5/2}}z^n
-\frac12 \sum_{n=2}^\infty \sum_{m=0}^{n-2} (-1)^{n-2}  z^{n}
\intop\!\frac{2\rmd^3 p}{(2\pi)^3}\frac{2\rmd^3 p'}{(2\pi)^3}
\overline{\Umom^{(S)}}\big(\frac{\vec{p}-\vec{p\,}'}{2},\frac{\vec{p}-\vec{p\,}'}{2}\big)
e^{-\frac{(m+1)\epsilon_{\vec{p\,}'}+(n-m-1)\epsilon_{\vec{p}}}{T}}\,.
\label{P-linU-1}
\end{align}
\end{widetext}
Changing variables $\vec{p\,}' \to \vec{P}/2 - 2\frac{(n - m - 1)}{n}\vec{q}$  and $\vec{p}\to \vec{P}/2 + 2\frac{(m + 1)}{n}\vec{q}$ in the integral, we can separate the center of mass motion, since $(m+1)\epsilon_{\vec{p\,}'}+(n-m-1)\epsilon_{\vec{p}}=
n\frac{\vec{P}^2}{8m_f}
+\frac{2}{n} (n-1 - m) (1 + m) \frac{q^2}{m_f}$. After integration over $\vec{P}$, we obtain
\begin{align}
&\intop\!\frac{2\rmd^3 p}{(2\pi)^3}\frac{2\rmd^3 p'}{(2\pi)^3}
\overline{\Umom^{(S)}}\big(\textstyle{\frac{\vec{p}-\vec{p\,}'}{2},
\frac{\vec{p}-\vec{p\,}'}{2}}\big)
e^{-\frac{(m+1)\epsilon_{\vec{p\,}'}+(n-m-1)\epsilon_{\vec{p}}}{T}}
\nonumber\\
&\quad =-\frac{2T}{\lambda^3}\frac{2^{3/2}}{n^{3/2}}
b_{2,\Phi}^{\rm (int, TS)}\Big(\frac{nT}{2 (n -1 - m) (1 + m)}\Big)\,,
\end{align}
where Eq.~(\ref{b2-2}) is used.
Substituting this result in the expression for the pressure (\ref{P-linU-1}) we obtain
\begin{align}
P^{\rm (TS)} &\approx \frac{T}{\lambda^3}\Big[2z
+  \sum_{n=2}^\infty 2 z^n \frac{(-1)^{n-1}}{n^{5/2}}
\nonumber\\
& +  \sum_{n=2}^\infty \sum_{m=1}^{n-1} (-1)^{n-2}  z^{n}
\frac{2^{3/2}}{n^{3/2}}
b_{2,\Phi}^{\rm (int,TS)}\Big(\frac{nT}{2 (n - m) m}\Big)\Big]\,.
\end{align}
Finally, we find  exact expressions for all linear contributions to the virial coefficients expressed through the second virial coefficient
\begin{align}
b_{n\ge 2}^{\rm (lin, U)} &= \frac{2(-1)^{n-1}}{n^{5/2}}
\nonumber\\
&+ (-1)^{n}
\Big(\frac{2}{n}\Big)^{3/2}\sum_{m=1}^{n-1}
b_{2,\Phi}^{\rm (int, TS)}\Big(\frac{nT}{2 (n - m) m}\Big)\,.
\label{bn-sum-rules}
\end{align}
Several terms are
\begin{align}
b_{2}^{\rm (lin, U)}(T) &= -\frac{1}{2^{3/2}} + b_{2,\Phi}^{\rm (int, TS)}(T)\,,
\\
b_{3}^{\rm (lin, U)}(T) &=\frac{2}{3^{5/2}}-\frac{2^{5/2}}{3^{3/2}} b_{2,\Phi}^{\rm (int, TS)}(3 T/4)\,,
\nonumber\\
b_{4}^{\rm (lin, U)}(T) &=-\frac{1}{16}+
\frac{1}{2^{3/2}} b_{2,\Phi}^{\rm (int, TS)}(T/2) \nonumber \\
&+ \frac{1}{2^{1/2}} b_{2,\Phi}^{\rm (int, TS)}(2 T/3)\,,
\nonumber \end{align}\begin{align}
b_{5}^{\rm (lin, U)}(T) &=\frac{2}{5^{3/2}}
\nonumber \\
&-\frac{2^{5/2}}{5^{3/2}}
\left(b_{2,\Phi}^{\rm (int, TS)}(5 T/12)
+ b_{2,\Phi}^{\rm (int, TS)}(5T/8)\right)\,.\nonumber
\end{align}
We note that have not come across similar expressions in the literature. One can see that the coefficients of higher order depend on the effectively lower value of the temperature than $b_2$. We emphasize that Eq.~(\ref{bn-sum-rules}) establishes relations between only those contributions to $b_n$ coefficients, which depend linearly on the potential. As we argued above, other diagrams beyond the first two in Fig.~\ref{fig:Phi-U-ser} included in the TS approximation are of the higher order in the potential (starting with $\Umom^2$), so they will not contribute to Eq.~(\ref{bn-sum-rules}).

Concluding this part we would like to stress once more that restricting ourselves by only two (TS) diagrams calculated beyond the linear in $\Umom$ approximation we treat the problem within the full self-consistent $\Phi$-derivable approximation scheme. Although such an approximation is thermodynamically consistent and exactly energy-momentum and current conserving, the result is certainly not the same as would be in the full theory including infinite number of diagrams in Fig.~\ref{fig:Phi-U-ser}. In case of a dilute gas at a finite temperature, one can restrict himself with a controlled  accuracy (up to $O(z^3)$ terms) considering only two virial coefficients, $b_1$ and $b_2$.
Calculated with only the TS diagrams, the $b_2$ coefficients  contain only terms linear in $\Umom$ that corresponds to the Born approximation for the fermion-fermion scattering amplitude. All other $\Phi$ diagrams in Fig.~\ref{fig:Phi-U-ser}  produce only higher order terms in $U$ and do not contribute within the Born approximation. The applicability of the Born approximation requires a smallness of typical magnitude of the pair potential, see conditions (\ref{slowB}), (\ref{fastB}), and (\ref{delt-Born-cond}) in Appendix~\ref{app:scattering}, that does not hold in case of realistic nucleon-nucleon pair potentials, see Fig.~\ref{fig:F2rat} below and the discussion in Section~\ref{ssec:effrange}.

\subsection{Application to the nuclear potential}\label{testTSnuclear}

As an illustration of an application of the derived expressions, we consider a system of neutrons interacting with the Yukawa potential (\ref{U-pot})
% with the Walecka parameters (\ref{Uomsig-par-W})
and we put $m_f=m_N=938$\,MeV.

In Fig.~\ref{fig:Phi-U-bs-P}a we show the values of coefficients $b_{2,\Phi}^{(\rm int, TS)}$ and $b_{3,\Phi}^{(\rm int, TS)}$ as functions of the temperature. Solid line, $b_{2,\Phi}^{(\rm int, TS)}$,  is calculated by Eq.~(\ref{b2-2a}), and dashed line, $b_{3,\Phi}^{(\rm int, TS)}$, by Eq.~(\ref{b3-U}) for the nucleon-nucleon potentials (\ref{U-pot}) with Walecka parameters (\ref{Uomsig-par-W}). We see that the coefficient  $b_{2,\Phi}^{(\rm int, TS)}$ is positive and  it represents steadily increasing function of $T$, whereas
$b_{3,\Phi}^{(\rm int, TS)}$ changes the sign, being negative for $T<90$\,MeV and positive at higher temperatures.  The dash-dotted line in Fig.~\ref{fig:Phi-U-bs-P}a shows  separate contribution
of the  $U^2$  term in (\ref{b3-U}), i.e.  in $b_{3,\Phi}^{(\rm int, TS)}$. This contribution to $b_{3,\Phi}^{(\rm int, TS)}$ is positive and sharply rising function of $T$. A strong attraction is produced by the second linear in $U$ term in the curly brackets in Eq.~(\ref{b3-U}).

We stress  that the virial coefficients $b_n$ enter  expressions for thermodynamic quantities,  each being multiplied by the corresponding $z^{n}$ pre-factor, cf. (\ref{P_TS}).
We find that the ratio $zb_{3,\Phi}^{\rm (int, TS)}[U^2 {\rm terms}]/b_{2,\Phi}^{\rm (int, TS)}\simeq 0.2$ for $T\simeq 20$ MeV and $n=0.05n_0$ and $b_{3,\Phi}^{\rm (int, TS)}[U^2 {\rm terms}]z/b_{2,\Phi}^{\rm (int, TS)}\simeq 0.4$ for $T\simeq 20$ MeV and $n=0.1n_0$.
Thereby  the linear in $U$ Born approximation applied to the Yukawa potential with the Walecka parameters (\ref{Uomsig-par-W}) is valid in this example only for rather  low temperatures and densities.

\begin{figure}
\centering
\includegraphics[width=6cm]{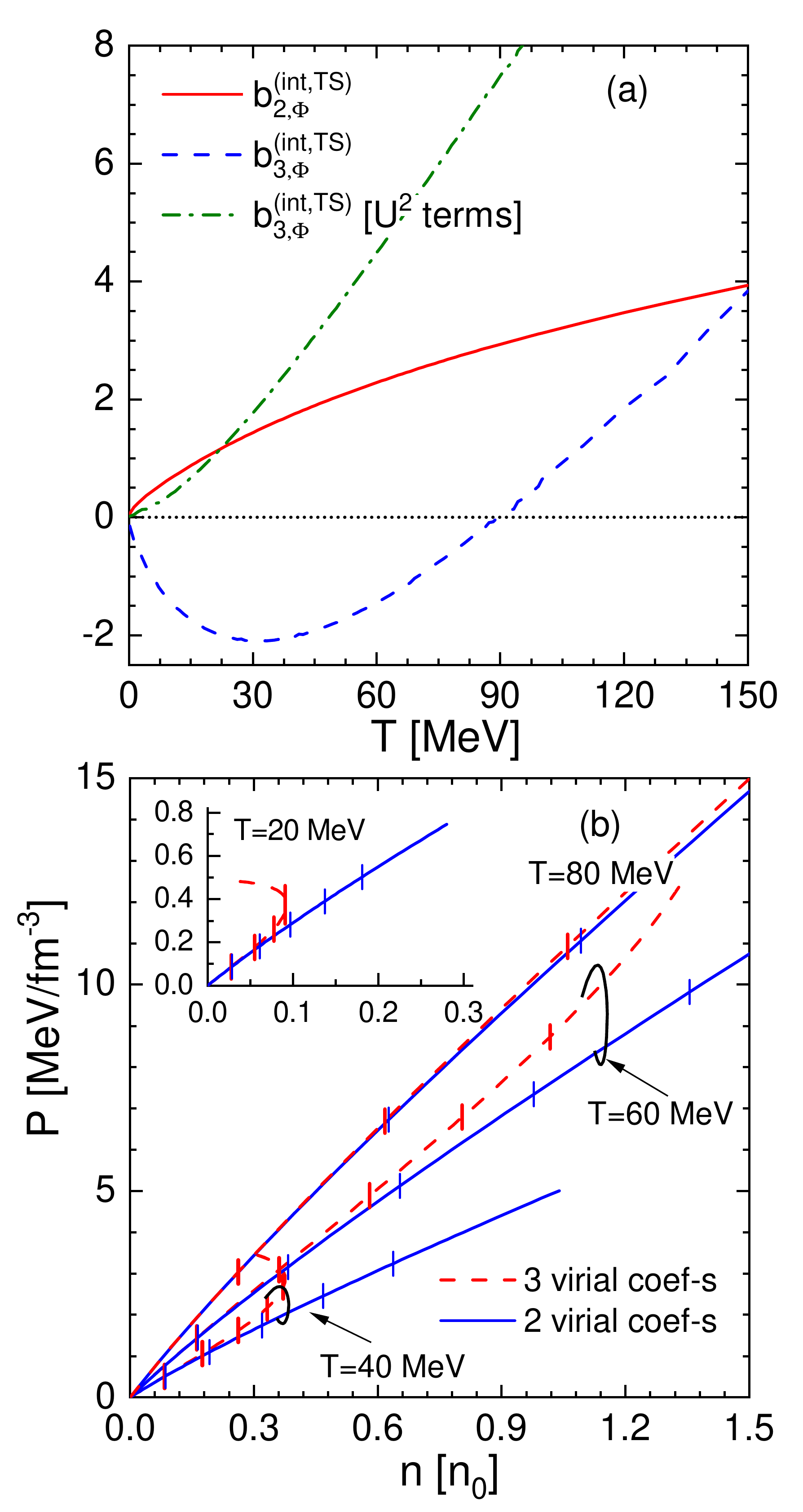}
\caption{Panel (a): The second and third coefficients in the expansion (\ref{P_TS}) of the pressure obtained within the $\Phi$-derivable approach at the TS approximation for the nucleon-nucleon Yukawa potential (\ref{U-pot}) with Walecka parameters (\ref{Uomsig-par-W}) shown as a function of temperature.
Solid line is calculated by Eq.~(\ref{b2-2}), and dashed line by Eq.~(\ref{b3-U}) . Dash-dotted line shows the contribution of the first $U^2$ term in curly brackets in Eq.~(\ref{b3-U}).\\
Panel (b): Pressure in TS approximation  as a function of density (in the units of normal atomic nucleus density $n_0=0.16\,{\rm fm}^{-3}$) calculated  for various temperatures with the help of expansion (\ref{P_TS}) keeping $z^2$ (solid lines) and $z^3$  terms (dashed lines). Vertical ticks on each line mark values of the pressure and density corresponding to $z$ changing from $z=0.1$ (the left tick) with a step 0.1. Thick ticks correspond to dashed lines, and thin ticks to solid lines.}
\label{fig:Phi-U-bs-P}
\end{figure}

In Fig.~\ref{fig:Phi-U-bs-P}b we plot  the pressure as a function of the density, $P^{\rm (TS)}(n)$, for several values of the temperature, as a parametric function of $z$ using the expansion (\ref{P_TS}) and the corresponding function of  $z$ for the density following from the thermodynamic relation (\ref{dPdmu-nf}). Vertical ticks on the lines in Fig.~\ref{fig:Phi-U-bs-P}b indicate the values of $z$. Solid lines show the result obtained when we keep two  terms in the expansion and dashed lines when we keep three terms. We see that for   $T\lsim 40$\,MeV the account of the third term  changes the equation of state already for $z\sim 0.3$. Moreover, for $T\sim 20-40$\,MeV at $z\sim 0.2-0.3$ the pressure calculated with three virial coefficients becomes a not single-valued function of density with increasing $z$ (see ``wiggles'' in dashed lines for $T=20$ and 40\,MeV) indicating invalidation of the virial   expansion in this case.
Thereby we conclude that for the typical nuclear potential (\ref{U-pot}) with Walecka parameters the expansion (\ref{P_TS}) can be used only for $z\lsim 0.3$ at temperatures $T=20$--40\,MeV.
At higher temperatures the expansion (\ref{P_TS}) can be used up to higher values of $z$: at $T=60$\,MeV for $z\leq 0.5$ and at $T=80$\,MeV for $z\leq  0.6$. Note also that within the second virial approximation the wiggles do not appear. The problem appears due to an anomalous behavior of the third virial coefficient with increasing $T$.

The Hartree contribution in Eq.~(\ref{b2-2a}) is given by
\begin{align}
&b_{2,\Phi}^{\rm (int,H)}(T) =\frac{2}{T\lambda^3}\Big(\frac{g_\sigma^2}{m_\sigma^2} - \frac{g_\om^2}{m_\om^2}\Big) =\frac{2}{T\lambda^3m_N^2}\Big(C_\sigma^2 - C_\om^2\Big)
\,.
\label{b2Phi-H}
\end{align}
The sign of the Hartree contribution, which does not change with the temperature, depends on the model parameters. For the Walecka choice with the parameter set (\ref{Uomsig-par-W}) we have $b_{2,\Phi}^{\rm (int,H)}(T)>0$, whereas  for the parameters of the modified Walecka models \cite{Boguta-Bodmer,Boguta,Cubero} and for  the CD Bonn central potential fit  (\ref{Uomsig-par-Bonn}) we find $b_{2,\Phi}^{\rm (int,H)}(T)<0$.
From the phase shifts of the nucleon-nucleon scattering it was found in~\cite{Horowitz2006_2} that for the neutron gas $b_{2}^{\rm (int)}>0$ at least up to $T<50$\,MeV. This gives us an argument in favour of the Walecka parameterization compared to the CD Bonn fit one within our simplified parameterizations of the nucleon-nucleon potential. On the other hand this circumstance may also indicate the importance of non-central $NN$ forces, which are disregarded in our consideration. A more detailed comparison will be done below.

For the Fock contribution  using the explicit form of the potential (\ref{U-pot}) we find
\begin{align}
&b_{2,\Phi}^{\rm (int,F)}(T) = \frac{2^{3/2}}{T}\!\!\intop\!\!\frac{\rmd^3 q}{(2\pi)^3}
e^{-\frac{\vec{q\,}^2}{m_f T}}\Umom(2q\,)
\nonumber\\
&\qquad =\frac{1}{4\pi\lambda T}\Big[g_\om^2-g_\sigma^2 -
g_\om^2\frac{\lambda m_\om}{\sqrt{8\pi}} \sqrt{\pi} e^{\frac{\lambda^2m_\om^2}{8\pi}}{\rm erfc}\Big(\frac{\lambda m_\om}{\sqrt{8\pi}}\Big)
\nonumber\\
&\qquad + g_\sigma^2\frac{\lambda m_\sigma}{\sqrt{8\pi}}\sqrt{\pi}e^{\frac{\lambda^2m_\sigma^2}{8\pi}}{\rm erfc}\Big(\frac{\lambda m_\sigma}{\sqrt{8\pi}}\Big)
\Big]
\,.
\label{b2Phi-F}
\end{align}
Here ${\rm erfc}$ is the complementary error function, ${\rm erfc}(x)=2\intop_x^\infty \exp(-t^2)\rmd t/\sqrt{\pi}$. Taking into account that for $x>0$ this function can be written as ${\rm erfc}(x)=\frac{2}{\sqrt\pi} e^{-x^2}/(x+\sqrt{x^2+\zeta(x)})$, where $\zeta(x)$ slowly varies with $x$ within interval $4/\pi\le \zeta(x)<2$, and $\zeta\approx 2-x^{-2}+O(x^{-4})$ for $x\gg 1$ and $\zeta\approx \frac{4}{\pi }+\frac{4 (4-\pi) x}{\pi ^{3/2}}+O(x^2)$ for $x\ll 1$, we have
\begin{align}
b_{2,\Phi}^{\rm (int,F)}(T) &=
 \frac{g_\sigma^2}{4\pi \lambda T}
\frac{1 - \sqrt{1+\zeta_\sigma\frac{8\pi}{\lambda^2m_\sigma^2} }}
     {1 + \sqrt{1+\zeta_\sigma\frac{8\pi}{\lambda^2m_\sigma^2} } }
\nonumber\\
&- \frac{g_\om^2}{4\pi\lambda T}
\frac{1 - \sqrt{1+\zeta_\om\frac{8\pi}{\lambda^2m_\om^2} }}
     {1 + \sqrt{1+\zeta_\om\frac{8\pi}{\lambda^2m_\om^2} } } \,,
\nonumber\\
\zeta_{\sigma,\om} &=\zeta\Big(\frac{\lambda m_{\sigma,\om}}{\sqrt{8\pi}}\Big)\,.
\end{align}
For low temperatures, $T\ll m_{\sigma}^2/(2m_N)\simeq 130$\,MeV with $m_{\sigma}\simeq 500$MeV,  we obtain
\begin{align}
b_{2,\Phi}^{\rm (int,F)}(T)
&\approx -\frac{1}{2}b_{2,\Phi}^{\rm (int,H)}(T) + O\Big(\frac{2 m_N T}{{m_\sigma}^2}\Big)
\,.
\label{b2Phi-F-2}
\end{align}
Thus, $b_{2,\Phi}^{\rm (int)}(T \ll m_{\sigma}^2/(2m_N))=b_{2,\Phi}^{\rm (int,H)}/2$.

In the limit of high temperatures, $T\gg m_{\om}^2/(2m_N)\sim 330$\,MeV,
we have
\begin{align}
b_{2,\Phi}^{\rm (int,F)}(T)
&\approx  \frac{g_\om^2-g_\sigma^2}{4\pi \lambda T}\,.
\label{b2Phi-F-3}
\end{align}
We see that within the $\sigma$-$\om$ model for nuclear potentials with a repulsion at small distances (i.e. $g_\om>g_\sigma$) the Fock contribution to $b_{2,\Phi}^{\rm (int)}$ is always positive at very high temperatures. The ratio of the Fock and Hartree contributions in this case is
\begin{align}
\frac{b_{2,\Phi}^{\rm (int, F)}(T)}{b_{2,\Phi}^{\rm (int, H)}(T)}
\approx\frac{m_N\,(g_\om^2-g_\sigma^2)}{2T\Big(C_\sigma^2 - C_\om^2\Big)}\,.\label{TCTS}
\end{align}
This ratio equals to $1.3(300\,{\rm MeV}/T)$ in case of the Walecka parameterization of the $nn$ potential  (\ref{Uomsig-par-W}), and to $-26(300\,{\rm MeV}/T)$ when one uses the central part of the CD Bonn potential, and thereby in the latter case there is a region of temperatures, where the Fock term essentially exceeds the Hartree one. Nevertheless we point out that  at asymptotically high temperatures the Fock term vanishes $\propto 1/\sqrt{T}$, cf.~(\ref{b2Phi-F-3}), whereas the Hartree term increases in modulus as $\propto \sqrt{T}$, cf. (\ref{b2Phi-H}).
Asymptotic expressions, which we have derived, also show that in the CD Bonn fit case the second virial coefficient at least once changes the sign at a certain value of the temperature. It occurs however at extremely large value, at which hadronic degrees of freedom would not survive.

The Hartree contribution to the third coefficient $b_{3,\Phi}^{\rm (int)}$ is given in Eq.~(\ref{b3Phi-H}). The classical part exceeds the quantum corrections for $T\gg  \pi^3/[\Umom(0)]^2 m_N^3]$ as we have estimated. For the Yukawa potentials, this inequality produces the constraint $T\gg \pi^3 m_N/(C^2_\om -C^2_\sigma)^2$, i.e. in case of the Walecka parameters the $U^2$ term is dominant for  $T> 6$ MeV,  whereas for the CD Bonn fit parameters the linear term remains dominant  at all temperatures of our interest.

Now we evaluate $z=n\lambda^3/2\simeq 0.2 (n/n_0)(m_\pi/T)^{3/2}$. Thus $z\ll 1$ for $n\ll n_0$ at all temperatures $T\lsim (1-2) m_\pi$ of our interest. From (\ref{b2Phi-H}), (\ref{b3Phi-H}) we  estimate $b_{3,\Phi}^{(\rm int,H)}[U \,{\rm{terms}}]z/b_{2,\Phi}^{(\rm int,H)}=-z/\sqrt{2}$,
$b_{3,\Phi}^{(\rm int,H)}[U^2 \,{\rm{terms}}]z/b_{2,\Phi}^{(\rm int,H)}=-n\Umom(0)/T$ that for the Yukawa potential with the Walecka parameters yields $b_{3,\Phi}^{(\rm int,H)}[U^2 \,{\rm{terms}}]/b_{2,\Phi}^{(\rm int,H)}\simeq -0.8 (n/n_0)(m_\pi/T)$ in  agreement with that is shown in  Fig \ref{fig:Phi-U-bs-P}a.

Also, using Eqs.~(\ref{b2Phi-H})
we find for the Hartree term that
$\big|b_{2,\Phi}^{\rm (int, H)}(T)\big|>|b_{2}^{(0)}|$ if $T>\pi^3 m_N/[4|C_\sigma^2-C_\om^2|^2]\simeq 1.4$\,MeV for the Walecka parameters (\ref{Uomsig-par-W}). For the CD Bonn fit potential the difference $|C_\sigma^2-C_\om^2|$ is much smaller than for the Walecka parameters and therefore the corresponding value of the  temperature,  $\lsim 140$\,MeV, proves to be two orders of magnitude higher.

In the next Section we will demonstrate that the Hartree part in Eq.~(\ref{b2-2a}) and the first term in (\ref{b3Phi-H}) coincide with the classical limit of the second and third virial coefficients in the virial equation of state.

\subsection{Application to the square barrier-well potentials}\label{testTSbarrier}

Substituting Eq.~(\ref{Usw-mom}) in Eq.~(\ref{b2-2}) and separating contribution from the Hartree and Fock diagrams we derive for the Hartree term
\begin{align}
b_{2,\Phi}^{\rm (int,H)}(T) &= -2\frac{\Umom_{\rm sw}(0)}{T\lambda^3}\,,
\end{align}
where $\Umom_{\rm sw}(0)$ is given by Eq.~(\ref{Ufourier}), and for the Fock term
\begin{align}
b_{2,\Phi}^{\rm (int, F)}(T) &=
\frac{(U_{\rm r}+U_{\rm a})}{T} \phi_{\rm sw}(T/T_{\rm cr,r})
-\frac{U_{\rm a}}{T} \phi_{\rm sw}(T/T_{\rm cr,a})\,,
\nonumber\end{align}\begin{align}
&T_{\rm cr,r}=1/(16r_0^2m_f)> T_{\rm cr,a}=1/(4r_U^2m_f)\,,\nonumber\\
\phi_{\rm sw}(a) &= \frac{1}{\sqrt{2}\pi} \intop_0^\infty \rmd x x^2 e^{-x^2/a} f_{\rm sw}(x)
\nonumber\\
&= 2^{-3/2}{\rm erf}\Big(\frac{\sqrt{a}}{2 }\Big)
- 2^{-3/2}\sqrt{\frac{a}{\pi}}
e^{-\frac{a}{4}} \,,
\nonumber\end{align}\begin{align}
\phi_{\rm sw}(a) &\approx
\sqrt{\frac{2}{\pi}}\frac{a^{3/2}}{24},\,\, a\ll 1\,,
\nonumber\\
\phi_{\rm sw}(a) &\approx 2^{-3/2} -2^{-3/2}\sqrt{\frac{a}{\pi}} e^{-\frac{a}{4}}
\,,\,\,
a\gg 1
\,.
\end{align}
The case $a\ll 1$ constitutes the limit of low temperatures when $T\ll T_{\rm cr,a}$. In this limit we find $b_{2,\Phi}^{\rm (int, F)}(T) \approx -\frac12 b_{2,\Phi}^{\rm (int),H}(T)$, cf. Eq.~(\ref{b2Phi-F-2}) above. Taking $r_U\simeq 1/m_\pi$ we estimate $T_{\rm cr,a}\simeq 5$\,MeV. For $r_U\simeq 1/(2m_\pi)$ we get $T_{\rm cr,a}\simeq 20$\,MeV. Taking $2r_0\simeq 0.6$\,fm, we estimate $T_{\rm cr,r}\simeq 30$\,MeV.

In the case $a\gg 1$, i.e. for  $T\gg T_{\rm cr,r}$, we deal with a high temperature limit. Then  the Fock term is determined by the repulsive part of the potential
$b_{2,\Phi}^{\rm (int, F)}(T)\approx U_{\rm r}/(2^{3/2}T)$. In this  limit
\begin{align}
\frac{b_{2,\Phi}^{\rm (int, F)}(T)}{b_{2,\Phi}^{\rm (int, H)}(T)}
\approx -\frac{\pi^{3/2}}{2(m_f T)^{3/2}}\frac{U_{\rm r}}{\Umom_{\rm sw}(0)}.
\end{align}
We further estimate
\begin{align}
\frac{b_{2,\Phi}^{\rm (int, F)}(T)}{b_{2,\Phi}^{\rm (int, H)}(T)}
\approx \Big(\frac{3T_{\rm cr,a}}{T}\Big)^{3/2}
\frac{U_{\rm r}}{U_{\rm a}}
\end{align}
for $U_{\rm r}\lsim U_{\rm a}$ and
\begin{align}
\frac{b_{2,\Phi}^{\rm (int, F)}(T)}{b_{2,\Phi}^{\rm (int, H)}(T)}
\approx -\Big(\frac{3T_{\rm cr,r}}{T}\Big)^{3/2}
\end{align}
for $U_{\rm r}\gg U_{\rm a}$ and $r_U\sim$ several $r_0$. Thus, in both cases the Hartree term exceeds the Fock one in the magnitude at least  for $T\lsim T_{\rm cr,a}\simeq 5$\,MeV at $r_U\simeq 1/m_\pi$ and for $T\lsim T_{\rm cr,a}\simeq 20$\,MeV at $r_U\simeq 1/(2m_\pi)$, and for $T>T_{\rm cr,r}\simeq 90$\,MeV (at $2r_0\simeq 0.6$\,fm). Compare with estimates (\ref{b2Phi-F-2}) -- (\ref{TCTS}) performed above for the Yukawa potentials.

Summarizing, in this Section we derived explicit expressions for all thermodynamic quantities in case of $\Phi$ described by the TS diagrams at the pair potential interaction. In this case we explicitly derived all thermodynamic quantities
and obtained the compact relation for the entropy (\ref{TS-short}).
Then, the fugacity expansion for $\Phi$ has been performed. We derived closed expressions for the first, second and third virial coefficients and we separated the Hartree and Fock contributions, which are quantified at hand of the specific potentials.  We found that our Eq. (\ref{b2-U-Born-delt}) for the $\Phi$ derivable second virial coefficient  has formally the same form as the Beth-Uhlenbeck  virial expansion one. Our results highlight the relationship between Hartree and Fock contributions.   Also within the Born approximation, i.e., keeping only terms linear in the potential, we found the expression for the virial coefficients of the $n$th order.

\section{Virial expansion  scheme for the equation of state}\label{virial}

\subsection{Density and fugacity expansions}\label{Densfug}
Kamerlingh Onnes in Ref.~\cite{KO1901} proposed to express the  pressure of one-component Boltzmann gas as series in powers of the density,
\begin{align}
P= n T \big(1+B\,n +C\,n^2 + \dots\big),
\label{P-class-cor}
\end{align}
where coefficients $B(T)$, $C(T)$, $\dots$  are  called the second, third, etc. virial coefficients in the density expansion. These coefficients indicate
the deviation of a real equation of state from that of the ideal gas.
The coefficients of the virial expansion (\ref{P-class-cor}) are measured and tabulated for many pure  gases, see e.g.~\cite{Landolt-Virial}.
On the theoretical side they can be calculated in terms of potentials acting among gas constituents, e.g.,
for  the classical Boltzmann gas of particles interacting via a binary potential $\Ucor(|\vec{r}\,|)$.

The derivation of (\ref{P-class-cor}) in  microscopic models is usually accomplished through an expansion of the pressure (or grand-thermodynamic potential) in powers of an effective parameter $z=\exp(\mu_f/T)$ called fugacity~\cite{Ursell,Mayer},
\begin{align}
P(z,T)=\frac{T}{\lambda^3}\sum_{k=1}^\infty b_k(T) z^k\,,
\label{P-class-z}
\end{align}
where we introduced dimensionless virial coefficients\footnote{In order to operate with the dimensionless virial coefficients one has to introduce in expansion~(\ref{P-class-z}) an overall factor with dimensionality of a volume. In  applications to nuclear physics for this aim one usually uses the cube of the thermal wavelength (\ref{lambda-def}), see, e.g., Refs.~\cite{Horowitz2006_2} and \cite{Voskresenskaya}. However  $\lambda\propto\hbar$, whereas the virial expansion method is also applicable in purely classical limit. So, as the result only combinations $\lambda^{3\,(n-1)} b_n$ have the proper classical limit, see Eq.~(\ref{B-b-rel}) and discussion in Sect.~\ref{Virsemicl} below.} $b_k$ in the fugacity expansion and the effective volume parameter defined through the thermal wavelength of the particle (\ref{lambda-def}). Differentiation of this series yields, in turn, the density as a power series of the fugacity\footnote{As before, we assume that there are no   bound states, as in case of neutron matter. In the presence of the bound state and clusters the relation for the density is more complicated, see, e.g., Refs.~\cite{RMS-NPA379,Ropke:2012qv}.}.
The relation between the density $n$ and the fugacity $z$ is as follows,
\begin{align}
n=\frac{\partial P}{\partial \mu_f}
=\frac{z}{T}\frac{\partial P}{\partial z}= \frac{1}{\lambda^3}\sum_{k=1}^\infty k b_k(T) z^k\,.
\label{dens-virial}
\end{align}
The virial density expansion (\ref{P-class-cor}) is
obtained, by formal inverting the last series to obtain the fugacity as a power series in the
density, and composing the latter with the expansion for the pressure.
The virial expansion of the pressure in powers of density is obtained with the help of the formal inverting of the fugacity series (\ref{dens-virial}), i.e., presenting  the fugacity as a power series in the
density and setting it in the expansion for the pressure (\ref{P-class-z}).
The first coefficients of the density expansion are
\begin{align}
B &= -\frac{\lambda^3}{b_1^2} b_2\,,\qquad
C = -2\frac{\lambda^6}{b_1^4} ( b_3 b_1 -2 b_2^2 ),
\label{B-b-rel}
\end{align}
and so on.

It should be noted that Eq.~(\ref{P-class-z}) is an exact representation of the grand-partition function provided that the {whole set of parameters $\{b_k\}$ is  known.} Formal questions of the convergence of the series  (\ref{P-class-z}) and (\ref{dens-virial}) and its applicability to various potentials were discussed intensively in the literature over the years, cf.~\cite{virial-converg,virial-converg-2}.
%For example, from Eq.~(\ref{b2clas}) we immediately conclude that the potential must have a repulsive core to prevent the system from collapse.
Once all coefficients $b_k$ are known, the set of equations (\ref{P-class-z}) and (\ref{dens-virial}) provides a parametrical definition of the pressure as a function of the density for a given temperature. The inversion of the series (\ref{dens-virial}) leading to Eq.~(\ref{P-class-cor}) requires that the densities and temperatures are at least such that $\lambda^3 n\ll b_1^2/|b_2|$.
The coefficient $b_1$ does not depend on the interparticle potential and temperature, being determined by the number of internal degrees of freedom of the gas constituents. For a one component gas of spinless particles $b_1=1$ and for a  one component gas of spin $1/2$ fermions (e.g. for the neutron gas) $b_1=2$.

%As one can see in Eq.~(\ref{b2clas}), t
The higher order virial coefficients depend on the inter-particle potential in a non-trivial way. Calculation of these virial coefficients, $b_k$ for $k>2$,  with an increase of $k$ becomes a  complicated numerical task even in the classical limit. The rigorous quantum mechanical approach for calculation of the virial coefficients has been formulated only for the second and third coefficients~\cite{Uhlenbeck-Beth,Beth,Pais-Uhlenbeck,Larsen,Larsen-Mascheroni}. Truncating the series and retaining in Eqs.~(\ref{P-class-z}) and (\ref{dens-virial}) the first $k_{\rm max}$ terms one has
\begin{align}
P(z,T) &=\frac{T}{\lambda^3}\sum_{k=1}^{k_{\rm max}} b_k(T) z^k,
\label{P-class-z-cut}\\
\lambda^3 n &=\sum_{k=1}^{k_{\rm max}} k b_k(T) z^k\,.
\label{dens-virial-cut}
\end{align}
Working within perturbation theory we may resolve Eq.~(\ref{dens-virial-cut}) as a series up to the order $(\lambda^3 n)^{k_{\rm max}}$,
\begin{align}
z=\sum_{k=1}^{k_{\rm max}} a_k (\lambda^3 n)^k+ O\big((\lambda^3 n)^{k_{\rm max}+1}\big)\,,
\label{z-n-exp}
\end{align}
where each coefficient $a_k$ is a function of coefficients $\{b_{i\le k}\}$,
and substituting it in Eq.~(\ref{P-class-z-cut}) we obtain pressure valid also up to order $O\big((\lambda^3 n)^{k_{\rm max}+1}\big)$.
Other thermodynamic quantities  follow from $P=P(z(\mu),T)$.   For example, the  entropy density is
\begin{align}
\frac{S}{V}=\frac{\partial P}{\partial T}\Big|_{\mu}=\frac{P-\mu n}{T}+ \frac32\frac{P}{T}+ \frac{T}{\lambda^3}\sum_{k=1}^{k_{\rm max}} z^k \partial_T b_k(T)\,,
\label{entropy-virial}
\end{align}
and the energy density is
\begin{align}
\frac{E}{V} &= \frac32 P +  \frac{T}{\lambda^3}\sum_{k=1}^{k_{\rm max}} z^k T\partial_T b_k(T)
\nonumber\\
&= \frac{T}{\lambda^3}\sum_{k=1}^{k_{\rm max}} z^k \Big(\frac32+T\partial_T\Big) b_k(T) \,,
\label{energy-virial}
\end{align}
where we used thermodynamic relation (\ref{thermoidentity}). If we substitute expansion (\ref{z-n-exp}) in Eqs.~(\ref{P-class-z-cut}),  (\ref{entropy-virial}) and (\ref{energy-virial}), we obtain pressure, entropy and energy as functions of density ant temperature, which satisfy thermodynamic relations perturbatively up to the order $O\big((\lambda^3 n)^{k_{\rm max}+1}\big)$.

\subsection{Thermodynamically-consistent approximations beyond a perturbative approach}\label{ssec:selfcons-virial}

Although a perturbative approach is satisfactory in general for $z<1$, it can be inconvenient for some repeated calculations, e.g., within numerical codes for heavy-ion collisions, where exact thermodynamic consistency is mandatory on each step of calculations, otherwise errors would be accumulating uncontrollably. To avoid the problem and impose thermodynamic consistency one should use directly the solution $z=z(n,T)$ of Eq.~(\ref{dens-virial-cut}) (the numerical one, if necessary) instead of the expansion (\ref{z-n-exp}). Then, thermodynamical quantities $P$, $S$, $E$ are obtained after substituting this solution in place of $z$, e.g. $P(n,T)=P(z=z(n,T),T)$. The thermodynamic relations hold then with the accuracy, with which we can solve  Eq.~(\ref{dens-virial-cut}). This treatment of the virial approximation to the equation of state is in a sense similar to the $\Phi$-derivable approach considered in Sec.~\ref{Phi}, when we restrict ourselves by consideration of finite number of $\Phi$ diagrams. Such an extension of the virial approach is  used in many applications, cf. ~\cite{Horowitz2006,Horowitz-response}.

Frequently one performs expansion of the pressure up to the second order in $z\ll 1$ dealing thereby with first two virial coefficients, $b_1$ and $b_2$, where the second one, $b_2$, can be reliably calculated either classically, if the potential is known, or quantum-mechanically, if the scattering experiments allow to determine sufficient number of partial scattering waves, cf. discussion below. In the latter case one may treat the obtained equation of state reproduced up to second order in  $z\ll 1$  as the model-independent equation of state.
 At this $z^2$-order the solution of Eq.~(\ref{dens-virial-cut}) yields
\begin{align}
z(n,T)=\frac{b_1}{4b_2(T)}\Big[\sqrt{1+8\frac{b_2(T)}{b_1^2}\lambda^3 n} -1 \Big],
\label{z-scv}
\end{align}
and in such a thermodynamically-consistent virial (tcv) treatment of the problem, the pressure $P_{\rm tcv} (n,T)$ is equal to
\begin{align}
P_{\rm tcv}(n,T)=\frac{n T}{2}
+\frac{b_1^2 T }{8b_2 \lambda^3}\Big[\sqrt{1+8 \frac{b_2}{b_1^2} \lambda^3 n}
-1\Big].
\label{p-scv}
\end{align}

Concluding, in this Section we introduced the density and fugacity expansions for the pressure --- the virial equation of state, and considered a thermodynamically-consistent extension of the virial treatment beyond the perturbative approach.

\section{Classical virial coefficients}\label{ssec:class-virial}
\subsection{Mayer's function and virial coefficients}\label{MayerFunc}

In classical statistical physics derivation of Eq.~(\ref{P-class-z}) starts with the expansion
of the grand-canonical statistical sum
%$\mathcal{Z}=\Sp[\exp(-(\hat{H}-\mu\hat{N})/T)]$
in powers of fugacities as~\cite{Huang,Reichl},
\begin{align}
\mathcal{Z}^{\rm (cl)}=\sum_{N=0}^{\infty} z^N W_N^{\rm (cl)} \,,
\end{align}
where $W_0^{\rm (cl)}=1$ and $W_N^{\rm (cl)}$ is the classical $N$-body canonical partition function
\begin{align}
W_N^{\rm (cl)} = \frac{1}{N!}\int\prod_{i=1}^N\frac{d_f\rmd^3 r_i\rmd^3p_i}{(2\pi)^3} e^{-\frac{1}{T}H^{(N)}(\vec{r}_1,\dots,\vec{r}_N,\vec{p}_1,\dots,
\vec{p}_N)}
\label{W_N-cl}
\end{align}
with $N$-body Hamiltonian, $H^{(N)}$,
\begin{align}
H^{(N)}=\sum_{i=1}^N\frac{\vec{p\,}_i^2}{2m_f}
+U_N(\vec{r}_1,\dots,\vec{r}_N)
\label{H-cl}
\end{align}
containing the sum of kinetic energies of particles and the potential energy of the inter-particle interaction depending on the particle position. As in previous Sections, we assume that two fermions interact only through the spin-independent two-body potential $\Ucor(|\vec{r}_{ij}|)$ depending on the distance between them, $\vec{r}_{ij}=\vec{r}_j-\vec{r}_i$, then
$U_N= \frac12\sum_{i,j=1}^N \Ucor(|\vec{r}_{ij}|)$. Degeneracy factor $d_f$ in (\ref{W_N-cl}) takes into account possible internal degrees of freedom, e.g., spin $S_f$.
Although spin is of the purely quantum origin, it enters the phase-space volume  through the degeneracy factor $d_f=2S_f +1$.

The integral over momenta can be easily evaluated in Eq.~(\ref{W_N-cl}) with the result
\begin{align}
W_N^{\rm (cl)} = \frac{1}{N!}
\int\prod_{i=1}^N\Big(d_f\frac{\rmd^3 r_i}{\lambda^3}\Big) e^{-\frac{1}{2T}\sum_{i,j=1}^N \Ucor(|\vec{r}_{ij}|)}\,.
\end{align}
The grand-canonical potential, being the logarithm of the partition function, can be re-expanded in terms of the $N$-body cluster-integrals expressible in terms $W_N$~\cite{Ursell,Mayer}.  The expansion of the grand-canonical potential becomes
\begin{align}
\Omega^{\rm (cl)}(T,\mu) &= -T\ln \mathcal{Z}^{\rm (cl)}=-\frac{TV}{\lambda^3}\sum_{N=1}^{\infty} z^N b_N^{\rm (cl)}\,,\,\,
\label{Omega-virial-cl}
\end{align}
where the virial coefficients are
\begin{align}
b_1^{\rm (cl)} &=\frac{\lambda^3}{V}W_1^{\rm (cl)}\,,
\quad
b_2^{\rm (cl)} =\frac{\lambda^3}{2V}\big(2W_2^{\rm (cl)}-[W_1^{\rm (cl)}]^2 \big)\,,
\nonumber\\
b_3^{\rm (cl)} &=\frac{\lambda^3}{6V}\big(6W_3^{\rm (cl)} - 6 W_2^{\rm (cl)}\, W_1^{\rm (cl)} + 2\, [W_1^{\rm (cl)}]^3\big) \,,
\label{bN-W-cl}
\end{align}
and so on. Writing out the statistical sums $W_N^{\rm(cl)}$ explicitly, for the several first coefficients we obtain
\begin{align}
b_1^{\rm (cl)} &= d_f,
\nonumber\\
b_2^{\rm (cl)}  &=\frac{d_f^2\lambda^3}{2V}
\Big(\int\frac{\rmd^3r_1\rmd^3 r_2}{\lambda^6}e^{-\Ucor(|\vec{r}_{12}|)/T}-\frac{V^2}{\lambda^6}\Big)
\nonumber\\
&=\frac{d_f^2}{2}\int\frac{\rmd^3r}{\lambda^3}\Big(e^{-\Ucor(|\vec{r\,}|)/T}-1\Big)
\,,
\label{bN-cl}\end{align}\begin{align}
b_3^{\rm (cl)}  &=\frac{d_f^3}{6}
\int\frac{\rmd^3r_1}{\lambda^3}\frac{\rmd^3 r_2}{\lambda^3}
\Big(e^{-\big(\Ucor(|\vec{r}_1|)+\Ucor(|\vec{r}_2|)+
\Ucor(|\vec{r}_{12}|)\big)/T}
\nonumber\\
&- e^{-\Ucor(|\vec{r}_1|)/T} - e^{-\Ucor(|\vec{r}_2|)/T}
- e^{-\Ucor(|\vec{r}_{12}|)/T} + 2
\Big)
\,.
\nonumber
\end{align}
Substituting these relations in Eq.~(\ref{B-b-rel}) we obtain  expressions for the coefficients of the density expansion of the pressure (\ref{P-class-cor}) in the classical limit. The first two coefficients  are
\begin{align}
B^{\rm (cl)}(T) &= -\frac{1}{2}\intop\rmd^3 r f_{\rm M}(r)\,,
%\big(1-e^{-\Ucor(|\vec{r}\,|)/T}\big)\,,
\label{b2clas}\\
%\nonumber\\
C^{\rm (cl)}(T) &= -\frac{1}{3}\intop\rmd^3 r_1 \rmd^3 r_2 f_{\rm M}(r_1)f_{\rm M}(r_2)f_{\rm M}(|\vec{r_1}-\vec{r_2}|)\,,
%\big(1-e^{-\Ucor(|\vec{r}_1\,|)/T}\big)
\nonumber
\\
%&\times
%\big(1-e^{-\Ucor(|\vec{r}_2\,|)/T}\big)
%\big(1-e^{-\Ucor(|\vec{r}_2-\vec{r}_1|)/T}\big).
f_{\rm M}(|\vec{r}\,|)&=e^{-\Ucor(|\vec{r}\,|)/T}-1\,.\nonumber
\end{align}
As it is seen on the example of the first coefficients $B^{(\rm cl)}$ and $C^{(\rm cl)}$, their main building block is the coordinate function $f_{\rm M}(|\vec{r}\,|)$, %$$f(|\vec{r}\,|)=e^{-\Ucor(|\vec{r}\,|)/T}-1\,,$$
 which is called the Mayer's f-function.
 In the ideal gas case,  the Mayer's f-function
vanishes everywhere. For a finite-range potential the Mayer's f-function takes finite values only in the interaction region and it becomes zero at large distances.
In this sense, the Mayer's f-function may serve as a smallness  parameter  in the perturbative expansion starting from the ideal gas in the zero-perturbation order.  Such an approximation scheme is behind the so called cluster expansion~\cite{Mayer,Ursell}.

Note that, although we  work here in the framework of classical physics,  expressions for the virial coefficients (\ref{bN-cl}) include explicitly the thermal wave length, $\lambda$, which is $\propto\hbar$. This is because of our choice to deal with dimensionless coefficients $b_j^{\rm (cl)}$ in the virial fugacity expansion (\ref{P-class-z}). Then $\lambda^3$ is the only scale to render the  coordinate integrals dimensionless. The coefficients (\ref{b2clas}) of the density virial expansion (\ref{P-class-cor}) are free of the $\lambda$ parameter. Also note that coefficients (\ref{b2clas}) entering $P(n,T)$ expansion  do not contain the degeneracy factor $d_f$ related to the internal degrees of freedom being absorbed in  $n$.

For potentials demonstrating  repulsion at short distances, $r<2r_0$ with   $U(2r_0)=0$, and attraction at longer distances, $f_{\rm M}$ consists of two contributions: one from  $r<2r_0$, where potential is  repulsive, and another one  from $r>2r_0$, where the potential is attractive.  Therefore we can write
\begin{align}
&B^{\rm (cl)}=B_{<}^{\rm (cl)} + B_{>}^{\rm (cl)}\,,
\label{B2cl-decomp}\\
&B_{<}^{\rm (cl)}= -2\pi\int_0^{2r_0} r^2\rmd r f_{\rm M} (r)=v_{\rm r}/2\,,
%\Big(1-e^{-U(r)/T}\Big)\,,
\label{B1cl-def}\\
&B_{>}^{\rm (cl)}=- 2\pi\int_{2r_0}^{\infty} r^2\rmd r f_{\rm M} (r)\,.
%\left(1-e^{-U(r)/T}\right)\,,
\label{B2cl-def}
\end{align}
The virial coefficient has dimensionality of volume, therefore, terms $B_{<}^{\rm (cl)}$ and $B_{>}^{\rm (cl)}$ can serve as characteristics of contributions of repulsive and attractive parts of the potential to the pressure. We can introduce the
attraction and repulsion volumes as
\begin{align}
v_{\rm r} &= 2 B_{<}^{\rm (cl)} = 4\pi\int^{2r_0}_{0} r^2\rmd r \left(1-e^{-U(r)/T}\right)\,,
\label{vr-gen}\\
v_{\rm a} &= 2 B_{>}^{\rm (cl)} = 4\pi\int_{2r_0}^{\infty} r^2\rmd r \left(1-e^{-U(r)/T}\right).
\end{align}
Factor 2 takes here into account that the volume $B$ is defined per particle, whereas the potential depends on the inter-particle distance.
The quantity can be also interpreted as the excluded volume, within which the probability to find a particle is strongly reduced.

The coefficient $B$ (and $b_2$), may vanish at the temperature $T_{\rm B}$ called the {\rm Boyle temperature}. For $T<T_{\rm B}$  attraction dominates and $B^{\rm (cl)}<0$, whereas for $T>T_{\rm B}$  repulsion wins and $B^{\rm (cl)}>0$. In the vicinity of the Boyle temperature $T\sim T_{\rm B}$, the virial coefficient $B$ is smaller than in case, when the Boyle temperature is absent. In the former case the gas behaves similarly to the ideal gas, if at such temperatures the higher-order term in the pressure is small, i.e., if $Cn^2\ll 1$.

\subsection{Classical virial coefficients for specific potentials}\label{classicalspec}

\subsubsection{Yukawa and LJ potentials}

\begin{figure}
\centering
\includegraphics[width=6cm]{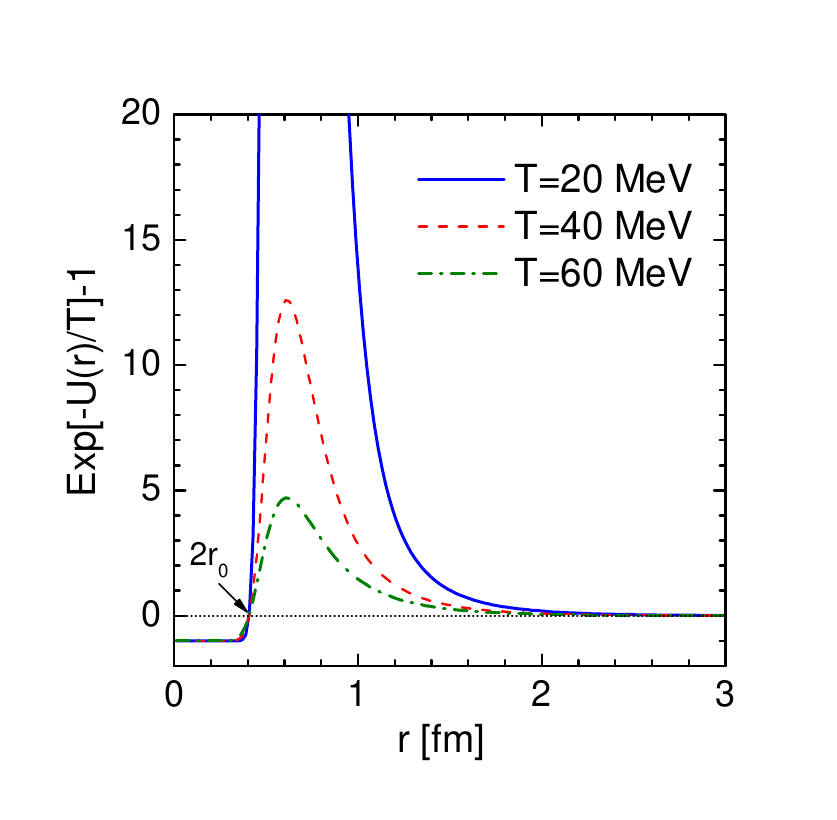}
\caption{The Mayer's $f_{\rm M}$-function, which is the integrand of the expression for the classical limit of the virial coefficient $B^{(\rm cl)}$, see Eq.~(\ref{b2clas}), as a function of the coordinate for several values of the temperature for the Yukawa potential (\ref{U-potR}) with Walecka parameters (\ref{Uomsig-par-W}).}
\label{fig:Binteg}
\end{figure}

The coordinate dependence of the Mayer's $f_{\rm M}$-function for the Yukawa potential with parameters (\ref{Uomsig-par-W})
is shown in Fig.~\ref{fig:Binteg} for three values of the temperature. The figure demonstrates two regimes in  the behavior of the Mayer's function for the repulsive ($r<2r_0$) and attractive ($r>2r_0$) parts of the potential. In the repulsion region the Meyer's function is almost constant and $\simeq 1$, whereas in the attraction region it is  a rapidly rising and then decreasing function strongly dependent on the temperature. The sharpness of $f_{\rm M}$ increases with a temperature decrease.

In Fig.~\ref{fig:V-repl} we show the temperature dependence of the effective repulsion volume defined in Eq.~(\ref{vr-gen}) for different potentials. We see that in general for the Yukawa-type potentials $v_{\rm r}$ decreases with a temperature increase faster than for the LJ-type potentials, since the latter ones increase with a decrease of $r$ much faster than the former ones, see Fig.~\ref{fig:U-pot}.

\begin{figure}
\includegraphics[width=7cm]{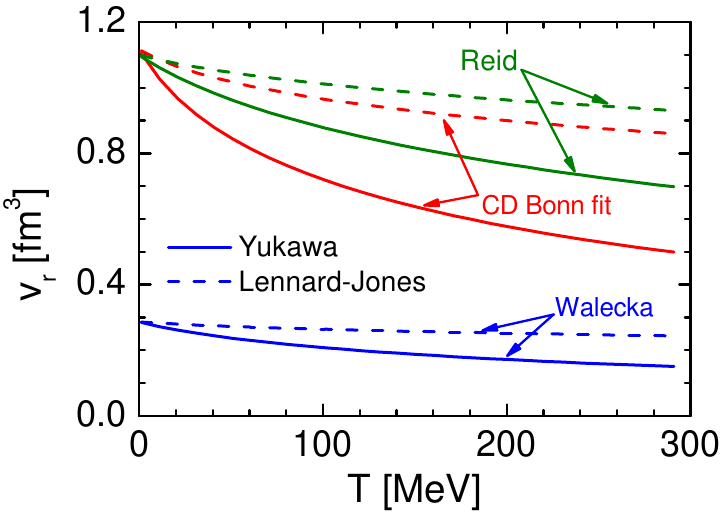}
\caption{ Dependence of the effective repulsive volume for  the potential, Eq.~(\ref{vr-gen}), on the temperature for the Yukawa-type potentials (\ref{U-potR}) shown by solid lines and for the  Lennard-Jones-type potentials (\ref{LJ-pot}) shown by dashed lines. Parameters of the Yukawa-type potentials are taken as in Eq.~(\ref{Uomsig-par-W}) for the Walecka set, Eq.~(\ref{Uomsig-par-Bonn}) for CD Bonn fit set, and Eq.~(\ref{U-Reid}) for the Reid set. In the figure we use the same labels for the LJ potentials taken with parameters  fitted to compare with the corresponding Yukawa-type potentials.}
\label{fig:V-repl}
\end{figure}

The repulsion volume, as we defined in Eq.~(\ref{vr-gen}), depends on the temperature because the typical size of the core is determined by the condition $U(r_c(T))=T$ and only for a infinitely hard core its radius, $r_c=2r_0$, is independent on the temperature.
For low temperatures the term $B_{<}^{\rm (cl)}$ can be evaluated as
\begin{align}
B_{<}^{\rm (cl)} &\approx \frac{16\pi}{3}r_0^3\Big(1 -\frac32 \frac{T}{r_0|U'(2r_0)|}\Big) +
O\Big(\frac{T^2}{r_0^2|U'(2r_0)|^2}\Big) \,,
 \label{B1cl-const}
\end{align}
where we used expansion of the potential near the point $r= 2r_0$. The temperature-independent term is the quadruplicated volume of the fermion. Its account in the virial expansion Eq.~(\ref{P-class-cor}) increases the pressure. The temperature dependence of the coefficient $B_{<}^{\rm (cl)}$  is controlled by the steepness of the potential at the point $2r_0$. The derivative of the potential heaving a repulsive core is necessarily negative at the point $2r_0$. For the potential (\ref{U-potR}) we have
\begin{align}
r_0 U'(2r_0)=-\frac{g_\om^2}{8\pi}(m_\om-m_\sigma) \Big(\frac{g_\om^2}{g_\sigma^2}\Big)^{\frac{m_\om}{m_\sigma-m_\om}}\,,
\end{align}
that for parameters (\ref{Uomsig-par-W}) gives $r_0 |U'(2r_0)|\simeq 300$\,MeV, and for the Reid potential we get $r_0 |U'(2r_0)|\simeq 430$\,MeV, whereas  with the CD Bonn fit parameters (\ref{Uomsig-par-Bonn}) we obtain $\simeq 170$\,MeV. The LJ potentials of the same core size and the depth produce larger values of the steepness parameter  $r_0 |U'(2r_0)|=12U_{0}=12|U_m|$, see the corresponding curves in Fig.~\ref{fig:U-pot}. For the Walecka parameter set we get $r_0 |U'(2r_0)|=12U_{0}=12|U_m|\simeq 1300$\,MeV, for the CD Bonn fit parameters $\simeq 500$\,MeV,  and for the Reid parameterization, $\simeq 1200$\,MeV. Therefore, the temperature dependence of $B_{<}^{\rm (cl)}$ is much weaker for the LJ potential, then for the Yukawa-type  potentials, compare solid and dash-dotted lines in Fig.~\ref{fig:V-repl}.
At $T\sim 150\mbox{--}200$\,MeV the transition to sub-nucleonic degrees of freedom is expected to occur\footnote{Application of the $\Phi$ and virial approaches to quark degrees of freedom was discussed, e.g., in \cite{Fischer2018,Blaschke2021}}. The temperature scale in Figs.~\ref{fig:V-repl}  and ~\ref{fig:Bclass}, ~\ref{fig:Bquant} below is extended to unrealistically high values only for illustration properties. It will further help us to better demonstrate relation between different asymptotic regimes.

\begin{figure}
\centering
\includegraphics[width=7cm]{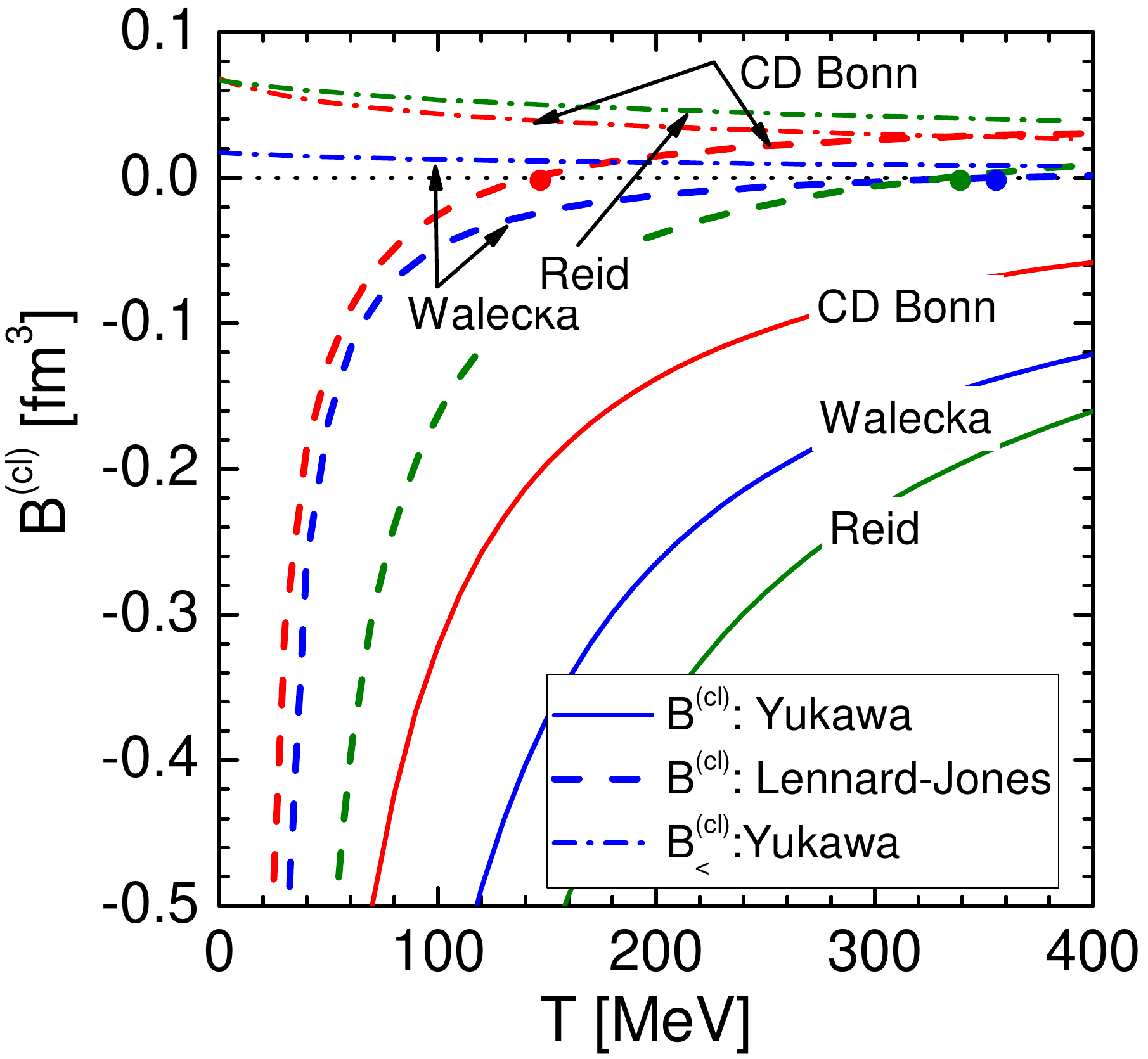}
\caption{Classical limit of the second virial coefficient $B^{\rm (cl)}$, Eq.~(\ref{B2cl-decomp}), for a  neutron  gas ($m_f=m_N=938$\,MeV) as a function of the temperature for  various potentials.
Solid lines show  results for the Yukawa-type potentials (\ref{U-potR}) with parameters corresponding to the Walecka model (labeled by W), see Eq.~(\ref{Uomsig-par-W}),  fitted to the CD Bonn potential (labeled by CD Bonn), see Eq.~(\ref{Uomsig-par-Bonn}), and to the  Reid potential (\ref{U-Reid}) (labeled by R).
Dashed lines show coefficients $B^{\rm (cl)}$ calculated for the Lennard-Jones potential (\ref{LJ-pot}) with the parameters $U_0$ and $r_0$ adjusted to reproduce the depths and  repulsion ranges  of the Yukawa-type potentials with the same labels as for the solid lines.
Dash-dotted lines show the separated contributions $B_{<}^{\rm (cl)}$, Eq.~(\ref{B1cl-def}) calculated for the Yukawa-type potentials. Bold dots indicate the Boyle temperatures for $B^{\rm (cl)}$ calculated with the LJ potentials.
}
\label{fig:Bclass}
\end{figure}

In Fig.~\ref{fig:Bclass} we compare classical coefficients $B^{\rm (cl)}$ calculated for various potentials discussed in Section~\ref{ssec:test-pot}. Solid curves are calculated for the potentials based on the Yukawa interaction, Eq.~(\ref{U-potR}) with Walecka parameters (\ref{Uomsig-par-W}), the Reid potential (\ref{U-Reid}), and the CD Bonn fit potential with parameters (\ref{Uomsig-par-Bonn}). In these cases, the quantity $B^{\rm (cl)}$ does not cross zero, i.e., there is no Boyle temperature. On the other hand, for the LJ potentials shown by dashed lines with the depths and repulsion ranges adjusted to the values for the Walecka, CD Bonn fit, and Reid potentials, the Boyle temperatures exist. For the Walecka and Reid potentials, the Boyle temperature proves to be unrealistically high, $>300$\,MeV,  the transition to quark matter should occur already at much lower temperatures, as we have mentioned. The smallest value of $T_{\rm B}$ corresponds to the potential adjusted to the values for the CD Bonn potential. Indeed, in this case $2r_0$ is the largest. The absence of the Boyle temperature for the Yukawa potentials is associated with very sharp decrease of the repulsive contribution $B_{<}^{\rm (cl)}$ with the temperature as illustrated in Fig.~\ref{fig:Bclass}.

The term $B_{>}^{\rm (cl)}$ quantifies the attractive part of the potential. It varies rapidly with temperature and depends strongly on the form of the potential. However, in the limit of high temperatures, $T\gg |U_m|$, its temperature dependence simplifies as we can expand $e^{-U(r)/T} \simeq 1-U(r)/T$ and obtain
\begin{align}
B_{>}^{(\rm cl)}(T\gg|U_m|)\approx \frac{2\pi}{T}\intop_{2r_0}^\infty r^2 U(r)\rmd r = -\frac{|U_m|}{T}\frac{v_{\rm a}^{\rm (eff)}}{2}\,,
\label{Bcl-2}
\end{align}
where temperature is related to the potential depth $|U_m|$ and $v_{\rm a}^{\rm (eff)}$ characterizes the spatial extension of the attractive potential pocket
\begin{align}
v_{\rm a}^{\rm (eff)}=\int_{2r_0}^\infty \rmd^3 r \Ucor(|\vec{r}\,|)/U_m\,.
\label{va-gen}
\end{align}
In the case of the Yukawa-type potential (\ref{U-potR}) this quantity is equal to
$v_{\rm a}^{\rm (eff)}\approx 5.40\,{\rm fm^3}$ for the Walecka model   (\ref{Uomsig-par-W}), $v_{\rm a}^{\rm (eff)}\approx 10.14\,{\rm fm^3}$ for the Reid potential (\ref{U-Reid}) and
$v_{\rm a}^{\rm (eff)}\approx 10.6\,{\rm fm^3}$ in case of parameters (\ref{Uomsig-par-Bonn}) adjusted to reproduce the CD Bonn potential. We see that typical values of the attraction volumes vary strongly. For the LJ potentials with the adjusted parameters ($U_m$ and $r_0$) the attraction volumes are smaller by a factor varying from 3 to 7: $v_{\rm a}^{\rm (eff)} \approx0.76\,{\rm fm^3}$ for the Walecka parameters and $v_{\rm a}^{\rm (eff)}\approx 3.0\,{\rm fm^3}$ for  the Reid potential and for the CD Bonn fit parameters.

From Fig.~\ref{fig:Bclass} (see solid and correspondingly marked dash-dotted lines) we conclude that in the classical limit the attractive part of the Yukawa-based nucleon potentials, $|B_>|$, exceeds the repulsive term, $B_< >0$ for all temperatures of our interest. It is not the case for  the LJ-based potentials (see dashed  lines in Fig.~\ref{fig:Bclass}) when there exist Boyle temperatures and attraction wins for $T<T_{\rm B}$ the repulsion wins for $T>T_{\rm B}$.

\subsubsection{Square barrier-well potential}

To get a feeling about a role played by  the repulsive and attractive parts of the potential in expressions for  the classical virial coefficients, let us consider a square barrier-well potential (\ref{Usw}).
For the coefficient $B^{\rm (cl)}$, see Eqs.~(\ref{B1cl-def}) and (\ref{B2cl-def}), we find
\begin{align}
B_<^{\rm (cl)} &= \frac12
\big( 1-e^{-U_{\rm r}/T} \big)  v^{\rm (sw)}_{\rm r} \,,
\nonumber\\
B_>^{\rm (cl)} &=
-\frac12\big(e^{U_{\rm a}/T} -1\big)v^{\rm (sw)}_{\rm a} \,,
\label{Bsw}
\end{align}
where $v_{\rm r}^{\rm (sw)}$ and $v_{\rm a}^{\rm (sw)}$ are the repulsion and attraction volumes defined in Eq.~(\ref{vr-va-def}).

In the literature one often considers the specific limit $U_{\rm r}\to \infty$, i.e., the potential with a impenetrable hard core.  From Eq.~(\ref{Bsw})  for the hard core potential ($U_{\rm r}\to\infty$) we find
\begin{align}
T_{\rm B}=\frac{U_{\rm a}}{\ln \Big[1+\frac{v^{\rm (sw)}_{\rm r}}{v^{\rm (sw)}_{\rm a}}\Big]}\,.\label{BoylHard}
\end{align}
For $U_{\rm r}<\infty$  the Boyle temperature increases compared to the hard-sphere case, since then $v^{\rm (sw)}_{\rm r}$ in Eq. (\ref{BoylHard})  should be replaced to $v^{\rm (sw)}_{\rm r}(1-\exp(-U_{\rm r}/T))$.

If repulsion is very strong and the attraction is weak in comparison with the temperature, i.e. $U_{\rm r}\gg T$ and $U_{\rm a}\ll T$, then using (\ref{Bsw}) we obtain the following second-order virial (v) equation of state:
\begin{align}
P_{\rm v} &= Tn(1 +Bn) +O(n^3)
\nonumber\\
  &= Tn\big(1+\half v^{\rm (sw)}_{\rm r}n\big)-n^2\frac{ {U_{\rm a}v^{\rm (sw)}_{\rm a}}}{2} + O(n^3)\,.
\label{PvW}
\end{align}
In this approximation the repulsion always wins over the attraction and the Boyle temperature is
\begin{align}
T_{\rm B}=U_{\rm a} v^{\rm (sw)}_{\rm a}/v^{\rm (sw)}_{\rm r}\,.\label{PvW1}
\end{align}
The relation $T\gg U_{\rm a}$, which is used in derivation of (\ref{PvW1}), holds here only for $v^{\rm (sw)}_{\rm a}\gg v^{\rm (sw)}_{\rm r}$.
Also, Eq.~(\ref{PvW1}) follows from Eq.~(\ref{BoylHard}) for $v^{\rm (sw)}_{\rm a}\gg v^{\rm (sw)}_{\rm r}$.
Note that taking the values of the attraction and repulsion volumes for the LJ potentials with parameters $U_0$ and $r_0$ adjusted to reproduce the depths and repulsion ranges of W, CD Bonn fit and Reid potentials ($v_{\rm r}/v_{\rm a}^{\rm (eff)}\simeq 0.4$ for the W potential and $\simeq 0.3$ for the CD Bonn fit and Reid potentials), and taking $U_{\rm a}\sim -U_m$ in (\ref{PvW1}) we estimate corresponding values of the Boyle temperature $T_{\rm B}[{\rm W}]\simeq 260$\,MeV, $T_{\rm B}[{\rm R}]\simeq 320$\,MeV, and $T_{\rm B}[{\rm W}]\simeq 140$\,MeV. These values are close to that we see in Fig.~\ref{fig:Bclass}.

\subsection{Extrapolated virial equations of state}\label{ssec:extrapol-vEoS}

Calculations of higher virial coefficients, $C$, $D$, $\dots$, become a  complicated numerical problem even for simplest potentials. Therefore, the density expansion (\ref{P-class-cor}) and the fugacity expansion (\ref{P-class-z}) are usually limited to several first coefficients. To enlarge the applicability range of the density expansion, one can  extrapolate the density dependence of the pressure by some methods (e.g. using  Pad\'e scheme). For example by such  a  modification of the virial expansion it proves to be possible to include a variety of new phenomena lying beyond the perturbative scheme, e.g. those associated with  presence of the critical point of the first-order phase transition. So, let us demonstrate abilities of such extrapolations.

The classical example of such an approach is the van der Waals equation of state, when in Eq.~(\ref{PvW}) one goes beyond a strict density expansion doing the replacement
\begin{align}
1+v^{\rm (sw)}_{\rm r}n/2\to
(1-v^{\rm (sw)}_{\rm r}n/2)^{-1}\,.
\label{repl}
\end{align}
Then one immediately recovers the renowned van der Waals ({\rm vdW}) form of the equation of state,
\begin{align}(P_{\rm vdW}+ n^2\frac{U_{\rm a}v^{\rm (sw)}_{\rm a}}{2})(1-\frac{v^{\rm (sw)}_{\rm r}n}{2})=Tn\,.
\label{PvdW}
\end{align}
The replacement leading to Eq.~(\ref{PvdW}) generates high order virial coefficients, which are absent in the original  Eq.~(\ref{PvW}), e.g. $C^{\rm (vdW)}=(v_{\rm r}^{\rm (sw)}/2)^2$ and also infinite tower of higher-order coefficients, so the $n$-th order coefficient is $(v_{\rm r}^{\rm (sw)}/2)^{n-1}$.

For further unified treatment of various forms of the equation of state it is convenient to introduce the dimensionless quantities
\begin{align}
&\widetilde{T}=\frac{T}{U_{\rm a}}\,,\,\,\widetilde{n}=\frac12v_{\rm r}^{\rm (sw)} n\,,\,\, \eta =\frac{v_{\rm a}^{\rm (sw)}}{v_{\rm r}^{\rm (sw)}}\,,\quad  \widetilde{P}=P\frac{ v_{\rm r}^{\rm (sw)}}{2 U_{\rm a}}\,.
\label{pvdw-param}
\end{align}
The  second-order virial equation of state (\ref{PvW}) can be written as
\begin{align}
\widetilde{P}_{\rm v}=\widetilde{T}{\widetilde{n}} + (\widetilde{T}-\eta)\widetilde{n}^2\,.
\label{PvEoS}
\end{align}
Thus, the dimensionless second virial coefficient is $\widetilde{B}=2B/v_{\rm r}^{\rm (sw)}=(\widetilde{T}-\eta)/\widetilde{T}$.
We notice that following (\ref{Bsw}), (\ref{BoylHard}) for $U_{\rm r}\gg T$ and $U_{\rm a}\ll T$ there is a simple relation between
the parameter $\eta$ and the dimensionless Boyle temperature, $\widetilde{T}_{\rm B}=\eta$, where  $\widetilde{T}_{\rm B}=T_{\rm B}/U_{\rm a}$.

In terms of (\ref{pvdw-param}) the van der Waals equation of state (\ref{PvdW}) has the form
\begin{align}
&\widetilde{P}_{\rm vdW}=\frac{\widetilde{T}{\widetilde{n}}}{1-\widetilde{n}}
-\eta \widetilde{n}^2\,.
\label{pvdw-dl}
\end{align}
Starting with the information about the two virial coefficients in Eq.~(\ref{PvW}), one can follow a different path, which we outlined in Section~\ref{ssec:selfcons-virial}, and fix the relation between $z$ and density, Eq.~(\ref{z-scv}), and the pressure $P_{\rm tcv}$  (\ref{p-scv}). Replacing in Eq.~(\ref{p-scv}) the combination $B=-b_2\lambda^3/b_1^2$ by $(v_{\rm r}^{\rm (sw)}/2)(\widetilde{T}-\eta)/\widetilde{T}$ and expressing everything in dimensionless parameters (\ref{pvdw-param}) we obtain
\begin{align}
&\widetilde{P}_{\rm tcv}=\frac{\widetilde{T}{\widetilde{n}}}{2}
-\frac18
\frac{\widetilde{T}^2}{\widetilde{T}-\eta}\Big[\sqrt{1-8\widetilde{n} \frac{\widetilde{T}-\eta}{\widetilde{T}}}-1\Big] \,.
\label{Pscv}
\end{align}
Both extrapolations (\ref{PvdW}) and (\ref{Pscv}) are to be compared with the original expression (\ref{PvEoS}).
Then, expanding Eqs.~(\ref{pvdw-dl}) and (\ref{Pscv}) we obtain
\begin{align}
\widetilde{P}_{\rm vdW}&=\widetilde{P}_{\rm v}+\widetilde{T}\,\big(\widetilde{n}^3+\widetilde{n}^4+O(\widetilde{n}^5)\big)\,,
\\
\widetilde{P}_{\rm tcv}&=\widetilde{P}_{\rm v} + 4\frac{(\widetilde{T}-\eta)^2}{\widetilde{T}}\widetilde{n}^3 + 20\frac{(\widetilde{T}-\eta)^3}{\widetilde{T}^2}\widetilde{n}^4+O(\widetilde{n}^5)\,.\nonumber
\end{align}
We observe that at the Boyle temperature both ``tcv''  and ``v'' pressures coincide with  the ideal gas pressure,
$P_{\rm tcv}(\widetilde{T}_{\rm B}) =P_{\rm v}(\widetilde{T}_{\rm B})= P_{\rm id}(\widetilde{T}_{\rm B}) = \widetilde{T}_{\rm B} \widetilde{n}$, whereas for the ``vdW'' case there are corrections of the order $\widetilde{n}^3$ and higher.

In Fig.~\ref{fig:vdW-com} the solid lines show $\widetilde{P}_{\rm vdW}$ as  a function of $\widetilde{n}$ for various temperatures, $\widetilde{T}$, and for $\eta=1$. Dashed lines depict $\widetilde{P}_{\rm v}$ given by Eq.~(\ref{PvEoS}), and dash-dotted lines demonstrate behavior of $\widetilde{P}_{\rm tcv}$ given by (\ref{Pscv}) for the same values of parameters $\widetilde{T}$ and $\eta$. All three equations of state agree at low densities and deviate at $\widetilde{n}>0.1$, whereby $\widetilde{P}_{\rm v}$ demonstrates presence of
a strong attractive interaction (the dashed curves bend down with increasing $\widetilde{n}$) and $\widetilde{P}_{\rm tcv}$ shows the  behavior typical for the repulsive interactions (the dash-dotted curves monotonously grow with increasing $\widetilde{n}$).

Note that all considered equations of state possess the scaling property: the plots do not change under the replacement $\eta\to s\,\eta$, $\widetilde{T}\to s\,\widetilde{T}$, and $\widetilde{P}\to s\,\widetilde{P}$, i.e., $\widetilde{P}(\widetilde{T},\widetilde{n},\eta)=\widetilde{P}(s\widetilde{T},\widetilde{n},s\eta)/s$.

In the ``tcv'' case  $P(n)$ demonstrates absence of an extremum. The density is limited for $\widetilde{T}>\eta$ by the condition $\widetilde{n}<\widetilde{n}_{\rm l}=\frac{1}{8(1 - \eta/\widetilde{T})}$ and the pressure becomes complex at larger $\widetilde{n}$.
Close to the end-point we can write the pressure as
\begin{align}
\widetilde{P}_{\rm tcv}\approx\frac{3}{2}\widetilde{T} \widetilde{n}_{\rm l}
-\widetilde{T} \widetilde{n}_{\rm l}^{1/2}(\widetilde{n}_1-\widetilde{n})^{1/2}\,.
\end{align}

In the ``v''-case the pressure is an increasing function of the density for $\widetilde{T}\ge\eta$ but it demonstrates  presence of instability for $ \widetilde{T}<\eta$ and for density $\widetilde{n}> \widetilde{T}/[2(\eta-\widetilde{T})]$ .

For the ``vdW'' equation of state
the pressure $\widetilde{P}_{\rm vdW}$ becomes infinite at $\widetilde{n}=1$.
Among the considered forms of the extrapolations of the virial equation of state, only the van der Waals form (\ref{PvdW}) demonstrates the presence of a critical end-point at $T=T_{\rm c}$, when $\rmd P_{\rm vdW}(T_{\rm c},n)/\rmd n =0$ and $\rmd^2 P_{\rm vdW}(T_{\rm c},n)/\rmd^2n =0$, and we deal with the first-order phase transition  for $T<T_{\rm c}$. We have
\begin{align}
\widetilde{T}_{\rm c}=\frac{8}{27}\eta\,,\quad \widetilde{n}_{\rm c}=\frac13\,,\quad \widetilde{P}^{\rm vdW}_{\rm c}=\frac{\eta}{27}\,.
\label{crit-point}
\end{align}
From these relations we can express parameters of the potential through the parameters of the critical point
\begin{align}
v_{\rm r}^{\rm (sw)}=\frac{2}{3n_{\rm c}}\,,\quad U_{\rm a} v_{\rm a}^{\rm (sw)}=\frac94\frac{T_{\rm c}}{n_{\rm c}}\,.\label{estimateEV}
\end{align}
Applying these relations to the liquid-gas phase transition in nuclear matter with empirical parameters~\cite{Elliot2013}  $T_{\rm c}=(17.9\pm 0.4)$\,MeV,  $n_{\rm c} = 0.06\pm 0.01\,{\rm fm}^{-3}$ and the critical pressure $P_{\rm c}=0.31\pm 0.07\,{\rm MeV/fm^{3}}$, using central values we get
\begin{align}
v_{\rm r}^{\rm (sw)}\simeq 11.1\,{\rm fm}^3\,,\quad U_{\rm a} v_{\rm a}^{\rm (sw)}\simeq 671.3\,{\rm MeV \cdot fm^3}\,.
\label{LG-U-par}
\end{align}
The obtained repulsion volume is much larger than those for the Walecka potential ($v_{\rm r}^{(\rm W)}\simeq 0.3\,{\rm fm}^3$) and for
the CD Bonn fit potential ($v_{\rm r}^{(\rm CD Bonn)}\simeq 1\,{\rm fm}^3$). Also, the attraction potential parameter, $U_{\rm a} v_{\rm a}^{\rm (eff)}$,  is larger than the values calculated using the Yukawa potential: $U_{\rm a} v_{\rm a}^{\rm (eff)}\sim 540\,{\rm MeV \cdot fm^3}$  for the Walecka parameters  and  $\sim 420\,{\rm MeV \cdot fm^3}$ for the CD Bonn choice.

Realistic equations of state of the nuclear matter show a much more complicated temperature dependence than that for the classical van der Waals equation of state. To take this into account for $|T-T_{\rm c}|\ll T_{\rm c}$, Ref.~\cite{Skokov:2009yu,Skokov:2010dd} introduced a modified van der Waals (mVW) equation of state with $P_{\rm mvdW}= f(T)P_{\rm vdW}$ with $f(T)=(1-2(T-T_{\rm c})^2)/(1+4(T-T_{\rm c})^2)$ chosen to fulfill condition $dP(n_{\rm c}, T_{\rm c})/dT=0$. Then, the general system of equations of non-ideal non-relativistic hydrodynamics was numerically
solved for densities and temperatures nearby the critical end-point. The extension of the van der Waals equation of state to the quantum statistics was considered Ref.~\cite{Redlich-vdW-quant}.

%
%\begin{align}
%\frac{\rmd P_{\rm v}}{\rmd n}   &= \widetilde{T} + 2(\widetilde{T}-\eta)\widetilde{n}
%\nonumber\\
%\frac{\rmd P_{\rm scv}}{\rmd n} &= \frac{\widetilde{T}}{2} + \frac{\widetilde{T}}{2 \sqrt{1-8 \widetilde{n}\frac{(\widetilde{T} - \eta )}{\widetilde{T}}}}
%\end{align}

\begin{figure}
\centering
\includegraphics[width=7cm]{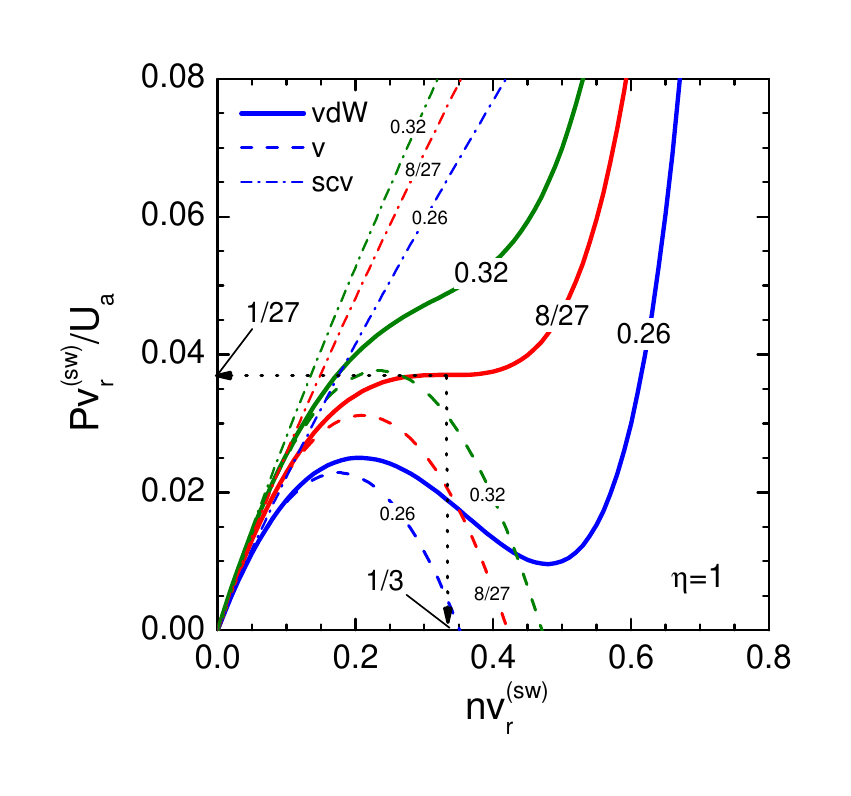}
\caption{Comparison of different extrapolations of the virial equation of state: the second-order virial equation of state (``v'', dashed lines) given by Eq.~(\ref{PvEoS}), the van der Waals form (``vdW'', solid lines), Eq.~(\ref{pvdw-dl}), and the thermodynamically consistent viral equation of state (``tcv'', dash-dotted lines), Eq.~(\ref{Pscv}). Pressure, density and temperature are measured in dimensionless units introduced in Eq.~(\ref{pvdw-param}). Calculations are performed with the parameter $\eta={v_{\rm a}^{\rm (sw)}}/{v_{\rm r}^{\rm (sw)}}=1$. Numbers near curves show  values of $\widetilde{T}=T/U_{\rm a}$. Line with label  $8/27$ corresponds to the critical temperature, $\widetilde{T}_{\rm c}$
and the critical pressure and density are given in Eq.~(\ref{crit-point}).
}
\label{fig:vdW-com}
\end{figure}

Note that the  replacement (\ref{repl}) leading to phenomenologically successful van der Waals model can be interpreted, as reduction of the volume allowed for the particle motion, $V\to V-v_{\rm p}N$, because of the particle finite volume, $v_{\rm p}$. This observation was used as motivation of \emph{excluded volume} (``ev'') model, in which the canonical partition function is taken in the form $Z_{\rm ev}(T,N,V)=Z_{\rm id}(T,N,V-v_{\rm p} N)\theta(V-v_{\rm p} N)$, with $Z_{\rm id}$, being the partition function of the ideal gas. The attractive part in such a model of the interaction is neglected.  The corresponding grand-canonical partition function is $\mathcal{Z}_{\rm ev}(T,\mu,V)=\sum_{N=0}^\infty \exp(\mu_f N/T) Z_{\rm ev}(T,N,V)$. In Refs.~\cite{EV-refs,EV-refs1} it was shown that the pressure $P_{\rm ev}=T\ln(\mathcal{Z}_{\rm ev})/V$ can be expressed as the pressure of the ideal gas with the chemical potential shifted as $\mu\to \mu - v_{\rm p} P_{\rm ev}$, i.e.,
\begin{align}
P_{\rm ev}(T,z)=\intop \frac{2\rmd^3 p}{(2\pi)^3}\frac{T}{\alpha}\ln
\Big(1+\alpha z e^{-\frac{e_p+v_{\rm p} P_{\rm ev}}{T}}\Big)\,.
\end{align}
Expanding this equation in $z$  we obtain the following relations for the virial coefficients:
\begin{align}
b_1^{\rm (ev)}&=2\,,\quad
b_2^{\rm(ev)} =-\frac{\alpha}{2^{3/2}} - 4 \frac{v_{\rm p}}{\lambda^3}\,,
\nonumber\\
b_3^{\rm(ev)} &=\frac{2 \alpha^2}{3^{5/2}}
+\frac{3\alpha v_{\rm p}}{2^{1/2}\lambda^3}+12\frac{v_{\rm p}^2}{\lambda^6}\,,
\nonumber\\
b_4^{\rm(ev)} &= -\frac{\alpha^3}{2^4}
-\frac{2^4 \alpha^2 v_{\rm p}}{3^{5/2} \lambda^3}
-\frac{\alpha^2 v_{\rm p}}{4 \lambda^3}
-\alpha\frac{8 \sqrt{2}   v_{\rm p}^2}{\lambda^6}
-\frac{128 v_{\rm p}^3}{3 \lambda^9}\,,
\label{bs-EV}
\end{align}
where, as it was used above,  $\alpha=1$ takes into account the Fermi-Dirac statistics and  $\alpha=0$ corresponds to the Boltzmann statistics. The values of the virial coefficients in the density expansions in the classical limit, i.e. $\alpha=0$, are
\begin{align}
B^{\rm (ev)} =v_{\rm p}\,,\quad
C^{\rm (ev)} =v_{\rm p}^2\,.
\label{BC-EV}
\end{align}
The next term is equal to $v_{\rm p}^3$. Hence the virial coefficients of the excluded volume model coincide with the coefficients $B^{\rm (vdW)}$, $C^{\rm (vdW)}$ for the van-der-Waals equation of state, if we drop attractive part of the potential $U_{\rm a}\to 0$ and identify $v_{\rm p}=v_{\rm r}^{\rm (sw)}/2=\frac{16\pi}{3}r_0^3$. The coefficient $1/2$ takes into account that the particle excluded volume $v_{\rm p}$ is calculated for one particle, whereas $v^{\rm (sw)}_{\rm r}$ is the repulsion volume for a pair of particles.
We conclude that the often used excluded volume model is nothing else than the van-der-Waals model, where the inter-particle attraction is omitted. However, within the former approach the effects of the quantum statistics (terms with $\alpha$ in Eq.~(\ref{bs-EV})) can be naturally taken into account. These effects could be neglected for $v_{\rm p}\gg \lambda^3$, or more precisely for temperatures $T\gg 2\pi/(2^{7/3} m_f v_p^{2/3})\approx 2/(m_f [v_{\rm r}^{\rm (sw)}]^{2/3})$, that for nucleons, $m_f=m_N$, and for $v_{\rm r}^{\rm (sw)}$ from (\ref{LG-U-par}) yields $T\gg 10$\,MeV. Thus, we conclude that the classical limit of the van der Waals equation of state can be marginally applicable for the description of the nuclear liquid-gas phase transition.

A van-der-Waals--type model for the nuclear matter equation of state, which takes into account not only the excluded volume effects but also attractive part of the $NN$ interaction and the Fermi statistics was developed in Ref.~\cite{Vovchenko-Anchshkin-MG}. The full account of the Fermi statistics allowed to use the model at $T=0$ and fit the model parameters to the properties of nuclear matter at saturation. The parameters are $v_{\rm r}^{\rm (sw)}=6.8$\,fm$^3$ and $U_{\rm a} v_{\rm a}^{\rm (sw)}=659$\,MeV$\cdot$fm$^3$, where the excluded volume becomes smaller compared to Eq.~(\ref{LG-U-par}) and the attractive part $U_av_{\rm a}^{(sw)}$ bigger. With these parameters the critical temperature of the liquid -- gas phase transition was found to be about 20\,MeV. If extended to high temperatures, this model shows that the attraction at low temperature changes to repulsion at the Boyle temperature of 100\,MeV~\cite{Vovchenko-MGS}.

The main assumption of the original excluded volume model is that the volume $v_{\rm p}$ is constant. The latter assumption holds only for the hard sphere potential, when $U(r<2r_0)\to \infty$. Any realistic potential has a finite stiffness  of the core, e.g.,  for the hadronic potentials based on the Yukawa one-boson exchange model the  potential behaves $\propto 1/r$ at small distances. Dependence of the repulsion volume on the temperature was illustrated in Fig.~\ref{fig:V-repl} for several nucleon-nucleon potentials (solid lines). We see that in the range of temperatures $50\,{\rm  MeV}\lsim T \lsim 150\,{\rm MeV}$ the repulsion volume is reduced by (10--25)\%.

Assuming that $v_{\rm p}$ is constant we can compare the virial coefficients for the excluded volume model with the results of calculations for the hard-sphere (hs) gas done over the years~\cite{HS-ref}. The analytic calculations of the first two viral coefficients yield
\begin{align}
B^{\rm (hs)}= \frac12 v_{\rm r}^{\rm (sw)}= v_{\rm p}\,,\quad
C^{\rm (hs)}= \frac{5}{24} (v_{\rm r}^{\rm (sw)})^2=\frac56 v_{\rm p}^2.
\label{BC-hs}
\end{align}
We see that the second virial coefficients in the ``hs'' and the ``ev'' models coincide, whereas the third virial coefficient of the ``hs'' model is $5/6$ times smaller than the coefficient of the ``ev'' model (\ref{BC-EV}). Thus we see that these models are not exactly equivalent, but give similar result for densities
\begin{align}
n\ll B^{\rm (hs)}/C^{\rm (hs)} \sim \frac{1}{v_{\rm p}} =\frac{2}{v_{\rm r}^{\rm (sw)}}\,,
\end{align}
when one can restrict himself to the second virial order.

\subsection{Limit of high temperatures ($T\gg |U_{\rm typ}|$)}\label{clashightemp}

Consider now the virial coefficients (\ref{bN-cl}) in the limit when the typical magnitude of the potential, $U_{\rm typ}$, is smaller than temperature, i.e. $T\gg |U_{\rm typ}|$. Then we find
\begin{align}
%b_2^{\rm (cl)}\approx
b_2^{\rm (cl)}(T\gg |U_{\rm typ}|)&\approx -\frac{d_f^2}{2T}\int\frac{\rmd^3r}{\lambda^3}\Ucor(|\vec{r\,}|)
\nonumber\\
&= -\frac{d_f^2}{2\lambda^3}\frac{\Umom(0)}{T}\,,
\label{b2cl-H}\\
b_3^{\rm (cl)}(T\gg |U_{\rm typ}|) &\approx \frac{d_f^3}{2T^2}\int\frac{\rmd^3r_1\rmd^3r_2}{\lambda^6}\Ucor(|\vec{r}_1|)\Ucor(|\vec{r}_2|)
\nonumber\\
&= \frac{d_f^3}{2\lambda^6}\frac{\Umom^2(0)}{T^2}\,.
\label{b3cl-H}
\end{align}
In this estimate we assume that the repulsive core is either absent or limited in the height. For nuclear Yukawa-type potentials the repulsive core is relatively soft and its contribution is rather small and decreases with a temperature increase, as discussed in Section~\ref{classicalspec}, see also Fig.~\ref{fig:Bclass}. Thus, at least for Yukawa-type potentials we have considered the repulsive core term can be indeed neglected.

Applying these results to the system of particles with the degeneracy factor $d_f=2$, as for the gas of spin 1/2 fermions considered in Section~\ref{ssec:fugacity-Phi}, we observe that they agree  with the corresponding Hartree contributions to the coefficients $b_{2,\Phi}^{\rm (int,H)}$ and $b_{3,\Phi}^{\rm (int,H)}$, see Eqs.~(\ref{b2Phi-H}) and (\ref{b3Phi-H}), i.e.,
\begin{align}
b_{2,\Phi}^{\rm (int,H)} &= b_2^{\rm (cl)}(T\gg |U_{\rm typ}|)\,,
\nonumber\\
b_{3,\Phi}^{\rm (int,H)} &= b_3^{\rm (cl)}(T\gg |U_{\rm typ}|) +O\Big(\frac{|\Umom(0)|}{\lambda^3 T}\Big)\,.
\end{align}
In the case of the Yukawa potential, $\Umom(0)$ is given by Eq. (\ref{U-pot}). We find out that the Hartree part of the coefficient $b_{2,\Phi}^{\rm (int, TS)}$ and $b_{3,\Phi}^{\rm (int, TS)}$ calculated within the $\Phi$-derivable approach with the TS diagrams agrees with
the classical expressions for the second and third virial coefficients, $b_2^{\rm (cl)}$ and $b_3^{\rm (cl)}$.
Thus, in the high temperature limit,
the pressure calculated within the $\Phi$ derivable approach tends to its classical limit determined by the first diagram (a) in Fig.~\ref{fig:Phi-U-ser}, which is linear in the potential and produces the Hartree contributions to all virial coefficients. The exchange diagrams and diagrams with loops are of quantum origin and their contributions die out in the classical limit.

Summarizing, in this Section we introduced the fugacity and density expansions of the pressure and discussed various truncation schemes of the expansion. We demonstrated that in the classical limit, one can separate contributions from the repulsive and attractive parts of the potential in the second virial coefficient. We showed that popular models, describing the effects of a first-order phase transition, e.g., the liquid--gas phase transition in nuclear matter, are based on the perturbative treatment of the attractive contribution, whereas some extrapolation scheme is applied to the repulsive contribution. As a result, the phenomenologically adjusted parameters of such models do not agree with those that follow from the typical nuclear potential.
We found that for all considered Yukawa-based nuclear potentials the repulsive part exceeds the attractive part only for unreasonably high values of the temperature ($T>400$~MeV).
This is an important point, as there are models that rely on purely repulsive excluded volume interaction to describe the high-temperature low-density behavior in heavy-ion collisions. Our finding, which indicates the absence of a Boyle temperature except for classical LJ potentials, shows a limitation of such  models. We also demonstrated that the classical limit of the virial coefficients $b_2$ and even $b_3$ can be reproduced within the $\Phi$ derivable approach with the account of the first Hartree diagram in $\Phi$.

\section{Quantum virial coefficients}
\label{sssec:cluster}

\subsection{General approach}\label{Generalap}

The quantum mechanical derivation of the virial expansion for the pressure starts with the similar formal expansion in terms of fugacities
\begin{align}
\mathcal{Z}=\sum_{N=0}^{\infty} z^N W_N \,,\,\,
W_N = \Tr_N[e^{-\hat{H}^{(N)}/T}]\,,\,\,%\hat{Q}_N=e^{-\hat{H}^{(N)}/T}
\end{align}
as in classical case,
with the only difference that the quantum $N$-body canonical partition function, $W_N$, is calculated as the trace of $e^{-\hat{H}^{(N)}/T}$ with the $N$-body Hamiltonian, $\hat{H}^{(N)}$ taken over the quantum $N$-body states~\cite{Slater}. For fermions  this means that wave-functions are anti-symmetrized. In the coordinate representation the trace can be written as
\begin{align}
&\Tr_N[\dots]=\frac{1}{N!}\sum_{s_1=\pm }\dots\sum_{s_N=\pm }\intop \rmd^3 r_1 \dots \rmd^3 r_N
\nonumber\\
&\quad\times
\bra{\{\vec{r}_1,s_1;\dots;\vec{r}_N,s_N\}} (\dots)
\ket{\{\vec{r}_1,s_1;\dots; \vec{r}_N,s_N\}}\,,
\label{Sp-def}
\end{align}
where $s_i$ indicate spin projections of the $i$-th particle.
The curly brackets in the in- and out-states indicate the full anti-symmetrization.
For fermions interacting through the spin-independent two-body potential, $\Ucor(|\vec{r}_{12}|)$, the single-, two- and three-body Hamiltonians
can be written as
\begin{align}
&\hat{H}^{(1)}=\hat{T}_1\,,\quad \hat{H}^{(2)}=\hat{T}_1 +\hat{T}_2 + \Ucor(|\vec{r}_{12}|)\,,
\\
&\hat{H}^{(3)}=\hat{T}_1 +\hat{T}_2 +\hat{T}_3
+ \Ucor(|\vec{r}_{12}|) + \Ucor(|\vec{r}_{13}|) + \Ucor(|\vec{r}_{23}|),
\nonumber\end{align}
where $\hat{T}_i=\frac{-1}{2m_f}\nabla^2_{\vec{r}_i}$ is the operator of kinetic energy of the $i$-th particle. Note that in contrast to Eq.~(\ref{H-cl}) we deal here with not commuting quantum operators of kinetic and potential energies.

%The grand-canonical potential, being the logarithm of the partition function, can be re-expanded in terms of the $N$-body cluster operator expressible in terms of operators $\hat{Q}_N=e^{-\hat{H}^{(N)}/T}$.

The cluster expansion of the grand-canonical potential similar to (\ref{Omega-virial-cl}) and (\ref{bN-W-cl}) can be made also for the quantum statistics~\cite{Kahn-Uhlenbeck} with the result
\begin{align}
\Omega(T,\mu) &= -T\ln \mathcal{Z}=-\frac{TV}{\lambda^3}\sum_{N=1}^{\infty} z^N b_N\,,\,\,
\label{Omega-virial}
\end{align}
and the virial coefficients expressed through the $N$ particle traces of the operators $\hat{Q}_N=e^{-\hat{H}^{(N)}/T}$, e.g.,
\begin{align}
b_1 &=\frac{\lambda^3}{V}\Tr_1[\hat{Q}_1]\,,
\nonumber\\
b_2 &=\frac{\lambda^3}{2V}\Big(2\Tr_2[\hat{Q}_2]-\Tr_1[\hat{Q}_1]^2\Big)\,,
\label{bN-def}\\
b_3 &=\frac{\lambda^3}{6V}\Big(6\Tr_3[\hat{Q}_3] - 6 \Tr_2[\hat{Q}_2]\, \Tr_1[\hat{Q}_1]+ 2\, \Tr_1[\hat{Q}_1]^3 \Big)\,,
\nonumber
\end{align}
and so on.

\subsubsection{First virial coefficient}

Such a genuine quantum effect, as spin, produces  differences between the classical and quantum calculations already in the first virial coefficient, for which we have
\begin{align}
b_1=\frac{\lambda^3}{V}\intop \rmd^3 r_1 \sum_{s_1=\pm }
\bra{\vec{r\,}_1,s_1}\hat{Q}_1\ket{\vec{r}_1,s_1} \,.
\label{b1-expl}
\end{align}
The matrix element of $\hat{Q}_1$ is diagonal in the spin space $\bra{\vec{r\,}'_1,s'_1}\hat{Q}_1\ket{\vec{r}_1,s_1} = \delta_{s'_1s_1}
\bra{\vec{r\,}'_1}\hat{Q}_1\ket{\vec{r}_1}$. To calculate the coordinate matrix element $\bra{\vec{r\,}'_1}\hat{Q}_1\ket{\vec{r}_1}$
we insert the complete sets of the momentum states (\ref{complete-pp-rr}),
%$\intop \rmd^3\vec{p}\,\ket{\vec{p}\,}\bra{\vec{p}\,}=1$ and
use the corresponding wave-function (\ref{rp-wf}) %$\braket{\vec{r}\,|\vec{p}\,}=e^{i\vec{p}\vec{r}}/(2\pi)^{3/2}$\,
and the scalar product of momentum states (\ref{norma-pp-rr}), see Appendix \ref{app:ME-NN-pot}. %$\braket{\vec{p}\,|\vec{p\,}'}=\delta^{(3)}(\vec{p}-\vec{p\,})$.
Thus, we can write
\begin{align}
&\bra{\vec{r\,}'_1}\hat{Q}_1\ket{\vec{r}_1} =
\bra{\vec{r\,}'_1}e^{-\hat{T}_1/T}\ket{\vec{r}_1}
\nonumber\\
&\quad
=
\intop\rmd^3 p_1 \intop\rmd^3 p_2
\frac{e^{i\vec{p}_1\vec{r\,}'_1/\hbar-\vec{p}_2\vec{r}_1/\hbar}}{(2\pi\hbar)^3}
 \bra{\vec{p}_1}e^{\beta \hat{T}_1}\ket{\vec{p}_2}
\nonumber\\
&\quad= \intop\frac{\rmd^3 p}{(2\pi\hbar)^3}
e^{-\frac{\vec{p\,}^2}{2m_fT}+i\vec{p}(\vec{r\,}'_1-\vec{r}_1)/\hbar}
 =\frac{1}{\lambda^3}
e^{-\pi(\vec{r}_1-\vec{r\,}'_1)^2/\lambda^2}\,.
\label{Q1}
\end{align}
Substituting this result in (\ref{b1-expl}) and integrating over the volume for $\vec{r\,}'_1=\vec{r}_1$, after summation over particle spins we obtain
\begin{align}
b_1=2\,.
\label{b1-q}
\end{align}
This value demonstrates the degeneracy factor for the particle with the spin $1/2$.
We see that the quantum result (\ref{b1-q}) corresponds to the classical result
(\ref{bN-cl}), if $d_f=2$. The $n$-th order virial coefficient contains also the same degeneracy factor but in the power $n$, as we see from the classical expansion (\ref{bN-cl}). This $d_f$ dependence does not influence the equation of state, since it is absorbed in the relation between fugacity $z$ and the density $n$, see Eq.~(\ref{dens-virial}).

\subsubsection{Second virial coefficient}

The quantum derivation of the second virial coefficient (\ref{bN-def}) is more involved, for general formalism see~\cite{Lee-Yang-I,Lee-Yang-II,Boyd-Larsen-Kilpatrick66}.

For evaluation of the trace in the expression for $b_2$ in Eq.~(\ref{bN-def}) we need the matrix element between the antisymmetrized states
\begin{align}
\ket{\{\vec{r}_1,s_1;\vec{r}_2,s_2\}}=\frac{1}{\sqrt{2}}\big(
\ket{\vec{r}_1,s_1;\vec{r}_2,s_2}-\ket{\vec{r}_2,s_2;\vec{r}_1,s_1}
\big)\,,
\label{antisym}
\end{align}
%\begin{widetext}
so that it can be written as
\begin{align}
b_2 &= \frac{\lambda^3}{2V}\intop \rmd^3 r_1 \rmd^3 r_2   \sum_{s_1,s_2=\pm }
\label{b2-expl}\\
&\quad\times
\bra{\{\vec{r\,}_1,s_1;\vec{r\,}_2,s_2\}}
\hat{Q}_2-\half\hat{Q}_1^2
\ket{\{\vec{r\,}_1,s_1;\vec{r\,}_2,s_2\}} \,.
\nonumber
\end{align}
The matrix element between  antisymmetrized states is
\begin{align}
&\bra{\{\vec{r\,}_1,s_1;\vec{r\,}_2,s_2\}}\hat{Q}_2-\half\hat{Q}_1^2
\ket{\{\vec{r}_1,s_1;\vec{r}_2,s_2\}}
\nonumber\\
&\quad
=
\bra{\vec{r}_1,s_1;\vec{r}_2,s_2}\hat{Q}_2
\ket{\vec{r}_1,s_1;\vec{r}_2,s_2}
\nonumber\\
&\qquad
-\bra{\vec{r}_1,s_1;\vec{r}_2,s_2}\hat{Q}_2
\ket{\vec{r}_2,s_2;\vec{r}_1,s_1}
\nonumber\\
&\qquad
-\bra{\vec{r}_1,s_1}Q_1\ket{\vec{r}_1,s_1}
  \bra{\vec{r}_2,s_2}Q_1\ket{\vec{r}_2,s_2}
\nonumber\\
&
=\delta_{s_1s_1}\delta_{s_2s_2}
\bra{\vec{r}_1;\vec{r}_2}\hat{Q}_2 \ket{\vec{r}_1;\vec{r}_2}
-\delta_{s_1s_2}\delta_{s_2s_1}\bra{\vec{r}_1;\vec{r}_2}\hat{Q}_2
\ket{\vec{r}_2;\vec{r}_1}
\nonumber\\
&\qquad
-\delta_{s_1s_1}\delta_{s_2s_2}\bra{\vec{r}_1}Q_1\ket{\vec{r}_1}
  \bra{\vec{r\,}_2}Q_1\ket{\vec{r}_2}\,.
\end{align}
Then the spin summation gives
\begin{align}
&\sum_{s_{1},s_2}
\bra{\{\vec{r}_1,s_1;\vec{r}_2,s_2\}}\hat{Q}_2-\half\hat{Q}_1^2\ket{\{\vec{r}_1,s_1;\vec{r}_2,s_2\}}
\nonumber\\
&\quad
=4\bra{\vec{r}_1;\vec{r}_2}\hat{Q}_2 \ket{\vec{r}_1;\vec{r}_2}
-2\bra{\vec{r}_1;\vec{r}_2} \hat{Q}_2 \ket{\vec{r}_2;\vec{r}_1}
\nonumber\\
&\qquad
-4\bra{\vec{r}_1}Q_1\ket{\vec{r}_1} \bra{\vec{r}_2}Q_1\ket{\vec{r}_2}
\nonumber\\
&\quad=4\bra{\vec{r}_1;\vec{r\,}'_2}e^{-\hat{H}^{(2)}/T}
- e^{-(\hat{T}_1+\hat{T}_2)/T} \ket{\vec{r}_1;\vec{r}_2}
\nonumber\\
&\qquad
-2\bra{\vec{r}_1;\vec{r}_2} e^{-\hat{H}^{(2)}/T} \ket{\vec{r}_2;\vec{r}_1}\,.
\label{U1-spinsum}
\end{align}
To evaluate two-particle matrix elements in the coordinate space we perform separation of the center-of-mass motion. In the new coordinates,
\begin{align}
\vec{R} &=\frac{\vec{r}_1+\vec{r}_2}{2},\quad\,\, \vec{r}=\vec{r}_1-\vec{r}_2,\,
\end{align}
 we have
 $\hat{T}_1+\hat{T}_2=\hat{T}_{\rm c.m.}+\hat{T}_{\rm rel}$, $\hat{H}^{(2)}=\hat{T}_{\rm c.m.} +\hat{H}^{(2)}_{\rm rel}$,
and $\hat{H}^{(2)}_{\rm rel}=\hat{T}_{\rm rel} + U(\vec{r}\,),$
where $\hat{T}_{\rm c.m.}=-\frac{1}{4 m_f}\nabla_{\vec{R}}^2$ and $\hat{T}_{\rm rel}=-\frac{1}{m_f}\nabla_{\vec{r}}^2$. Now the center-of-mass motion can be separated, since $\hat{T}_{\rm c.m.}$ commutes with $\hat{T}_{\rm rel}$ and $\hat{H}^{(2)}_{\rm rel}$, and we have for the first term in (\ref{U1-spinsum}),
\begin{align}
&\bra{\vec{r}_1;\vec{r}_2}e^{-\hat{H}^{(2)}/T} - e^{-(\hat{T}_1+\hat{T}_2)/T} \ket{\vec{r}_1;\vec{r}_2}
\nonumber\\
&\quad =\bra{\vec{R\,}}e^{-\hat{T}_{\rm c.m.}/T}\ket{\vec{R}}
\bra{\vec{r\,}} e^{-\hat{H}^{(2)}_{\rm rel}/T} - e^{-\hat{T}_{\rm rel}/T}\ket{\vec{r}\,} \,,
\end{align}
and for the second one,
\begin{align}
\bra{\vec{r}_1;\vec{r}_2} e^{-\hat{H}^{(2)}/T} \ket{\vec{r}_2;\vec{r}_1}
&=\bra{\vec{R}}e^{-\hat{T}_{\rm c.m.}/T}\ket{\vec{R}\,}\nonumber\\
&\times
\bra{-\vec{r}\,} e^{-\hat{H}^{(2)}_{\rm rel}/T} \ket{\vec{r}\,}\,.
\label{CMsepar}
\end{align}
Matrix elements between the center-of-mass coordinate can be evaluated
using the result (\ref{Q1}) as
\begin{align}
\bra{\vec{R\,}'}e^{-\hat{T}_{\rm c.m.}/T}\ket{\vec{R}}=\frac{2^{3/2}}{\lambda^3}
e^{-2\pi(\vec{R}-\vec{R\,}')^2/\lambda^2}\,,
\label{Tcm-ME}
\end{align}
where we took into account the difference in the particle masses in $\hat{T}_{\rm c.m.}$ and in $\hat{T}_1$  and consider different bra-  and ket-coordinate states for generality.

Combining Eqs.~(\ref{Tcm-ME}), (\ref{CMsepar}) and (\ref{U1-spinsum}) and substituting the result in (\ref{b2-expl}) we obtain
\begin{align}
b_2 &= 2^{1/2}\intop\rmd^3 \vec{r}\,\Big[ 4\bra{\vec{r}\,} e^{-\hat{H}^{(2)}_{\rm rel}/T} - e^{-\hat{T}_{\rm rel}/T}\ket{\vec{r}}
\nonumber\\
&-2\bra{-\vec{r\,}} e^{-\hat{H}^{(2)}_{\rm rel}/T} \ket{\vec{r}\,}
\Big]\,.
\label{b2-full}
\end{align}
This expression has to be contrasted with the classical result (\ref{bN-cl}). The striking feature is the presence of the second term, which appears because of  the anti-symmetrization of the wave-function (\ref{antisym}). Due to this, the term $b_2$ does not vanish, if we put $U\to 0$, while $b_2^{\rm (cl)}=0$ for $U=0$. Also, the first term in (\ref{b2-full}) cannot be reduced to $b_2^{\rm (cl)}$,
since the operators $\hat{H}^{(2)}_{\rm rel}$ and $\hat{T}_{\rm rel}$ do not commute.

It is convenient to separate from the second virial coefficient the term corresponding to the ideal gas ($\Ucor=0$) and the term induced by the particle interaction, so that
\begin{align}
b_2=b_2^{(0)}+b_2^{\rm (int)} \,.
\label{b2-decomp}
\end{align}
The virial coefficient for an ideal Fermi gas is
\begin{align}
b_2^{(0)} &= -2^{3/2}\intop\rmd^3 r\,\bra{-\vec{r\,}} e^{-\hat{T}_{\rm rel}/T} \ket{\vec{r}\,}
\nonumber\\
&= -\frac{1}{\lambda^3}\intop\rmd^3 r e^{-2\pi \vec{r\,}^2/\lambda^2}
=-\frac{1}{2^{3/2}},
\label{b0}
\end{align}
where again we used Eq.~(\ref{Q1}) and took into account the difference between the masses in $\hat{T}_{\rm rel}$ and $\hat{T}_1$.
The second term in Eq.~(\ref{b2-decomp}) dependent on the interaction is
\begin{align}
b_2^{\rm (int)} &= 2^{1/2}\intop\rmd^3 r\,\Big[ 4\bra{\vec{r}\,} e^{-\hat{H}^{(2)}_{\rm rel}/T} - e^{-\hat{T}_{\rm rel}/T}\ket{\vec{r}\,}
\nonumber\\
&-2
\bra{-\vec{r\,}} e^{-\hat{H}^{(2)}_{\rm rel}/T} -e^{-\hat{T}_{\rm rel}/T} \ket{\vec{r}\,}
\Big]\,.
\label{b2int-full}
\end{align}
This expression can be separated in two pieces according to possible values of the total spin of colliding particles, $S=0$ and $S=1$,
\begin{align}
b_2^{\rm (int)} &=
b_{2,S=0}^{\rm (int)} +b_{2,S=1}^{\rm (int)} \,,
\label{b2int-spin-sep} \\
b_{2,S=0}^{\rm (int)} &= 2^{1/2}\intop\rmd^3 \vec{r}\bra{\vec{r}_+} e^{-\hat{H}^{(2)}_{\rm rel}/T} - e^{-\hat{T}_{\rm rel}/T}\ket{\vec{r}_+}\,,
\label{b2int-S0}\\
b_{2,S=1}^{\rm (int)} &= 3\cdot 2^{1/2}\intop\rmd^3 \vec{r}\bra{\vec{r}_{-}} e^{-\hat{H}^{(2)}_{\rm rel}/T} - e^{-\hat{T}_{\rm rel}/T}\ket{\vec{r}_{-}}\,,
\label{b2int-S1} \\
&\ket{\vec{r}_\pm\,} =\frac{1}{\sqrt{2}}\big(\ket{\vec{r}\,}\pm\ket{-\vec{r}\,}\big)\,.\nonumber
\end{align}
Factor $3$ in  $b_{2,S=1}^{\rm (int)}$ corresponds to the degeneracy of the spin-one states.

We stress that the derived expression for the second virial coefficient (\ref{b2-full}), and its interaction-dependent part (\ref{b2int-full}) given also in (\ref{b2int-spin-sep}) in the spin representation, are the full quantum mechanical relations expressing the coefficients through the inter-particle potential.

\subsection{Beth-Uhlenbeck approach}
\label{ssec:fullquant-b2}
\subsubsection{Second virial coefficient and  scattering phase shifts}

Here we will show that the second virial coefficient $b_2$, Eq.~(\ref{b2int-full}), can be expressed through characteristics of a particle scattering, which can be directly measured in experiments.

In the derivation we shall not rely on the wave-function description of the scattering process, as in Refs.~\cite{Huang,LLV,Uhlenbeck-Beth,Beth}, where the change of the  density of states is inferred from the change of the wave-function asymptotic at a remote (fictitious) space boundary. We will straightly  operate in the momentum space, where the symmetry properties of the scattering states in the channels with various spins can be followed more evidently. Thus we start with rewriting Eqs.~(\ref{b2int-S1}) and (\ref{b2int-S0}) in the momentum space,
\begin{align}
b_{2,S}^{\rm (int)} &= (2S+1)\,2^{1/2}
\!\!\!\intop \rmd^3 p
%\nonumber\\
%&\quad\times
\bra{\vec{p},S} e^{-\hat{H}^{(2)}_{\rm rel}/T} - e^{-\hat{T}_{\rm rel}/T}\ket{\vec{p},S}
\,,
\nonumber\\
&\ket{\vec{p},S=0}={\textstyle\frac1{\sqrt 2}} (\ket{\vec{p}\,} +\ket{-\vec{p}\,})\,,
\nonumber\\
&\ket{\vec{p},S=1}={\textstyle\frac1{\sqrt 2}} (\ket{\vec{p}\,} -\ket{-\vec{p}\,})\,.
\label{b2-full-p}
\end{align}
Following the method of Refs.~\cite{Watson56,Baumgartl67} we represent
the operator exponents in (\ref{b2-full-p}) as the Laplace transformations of the resolvent operator  $\hat{\mathcal{G}}_0$ of the kinetic energy operator $\hat{T}_{\rm rel}$ and the resolvent $\hat{\mathcal{G}}$ of the full Hamiltonian $\hat{H}_{\rm rel}$, see definitions  (\ref{resolvent}) in Appendix \ref{app:scattering},
\begin{align}
e^{-\hat{T}_{\rm rel}/T} &=
-\intop_{C}\frac{\rmd z}{2\pi i} e^{-z/T} \hat{\mathcal{G}}_0(z),
\nonumber\\
e^{-\hat{H}_{\rm rel}^{(2)}/T} &=
-\intop_{C}\frac{\rmd z}{2\pi i} e^{-z/T} \hat{\mathcal{G}}(z),
\label{integ-resolvent}
\end{align}
where contour $C$ shown in Fig.~\ref{fig:int-contour} encircles all singularities of $\hat{\mathcal{G}}(z)$ or $\hat{\mathcal{G}}_0(z)$
and is closed with the circle of large radius.
%%%%%%%%%%%%%
\begin{figure}
\parbox{3cm}{\includegraphics[width=3cm]{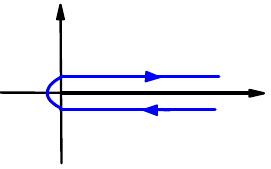} }
\caption{Integration contour in Eq.~(\ref{integ-resolvent}) }
\label{fig:int-contour}
\end{figure}
%%%%%%%%%%%%%
We assume that there are no bound states in the spectrum of the resolvent, i.e., there are no poles on the real axis for $\Re z<0$, only a cut for $\Re z\ge 0$. Properties of the resolvent operators are discussed in Appendix~\ref{app:scattering}, where it is shown that the difference $\hat{\mathcal{G}}(z)-\hat{\mathcal{G}}_0(z)$ can be expressed through the scattering operator $\hat{\mathcal{T}}$, which is defined in Eq.~(\ref{T-def}) and obeys the Lippmann-Schwinger equation (\ref{LSE}). Now, using  Eq.~(\ref{G-G0-TGG}) the difference of exponents in ~(\ref{integ-resolvent}) can be presented as
\begin{align}
 e^{-\frac{\hat{H}^{(2)}_{\rm rel}}{T}} - e^{-\frac{\hat{T}_{\rm rel}}{T}}=
 -\intop_{C}\frac{\rmd z}{2\pi i} e^{-\frac{z}{T}} \hat{\mathcal{G}}_0(z) \hat{\mathcal{T}}(z) \hat{\mathcal{G}}_0(z)\,.
\end{align}
The matrix element of this operator equation gives\footnotemark\footnotetext{In the presence of bound states, in r.h.s. of Eq.~(\ref{mel-T}) one has to add
terms $\sum_{n}e^{-\epsilon_n/T}\braket{\vec{p},S|n}\braket{n|\vec{p},S}$, where $\epsilon_n<0$ is the binding energy of the state $\ket{n}$.}
\begin{align}
\bra{\vec{p},S} e^{-\frac{\hat{H}^{(2)}_{\rm rel}}{T}} - e^{-\frac{\hat{T}_{\rm rel}}{T}}
\ket{\vec{p},S} =
 -\intop_{C}\frac{\rmd z}{2\pi i} e^{-\frac{z}{T}}
 \frac{\bra{\vec{p},S}\hat{\mathcal{T}}(z) \ket{\vec{p},S} }{(z-E_{\vec{p}})^2}\,.
 \label{mel-T}
\end{align}
Now, performing the contour integration explicitly we can cast
the second virial coefficient defined in (\ref{b2-full-p}) as
\begin{align}
b_{2,S}^{\rm (int)} &= - (2S+1)\,2^{1/2} \intop\rmd^3 p\intop_0^\infty\frac{\rmd E}{\pi} e^{-E/T}
\nonumber\\
&\times\lim_{\eta\to 0}
\Im \frac{\bra{\vec{p},S}\hat{\mathcal{T}}(E + i\eta) \ket{\vec{p},S} }{(E+i\eta-E_{\vec{p}})^2}\,.
\label{b2-T-1}
\end{align}
Note that under the integral we have formally the forward scattering amplitude evaluated for off-shell energies, i.e. $E$ can differ from $E_{\vec{p}} = \vec{p\,}^2/ m_f$. However, using the optical theorem, we show in Appendix~\ref{app:onshell} that the imaginary part of the integrand can be reduced to the mass-shell.
Using the relation (\ref{int-onshell}), we can rewrite Eq.~(\ref{b2-T-1}) as
\begin{align}
b_{2,S}^{\rm (int)} &= - (2S+1)\,2^{1/2}   \intop\rmd^3 p \, e^{-E_{\vec{p\,}}/T}
\nonumber\\
&\times\Bigg\{
\Big( \frac1{2E_p}+ \frac{\rmd}{\rmd E_p}
\Big)\Re\bra{\vec{p},S}\hat{\mathcal{T}}(E_{\vec{p\,}} + i\eta) \ket{\vec{p},S}
\nonumber\\
& - \pi \intop\rmd^3 p'\delta(E_{\vec{p\,}'}-E_{\vec{p\,}})\Im\Big[
\bra{\vec{p\,}',S}\hat{\mathcal{T}}(E+i\eta) \ket{\vec{p\,},S}^*
\nonumber\\
&\times
 \frac{\rmd }{\rmd E}\bra{\vec{p\,}',S}\hat{\mathcal{T}}(E+i\eta) \ket{\vec{p},S}
 \Big]\Big|_{E=E_{\vec{p}}}
 \Bigg\}\,,
\label{int-onshell-2}
\end{align}
%\end{widetext}
superscript $*$ denotes the complex conjugation.

The matrix element of the scattering operator between states with momenta $\vec{p}$ and $\vec{p\,}'$, such that $|\vec{p}\,|=|\vec{p\,}'|$, taken at the energy $E_{\vec{p}}+i\eta$, defines the scattering amplitude in the center-of-mass reference frame,
\begin{align}
&F^{(S)}(E_p,(\vec{n}\vec{n\,}'))
= -2\pi^2 m_f \bra{\vec{p\,}',S} \hat{\mathcal{T}}(E_{\vec{p}}+i\eta) \ket{\vec{p}, S}
\nonumber\\
&\quad = f^{(S)}(E_p,(\vec{n}\vec{n\,}'))+(-1)^{S}f^{(S)}(E_p,-(\vec{n}\vec{n\,}'))\,,
\label{f-def}
\end{align}
which is the function of the energy in this reference frame and the angle between momenta of incoming and outgoing particles, $\vec{p}= \vec{n} |\vec{p}\,|$ and $\vec{p\,}'= \vec{n\,}' |\vec{p\,}|$.
The non-symmetrized amplitude
\begin{align}
f^{(S)}(E_{\vec{p}},(\vec{n}\vec{n\,}'))=-2\pi^2 m_f \bra{\vec{p\,}'}\hat{\mathcal{T}}(E_{\vec{p}}+i\eta) \ket{\vec{p\,}}
\label{f-nonident}
\end{align}
 describes the scattering of the non-identical particles of the same mass  interacting by the same potential. We keep here index $S$ for the non-symmetrized amplitude although the potential, which  we consider, does not depend on the spin.
Substituting (\ref{f-def}) in Eq.~(\ref{int-onshell-2}) we can express the second virial coefficient as
%$$
%2\pi \intop\rmd^3 p'\delta(E_{\vec{p\,}'}-E_{\vec{p\,}})\frac{1}{4\pi^4 m^2}
%=2\pi \intop\rmd\Omega\rmd p'^2\frac12 p'\delta(p'^2-p^2) m \frac{1}{4\pi^4 m^2}
%= \intop\frac{\rmd\Omega}{4\pi}  p  \frac{1}{\pi^2 m}
%$$
\begin{align}
&b_{2,S}^{\rm (int)} =  (2S+1)\,2^{1/2}
\intop\rmd^3 p\, e^{-E_{\vec{p\,}}/T}
\frac{\mathcal{F}_S(E_{\vec{p}})}{2\pi^2}\,,
\label{b2S-F}
\end{align}
where
\begin{align}
&\mathcal{F}_S(E_{\vec{p}}) =  \frac{1}{2p} \Big( \frac1{p}+ \frac{2p}{m_f}\frac{\rmd}{\rmd E_{\vec{p}}}
\Big)\Re F^{(S)}(E_{\vec{p}},1)\label{Fs-def}
\\
&\,\,+\frac12
 \intop\frac{ \rmd \Omega_{\vec{n\,}'}}{4\pi}
\Im\Big[
F^{(S)*}(E_{\vec{p}},(\vec{n}\vec{n\,}'))
 \frac{2p}{m_f}\frac{\rmd }{\rmd E_{\vec{p}}}F^{(S)}(E_{\vec{p}},(\vec{n}\vec{n\,}'))
 \Big]
\,.\nonumber
\end{align}

The standard tool for the extraction of the information from scattering experiments is the partial wave analysis, when one determines the part of the amplitude responsible for the scattering with a specific angular momentum. We remind here that we consider a simple case of spin-independent potential, when spin and angular momentum are both good quantum numbers. Then using Eqs. (\ref{F-pwa}), (\ref{F-delt-def}) of Appendix F the scattering amplitude $f^{(S)}(E_{\vec{p}},(\vec{n}\vec{n\,}'))$ can be presented as
\begin{align}
f^{(S)}(E_{\vec{p}},(\vec{n}\vec{n\,}')) &= \sum_{l}(2l+1) f^{(S)}_l(E_{\vec{p}}) P_l(\vec{n}\vec{n\,}')\,,
\label{f-pwa}
\end{align}
with the partial wave amplitudes
\begin{align}
f^{(S)}_l(E_{\vec{p}}) &= \frac{1}{p}e^{i\delta_l^{(S)}(E_{\vec{p}})}
\sin\delta_l^{(S)}(E_{\vec{p}})\,,
\label{delta-def}
\end{align}
$P_l(x)$ are the Legendre polynomials.
Using the relation (\ref{Pl-integ})
%\begin{align}
%\intop \frac{\rmd \Omega_{\vec{n\,}''}}{4\pi}
%P_l(\vec{n}\vec{n\,}'') P_m(\vec{n\,}''\vec{n\,}')=\frac{\delta_{lm}}{2l+1}
%P_l(\vec{n}\vec{n\,}')\,,
%\end{align}
we find ~\cite{LLV} a  compact expression for $\mathcal{F}_S$:
\begin{align}
\mathcal{F}_S(E_{\vec{p}}) &= 2
\sum_{l[S]}(2l+1)\Big\{ \frac{1}{2p} \Big( \frac1{p}+ \frac{\rmd}{\rmd p}
\Big)\Re f^{(S)}_l(E_{\vec{p}})
\nonumber\\
&\qquad +\Im\Big[ f^{(S)*}_l(E_{\vec{p}}) \frac{\rmd }{\rmd p}f^{(S)}_l(E_{\vec{p}}) \Big]
\Big\}
\nonumber\\
&= 2\sum_{l[S]}(2l+1)\frac{1}{2p^2}
 \frac{\rmd\delta_l^{(S)}(E_{\vec{p}})}{\rmd p}\,.
 \end{align}
For $S=0$ the sum goes over even values of $l$ and for $S=1$ over odd values.
Substituting the obtained expression in (\ref{b2S-F}) we can present  the second virial coefficient $b_2^{(\rm int)}=b_{2,S=0}^{(\rm int)}+
b_{2,S=1}^{(\rm int)}$ as
\begin{align}
b_2^{\rm (int)} &= 2^{3/2}\intop_0^\infty \frac{\rmd p}{\pi} e^{-E_{\vec{p}}/T}
\frac{\rmd \delta_{\rm tot}(E_{\vec{p}})}{\rmd p}\,,
\label{b2-phase}\\
\delta_{\rm tot}(E_{\vec{p}}) &=\sum_{S=0,1}\sum_{l[S]}(2S+1)(2l+1)\delta_l^{(S)}(E_{\vec{p}})\,,
\label{delta-tot}
\end{align}
where we introduced the sum of phase shifts including the degeneracy factors. In our case of the spin-independent potential, when both spin and angular momentum are appropriate quantum numbers, we have
\begin{align}
\delta_{\rm tot}=\delta_0^{(0)}+9\delta^{(1)}_{1}+5\delta_2^{(0)}+
21\delta_{3}^{(1)}+\dots
\end{align}
For a more complicated potential with the $\vec{L}\vec{S}$ coupling the
phases are classified with the help of the total momentum  $\vec{J}=\vec{S}+\vec{L}$. Then, introducing the  notation $\delta_{^{2S+1}L_{J}}$ we identify phase shifts
$\delta_0^{(0)}\to \delta_{^1S_0}$ and $\delta_2^{(0)}\to \delta_{^1D_2}$ and  nine equal p-wave amplitudes decouple in amplitudes with various $J=0,1,2$, so that  $9\delta^{(1)}_{1}\to  \delta_{^3P_0}+3\delta_{^3P_1}+5\delta_{^3P_2}$.
Similarly,  21 f-wave amplitudes decouple also in three groups,
$21\delta^{(3)}_{1}\to  5\delta_{^3F_2} + 7\delta_{^3F_3} + 9\delta_{^3F_4}$, and so on.

The merit of the  expressions (\ref{b2-phase}) obtained by Gropper~\cite{Gropper-36} and by Beth and Uhlenbeck~\cite{Beth} is that they do not rely on any modelling of a potential acting between particles and calculations can be performed  using the experimentally measured phase shifts.
%If the temperature is small enough, it is sufficient to take into account several first partial waves. {\bf otkuda eto utverzhdenie sleduet?}

\subsubsection{Phase shifts and effective range approximation}\label{ssec:effrange}

Employing that the function $p^{2l+1}\cot\delta_l^{(S)}(E_{\vec{p}})$ is an analytical function of the relative energy of colliding particles~\cite{Chew-Gold49}, see also Section 11.6 in~\cite{Joachain}, one can present it as a series of $E_{\vec p}$ or $p^2$,
\begin{align}
&p^{2l+1}\cot\delta_l^{(S)}(E_{\vec{p}})=-\frac{1}{a_l^{(S)}} + \frac12r_{{\rm ef},l}^{(S)} p^2 +O(p^4)\,,\,\,
\nonumber\\
&p=\sqrt{m_f E_{\vec{p}}}\,.
\label{delta-effr}
\end{align}
The quantity  $a_l^{(S)}$ for $l=0$ is called the scattering length  and for $l=1$, the scattering volume, and $r_{{\rm ef},l}^{(S)}$ is the scattering effective range. The latter quantity measures  the typical range of the interaction, $r_{\rm typ}$, in case of a finite-range potential, then for $r$ essentially larger than $r_{\rm typ}$  the potential can be considered as negligibly small. The coefficients $a_l^{(S)}$ and $r_{{\rm e},l}^{(S)}$ can be extracted from the  analysis of the experimental phase shifts. For $l=0$, expansion (\ref{delta-effr}) was derived and discussed in Refs.~\cite{Schwinger47,Bethe49,Blatt-Jackson49}. It proves to be that the  coefficients $a_0^{(S=0)}$ and $r_{{\rm ef},0}^{(S=0)}$ are expressed only through the integrated value of the potential, e.g. in the Born approximation $a_0^{(S=0)}=(m_f/(4\pi))\int U(r)d^3 r$, cf. Eq.~(\ref{lowp-param-Born}) in Appendix \ref{app:scattering}. Since the energy dependence of the  phase shifts can be extracted from the  particle scattering experiments, constructing appropriate pair potentials one should reproduce these quantities. General properties of the third coefficient in (\ref{delta-effr}) were investigated in \cite{Blatt-Jackson49}, and other extensions beyond the $p^2$ term are discussed in~\cite{Mathelitsch84}.

From (\ref{delta-effr}) we get the quantity
\begin{align}
\frac{\rmd \delta_l^{(S)}}{\rmd p} = p^{2l}
\frac{-\frac{1+2l}{a_l^{(S)}}-\frac{1-2l}{2}r_{{\rm ef},l}^{(S)} p^2}
{\big(-\frac{1}{a_{l}^{(S)}} + \frac12 r_{{\rm ef},l}^{(S)} p^2\big)^2+ p^{4l+2}}\,,
\label{ER-ddelt}
\end{align}
determining the second virial coefficient (\ref{b2-phase}).
We see that for $l=0$ and $a_0^{(0)}r_{{\rm ef},0}^{(0)}<0$  the function $\delta_{0}^{(0)}(p)$ has a maximum at $p_{\rm max}=\sqrt{2/ | a_0^{(0)}r_{{\rm ef},0}^{(0)}|}$.

Expanding the scattering phase shift in (\ref{delta-effr}) in series of $p$ we obtain for the s-wave ($l=0$)
\begin{align}
\delta_0^{(0)}\approx -a_0^{(0)} p \Big[1 + \frac12 a_0^{(0)}(r_{{\rm ef},0}^{(0)} - \frac23 a_0^{(0)})p^2\Big]\,,
\label{delta0-exp}
\end{align}
for  $p^2|a_0^{(0)}(r_{{\rm ef},0}^{(0)}-\frac23 a_0^{(0)})|\ll 1$, and $\delta_0^{(1)}=0$, whereas  for  $l\neq 0$ we have
\begin{align}
\delta_l^{(S)}\approx -a_l^{(S)} p^{2l+1}\Big[1+\frac12 a_l^{(S)}r_{{\rm ef},l}^{(S)} p^2\Big] \frac{1+(-1)^{S+l}}{2}\,,
\label{deltal-exp}
\end{align}
at $p^2|a_l^{(S)}r_{{\rm ef},l}^{(S)}|\ll 1$.

\begin{figure}
\centering
\includegraphics[width=6cm]{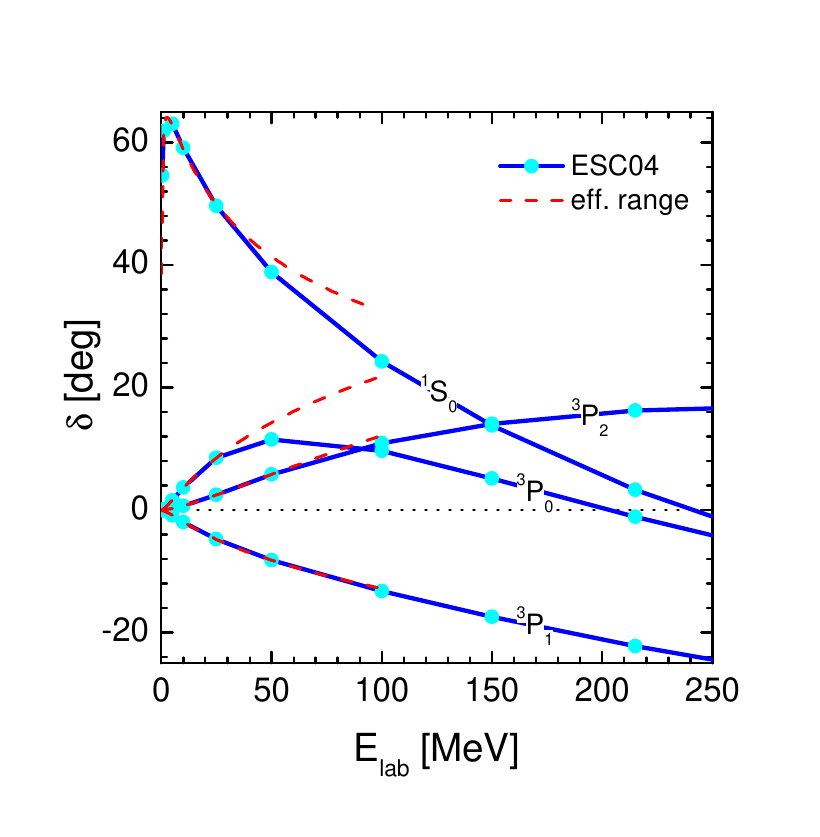}
\caption{ Scattering phase shifts versus laboratory energy $E_{\rm lab}=2E_{\vec{p}}$ for $^1{\rm S}_0$, $^3{\rm P}_0$, $^3{\rm P}_1$, and $^3{\rm P}_2$ partial waves. Dots connected by solid lines show the results of the ESC04 model~\cite{Rijken-06}. Dashed lines are low-energy interpolations within the effective range approach (\ref{delta-effr}) with the parameters (\ref{a-reff-param}).}
\label{fig:delta-NN}
\end{figure}

In Fig.~\ref{fig:delta-NN} we show the phase shifts $\delta^{(S)} (E_{\rm lab})$ of the nucleon-nucleon elastic scattering in the isospin~1 state for $^1{\rm S}_0$, $^3{\rm P}_0$, $^3{\rm P}_1$, and $^3{\rm P}_2$ (solid dots connected by solid lines to guide the eye) obtained from the analysis of the neutron-proton scattering within the ESC04 model~\cite{Rijken-06} by the Nijmegen group. The corresponding low-energy parameters of the $NN$ scattering calculated using Eq.~(\ref{delta-effr}) are
\begin{align}
&^1{\rm S}_0:\, a_{^1{\rm S}_0}= -23.7\,{\rm fm},\,\, r_{{\rm ef},^1{\rm S}_0}=2.76\,{\rm fm} \,,
\nonumber\\
&^3{\rm P}_0:\, a_{^3{\rm P}_0}= -2.68\,{\rm fm}^3,\,\, r_{{\rm ef},^3{\rm P}_0}=4.86\,{\rm fm}^{-1}\,,
\nonumber\\
&^3{\rm P}_1:\, a_{^3{\rm P}_1}= \phantom{-}1.42\,{\rm fm}^3,\,\, r_{{\rm ef},^3{\rm P}_1}=-8.50\,{\rm fm}^{-1}\,,
\nonumber\\
&^3{\rm P}_2:\, a_{^3{\rm P}_2}= -0.323\,{\rm fm}^3,\,\, r_{{\rm ef},^3{\rm P}_2}=5.05\,{\rm fm}^{-1}\,.
\label{a-reff-param}
\end{align}
Actually, the $^1{\rm S}_0$ parameters were obtained in Ref.~\cite{Rijken-06}. This required a careful analysis of the phase rapidly varying with the momentum. The p-wave parameters we found by fitting the low-energy phase shifts given in~\cite{Rijken-06}.

The phase shifts given by Eq.~(\ref{delta-effr}) calculated with the parameters (\ref{a-reff-param}) are shown in Fig.~\ref{fig:delta-NN} by the dashed lines. We see that the effective-range approximation gives appropriate fit of the presented  scattering phase shifts for the laboratory energy $0\le E_{\rm lab} \lsim 50$\,MeV. For higher energies, Eq.~(\ref{delta-effr}) can be used only as an extrapolation. The analysis of the $NN$ scattering~\cite{Bethe49,Blatt-Jackson49,Mathelitsch84} shows that the next term in the expansion~(\ref{delta-effr}) is rather small and becomes dominant at momenta $p\gsim 160$\,MeV.

The low-energy parameters of the nucleon-nucleon scattering can be easily estimated in the Born (subscript B) approximation for the central Yukawa-type potential~(\ref{U-potR}). With the help of Eq.~(\ref{lowp-param-L01}) in Appendix \ref{app:scattering} we find
\begin{align}
a_{{\rm B},0}^{(0)}
%&= \frac{m_N}{4\pi}\intop \rmd^3 r U(r)
&=\frac{m_N}{4\pi}\big(\frac{g_\om^2}{m_\om^2} - \frac{g_\sigma^2}{m_\sigma^2}\big), \,\,\,
%\nonumber\\
a_{{\rm B},1}^{(1)}
%&= \frac{m_N}{36\pi}\intop \rmd^3 r U(r)r^2
=\frac{m_N}{6\pi}\big(\frac{g_\om^2}{m_\om^4} - \frac{g_\sigma^2}{m_\sigma^4}\big),
\nonumber\\
r_{{\rm B, ef},0}^{(0)} &= -\frac{6a_{{\rm B},1}^{(1)}}{[a_{{\rm B},0}^{(0)}]^2}\,,\quad
r_{{\rm B, ef},1}^{(1)} =-\frac{4m_N}{3\pi [a_{{\rm B},1}^{(1)}]^2}\big(\frac{g_\om^2}{m_\om^6} - \frac{g_\sigma^2}{m_\sigma^6}\big),
\label{LE-Born-Yukawa}
\end{align}
that yields $a_{{\rm B},0}^{(0)}=-1.2$\,fm and $a_{{\rm B},1}^{(1)}=-0.33\,{\rm fm}^3$ for the Walecka parameters (\ref{Uomsig-par-W}) and
$a_{{\rm B},0}^{(0)}=0.12$\,fm and $a_{{\rm B},1}^{(1)}=-0.27\,{\rm fm}^3$ for the CD Bonn potential fit parameters (\ref{Uomsig-par-Bonn}). The range terms for the s-wave are, correspondingly,
$r_{{\rm B, ef},0}^{(0)}=  1.4$\,fm and $113$\,fm (the latter quantity is unrealistically large), and for the p-wave $r_{{\rm B, ef},1}^{(1)}= -4.8\,{\rm fm}^{-1}$ and 6.6\,${\rm fm}^{-1}$.

\begin{figure}
\centering
\parbox{5.5cm}{\includegraphics[width=5.5cm]{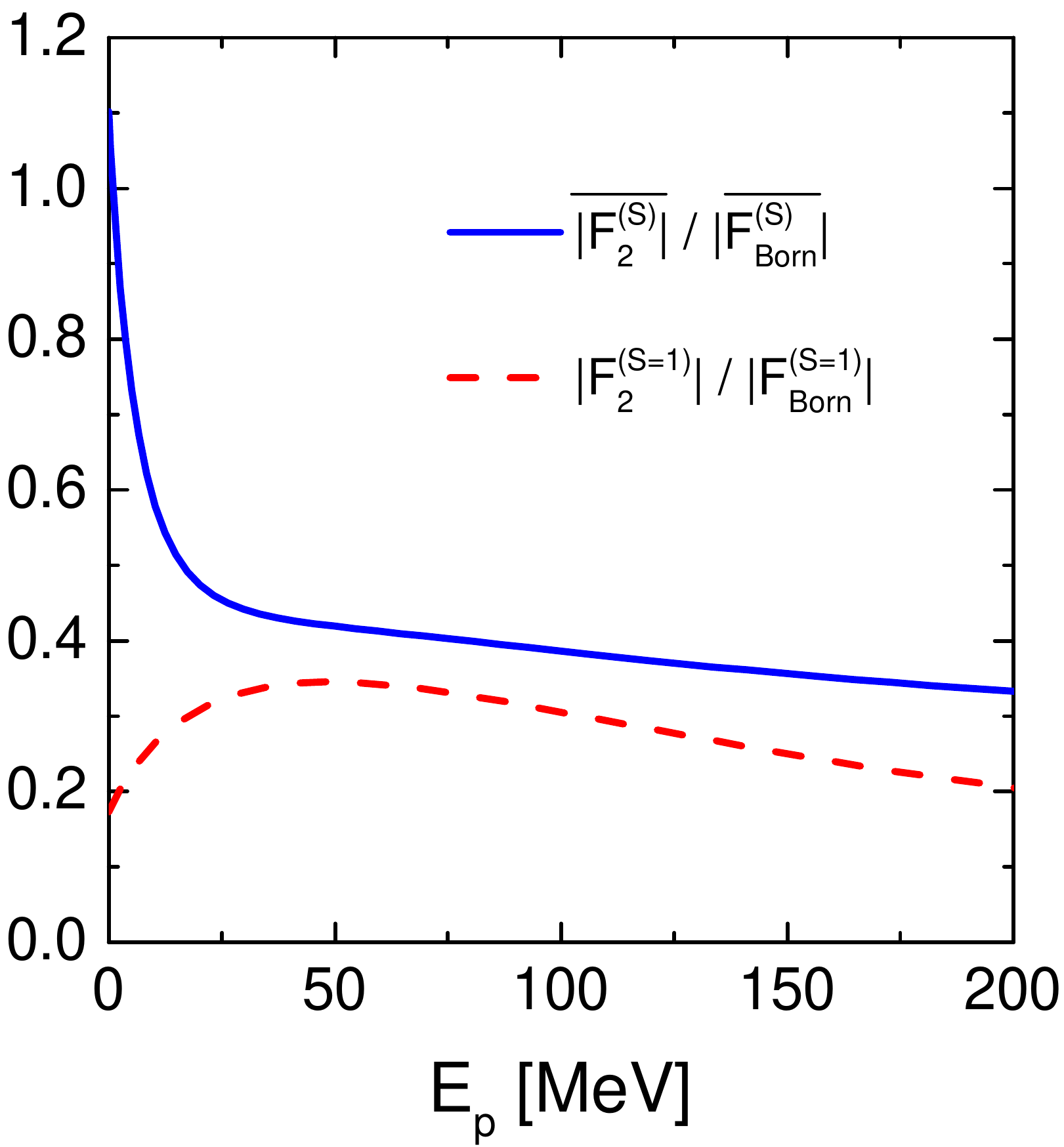}}
\caption{Ratio of the double scattering amplitude (\ref{FS2-def})
and the first-order Born scattering amplitude (\ref{f-Born}) as a function of the relative nucleon energy in the center of mass reference frame calculated for the Walecka parameterization (\ref{Uomsig-par-W}) of the Yukawa-type pair potential. Solid line demonstrates the ratio of isospin averaged amplitudes and dashed line shows the ratio of spin-one amplitudes.}
\label{fig:F2rat}
\end{figure}

To estimate the applicability of the Born approximation for the Yukawa potentials (\ref{U-potR}), we calculate in Appendix~\ref{app:double-scatt}
the second-order term in the expansion of the scattering operator in the value of the pair  potential (\ref{LSE-iter}). Fig.~\ref{fig:F2rat} demonstrates the ratio of the double scattering amplitude $F_2^{(S)}$ given by Eqs.~(\ref{FS2-def}),  (\ref{T-2}) and (\ref{Aij-def}) to the first-order amplitude (\ref{f-Born}). We use the Walecka parameterization of the Yukawa-type  potential. Solid line shows the ratio for the isospin averaged amplitudes. At the threshold ($p\to 0$) the ratio of the amplitudes is $\approx 1.1$, see analytical estimate (\ref{F22F1-thresh}). Within the energy range $ 0<E_{\vec{p}}\lsim 1$\,MeV, the ratio decreases sharply to the value $ |\overline{F_2^{(S)}}|/|\overline{F_{\rm Born}^{(S)}}|\simeq 0.5$ at $E_{\vec{p}}\simeq 20$\,MeV. With the further energy increase the ratio continues decreasing but very smoothly,
yielding $|\overline{F_2^{(S)}}|/|\overline{F_{\rm Born}^{(S)}}|\simeq 0.32$ at $E_{\vec{p}}=200$\,MeV. As we see from the figure the main contribution of the second-order scattering amplitude to the isospin averaged amplitude at low energies ($E_{\vec{p}}\lsim 20$\,MeV) is given by the spin-zero term.  Selecting the spin-one channel we see that the ratio of the amplitudes $ |\overline{F_2^{(S=1)}}|/|\overline{F_{\rm Born}^{(S=1)}}|$ does not exceed 0.35, cf. dashed line in Fig.~\ref{fig:F2rat}. In this channel the Pauli exclusion principle forbids the scattering of two-identical particles in the s-wave.  As we see, the scattering in the spin-one channel   can be approximately described within the Born approximation. At the end, note  that the pure spin-one scattering could be realized in the polarized neutron gas.

The above estimates of the low-energy parameters  were done for the central part of the pair nucleon potential in the Born approximation.
It is instructive to compare them with those following from the available experimental information on the nucleon-nucleon scattering. Comparing estimates of the $^1$S$_0$ scattering length performed above (\ref{LE-Born-Yukawa}) within the Born approximation   and the experimental value of the $^1$S$_0$ scattering length (\ref{a-reff-param}) we see that the Born approximation estimations yield an order of magnitude
smaller value for  the case of the  potential with the Walecka parameters (\ref{Uomsig-par-W}) and two order of magnitude smaller result (even of another sign) for the CD potential fit with parameters (\ref{Uomsig-par-Bonn}). As we have mentioned, for an adequate description of the nucleon scattering by the CD Bonn potential, one has to include non-central parts (tensor, spin-orbit, etc.) and relativistic corrections. Therefore it is not strange that employing only the central part of the CD potential fit with parameters (\ref{Uomsig-par-Bonn}) we obtain unrealistic result. At the same time, in spite of the value of the $^1$S$_0$ scattering length obtained with the Walecka parameters (\ref{Uomsig-par-W}) is an order of magnitude smaller than the experimental value,  this result has a physical meaning, as we will demonstrate  below.

The large value of the experimental $^1$S$_0$ scattering length signals  presence of a quasi-bound state close to the threshold generated in the solution of the Lippmann-Schwinger equation. Substituting expansion (\ref{delta-effr}) in Eq.~(\ref{F-delt-def}) we can separate the pole contribution, as it was done in Ref.~\cite{Lutz-NN},
\begin{align}
f_{^1{\rm S}_0}(E_{\vec{p}}) &\approx \frac{1}{-\frac{1}{a_{^1{\rm S}_0}} + \frac12r_{{\rm ef},^1{\rm S}_0} p^2 -i p}
\nonumber\\
&\approx \frac{iw}{p+ip_{\rm qb}} -a_{^1{\rm S}_0}^{\rm (ef)} -ib_{^1{\rm S}_0}^{\rm (ef)} p\,,
\label{LE-epxansion}
\end{align}
with the pole parameters
\begin{align}
w &=(1-2r_{{\rm ef},^1{\rm S}_0}/a_{^1{\rm S}_0})^{-1/2}\simeq 0.90\,,
\nonumber\\
p_{\rm qb} &=\frac{1-w}{w\,r_{{\rm ef},^1{\rm S}_0}}\simeq 0.04\,{\rm fm}^{-1}\,,
\label{qb-parametres}
\end{align}
and the residual effective scattering length is
\begin{align}
a_{^1{\rm S}_0}^{\rm (ef)} &= a_{^1{\rm S}_0} +\frac{r_{{\rm ef},^1{\rm S}_0} w^2}{1-w}\simeq -1.18\,{\rm fm},\,\,
\nonumber\\
b_{^1{\rm S}_0}^{\rm (ef)} &=\frac{w}{p_{\rm qb}^2}-a_{^1{\rm S}_0}^2\simeq 1.54\,{\rm fm}^2\,,
\label{a-eff}
\end{align}
where we used the numerical values of the empirical parameters for the $^1$S$_0$ wave (\ref{a-reff-param}). Note that expression in the second line in (\ref{LE-epxansion}) assumes that $p\ll |a_{^1{\rm S}_0}^{\rm (ef)}/b_{^1{\rm S}_0}^{\rm (ef)}|$ and $|a_{^1{\rm S}_0}^{\rm (ef)}/b_{^1{\rm S}_0}^{\rm (ef)}|> p_{\rm qb}$. It is remarkable that the effective scattering length $a_{^1{\rm S}_0}^{\rm (ef)}$ is now of the same magnitude as the scattering length obtained in the Born approximation (\ref{LE-Born-Yukawa}) for the Yukawa-type potential (\ref{U-potR}) with the Walecka parameters (\ref{Uomsig-par-W}).

\begin{figure}
\centering
\includegraphics[width=7cm]{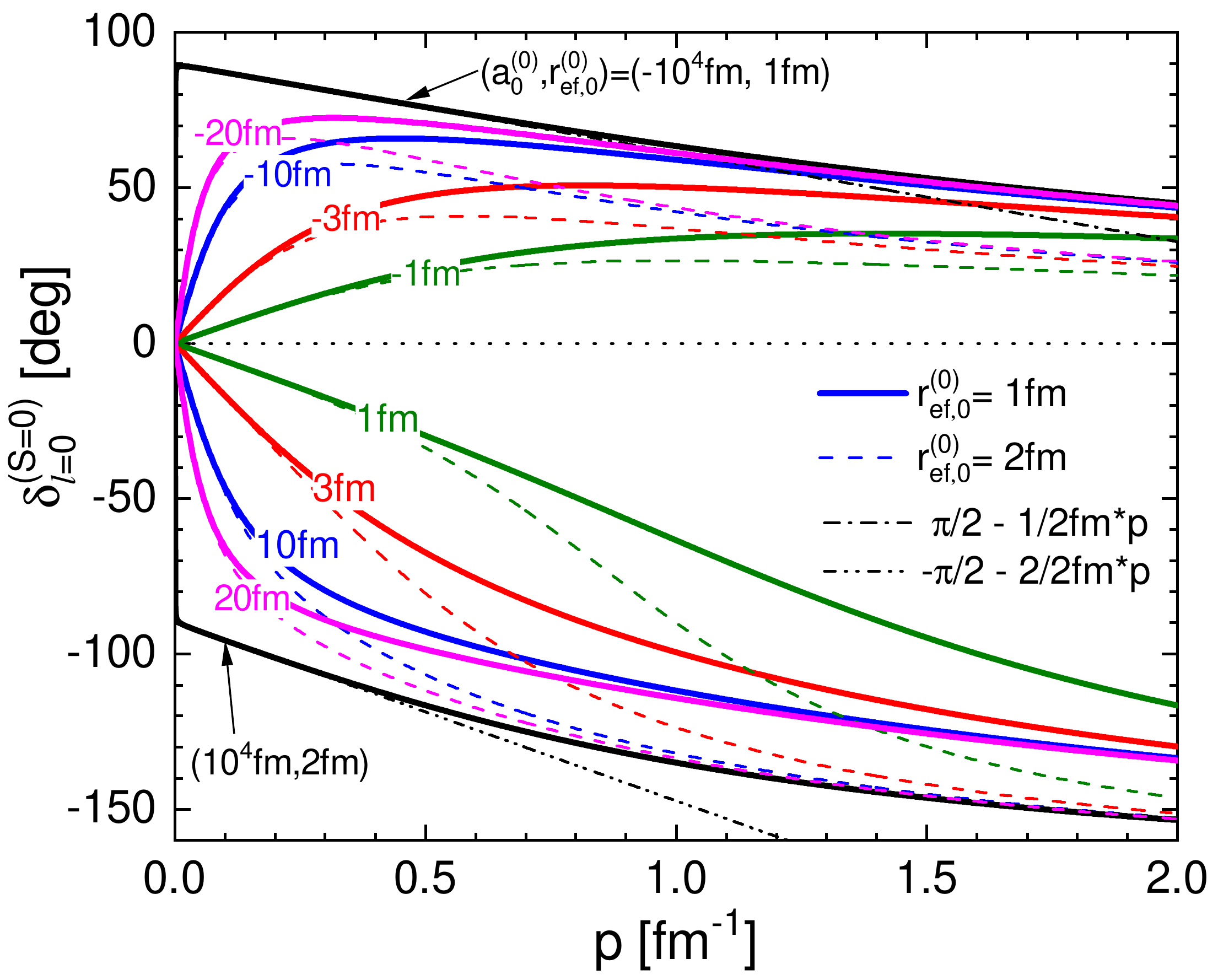}
\caption{Scattering phase shifts defined by Eq.~(\ref{delta-effr}) for $l=0$  as functions of relative scattering momentum for several values of the scattering length $a_0^{(0)}$ and the effective range $r_{{\rm  ef},0}^{(0)}$. Dash-dotted and dash-doubly-dotted lines represent the approximation (\ref{lin<}). }
\label{fig:delta-effr}
\end{figure}

In Fig.~\ref{fig:delta-effr} we illustrate the momentum dependence of the s-wave phase shift, $\delta_0^{(0)}$, given by Eq.~(\ref{delta-effr}) for various scattering lengths and range terms. Comparing the solid and dashed lines drawn for $r_{{\rm ef},0}^{(0)}=1$\,fm and 2\,fm, respectively, we conclude that the increase of the range term suppresses the magnitude of $\delta_0^{(0)}$ for the attractive scattering length ($a_0^{(0)}<0$), solid curves are situated above dashed ones, whereas for the repulsive scattering length ($a_0^{(0)}>0$) the increase of the range term leads to the increase of the phase shift magnitude, solid curves are situated below the dashed ones. We see also that with an increase of the scattering length magnitude, the interval of momenta, where Eq. (\ref{delta0-exp}) is applicable, shrinks dramatically. As follows from (\ref{delta-effr}), in the interval of momenta  $1/|a_0^{(0)} r_{{\rm ef},0}^{(0)}|\ll p^2\ll 1/(r_{{\rm ef},0}^{(0)})^2 $ the phase shift can be written as
\begin{align}
\delta_0^{(0)}\approx -{\rm sgn(a_0^{(0)})} \frac{\pi}{2}-\frac12 r_{{\rm ef},0}^{(0)}p\,.
\label{lin<}
\end{align}
For momenta $p^2\lsim 1/|a_0^{(0)} r_{{\rm ef},0}^{(0)}|$ there is a sharp transition from expansion (\ref{lin<}) to expansion
(\ref{delta0-exp}), so that the derivative  $\frac{\rmd \delta_0^{(0)}}{\rmd p}$ in (\ref{ER-ddelt}) has a $\delta$-function-like maximum at tiny momenta $p\sim 1/|a_0^{(0)}|$. Presence of this maximum leads to a constant contribution to the second virial coefficient, see Eq.~(\ref{b2-effr-large-a}) below.

In Fig.~\ref{fig:delta-effr} we see that dash-dotted and dash-double-dotted lines representing expansions (\ref{lin<}) for $a_0^{(0)}<0$ and $>0$
respectively coincide well at low momenta ($p\lsim 0.5$\,fm$^{-1}$) with the solid lines calculated according to Eq.~(\ref{delta-effr}) with parameters $|a_0^{(0)}|=10^4$\,fm and $r_{\rm ef,0}^{(0)}=1$ and 2\,fm. It is remarkable that for large values of $|a_0^{(0)}|\gg r_{{\rm ef}, 0}^{(0)}$ the scattering phase shift becomes practically independent on the scattering length in a broad range of the momenta, being mainly determined by the range term.

As it will be shown in the next Section,  estimates of the second virial coefficients at low temperatures crucially depend on the relative values of the scattering length and the effective range.

%%%%%%%%%%%%%%%%

\subsubsection{Second virial coefficients in low temperature limit}

A characteristic   momentum in the integral (\ref{b2-phase}) is $p\sim \sqrt{m_f T}$. Consider expression (\ref{b2-phase}) at a rather small temperature when we may use the momentum expansions derived above.
Substituting (\ref{delta0-exp}), (\ref{deltal-exp}) in Eq.~(\ref{b2-phase}) we obtain for small temperatures
$|a_l^{(S)}r_{{\rm ef},l}^{(S)}|m_f T\ll 1$, $ |a_l^{(S)}| (m_f T)^{l}\ll 1$, and $m_f T|a_0^{(0)}(r_{{\rm ef},0}^{(0)}-\frac23 a_0^{(0)})|\ll 1$,
\begin{align}
b_2^{\rm (int)} &\approx -2^{3/2}\intop_0^\infty \frac{\rmd p}{\pi} e^{-E_{\vec{p}}/T}
\Big[a_0^{(0)} + \frac32 a_0^{(0)}\big(r_{{\rm ef},0}^{(0)}-\frac23a_0^{(0)}\big) p^2
\nonumber\\
&+27 a_1^{(1)} p^2\Big]
\nonumber\\
&= -2\frac{a_{0}^{(0)}}{\lambda}\Big[1 +\frac{3\pi}{2\lambda^2}\Big( a_0^{(0)} r_{{\rm ef},0}^{(0)}-\frac23a_0^{(0)2} + 18 a_1^{(1)}\Big)\Big]{ \,.}
%=-\sqrt{\frac{2m_f T}{\pi}} a_0^{(0)}
\label{b2-lowT-a1}
\end{align}

In the Born approximation [see Eq.~(\ref{T-Born}) in Appendix \ref{app:scattering}], which is applicable, if conditions (\ref{slowB}) and (\ref{fastB})  are fulfilled, requiring a smallness of the potential, there exists a simple relation between the scattering length and the potential. Then neglecting the sub-leading terms ($\propto T^{3/2}$) in Eq. (\ref{b2-lowT-a1}) and using Eq.~(\ref{lowp-param-L01}) we obtain
\begin{align}
b_2^{\rm (int)}\approx -\sqrt{\frac{m_f^3 T}{8\pi^3}} \intop \rmd^3 r \Ucor(r)\,.
\label{b2-Born}
\end{align}
This approximation is valid, if $\delta_0(p\sim\sqrt{m_fT})\ll 1$,  that implies
$$
|\Ucor_{\rm typ}| r_{\rm typ}^2 m_f \sqrt{m_f T} r_{\rm typ}\ll 1 \,.
$$
Comparing Eq.~(\ref{b2-Born}) with Eq.~(\ref{b2-2a}) obtained within the $\Phi$ derivable approximation with the TS diagrams shown in Fig.~\ref{fig:Sig-U}, we see that (\ref{b2-Born}) coincides with the $2b_{2,\Phi}^{\rm int,H}$ term that
yields the full result provided $b_{2,\Phi}^{\rm int,F}=-b_{2,\Phi}^{\rm int,H}/2$. In case of the Yukawa nucleon-nucleon pair potential, the latter relation holds for low temperatures, $T\ll m_\sigma^2/2 m_N\simeq 130$ MeV,  cf. Eq.~(\ref{b2Phi-F-2}).
Note that for the nuclear potentials shown in Fig. \ref {fig:U-pot} we estimate  $|U_{\rm typ}|(m_N r_{\rm typ}^2)\sim 1$. With $r_{\rm typ}\sim 1/m_\pi$ the condition $\sqrt{mT}r_{\rm typ}\ll 1$ is realized for $T\ll 30$ MeV, and with $r_{\rm typ}\sim 1/m_\sigma$ it is satisfied for  $T\ll 130$ MeV.

To study the second virial coefficient within the effective range approximation for the scattering phase shifts, we
substitute Eq.~(\ref{ER-ddelt}) in Eq.~(\ref{b2-phase}) and obtain
\begin{align}
b_2^{\rm (int)} &= -\sum_{S,l[S]}(2S+1)(2l+1)
 \frac{a_{l}^{(S)}}{\lambda^{2l+1}}
J_l\big(\frac{a_l^{(S)}}{\lambda^{2l+1}},
\frac{a_l^{(S)}r_{{\rm ef},l}^{(S)}}{\lambda^2}\big),
%\nonumber\\
\end{align}
where
\begin{align}
J_l(x,y) &= 2^{3/2}\intop_0^\infty \frac{\rmd z}{\pi} e^{-\frac{z^2}{2\pi}}
z^{2l} \frac{(1+2l)+(1-2l)\frac{1}{2}  y z^2}
{\big(1 - \frac12  y z^2\big)^2+ x^2 z^{4l+2}}.
\label{Jl-def}
\end{align}
Consider first the s-wave contribution ($l=0$).
Then, the integrand in Eq.~(\ref{Jl-def}) can be represented as
\begin{align}
&\frac{1 + \frac12 y z^2}{\big(1 - \frac12  y z^2\big)^2+ x^2 z^2}  =
\frac{\tau}{(\rho_+ - \rho_-)}
\nonumber\\
&\quad\qquad\qquad\times\Bigg(
\frac{\frac{\tau}{\rho_-} - 1}{\frac{y\tau}{2\rho_-} z^2 + 1} - \frac{\frac{\tau}{\rho_+} - 1}{\frac{y\tau}{2\rho_+} z^2 + 1 }
\Bigg)\,,
\nonumber\\
&\rho_\pm  =
1-\tau \pm \sqrt{(1-\tau)^2-\tau^2}\,,\quad \tau=\frac{y}{x^2}\,.
\end{align}
As the consequence we can express $J_0(x,y)$ as follows
\begin{align}
J_0(x,y) = &\frac{\tau}{(\rho_+ - \rho_-)}\Big[
\Big(\frac{\tau}{\rho_-} - 1\Big)J_0\Big(\sqrt{\frac{y\tau}{2\rho_-}},0\Big) -
\nonumber\\
&\Big(\frac{\tau}{\rho_+} - 1\Big)J_0\Big(\sqrt{\frac{y\tau}{2\rho_+}},0\Big)
\Big].
\label{J0-tau}
\end{align}
where $J_0(x,y)$ is expressed via the $J_0$ function of only one argument. Three parameters, $\lambda$, $a_0^{(0)}$, and $r_{\rm ef,0}^{(0)}$ determining the s-wave contribution to the virial coefficient are combined in the temperature-independent scale
\begin{align}
\tau=y/x^2={r_{{\rm ef},0}^{(0)}}/{a_0^{(0)}},
\label{tau-scale}
\end{align}
and in the temperature-dependent scale $y={a_0^{(0)}r_{{\rm ef},0}^{(0)}}/{\lambda^2}$.
For the vanishing second argument, the function $J_l(x,0)$ can be evaluated easily as
\begin{align}
J_0(x,0)
&=\frac{2^{3/2}}{\pi|x|}\intop_0^\infty\frac{\rmd z e^{-\frac{z^2}{2\pi x^2}}}{1  + z^2}
=\frac{\sqrt{2}}{|x|}e^{\frac{1}{2\pi x^2}}{\rm erfc}\Big(\frac{1}{\sqrt{2\pi x^2}}\Big)
\nonumber\\
&=\left\{
\begin{array}{cc}
2 - 2\pi x^2 +6\pi^2x^4 \,, \quad & |x|\ll 1/\sqrt{2\pi},\\
\frac{\sqrt{2}}{|x|}-\frac{2}{\pi x^2}\,, \quad & |x|\gg 1/\sqrt{2\pi}\,.
\end{array}
\right.
\label{J0-exp}
\end{align}

As we have mentioned the experimental value of the nucleon-nucleon s-wave scattering length in the spin-singlet state is anomalously large
and $|\tau|\ll 1$. Hence we are interested in the small-$\tau$ expansion of Eq.~(\ref{J0-tau}). We observe that for $|\tau|\ll 1$ the arguments of two $J_0$ functions are
\begin{align}
\sqrt{\frac{y\tau}{2\rho_-}} &\approx
\sqrt{\frac{y}{\tau}}\Big(1 - \frac{\tau}{2} - \frac{\tau^2}{4} \Big)\,,\,\,
\nonumber\\
\sqrt{\frac{y\tau}{2\rho_+} } &\approx \frac{\sqrt{y\tau}}{2}\Big(1+\frac{\tau}{2}+\frac{\tau^2}{2}\Big),
\end{align}
and produce, thereby, two different temperature-dependent scales
\begin{align}
s_a &= \sqrt{\frac{y}{\tau}}=|x|=\frac{|a_0^{(0)}|}{\lambda}\simeq \frac{1}{\sqrt{2\pi}} \frac{|a_0^{(0)}|}{20\,{\rm fm}}\sqrt{\frac{T}{0.1\,{\rm MeV}}}
\label{sa-def}\\
s_r &= \sqrt{y\tau}=\Big|\frac{y}{x}\Big|=\frac{|r_{{\rm ef},0}^{(0)}|}{\lambda}\simeq \frac{1}{\sqrt{2\pi}}\frac{|r_{{\rm ef},0}^{(0)}|}{2\,{\rm fm}}\sqrt{\frac{T}{10\,{\rm MeV}}}
\label{sr-def}
\end{align}
and $\tau$ corrections to them. Expanding now Eq.~(\ref{J0-tau}) up to the first order in $\tau$ we have
\begin{align}
& J_0(x,y) \approx  \big(1-\half\tau\big)J_0\big(s_a-\half s_a\tau,0\big) + \half\tau J_0\big(\half s_r,0\big)
\nonumber\\
\qquad\approx & \big(1-\half\tau\big) J_0(s_a,0) - \half s_a\tau J'_0(s_a,0) + \half\tau J_0\big(\half s_r,0\big)\,.
\label{J0-tau-exp}
\end{align}
Now, using the identity for the derivative of the $J_0$ function,
\begin{align}
\partial_x J_0(x,0) &=\frac{2-J_0(x,0)}{\pi  x^3} -  \frac{J_0(x,0)}{x} \,,
\end{align}
we finally find
\begin{align}
 J_0(x,y) \approx \Big(1+ \frac{\tau}{2\pi s_a^2}\Big) J_0(s_a,0)  -  \frac{\tau}{\pi  s_a^2}
+ \frac{\tau}{2} J_0\big(\half s_r,0\big).
\label{J0-tau-exp-2}
\end{align}
Thus, in the limit cases $\tau\ll 1$, the contribution to the $b_2^{\rm (int)}$ from the s-wave scattering can be presented as
\begin{align}
 b_{2,l=0}^{\rm (int)} &= -\frac{a_{0}^{(0)}}{\lambda}
J_0\big(\frac{a_0^{(0)}}{\lambda}, \frac{a_0^{(0)}r_{{\rm ef},0}^{(0)}}{\lambda^2}\big)
\nonumber\\
&\approx -\Big(\frac{a_{0}^{(0)}}{|a_{0}^{(0)}|}  +  \frac{\lambda^2 \,r_{{\rm ef},0}^{(0)}}{2\pi |a_0^{(0)}|^3}\Big)
\sqrt{2} e^{\frac{ 1}{2\pi s_a^2 }}{\rm erfc}\Big(\frac{1}{\sqrt{2\pi}s_a}\Big)
\nonumber\\
& + \frac{\lambda r_{{\rm ef},0}^{(0)}}{\pi\,a_0^{(0)2}}
- \frac{ r_{{\rm e},0}^{(0)}}{ | r_{{\rm e},0}^{(0)}|}\sqrt{2}   e^{\frac{ 4}{2\pi s_r^2 }}{\rm erfc}\Big(\frac{2}{\sqrt{2\pi}s_r}\Big)
\,.
%, \quad |a_0^{(0)}r_{{\rm ef},0}^{(0)}|\ll 1/(m_f T)\,.
\label{b2-effr}
\end{align}
Within the zero-range approximation similar result was obtained in Ref.~\cite{Ho-Mueller-2004} and then it was extended for the finite range in~\cite{Liu-2013}. The result (\ref{b2-effr}) relies only on the smallness $\tau\ll 1$ and is valid for any temperature unless the effective range approximation can be still used for the description of the particle scattering at momenta $p\sim\sqrt{m T}$.

Now we can analyze temperature dependence of expression (\ref{b2-effr}). Since  $s_r\ll s_a$ for any temperature, we can consider three temperature intervals
\begin{align}
&{\rm I}.~~\,\frac{s_r}{2}\ll s_a\ll \frac{1}{\sqrt{2\pi}}\,,
\label{srsa-I}\\
&{\rm II}.~\frac{s_r}{2}\ll \frac{1}{\sqrt{2\pi}} \ll s_a\,,
\label{srsa-II}\\
&{\rm III}.~\frac{1}{\sqrt{2\pi}} \ll \frac{s_r}{2}\ll s_a\,.
\label{srsa-III}
\end{align}
Case~I is the low temperature limit that for the parameters of the nucleon-nucleon interaction corresponds to temperatures  $T\ll 0.1$\,MeV as follows from the condition $\sqrt{2\pi} s_a\ll 1$ and the estimate in (\ref{sa-def}). Case~III is the high temperature limit valid for temperatures $T\gg 40$\,MeV as given by condition $\sqrt{\pi/2} s_r\gg 1$. For the intermediate temperatures $1/(m_f a_0^{(0)2})\ll T\ll  4/(r_{{\rm ef},0}^{(0)2}m_f )$, when $0.1\,{\rm MeV}\ll T\ll 20\,{\rm MeV}$, case~II is realized.

For the low-temperature case (case I), using expansion (\ref{J0-exp}) we can rewrite Eq.~(\ref{b2-effr}), keeping the terms quadratic in $s_a$ and of the order $s_r^0$, as follows
\begin{align}
 b_{2,l=0}^{\rm (int)}& \approx -2\frac{a_{0}^{(0)}}{\lambda} +2\pi \frac{a_{0}^{(0)3}}{\lambda^3} -3\pi\frac{r_{{\rm ef},0}^{(0)} a_0^{(0)2}}{\lambda^3}
 \nonumber\\
& =-\sqrt{\frac{2m_fT}{\pi}} a_{0}^{(0)}\Big[1-  \frac{a_{0}^{(0)2}}{2} m_f T +\frac{3r_{{\rm ef},0}^{(0)} a_0^{(0)}}{4}m_f T
\Big]\,,
\label{b2-effr-small-a}
\end{align}
i.e., we recovered  expansion (\ref{b2-lowT-a1}).

For the intermediate temperatures (case~II), see Eq.~(\ref{srsa-II}), we expand Eq.~(\ref{b2-effr}) keeping terms of the order $s_a^{-1}$ and $s_r^0$,
\begin{align}
&b_{2,l=0}^{\rm (int)}
\approx -\frac{a_{0}^{(0)}}{|a_{0}^{(0)}|}\sqrt{2}  +  \frac{2\lambda}{\pi a_0^{(0)}} - \frac{ r_{{\rm ef},0}^{(0)}}{\lambda}
\nonumber\\
&\quad =-{\rm sgn}\big(a_{0}^{(0)}\big) \sqrt{2} +\frac{2^{3/2}}{\sqrt{\pi m_f T}} \frac{1}{a_0^{(0)}} -\sqrt{\frac{m_f T}{2\pi}}\,r_{{\rm ef},0}^{(0)}\,.
\label{b2-effr-large-a}
\end{align}
We see that $b_2^{\rm (int)}$ is now almost constant and its temperature dependence for  temperatures when $|r_{{\rm ef},0}^{(0)}|/\lambda\ll 2\lambda/(\pi |a_0^{(0)}|)$, i.e.  $ T\ll 4/(m_f |a_0^{(0)} r_{{\rm ef},0}^{(0)}|)\simeq 4\,{\rm  MeV}/(|a_0^{(0)}r_{{\rm ef},0}^{(0)}|/[40\,{\rm fm}^2])$, is  determined by the scattering-length--dependent term rather than the effective range term, whereas for $T\gg 4/(m_f |a_0^{(0)} r_{{\rm ef},0}^{(0)}|)$
the temperature dependence is determined by the effective range term.

For high temperatures $ T\gg 4/(r_{{\rm ef},0}^{(0)2}m_f )\simeq 20$\,MeV (as follows from the condition $\sqrt{\pi/2} s_r\gg 1$), we deal with case III, and the expansion (\ref{b2-effr}) reads
\begin{align}
b_{2,l=0}^{\rm (int)}
&\approx -\frac{a_{0}^{(0)}}{|a_{0}^{(0)}|}\sqrt{2}  +  \frac{2\lambda}{\pi a_0^{(0)}} - \frac{ r_{{\rm ef},0}^{(0)}}{|r_{{\rm ef},0}^{(0)}|}\sqrt{2} +\frac{4\lambda}{\pi r_{{\rm ef},0}^{(0)}}
\nonumber\\
& = - \big[{\rm sgn}\big(a_{0}^{(0)}\big)+{\rm sgn}\big(r_{{\rm ef},0}^{(0)}\big)\big] \sqrt{2}
\nonumber\\
 & + \frac{2^{3/2}}{\sqrt{\pi m_f T}} \frac{1}{a_0^{(0)}} + \frac{2^{5/2}}{\sqrt{\pi m_f T}} \frac{1}{r_{{\rm ef},0}^{(0)}}\,.
\label{b2-effr-large-T}
\end{align}
Since for the nucleon interaction $|r_{{\rm ef},0}^{(0)}|\ll |a_0^{(0)}|$, the last term exceeds the second ($a_0^{(0)}$-dependent) term at any temperature. Thus, we demonstrated that at all temperatures $T\gg 4/(m_f |a_0^{(0)}| r_{{\rm ef},0}^{(0)})$, i.e for $T\gg 3$\,MeV, the thermal effects almost completely wash out the information on the actual value of the anomalously large scattering length $a_0^{(0)}$ but not on its sign (indicating either attraction or repulsion). Note that for the case of the s-wave $nn$ scattering, $a_0^{(0)}<0$ and $b_{2,l=0}^{\rm (int)}>0$ for all temperatures. When the quantum correction (\ref{b0}) is included, the quantity $b_{2,l=0}$ changes sign at $T\simeq 0.02$\,MeV and remains positive at all temperatures of our interest.

To estimate the contribution from the p-wave to $b_2$ we rewrite function $J_l$ in Eq.~(\ref{Jl-def}) for  $l=1$   and separate terms linear in $y$ for $|y|\ll 1/({2\pi})$,
\begin{align}
J_1(x,y) &\approx\frac{2^{3/2}}{|x|}\Big[3\intop_0^\infty\frac{\rmd z}{\pi} e^{-\frac{z^2}{2 \pi  x^{2/3}}} \frac{ z^2}{(1+z^6)}
\nonumber\\
&-\frac{y}{2x^{2/3}}\intop_0^\infty\frac{\rmd z}{\pi}e^{-\frac{z^2}{2 \pi  x^{2/3}}} \frac{ z^4 (z^6-5)}{(1+z^6)^2}\Big]\,.
\end{align}
The low and high $|x|$ limits of this expression are
\begin{align}
J_1(x,y)\approx
\left\{\begin{array}{cc}
6 \pi  + 15 \pi^2 y &, |x|\ll 1/(2\pi)^{3/2},\\
\frac{\sqrt{2}}{|x|}- \frac{\sqrt{2}}{\pi |x|^{5/3}} +\frac{y}{\pi x^2} &, |x|\gg 1/(2\pi)^{3/2}\,.
\end{array}
\right.
\end{align}
Hence, the contribution from the p-wave to the virial coefficient $b_2^{\rm (int)}$ for $|a_1^{(1)}r_{{\rm ef},1}^{(1)}|\ll \lambda^2/(2\pi)$ is
\begin{align}
& b_{2,l=1}^{\rm (int)} = -9\frac{a_{1}^{(1)}}{\lambda^3}
J_1\big(\frac{a_1^{(1)}}{\lambda^3}, \frac{a_1^{(1)}r_{{\rm ef},1}^{(1)}}{\lambda^2}\big)
\nonumber\\
&\approx\left\{
\begin{array}{lc}
-9\frac{a_{1}^{(1)}}{\lambda^3}\Big[6\pi +15\pi^2\frac{a_1^{(1)}r_{{\rm ef},1}^{(1)}}{\lambda^2}\Big]  &, \frac{|a_1^{(1)}|}{\lambda^3}\ll  \frac{1}{(2\pi)^{3/2}} \\
-9\frac{a_{1}^{(1)}}{\lambda^3}\Big[\frac{\sqrt{2}\lambda^3}{|a_{1}^{(1)}|} -\frac{\sqrt{2}\lambda^5}{\pi |a_1^{(1)}|^{5/3}} +\frac{\lambda^4 r_{{\rm ef},1}^{(1)}}{\pi a_1^{(1)}}\Big] &, \frac{|a_1^{(1)}|}{\lambda^3}\gg \frac{1}{(2\pi)^{3/2}}
\end{array}
\right.
\nonumber\\
&\approx\left\{
\begin{array}{lc}
- \frac{(9 m_f T)^{3/2}}{(2\pi)^{1/2}} a_{1}^{(1)},
\quad |a_1^{(1)}|\ll   \frac{\lambda^3}{(2\pi)^{3/2}} &
  \\
-9\sqrt{2}\Big[{\rm sgn}(a_{1}^{(1)})
-\frac{2}{m_fT[a_1^{(1)}]^{2/3}}
+\sqrt{\frac{r_{{\rm ef},1}^{(1)2}}{\pi m_f T}} \Big], &\\
\phantom{x}\quad\qquad\qquad\qquad |a_1^{(1)}|\gg  \frac{\lambda^3}{(2\pi)^{3/2}} &
\end{array}
\right.
\label{b2-l1-effr-large-a}
\end{align}
The condition for the applicability of the small scattering-length limit ($|a_1^{(1)}|\ll \lambda^3/(2\pi)^{3/2}$) looks for the p-wave as follows, $T\ll \frac{1}{m_f [a_{1}^{(1)}]^{2/3}}\approx 9\, {\rm MeV}/(a_1^{(1)}/10\,{\rm fm}^3)^{2/3}$. Comparing this condition with that for the s-wave, Eq.~(\ref{srsa-I}), we see that the applicability range get broadened  and is less sensitive to the value of the scattering volume. Typical values of the p-wave scattering volumes are within the range  $0.3\,{\rm fm}^3\lsim |a_{^3{\rm P}_0, ^3{\rm P}_1, ^3{\rm P}_2}| \lsim 4\,{\rm fm}^3$~\cite{Dumbrais-83}. Thus, one can use the limit $|a_1^{(1)}|\ll \lambda^3$ for temperatures $T\ll 100$\,MeV.

Now let us evaluate s-wave and p-wave contributions to the virial coefficient $b_2^{\rm (int)}$. Substituting parameters (\ref{a-reff-param}) in Eq.~(\ref{Jl-def}) we can write
\begin{align}
b_2^{\rm (int)} &\approx b_{2,l=0}^{\rm (int)} + b_{2,l=1}^{\rm (int)}
\nonumber\\
b_{2,l=0}^{\rm (int)} &=  -\frac{a_{^1{\rm S}_0}}{\lambda}
J_0\big(\frac{a_{^1{\rm S}_0}}{\lambda}, \frac{a_{^1{\rm S}_0}r_{^1{\rm S}_0}}{\lambda^2}\big)
\nonumber\\
b_{2,l=1}^{\rm (int)} &=  -\sum_{J=0,1,2}(2J+1)\frac{a_{^3{\rm P}_J}}{\lambda^3}
J_1\big(\frac{a_{^3{\rm P}_J}}{\lambda^3}, \frac{a_{^3{\rm P}_J}r_{^3{\rm P}_J}}{\lambda^2}\big).
\label{b-effr-l}
%\nonumber\\
%&-3\frac{a_{^3{\rm P}_1}}{\lambda^3}
%J_1\big(\frac{a_{^3{\rm P}_1}}{\lambda^3}, \frac{a_{^3{\rm P}_1}r_{^3{\rm P}_1}}{\lambda^2}\big)
%\nonumber\\
%&-5\frac{a_{^3{\rm P}_2}}{\lambda^3}
%J_1\big(\frac{a_{^3{\rm P}_2}}{\lambda^3}, \frac{a_{^3{\rm P}_2}r_{^3{\rm P}_2}}{\lambda^2}\big)
\end{align}
These coefficients -- the sum, $b_2^{\rm (int)}$, and separate contributions from various partial waves, -- are shown in Fig.~\ref{fig:b2-phase} by solid lines as functions of the temperature for the low-energy parameters (\ref{a-reff-param}).  For comparison, by dashed lines we depict the result obtained with Eq.~(\ref{b2-phase}), where we substituted the interpolated phase shifts shown in Fig.~\ref{fig:delta-NN}. We see that for temperatures $T< 15$\,MeV the effective range approximation is quite adequate for calculations of the second virial coefficients. The s-wave contribution indeed corresponds to the large-scattering length limit of Eq.~(\ref{J0-exp}), see Eq.~(\ref{b2-effr-large-a}). The $^3{\rm P}_0$ and $^3{\rm P}_2$ waves give attractive contributions and $^3P_1$ wave gives a repulsive one, which essentially compensate the $^3{\rm P}_2$ part. To conclude about the potential importance of inclusion of the higher partial waves we compare our result for $b_2^{\rm (int)}$ with the calculations from Ref.~\cite{Horowitz2006_2}\footnote{In Ref.~\cite{Horowitz2006_2} the virial coefficients are defined by factor 2 smaller than in our work, cf. our Eq.~(\ref{P-class-z}) and Eq.~(3) there.},  where partial waves up to $l=6$ are included. Comparing dots with dashed line (labeled S+P) calculated with the interpolated phase shifts  we see that higher partial waves become sizable for $T\gsim 10$\,MeV. Solid lines show the results of calculation with the phase shifts given in the effective range approximation, which overestimate the results obtained with the interpolated phase shifts. We see that this approximation can be trusted for $T\lsim 10$\,MeV.

\begin{figure}
\centering
\includegraphics[width=6cm]{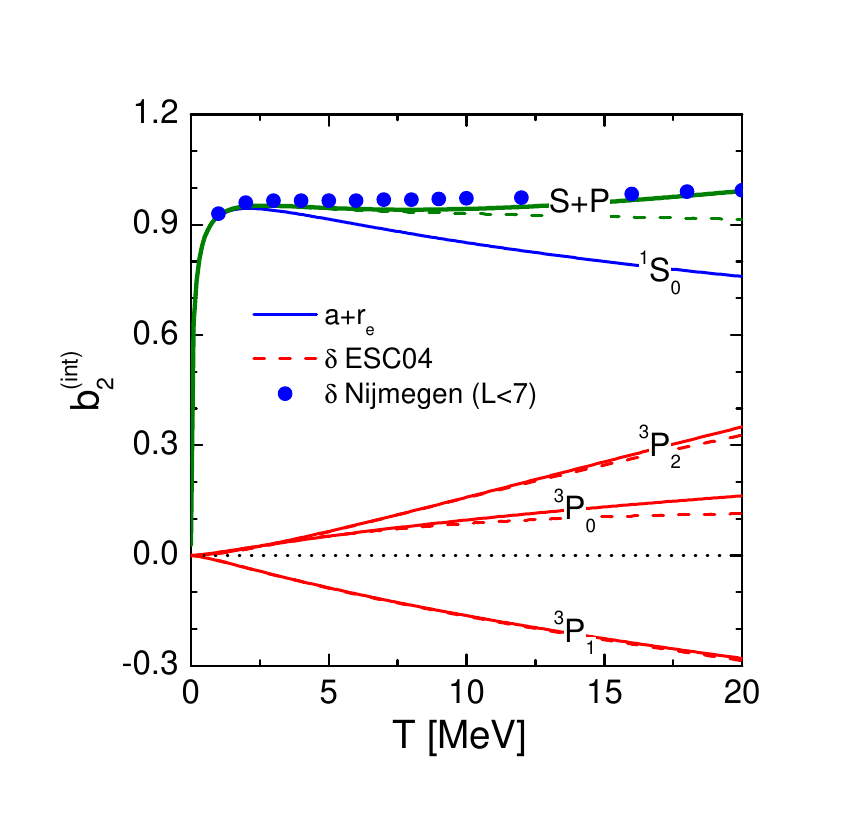}
\caption{Second virial coefficient and its partial-wave contributions versus temperature. Solid lines are calculated with Eq.~(\ref{b-effr-l}) and phase shifts with $l=0,1$ taken in the effective-range approximation (\ref{delta-effr}) with the low-energy parameters (\ref{a-reff-param}). Dashed lines are calculated with Eq.~(\ref{b2-phase}) and phase shifts with $l=0,1$ of ESC04 analysis~\cite{Rijken-06} interpolated in the energy range $0\le  E_{\rm lab}\le 200$\,MeV.
Dots show the results from Ref.~\cite{Horowitz2006_2} obtained with the Nimegen phase shifts for all partial waves with $l<7$.}
\label{fig:b2-phase}
\end{figure}

Thus, from the above analysis we see that the nucleon-nucleon interaction gives overall the attractive correction of the order $n^2$ to the equation of state for temperatures up to 20\,MeV. Analysis~\cite{Horowitz2006_2} confirms that $b_2^{\rm (int)}$ remains attractive at least up to $T=50$\,MeV.

Summarizing, in this Section, we first considered general expressions for the first three virial coefficients and then we derived a closed expression for the second virial coefficient through the T-matrix~(\ref{b2-T-1}) and then through the phase shifts without invoking any fictitious space boundary. We derived parameters of the effective-range approximation for the p-wave scattering amplitude obtained by the Nijmegen group~\cite{Rijken-06}, see Fig.~\ref{fig:delta-NN}. The experimental scattering parameters were compared with the results of Bonn approximation, which is demonstrated to be inapplicable at the hand of explicit calculation of the double-scattering contribution to the scattering amplitude, see Fig.~\ref{fig:F2rat}. Next, we studied the temperature dependence of the second virial coefficient using the effective range approximation and derived general analytical expressions for the s-wave and p-wave contributions in terms of the scattering lengths and the range terms.
We demonstrated in Fig.~\ref{fig:b2-phase} that for $E<50$ MeV the $b_2$ coefficient, calculated within the effective range approximation, closely agrees with the results obtained with the analysis~\cite{Horowitz2006_2} including experimental partial waves with $l<7$.
Because the experimental $nn$ scattering length is anomalously large and negative owing to the presence of the quasi-bound state close to the scattering threshold, the interacting part of the second virial coefficient proves to be nearly constant as a function of the temperature for 2\,MeV$<T<$20\,MeV and its magnitude is insensitive to the magnitude of the scattering length.

\section{Virial coefficients in semiclassics}\label{Virsemicl}

\subsection{ Second virial coefficients in semiclassical expansion}\label{semiclsecvircoef}

\subsubsection{Semiclassical expansion of Eqs.~(\ref{b2-full})-(\ref{b2int-full}) }

Let us now demonstrate how the full quantum expressions for the second virial coefficients (\ref{b2-full})-(\ref{b2int-full}) turn into its classical counterpart Eq.~(\ref{bN-cl}) at the degeneracy factor $d_f=2$ and find quantum correction terms.

Consider now the matrix elements entering the first term in Eq.~(\ref{b2int-full}). We name the first term with matrix elements between the same coordinate states as the \emph{direct} term, and the second term, as the \emph{exchange} term.
The matrix element of the relative kinetic energy can be evaluated using Eq.~(\ref{Q1}),
\begin{align}
\bra{\vec{r\,}}e^{-\hat{T}_{\rm rel}/T}\ket{\vec{r\,}'}=\frac{1}{2^{3/2}\lambda^3}
e^{-\pi(\vec{r}-\vec{r\,}')^2/(2\lambda^2)}\,,
\label{Tkin-exch}
\end{align}
where we take into account the difference in the particle mass in $\hat{T}_{\rm rel}$ and $\hat{T}_1$.

We evaluate the matrix element of $e^{\frac{-\hat{H}^{(2)}_{\rm rel}}{T}}$ inserting the complete sets of momentum states (\ref{complete-pp-rr}),
\begin{align}
&\bra{\vec{r\,}'}e^{\frac{-\hat{H}^{(2)}_{\rm rel}}{T}} \ket{\vec{r}\,}
=\intop\frac{\rmd^3 p' \rmd^3 p}{(2\pi\hbar)^3} \braket{\vec{p\,}'|\vec{p}\,}I_{\vec{p}\vec{p\,}'}(\vec{r},\vec{r\,}',\beta)\,,
\label{I-rist}
\end{align}
where
\begin{align}
&I_{\vec{p}\vec{p\,}'}(\vec{r},\vec{r\,}',\beta) =
(2\pi\hbar)^3\varphi^*_{\vec{p}}(\vec{r}\,)
e^{-\beta\hat{H}^{(2)}_{\rm rel}}\varphi_{\vec{p\,}'}(\vec{r\,}')\,,
\label{I-def}
\end{align}
and the operators act on the plane-wave function, $\varphi_{\vec{p}}(\vec{r}\,)=e^{i\vec{p\,}\vec{r}/\hbar}/(2\pi\hbar)^{3/2}$, see Eq.~(\ref{rp-wf}). We use the notation $\beta=1/T$.
Further on in this Section we will write out $\hbar$ explicitly for the sake of book-keeping.
Now, taking into account that
\begin{align}
I_{\vec{p}\vec{p\,}'}(\vec{r},\vec{r\,}',\beta) = (2\pi\hbar)^3\varphi^*_{\vec{p}}(\vec{r\,})
\varphi_{\vec{p}}(\vec{r\,}')I_{\vec{p}\vec{p\,}'}(\vec{r\,}',\vec{r\,}',\beta),
\end{align}
we can write the matrix element in the form
\begin{align}
&\bra{\vec{r\,}'}e^{\frac{-\hat{H}^{(2)}_{\rm rel}}{T}} \ket{\vec{r}\,}
=\intop \frac{\rmd^3 p}{(2\pi\hbar)^3}
e^{i\vec{p}(\vec{r\,}' - \vec{r})/\hbar}
I_{\vec{p\,}\vec{p}}(\vec{r\,}',\vec{r\,}',\beta)
\,,
\end{align}
where the translation invariance is evident. To find the function $I_{\vec{p\,}\vec{p}}(\vec{r\,},\vec{r},\beta)$ we use the method proposed in Ref.~\cite{Kirkwood33}, see also   \S\,33 in~\cite{LLV}. Differentiating Eq.~(\ref{I-def}) for $\vec{r\,}'=\vec{r}$ with respect to $\beta$ we obtain the differential equation for the function $I_{\vec{p}\vec{p}}=I_{\vec{p\,}\vec{p}}(\vec{r\,},\vec{r},\beta)$,
\begin{align}
&\frac{\partial I_{\vec{p}\vec{p}}}{\partial \beta}
= -(2\pi\hbar)^3 \varphi^*_{\vec{p}}(\vec{r}\,) \hat{H}^{(2)}_{\rm rel}
e^{-\beta\hat{H}^{(2)}_{\rm rel}}\varphi_{\vec{p}}(\vec{r}\,)
\label{Ipp-eq}\\
&= -(2\pi\hbar)^3 \varphi^*_{\vec{p}}(\vec{r}\,) \Big(-\frac{\hbar^2}{m_f}\nabla^2 + \Ucor(|\vec{r}\,|)\Big)
\varphi_{\vec{p}}(\vec{r}\,)
I_{\vec{p}\vec{p}}
\nonumber\\
&=-\big(\frac{p^2}{m_f}+\Ucor(|\vec{r}\,|)\big)I_{\vec{p}\vec{p}} + 2 i \frac{\hbar}{m_f} p_a \big(\nabla_aI_{\vec{p}\vec{p}}\big)
+\frac{\hbar^2}{m_f}\big(\nabla^2I_{\vec{p}\vec{p}}\big)\nonumber
\end{align}
with the boundary condition $I_{\vec{p}\vec{p}}=1$ for $\beta=0$.  Equation (\ref{Ipp-eq}) mixes various orders in $\hbar$, and we can search the solution as a series of successive approximations
\begin{align}
I_{\vec{p}\vec{p}}(\vec{r\,},\vec{r},\beta)=e^{-\beta(\frac{p^2}{m_f}+\Ucor(|\vec{r}\,|)}\sum_{k=0}^\infty\hbar^k \chi_k(\vec{r},\vec{p}\,;\beta)\,,
\label{I-exp-1}
\end{align}
where $\hbar^k$ indicates the order in the semiclassical expansion. Substituting expansion (\ref{I-exp-1}) in (\ref{Ipp-eq}) we obtain the following set of coupled differential equations for functions $\chi_k$ :
\begin{align}
\frac{\partial \chi_k}{\partial \beta} &=
-i\frac{2}{m_f}(\vec{p}\,\hat{\vec{\mathcal{D}}}_\beta) \chi_{k-1}
+\frac{1}{m_f}\hat{\vec{\mathcal{D}}}_\beta^2\chi_{k-2} \,,
\nonumber\\
\hat{\vec{\mathcal{D}}}_{\beta} &= \beta (\nabla \Ucor) - \nabla,
\label{I-eq}
\end{align}
where $\chi_0=1$ and boundary conditions for other functions are $\chi_{k}(\beta=0)=0$ for $k > 0$. Here each operator $\nabla$ standing outside  the bracket acts on everything to the right of it. Operator $\nabla$ within brackets, e.g. $(\nabla_a \Ucor)$, acts only on the nearest function. With the help of this recursion formula one can calculate  any term in the series (\ref{I-exp-1}) performing  operations of
differentiation and integration with  the two preceding terms. The first two functions $\chi_{k\le 2}$ are given in Appendix~\ref{app:QPexp}.
Using expansion (\ref{I-exp-1}) we can write
\begin{align}
&\bra{\vec{r\,}'}e^{\frac{-\hat{H}^{(2)}_{\rm rel}}{T}} \ket{\vec{r}\,}
= e^{-\frac{\Ucor(|\vec{r}\,|)}{T}}\intop \frac{\rmd^3 p}{(2\pi\hbar)^3}
e^{-\frac{p^2}{m_f T}+\frac{i}{\hbar}\vec{p}(\vec{r\,}' - \vec{r})}
\nonumber\\
&\times \big(1+\hbar\chi_1(\vec{r},\vec{p};\beta) + \hbar^2\chi_2(\vec{r},\vec{p};\beta)
+O(\hbar^3)\big) \,.
\label{h-exp}
\end{align}
Substituting this expansion in Eq.~(\ref{b2int-full}) and using the momentum integrals (\ref{p-aver}) and (\ref{chi12-paver}) we separate the direct and exchange terms stemming to the first and second matrix elements in (\ref{b2int-full}), respectively
\begin{align}
b_2 &= b_2^{\rm (dir)}+b_2^{\rm (exch)} ,
\label{b2-dir-exch}
\end{align}
where the direct term is
\begin{align}
\lambda^3 b_2^{\rm (dir)}&= 2^{5/2}\lambda^3\intop\rmd^3 r\, \bra{\vec{r}\,} e^{-\hat{H}^{(2)}_{\rm rel}/T} - e^{-\hat{T}_{\rm rel}/T}\ket{\vec{r}\,}\nonumber\\
&=\lambda^3 b_2^{\rm (cl)}+\lambda^3\delta b_2^{\rm (dir)}
\nonumber\\
&\approx 2\intop\rmd^3 r\big(e^{-\frac{\Ucor(|\vec{r}\,|)}{T}} -1 \big)
\nonumber\\
&+ \frac{\lambda^2}{12\pi} \intop\rmd^3 r
e^{-\frac{\Ucor(|\vec{r}\,|)}{T}}\Big( \frac{(\nabla U)^2}{T^2}  -2\frac{\nabla^2 U}{T}\Big)\,.
\label{b2-dir}
\end{align}
We remark that only terms with even powers of $\hbar$ from expansion (\ref{h-exp}) can contribute to the direct term, since all  momentum integrals of the odd terms $\chi_{2k+1}$ in Eq.~(\ref{h-exp}) are proportional to $\vec{r}-\vec{r'}$ and vanish when we put $\vec{r\,}'=\vec{r}$ in the direct term. However these terms will contribute to the exchange term where $\vec{r\,}'=-\vec{r}$.
Performing integrations by parts and using the relations (\ref{byparts-1}) we identify in the direct term (\ref{b2-dir}) the classical limit of the virial coefficient (\ref{bN-cl})
%(\ref{b2-class})
and the quantum correction
\begin{align}
\lambda^3 \delta b_{2}^{\rm (dir)} &=- \frac{\lambda^2}{12\pi} \intop\rmd^3 r
e^{-\frac{\Ucor(|\vec{r}\,|)}{T}} \frac{(\nabla U)^2}{T^2}\,.
\label{b2-dir-2}
\end{align}
 The correction term was first derived in Refs.~\cite{Gropper-36,Uhlenbeck-Beth}.
It is formally of the order $\lambda^2\propto \hbar^2$ or, saying  more accurately,  the effective expansion parameter of the semiclassical expansion here is $\lambda^2/r_{\rm typ}^2$, where $r_{\rm typ}$ is, as before, the typical distance for the change of the potential.
The next term in $\lambda^3 b_2^{\rm (dir)}$, which we disregarded in (\ref{b2-dir}), is proportional to $\lambda^4/r_{\rm typ}^4$, see Ref.~\cite{Gropper-36}.

Combining the virial coefficient of the ideal Fermi gas $b_2^{(0)}$, Eq.~(\ref{b0}), and the second term in Eq.~(\ref{b2int-full}) we can write the exchange contribution to the full second virial coefficient (\ref{b2-decomp}) as
\begin{align}
&\lambda^3 b_2^{\rm (exch)} = -\frac{\lambda^3}{2^{3/2}}
\nonumber\\
&\quad -2^{3/2}\lambda^3\intop\rmd^3 r\,
\bra{-\vec{r\,}} e^{-\hat{H}^{(2)}_{\rm rel}/T} -e^{-\hat{T}_{\rm rel}/T} \ket{\vec{r}\,}
\Big].
\end{align}
Substituting the matrix elements (\ref{Tkin-exch}) and (\ref{h-exp}) with functions $\chi_{1,2}$ from Eq.~(\ref{chi12-paver}) of Appendix \ref{app:QPexp} we find
\begin{align}
\lambda^3 b_2^{\rm (exch)} &= \lambda^3 b_{2,{\rm mf}}^{\rm (exch)}+\lambda^3 b_{2,\nabla}^{\rm (exch)}\,,
\label{b2-exch}
\\
\lambda^3 b_{2,{\rm mf}}^{\rm (exch)}& = -\frac{\lambda^3}{2^{3/2}} -\intop \rmd^3 r
e^{-2\pi r^2/\lambda^2}\big(e^{-U(r)/T} -1\big)\,,
\label{b2-exch-mf}
\\
\lambda^3 b_{2,\nabla}^{\rm (exch)} &\approx  - \intop \rmd^3 r
e^{-2\pi r^2/\lambda^2} e^{-U(r)/T} \Big[ \frac{(\vec{r}\,\nabla U)}{ T}
\nonumber\\
&+
\frac{\lambda^2}{4\pi}\frac16\Big( \frac{(\nabla U)^2}{T^2}  -2\frac{\nabla^2 U}{T}\Big)
  \nonumber\\
  &+\frac{1}{2}
  \Big( \frac{(\vec{r}\, \nabla U)^2}{T^2} - \frac{4}{3} \frac{r_a r_b\nabla_a\nabla_bU}{T} \Big) \Big]\,.
\label{b2-exch-grad}
\end{align}
Here we separated  explicitly the `mean-field' term independent on the gradient of the potential, $\lambda^3 b_{2,{\rm mf}}^{\rm (exch)}\propto \lambda^3$, since it is determined by typical $r\sim \lambda/\sqrt{2\pi}$. The first term in expression Eq.~(\ref{b2-exch-grad}) for $\lambda^3 b_{2,\nabla}^{\rm (exch)}$ is $\propto \lambda^4/r_{\rm typ}$ and other terms $\propto \lambda^5/r_{\rm typ}^3$. Note that the part of the  exchange term (\ref{b2-exch-mf}) and the first term in Eq.~(\ref{b2-exch-grad}) were obtained in Ref.~\cite{Gropper-36}.
All the gradient terms $\lambda^3 b_{2,\nabla}^{\rm (exch)}$ are formally of order  $\lambda^{3+n}$, $n\ge 1$, and hence must be dropped with our accuracy, since the correction terms $\propto \lambda^4$ have been disregarded in the direct terms, cf.  (\ref{b2-dir-2}).

With the help of Eqs.~(\ref{byparts-2}) and (\ref{byparts-3}) we can reduce the order of derivatives acting on the potentials in Eq.~(\ref{b2-exch-grad}) and cast it in the  form
\begin{align}
\lambda^3 b_{2,\nabla}^{\rm (exch)}
&\approx  - \intop \rmd^3 r
e^{-2\pi r^2/\lambda^2} e^{-U(r)/T}
\nonumber\\
&\times
\Big[\frac{10}{3}\Big(1 - \frac{4}{5}\frac{\pi r^2}{\lambda^2}\Big) \frac{(\vec{r}\,\nabla U)}{ T}
\nonumber\\
&
-\frac16 \frac{(\vec{r}\, \nabla U)^2}{T^2}-
\frac{\lambda^2}{24\pi}\frac{(\nabla U)^2}{T^2}
 \Big].
  \label{b2-exch-2}
\end{align}

From Eqs.~(\ref{b2-exch-mf}) and (\ref{b2-exch-grad}) we see that the limits $\lambda\to 0$ and  $U\to 0$ are not interchangeable in the exchange terms:
$\lim_{U\to 0}\lim_{\lambda\to 0}\big[\lambda^3 b_2^{\rm (exch)}\big]=0$, whereas
$\lim_{\lambda\to 0}\lim_{U\to 0}\big[\lambda^3 b_2^{\rm (exch)}\big] = \lambda^3 b_2^{(0)}$.
Thus we conclude that for a fixed potential the full quantum mechanical expression for $b_2$ reduces to the classical limit
\begin{align}
b_2(\lambda \to 0)\to b_2^{(\rm cl)}
\end{align}
and the exchange term vanishes in this limit.

To clarify the expansion orders of various terms in semiclassical expansion of the second virial coefficients,
we will apply the derived formulas for the smoothed square barrier-well potential (\ref{Usw-smooth}) and to the nuclear potentials introduced in Section~\ref{app:NNpot}.

\subsection{Estimations for test potentials}\label{semicltest}
\subsubsection{Smoothed square barrier-well potential }

For the square barrier-well potential the classical part of the second virial coefficient (\ref{bN-cl})
can be expressed with the help of Eq.~(\ref{Bsw}) as follows
\begin{align}
\lambda^3 b_2^{\rm (cl)}=-
2\big( 1-e^{-U_{\rm r}/T} \big)  v^{\rm (sw)}_{\rm r}
-2\big( 1-e^{U_{\rm a}/T} \big)  v^{\rm (sw)}_{\rm a},
\label{b2cl-sw}
\end{align}
where we used the repulsion and attraction volumes, $v^{\rm (sw)}_{\rm r,a}$, for the square barrier-well potentials defined in Eq.~(\ref{vr-va-def}).

The semiclassical approximation is valid only for smooth potentials. Therefore, for the estimation of the quantum corrections, which contain gradients of the potential, we should use the smoothed version of the square barrier-well potential~(\ref{Usw-smooth}).
The gradient of the potential is
\begin{align}
\frac{\rmd U_{{\rm sw},\sigma}}{\rmd r}=
-\frac{U_{\rm a}+U_{\rm r}}{2\sigma\cosh^2\frac{r-2r_0}{\sigma}}
+\frac{U_{\rm a}}{2\sigma\cosh^2\frac{r-r_U}{\sigma}}\,.
\label{DUsw}
\end{align}
We  assume that $\sigma\ll \min\{2r_0,r_U-2r_0\}$. Then
\begin{align}
\Big(\frac{\rmd U_{{\rm sw},\sigma}}{\rmd r}\Big)^2 &\approx
\frac{(U_{\rm a}+U_{\rm r})^2}{4\sigma^2\cosh^4\frac{r-2r_0}{\sigma}}
+\frac{U_{\rm a}^2}{4\sigma^2\cosh^4\frac{r-r_U}{\sigma}}\nonumber\\
&+O(e^{-2r_0/\sigma}, e^{-r_U/\sigma}, e^{-(r_U-2r_0)/\sigma})\,,
\label{DDUsw}
\end{align}
i.e. we exploit that the derivative of the potential is well localized at distances $r\sim 2r_0$ and $r\sim r_U$. Although $\sigma$ is much shorter than $2r_0,r_U-2r_0$, within a semiclassical expansion scheme we should choose it to be much larger than $\lambda$, i.e., $\sigma\gg\lambda$. Thus terms  $\propto e^{-2r_0/\lambda}$ and
$e^{-2r_U/\lambda}$ will be further disregarded as well as terms  $O(e^{-2r_0/\sigma}, e^{-r_U/\sigma}, e^{-(r_U-2r_0)/\sigma}).$

When we substitute Eq.~(\ref{DDUsw}) in Eq.~(\ref{b2-dir-2}) we obtain
\begin{align}
&\lambda^3 \delta b_{2}^{\rm (dir)} \approx - \frac{\lambda^2}{12T^2\sigma^2} \Bigg[ \intop_0^{2r_0}\rmd r r^2
e^{-\frac{U_{\rm r}}{T}}
\frac{(U_{\rm a}+U_{\rm r})^2}{\cosh^4\frac{r-2r_0}{\sigma}}
\nonumber\\
&\qquad + \intop_{2r_0}^{r_U}\rmd r r^2
e^{\frac{U_{\rm a}}{T}}
\frac{(U_{\rm a}+U_{\rm r})^2}{\cosh^4\frac{r-2r_0}{\sigma}}
\nonumber\\
&\qquad + \intop_{2r_0}^{r_U}\rmd r r^2
e^{\frac{U_{\rm a}}{T}} \frac{U_{\rm a}^2}{\cosh^4\frac{r-r_U}{\sigma}}
+\intop_{r_U}^\infty\rmd r r^2
\frac{U_{\rm a}^2}{\cosh^4\frac{r-r_U}{\sigma}}
\Bigg].
\label{db2-dir}
\end{align}
In the limit $2r_0,r_U-2r_0\gg \sigma$, which we consider, we find
\begin{align}
\lambda^3 \delta b_{2}^{\rm (dir)} \approx & -\frac{1}{24\pi}
\Big[v_{\rm r}^{\rm (sw)}\frac{(U_{\rm a}+U_{\rm r})^2}{T^2}
\big( e^{-\frac{U_{\rm r}}{T}} + e^{\frac{U_{\rm a}}{T}}\big)
\frac{\lambda^2}{\sigma 2r_0 }
\nonumber\\
&+ v_U^{\rm (sw)}\frac{U_{\rm a}^2}{T^2} \big(1 + e^{\frac{U_{\rm a}}{T}}\big)
\frac{\lambda^2}{\sigma r_U }
\Big],
\label{b2dir-sw}
\end{align}
where we used that $\intop_0^\infty \rmd x/\cosh^4x=2/3$. Notations for the repulsion volume of the potential, $v_{\rm r}$, and the total potential volume, $v_U$, were introduced in  Eq.~(\ref{vr-va-def}).

Consider now the exchange term (\ref{b2-exch-mf}), where we separate the integration range in the regions with repulsive, attractive and zero potentials,
\begin{align}
&\lambda^3 b_{2,{\rm mf}}^{\rm (exch)} = - 4\pi e^{-\frac{U_{\rm r}}{T}}
\int_0^{2r_0} \rmd r r^2 e^{-\frac{2\pi r^2}{\lambda^2}}
\nonumber\\
& -
4\pi e^{\frac{U_{\rm a}}{T}}
\int_{2r_0}^{r_U} \rmd r r^2 e^{-\frac{2\pi r^2}{\lambda^2}}
 - 4\pi\int_{r_U}^\infty \rmd r r^2
e^{-\frac{2\pi r^2}{\lambda^2}}.
\end{align}
\begin{widetext}
Integrations can be performed in the closed form. We have
\begin{align}
\lambda^3 b_{2,{\rm mf}}^{\rm (exch)} &=
- e^{-\frac{U_{\rm r}}{T}}\frac{\lambda^3}{2^{3/2}}
- \Big(e^{\frac{U_{\rm a}}{T}}-e^{-\frac{U_{\rm r}}{T}}\Big)
\Big[\lambda^2 2r_0 e^{-\frac{8 \pi r_0^2}{\lambda^2}}
+\frac{\lambda^3}{2^{3/2}} {\rm erfc}\Big(\frac{\sqrt{2\pi} (2r_0)}{\lambda}\Big)
\Big]
\nonumber\\
&-\Big(1-e^{\frac{U_{\rm a}}{T}}\Big)
\Big[\lambda^2 r_U e^{-\frac{2 \pi r_U^2}{\lambda^2}}+
\frac{\lambda^3 }{2^{3/2}}{\rm erfc}\Big(\frac{\sqrt{2 \pi } r_U}{\lambda}\Big)
\Big]\,.
\end{align}
Taking into consideration asymptotic of the complimentary error function, as in derivation of Eq.~(\ref{b2Phi-F-2}), we obtain
\begin{align}
\lambda^3 b_{2,{\rm mf}}^{\rm (exch)} &=
-e^{-\frac{U_{\rm r}}{T}}\frac{\lambda^3}{2^{3/2}}
- \frac{3v_{\rm r}^{\rm (sw)}}{4\pi}\Big(e^{\frac{U_{\rm a}}{T}}-e^{-\frac{U_{\rm r}}{T}}\Big)
\Big[ \frac{\lambda^2}{(2r_0)^2} + \frac{\lambda^4}{2\pi (2r_0)^4}
\frac{1}{1 + \sqrt{1+\frac{\zeta\lambda^2}{8\pi r_0^2}}}
\Big] e^{-\frac{8 \pi r_0^2}{\lambda^2}}
\nonumber\\
&-\frac{3v_U^{\rm (sw)}}{4\pi}\Big(1-e^{\frac{U_{\rm a}}{T}}\Big)
\Big[\frac{\lambda^2}{r_U^2} + \frac{\lambda^4}{2\pi r_U^4}
\frac{1}{1 + \sqrt{1+\frac{\zeta\lambda^2}{2\pi r_U^2}}}
\Big] e^{-\frac{2 \pi r_U^2}{\lambda^2}} \simeq -e^{-\frac{U_{\rm r}}{T}}\frac{\lambda^3}{2^{3/2}} +  O\left(e^{-\frac{8 \pi r_0^2}{\lambda^2}}, e^{-\frac{2 \pi r_U^2}{\lambda^2}}\right) \,.
\label{b2exch-mf-sw}
\end{align}
%{\bf In the last equality  exponentially small  terms $e^{-\frac{8 \pi r_0^2}{\lambda^2}}$, $e^{-\frac{8 \pi r_U^2}{\lambda^2}}$  are omitted, since we have dropped larger  contributions $\propto e^{-2r_0/\lambda}$ and $e^{-r_U/\lambda}$.  gde vikinuti? i po kakoi prichine? Mozhet imeli vvidu $e^{-2r_0/\sigma}$, $e^{-r_U/\sigma}$, inache vse ozenki dalee ne godyatsia?}

The gradient exchange term is estimated with the help of the separation  of the integration range on two regions, as in Eq.~(\ref{db2-dir}). Employing (\ref{DUsw}) and (\ref{DDUsw}) we find
\begin{align}
\lambda^3 b_{2,\nabla}^{\rm (exch)}
\approx & - 4\pi v_{\rm r}^{\rm (sw)}
e^{-\frac{8\pi r_0^2}{\lambda^2}}
\big(e^{-\frac{U_{\rm r}}{T}} + e^{\frac{U_{\rm a}}{T}} \big)\frac{(U_{\rm a}+U_{\rm r})}{ T}
\Big[\frac{(2r_0)^2}{\lambda^2} - \frac{5}{4\pi}
-\frac{(U_{\rm a}+U_{\rm r})}{48T} \frac{(2r_0)}{\pi\sigma}
\Big(1 + \frac{1}{4\pi}\frac{\lambda^2}{(2r_0)^2} \Big) \Big]
\nonumber\\
&- 4\pi v_U^{\rm (sw)}
e^{-\frac{2\pi r_U^2}{\lambda^2}}
\big(1 + e^{\frac{U_{\rm a}}{T}} \big)
\frac{U_{\rm a}}{ T}
\Big[\frac{r_U^2}{\lambda^2} - \frac{5}{4\pi}
+\frac{U_{\rm a}}{48T} \frac{r_U}{\pi\sigma}
\Big(1 + \frac{1}{4\pi}\frac{\lambda^2}{r_U^2} \Big) \Big] \simeq  O\left(e^{-\frac{8 \pi r_0^2}{\lambda^2}}, e^{-\frac{2 \pi r_U^2}{\lambda^2}}\right) \,.
\label{b2exch-grad-sw}
\end{align}
\end{widetext}
Further,  simplifying consideration we shall assume $e^{-\frac{8 \pi r_0^2}{\lambda^2}},e^{-\frac{2 \pi r_U^2}{\lambda^2}}\ll 1$ and  drop the exchange-gradient correction. Certainly, as we see, it cannot be done at least at very low temperature.

Now we are able to compare the classical result (\ref{b2cl-sw}) with the direct and exchange terms derived within semiclassical expansion, (\ref{b2dir-sw}), (\ref{b2exch-mf-sw}), and (\ref{b2exch-grad-sw}). For simplicity  consider separately the cases of purely repulsive and purely attractive potentials.

For the purely repulsive square barrier-well potential we put $U_{\rm a}=0$ and get
\begin{align}
&\frac{\delta b_{2}^{\rm (dir)} }{b_2^{\rm (cl)}} \approx
\frac{1}{48\pi}\frac{\lambda^2}{2r_0\sigma}
\frac{U_{\rm r}^2}{T^2}\coth\frac{U_{\rm r}}{2T}\,,
\nonumber\\
&\frac{ b_{2,{\rm mf}}^{\rm (exch)} }{ b_2^{\rm (cl)} } \approx
\frac{3}{8\pi}
\frac{2^{-3/2}}{\big(e^{U_{\rm r}/T}-1
\big)  }\frac{\lambda^3}{(2r_0)^3}\,,
\nonumber\\
&\frac{b_{2,\nabla}^{\rm (exch)} }{ b_2^{\rm (cl)} } \approx 0\,.
\label{RepulsUcor}
\end{align}
In the high temperature limit, $T\gg U_{\rm r}$, the first term
${\delta b_{2}^{\rm (dir)} }/{b_2^{\rm (cl)}}$ is a positive (attractive) correction
$\frac{\delta b_{2}^{\rm (dir)} }{b_2^{\rm (cl)}} \approx
\frac{1}{48\pi}\frac{\lambda^2}{2r_0\sigma}
\frac{U_{\rm r}}{T}\propto 1/T^2$, i.e. it decreases with an increasing temperature. Notice the presence of a small numerical coefficient $\frac{1}{48\pi}\simeq \frac{1}{150}$ in this expression. Next correction is also positive, $\frac{ b_{2,{\rm mf}}^{\rm (exch)} }{ b_2^{\rm (cl)} } \approx
\frac{1}{24}\frac{T}{U_{\rm r}}\frac{\lambda^3}{(2r_0)^3}\propto T^{-1/2}$. Thus, both corrections decrease with a temperature increase demonstrating validity of the classical result.
We note that the relative correction $b_{2,{\rm mf}}^{\rm (exch)} /\delta b_{2}^{\rm (dir)}\simeq 3\frac{\lambda\sigma}{2r_0^2}\frac{T^2}{U^2_{\rm r}}\propto T^{3/2}$ increases with an increasing temperature, but its value essentially depends on the relation between the  parameters entering this expression. This fact demonstrates nontrivial dependence of the actual value of the parameter of the semiclassical expansion on the temperature. In the low temperature limit, $T\ll U_{\rm r}$, the correction $\frac{\delta b_{2}^{\rm (dir)} }{b_2^{\rm (cl)}} \approx \frac{1}{48\pi}\frac{\lambda^2}{2r_0\sigma} \frac{U^2_{\rm r}}{T^2} \propto T^{-3}$ strongly increases with decreasing $T$, whereas $\frac{ b_{2,{\rm mf}}^{\rm (exch)} }{ b_2^{\rm (cl)} } \approx \frac{1}{24}e^{-U_{\rm r}/T}\frac{\lambda^3}{(2r_0)^3}$ is exponentially small and can be safely neglected. Thus, at low temperature the classical approximation becomes invalid due to the grows of the correction $\delta b_{2}^{\rm (dir)}$.

%{\bf For the fixed ratio $U_{\rm r}/T$, the first term in (\ref{RepulsUcor}) is a positive (attraction) correction proportional to $\lambda^2/(2r_0 \sigma)$. In the case we consider this factor is much larger than $\lambda^2/(2r_0)^2$ and much smaller than $\lambda/(2r_0)$. Next correction in (\ref{RepulsUcor}) is also positive and   of the order $(\lambda^3/(2r_0)^3)$ for the fixed ratio $U_{\rm r}/T$. On the other hand for $U_{\rm r} \ll T$ the correction ${\delta b_{2}^{\rm (dir)} }/{b_2^{\rm (cl)}}$ is $\propto U_{\rm r}/T\ll 1$ and the correction ${ b_{2,{\rm mf}}^{\rm (exch)} }/{ b_2^{\rm (cl)} } $  is $\propto T/U_{\rm r}$. Thus the ratio
%$b_{2,{\rm mf}}^{\rm (exch)} /\delta b_{2}^{\rm (dir)}\sim \frac{\lambda\sigma}{(2r_0)^2}\frac{T^2}{U^2}$ in this limit and thereby is it small or large depends on relation between various parameters.
%For $U_{\rm r} \gg T$ the correction ${\delta b_{2}^{\rm (dir)}}/{b_2^{\rm (cl)}}$ grows $\propto (U_{\rm r}/T)^2$ with increasing $U_{\rm r}$, whereas the  correction ${ b_{2,{\rm mf}}^{\rm (exch)} }/{ b_2^{\rm (cl)} }$ becomes exponentially small, $\propto e^{-U_{\rm r}/T}$, and can be safely neglected.}

Analogous expressions follow also for the purely attractive potential after the replacement $U_{\rm r}\to -U_{\rm a}$ and $2r_0\to r_U$,
\begin{align}
\frac{\delta b_{2}^{\rm (dir)} }{b_2^{\rm (cl)}} &\approx
- \frac{1}{48\pi} \frac{\lambda^2}{r_U\sigma }\frac{U_{\rm a}^2}{T^2}\coth\frac{U_{\rm a}}{2T}\,,
\nonumber\\
\frac{ b_{2,{\rm mf}}^{\rm (exch)} }{ b_2^{\rm (cl)} } &\approx
-\frac{3}{8\pi}
\frac{2^{-3/2}}{\big(1-e^{-U_{\rm a}/T}
\big)  }\frac{\lambda^3}{r_U^3}\,,
\nonumber\\
\frac{b_{2,\nabla}^{\rm (exch)} }{ b_2^{\rm (cl)} } &\approx 0 \,.
\label{AttractUcor}
\end{align}
In the high temperature limit, $T\gg U_{\rm a}$, the first term
${\delta b_{2}^{\rm (dir)} }/{b_2^{\rm (cl)}}$ is a negative (repulsive) correction
$\frac{\delta b_{2}^{\rm (dir)} }{b_2^{\rm (cl)}} \approx -
\frac{1}{48\pi}\frac{\lambda^2}{r_U\sigma} \frac{U_{\rm a}}{T}\propto - T^{-2}$, i.e. it decreases with increasing temperature. Again we observe the presence of a small numerical coefficient $\frac{1}{48\pi}\simeq 1/150$. Next correction is also negative (repulsive), $\frac{ b_{2,{\rm mf}}^{\rm (exch)} }{ b_2^{\rm (cl)} } \approx -\frac{1}{24}\frac{T}{U_{\rm a}}\frac{\lambda^3}{r_U^3}\propto -T^{-1/2}$. Thus, both corrections decrease in the magnitude with an increasing temperature demonstrating validity of the classical result.
In the low temperature limit, $T\ll U_{\rm a}$, the corrections $\frac{\delta b_{2}^{\rm (dir)} }{b_2^{\rm (cl)}} \approx - \frac{1}{48\pi} \frac{\lambda^2}{r_U\sigma} \frac{U^2_{\rm a}}{T^2}\propto - T^{-3}$ and
$\frac{ b_{2,{\rm mf}}^{\rm (exch)} }{ b_2^{\rm (cl)} } \approx -\frac{1}{24}\frac{\lambda^3}{r_U^3}\propto - 1/T^{3/2}$ strongly increase in the magnitude with decreasing $T$. So we demonstrated that  at low temperature the classical approximation becomes invalid due to the grows of  both corrections $\frac{\delta b_{2}^{\rm (dir)} }{b_2^{\rm (cl)}}$ and $\frac{ |b_{2,{\rm mf}}^{\rm (exch)} |}{ b_2^{\rm (cl)} }$. Also notice that the correction $\delta b_{2}^{\rm (dir)}$ grows with decreasing $T$ more rapidly than $|b_{2,{\rm mf}}^{\rm (exch)}|$ and that the correction $\frac{ b_{2,{\rm mf}}^{\rm (exch)} }{ b_2^{\rm (cl)} }$ is independent on the potential $U_{\rm a}$ in this limit.

Summarizing, on example of the barrier-well potential we explicitly demonstrated that for both  the repulsive and attractive potentials the classical expression for the second virial coefficient holds in the high temperature limit and the low temperature limit is purely quantum, as it was expected.

\subsubsection{Estimations for the nuclear potential}

Now we will study  relations between various contributions in the semiclassical expansion of  the second virial coefficients, which we have derived in Section~\ref{semiclsecvircoef},  using the  test nuclear potentials introduced in Section~\ref{app:NNpot}. To be specific we focus on  the case of the Yukawa type potential (\ref{U-potR}) with Walecka parameters (\ref{Uomsig-par-W}).

In  Fig.~\ref{fig:Bquant} we show the classical limit of the second virial coefficient and the quantum corrections as functions of the temperature for the gas of neutrons ($m_f=m_N=940$\,MeV) interacting via the  potential (\ref{U-potR}) with Walecka parameters (\ref{Uomsig-par-W}).
Bold solid line shows the result obtained in the classical approximation, $b_2^{\rm (cl)}$ in  Eq.~(\ref{bN-cl}). Dash line depicts the quantum correction of the order $\hbar^2$ in expansion (\ref{h-exp}) from the direct term, $\delta b_{2}^{(\rm dir)}$ given by Eq.~(\ref{b2-dir-2}). Dash-dotted line stands for the quantum correction term $b_{2,{\rm mf}}^{(\rm exch)}$ of the order $\hbar^3$ given by Eq.~(\ref{b2-exch-mf}). We see that in the given case at all $T$ the $\hbar^2$ correction $|\delta b_{2}^{(\rm dir)}|$ exceeds the $\hbar^3$ correction $|b_{2,{\rm mf}}^{(\rm exch)}|$. Short-dash line shows the quantum correction term $b_{2,\nabla}^{(\rm exch)}\propto \lambda^4$ given by Eq.~(\ref{b2-exch-grad}). In our barrier-well example considered above we disregarded this correction assuming that it is exponentially small, cf. Eq.~(\ref{b2exch-grad-sw}). In the given example we see that the gradient correction blows off for $T<50$ MeV but can be neglected for $T>50$ MeV. Dash-double-dot line shows the sum $b_{2}^{(\rm cl)}+\delta b_{2}^{(\rm dir)}$ including only  corrections of the order of $\hbar^2$, whereas dotted line demonstrates the sum $b_{2}^{(\rm cl)}+\delta b_{2}^{(\rm dir)}+b_{2,{\rm mf}}^{(\rm exch)}$  including corrections of the order of $\hbar^3$. As we see from the figure,  the semiclassical approximation in the given example becomes completely invalid  for $T\lsim 50$ MeV, since the quantum corrections of the higher order give the contribution of the order or even  larger than the main classical term at such temperatures. Oppositely, for $T\gsim 100$ MeV semiclassical approach  becomes a rather good approximation. Note also that the sum $b_{2}^{(\rm cl)}+\delta b_{2}^{(\rm dir)}$ calculated in the order of $\hbar^2$  is closer to the classical term $b_{2}^{(\rm cl)}$ than the sum $b_{2}^{(\rm cl)}+\delta b_{2}^{(\rm dir)}+b_{2,{\rm mf}}^{(\rm exch)}$  including corrections of the order of $\hbar^3$. This reflects the fact that the semiclassical series  is  asymptotic series. Due to that in the ordinary quantum mechanics one usually keeps only two terms in the semiclassical expansion.

Figure~\ref{fig:Bquant} clearly shows that at rather high temperature the virial coefficients can be considered within the semiclassical approach.

\begin{figure}
\centering
\includegraphics[width=8cm]{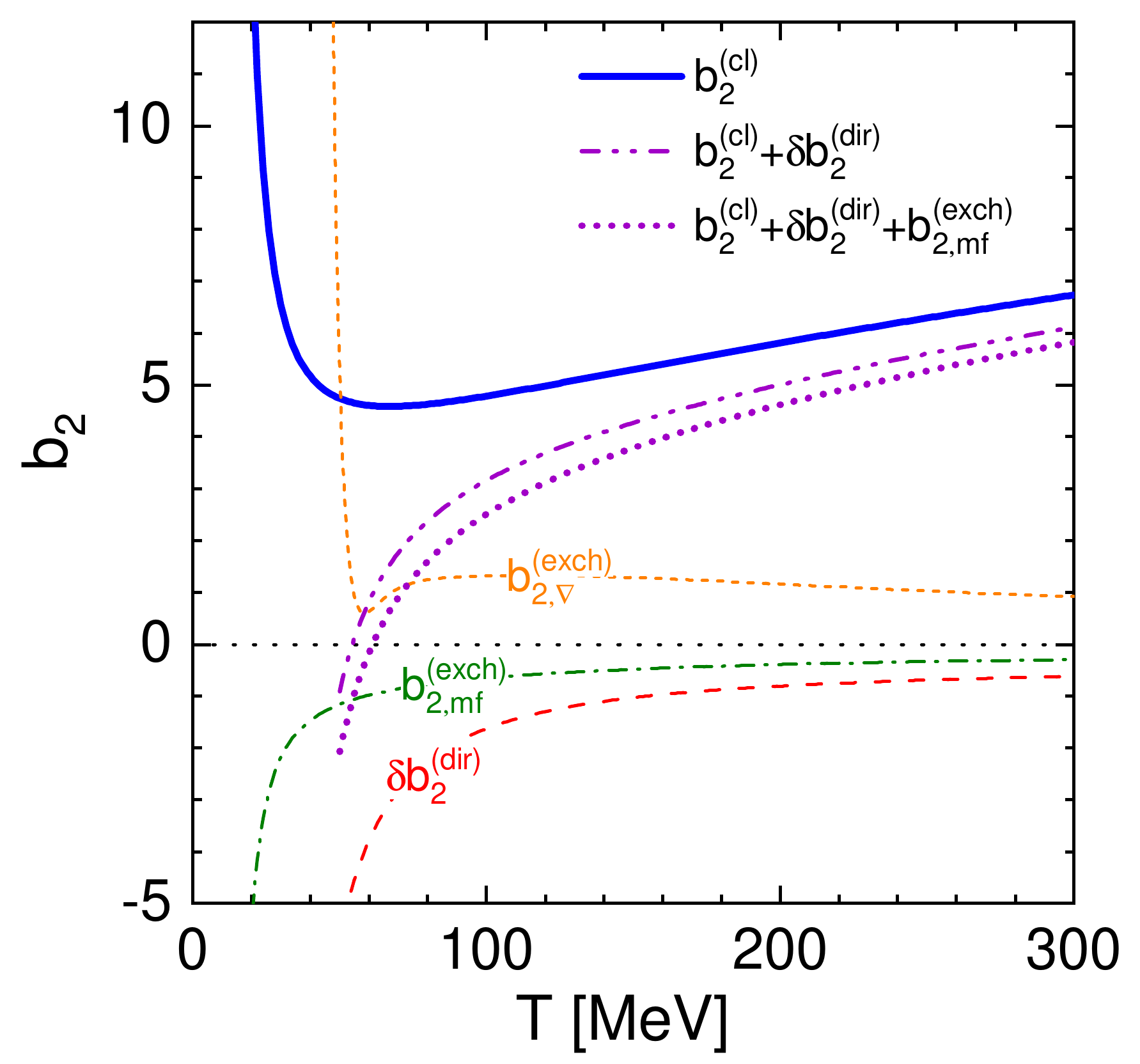}
\caption{Second virial coefficient $b_2$ in the classical limit and quantum corrections to it as functions of the temperature for the system of neutrons interacting via the  potential (\ref{U-potR}) with Walecka parameters (\ref{Uomsig-par-W}). Solid line shows the full classical expression (\ref{bN-cl}). Dashed line demonstrates the  quantum correction $\delta b_{2}^{(\rm dir)}$ of order $\hbar^2$ in expansion (\ref{h-exp}) from the direct term (\ref{b2-dir-2}). Dash-dotted line shows the quantum correction term $b_{2,{\rm mf}}^{(\rm exch)}$ given by Eq.~(\ref{b2-exch-mf}) that yields a correction of order $\hbar^3$.  Short-dash line shows the quantum correction term $b_{2,\nabla}^{(\rm exch)}\propto \lambda^4$ given by Eq.~(\ref{b2-exch-grad}).
Dash-double-dot line demonstrates the term $b_{2}^{(\rm cl)}+\delta b_{2}^{(\rm dir)}$, i.e. it includes corrections of the order of $\hbar^2$.
Dotted line demonstrates the sum of terms including corrections of the order of $\hbar^3$:
$b_{2}^{(\rm cl)}+\delta b_{2}^{(\rm dir)}+b_{2,{\rm mf}}^{(\rm exch)}$.
%Dash-double-dotted line shows calculation of $b^{(\rm int, TS)}_{2,\Phi}+b_{2}^{(0)}$ done beyond the Born approximation {\bf tak?} with tadpole-sandwich diagrams in the $\Phi$ derivable approach.
}
\label{fig:Bquant}
\end{figure}

Summarizing, in this Section we demonstrated how the full quantum expression for the second virial coefficient turns into its classical counterpart and quantum corrections. Then, we applied the derived semiclassical results to evaluate the second virial coefficient analytically for the case of the smoothed square barrier-well potential and numerically for the Yukawa-type Walecka central pair potential. Both considered examples demonstrate validity of the semiclassical approximation for rather large temperatures and its nonapplicability in low temperature limit. The magnitudes of different contributions in $b_2$ derived in the present manuscript are shown in Fig.~\ref{fig:Bquant} for the potential  (\ref{U-potR}) with Walecka parameters (\ref{Uomsig-par-W}).

\section{Comparison of the fugacity expansions within $\Phi$ derivable and virial approaches}\label{sec:comp}

\subsection{Second virial coefficient}

In Section \ref{ssec:fugacity-Phi} within the framework of the $\Phi$ derivable approach for the set of the TS diagrams, we obtained the expression (\ref{b2-2}) for the coefficient $b_{2,\Phi}^{(\rm int,TS)}$ of the term $\propto z^2$ in the $z$ expansion (\ref{P_TS}) of the pressure (\ref{P-U-fin}). This term is linear in the interaction potential. In Eq.~(\ref{b2-U-Born-delt}) we showed that it can be written through the derivative of the scattering phase shift in the Born approximation that resembles the full quantum expression for the second virial coefficient (\ref{b2-phase}) through the full scattering amplitude obtained by Gropper~\cite{Gropper-36} and Beth and Uhlenbeck~\cite{Uhlenbeck-Beth,Beth}.

Let us consider what the exact expression (\ref{b2int-full}), from which Eq.~(\ref{b2-phase}) was obtained, gives for the term linear in the potential.
We expand Eq.~(\ref{b2int-full}) formally for small $\hat{U}$ using the relation from Ref.~\cite{Lee-Yang-I}
\begin{align}
&e^{-\frac{\hat{H}^{(2)}_{\rm rel} }{T}}
\approx
e^{-\frac{\hat{T}_{\rm rel} }{T} } -\intop_0^1\rmd \xi e^{-(1-\xi)\frac{\hat{T}_{\rm rel} }{T} }\frac{\hat{U}}{T}
e^{-\xi\frac{\hat{T}_{\rm rel} }{T} }\,.
\end{align}
Taking into account that for the integrated matrix element the operators can be cyclicly exchanged, e.g., for an operator $\hat{O}$:
\begin{align}
\intop\rmd^3 r \bra{\pm\vec{r}\,} e^{-\frac{\hat{T}_{\rm rel}}{T}}\hat{O}\ket{\vec{r}\,}=
\intop\rmd^3 r \bra{\pm\vec{r}\,} \hat{O} e^{-\frac{\hat{T}_{\rm rel}}{T}}\ket{\vec{r}\,}\,,
\end{align}
we can write  Eq.~(\ref{b2int-full}) as
\begin{align}
b_2^{\rm (int)} &\approx -\frac{2^{1/2}}{T}\intop\rmd^3 \vec{r}\,\Big[
4\bra{\vec{r}\,} e^{-\hat{T}_{\rm rel}/T}\hat{U}(\vec{r}\,)\ket{\vec{r}\,}
\nonumber\\
&-2
\bra{-\vec{r\,}} e^{-\hat{T}_{\rm rel}/T}\hat{U}(\vec{r}\,) \ket{\vec{r}\,}
\Big]
=-\frac{2^{1/2}}{T}\intop\rmd^3 \vec{r} \intop \rmd^3p_1\rmd^3 p_2
\nonumber\\
&\times\Big[
4\braket{\vec{r}\,|\vec{p}_1}\bra{\vec{p}_1} e^{-\hat{T}_{\rm rel}/T}\ket{\vec{p}_2}\braket{\vec{p}_2\,|\vec{r}\,}U(\vec{r}\,)
\nonumber\\
&-2
\braket{-\vec{r}\,|\vec{p}_1}\bra{\vec{p}_1} e^{-\hat{T}_{\rm rel}/T}
\ket{\vec{p}_2}\braket{\vec{p}_2\,|\vec{r}\,} U(\vec{r}\,) \Big].
\end{align}
Taking into account that $\bra{\vec{p}_1} e^{-\hat{T}_{\rm rel}/T}\ket{\vec{p}_2}
=e^{-p_1^2/mT}\delta^{(3)}(\vec{p}_1-\vec{p}_2)$ we obtain
\begin{align}
b_2^{\rm (int)} &\approx-\frac{2^{3/2}}{T}\!\!\! \intop\!\!\frac{\rmd^3p}{(2\pi)^3}
e^{-\frac{p^2}{mT}}
\intop\rmd^3 \vec{r} \big( 2 \Ucor(\vec{r}\,) -  e^{-i2\vec{p}\vec{r}} \Ucor(\vec{r}\,) \big)
\nonumber\\
&=-\frac{2^{3/2}}{T}\!\!\! \intop\!\!\frac{\rmd^3p}{(2\pi)^3}
e^{-\frac{p^2}{mT}}\big( 2 \Umom(0) -  \Umom(2\vec{p}\,) \big) \,,
\label{b2-exp}
\end{align}
where we define the Fourier transform of the potential as  $\Umom(\vec{q}\,)=
\intop \rmd^3 r e^{-i\vec{q}\,\vec{r}} \Ucor(\vec{r}\,)$, cf. Eq.~(\ref{U-matrix-p}) in Appendix \ref{app:ME-NN-pot}. The obtained relation exactly coincides with $b_{2,\Phi}^{\rm (int, TS)}$ given by  Eq.~(\ref{b2-2}).

Thus, we find that the exact result for the second virial coefficient (\ref{bN-def}) or (\ref{b2-full}) and the coefficient $b_{2,\Phi}^{\rm (int,TS)}$  calculated within the $\Phi$ derivable approach with the TS diagrams (the first two in Fig.~\ref{fig:Phi-U-ser}) differ only in the second order of the interaction potential, i.e.,
\begin{align}
b_2^{(\rm int)} - b_{2,\Phi}^{\rm (int,TS)}= O(\Umom^2)\,.
\label{b-bPhi-diff}
\end{align}
Including in the $\Phi$ functional diagrams of higher order, e.g., diagrams c) and d) in Fig.~\ref{fig:Phi-U-ser}, the difference between $b_2^{(\rm int)}$ and the virial coefficient calculated in the $\Phi$ derivable approach, $b_{2,\Phi}^{(\rm int)}$ will become of higher orders of $\Umom$.
Finally, for the $\Phi$ functional with the infinite set of the so-called ring diagrams, see  Fig,~\ref{fig:ring-diag}, the difference will vanish completely since in this case the nucleon self-energy will be determined by the T-matrix satisfying the Lippmann-Schwinger equation (\ref{LSE-me}) instead of the amplitude in the Born approximation. For the pair-potential interaction, all other not-ring diagrams do not contribute to the second virial coefficient, see Section 16 in Ref.~\cite{LL9}, but contribute in part to the modification of the scattering amplitude in nuclear matter.

\begin{figure}
\centering
\parbox{8cm}{\includegraphics[width=8cm]{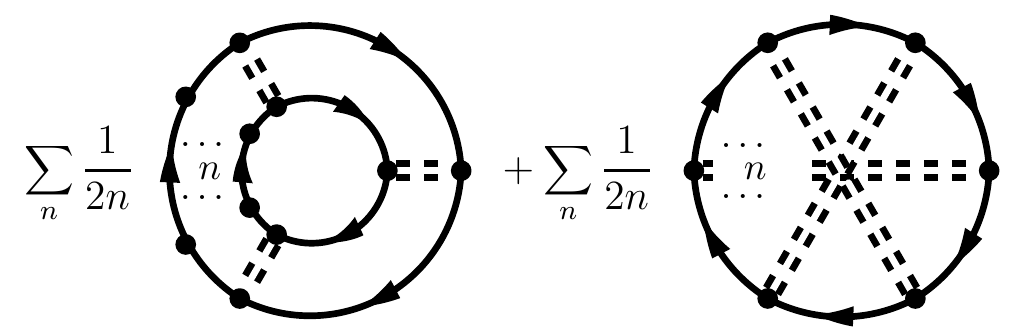}}
\caption{
The set of the ring diagrams in the $\Phi$ functional which produce the coefficient $b_{2,\Phi}^{(\rm int,TS)}$ of the term $\propto z^2$ in the $z$ expansion (\ref{P_TS}) of the pressure (\ref{P-U-fin}) expressed through the phase shifts corresponding to the full scattering amplitude satisfying the Lippmann-Schwinger equation (\ref{LSE-me}).}
\label{fig:ring-diag}
\end{figure}

The results of the leading order in $\Umom$ cancelling in Eq.~(\ref{b-bPhi-diff}) correspond to Born approximation for the scattering amplitude.  As was discussed in Section~\ref{ssec:effrange}, this approximation is not satisfied for nuclear potentials, which produce quasi-bound states below threshold and, consequently, a very large scattering length. Indeed, for the neutron-neutron scattering, we have $a^{\rm nn}_0 \simeq -19$\,fm, whereas the effective range parameter is much smaller, $r^{\rm nn}_0\simeq 2.3$\,fm.
As the result, the temperature dependence of the coefficient $b_2^{(\rm int)}$ calculated with the empirical low-energy parameters of the
nucleon-nucleon scattering and the coefficient $b_{2,\Phi}^{\rm (int,TS)}$ given by Eq.~(\ref{b2-2}) are essentially different. It is illustrated in Fig.~\ref{fig:F2rat-2}. The coefficient $b_2^{\rm (int)}$ calculated with  Eq.~(\ref{b-effr-l}) using the effective range representation (\ref{delta-effr}) of the s-wave and p-wave phase shifts with parameters (\ref{a-reff-param}) shows a peculiar temperature dependence, see solid line in Fig.~\ref{fig:F2rat-2}. It rises sharply at $0<T<(1\mbox{--}2)$\,MeV and then stays approximately constant for $T<20$\,MeV in accordance with expansions (\ref{b2-effr-large-a}) and (\ref{b2-l1-effr-large-a}). This differs drastically from the temperature dependence of the coefficient $b_{2,\Phi}^{\rm (int,TS)}$ shown by dash-dotted line even at low temperatures, where $b_{2,\Phi}^{\rm (int,TS)}\propto T^{1/2}$.
The coincidence of the solid line at $T\lsim 20$ MeV with the bold dots corresponding to the results of Ref.~\cite{Horowitz2006_2}
shows that for such temperatures the main contribution to the phase shifts is given by the s- and p-waves.

Employing the fact that $|a^{\rm nn}_0|\gg r^{\rm nn}_0$, Refs.~\cite{HammerSon2021,SchaferBaym2021} suggested to consider a broad region of inter-particle distances between $r^{\rm nn}_0\sim 2\,{\rm fm}$ and $a_0^{\rm nn}\sim$20\,fm, as a region of ``unnuclear'' physics, where all relevant physics is described by universal scaling relations. This range of distances corresponds to densities $n_0\gg n\gsim 1/|a_0^{\rm nn}|^3\sim 0.001n_0$ at $T=0$, and momenta and temperatures $1/ r^{\rm nn}\gg p\sim \sqrt{mT}\gg 1/|a_0^{\rm nn}|$, i.e 10\,MeV$\gg T\gg 0.1$\,MeV. Would such universal scaling relations characteristic to the unitarity limit of the neutron-neutron interaction indeed exist in the relatively broad range of densities and temperatures, there would be no place for many non-trivial physical phenomena like, e.g., the Cooper pairing or clustering occurring actually in the same range of densities and temperatures. In reality, many new relevant scales can be formed in the medium.

Results exploiting the fact that scattering length is anomalously large, being applicable at the consideration of the very low densities and temperatures, do not take into account that the quasi-bound state located close to the threshold should be very sensitive to the Pauli-blocking and finite temperature effects. As was pointed out in Refs.~\cite{ZS,SRS1990,Ropke:2012qv,RopkeSchnell}, for densities with $p_{\rm F}\gsim \pi/|a_0^{\rm nn}|$ (the perturbation theory parameter for the low density Fermi gas is $p_{\rm F}|a_0^{\rm nn}|/\pi$, cf.~\cite{LL9}) the quasi-bound state should be  destroyed because of the action of the Pauli blocking. For instance, even the stronger bound deuteron is dissolved in the symmetric nuclear matter for $n\gsim n_0/20$, see Ref.~\cite{RopkeSchnell}. Rough estimates and calculations in Ref.~\cite{RopkeSchnell}, cf. Fig.~3 there, suggest that the quasi-bound state can be dissolved by the Pauli blocking already for $n\lsim 0.01\, n_0$ at $T=10$\,MeV.

If we drop the contribution of the quasi-bound state (put $w\to 0$ in second line of Eq. (\ref{LE-epxansion})), the second virial coefficient $b_2^{\rm (int)}$ given by Eq.~(\ref{b2-phase}) in the effective range approximation (\ref{b-effr-l}) can be estimated as
\begin{align}
b_{2,l=0}^{\rm (int)} \approx \widetilde{b}_{2,l=0}^{\rm (int)} \equiv -2 \frac{a_{^1{\rm S}_0}^{\rm (ef)}}{\lambda}+
\frac{4\pi}{\lambda^3}a_{^1{\rm S}_0}^{\rm (ef)} b_{^1{\rm S}_0}^{\rm (ef)},
\label{b2-no-pole}
\end{align}
where we used expansion (\ref{b2-effr-small-a}) and the effective scattering parameters, $a_{^1{\rm S}_0}^{\rm (ef)}$ and $b_{^1{\rm S}_0}^{\rm (ef)}$, defined in Eq.~(\ref{a-eff}). The expansion is valid when $2\pi b_{^1{\rm S}_0}^{\rm (ef)}/\lambda^2 \ll 1$ that corresponds to $T\ll 30$\,MeV. The coefficient $b_2^{\rm (int)}$ from Eq.~(\ref{b2-no-pole}) is shown in Fig.~\ref{fig:F2rat-2} by dotted line, which for $T\lsim 2$ MeV well coincides with the dash-dotted line depicting the coefficient $b_{2,\Phi}^{\rm (int,TS)}$. This reflects the fact that the scattering length calculated in the Born approximation, Eq.~(\ref{LE-Born-Yukawa}), with the Walecka parameters (\ref{Uomsig-par-W}) is equal to $-1.2$\,fm that is close to the value of $a_{^1{\rm S}_0}^{\rm (ef)}=-1.18$\,fm from Eq.~(\ref{a-eff}).

\begin{figure}
\centering
\parbox{8cm}{\includegraphics[width=8cm]{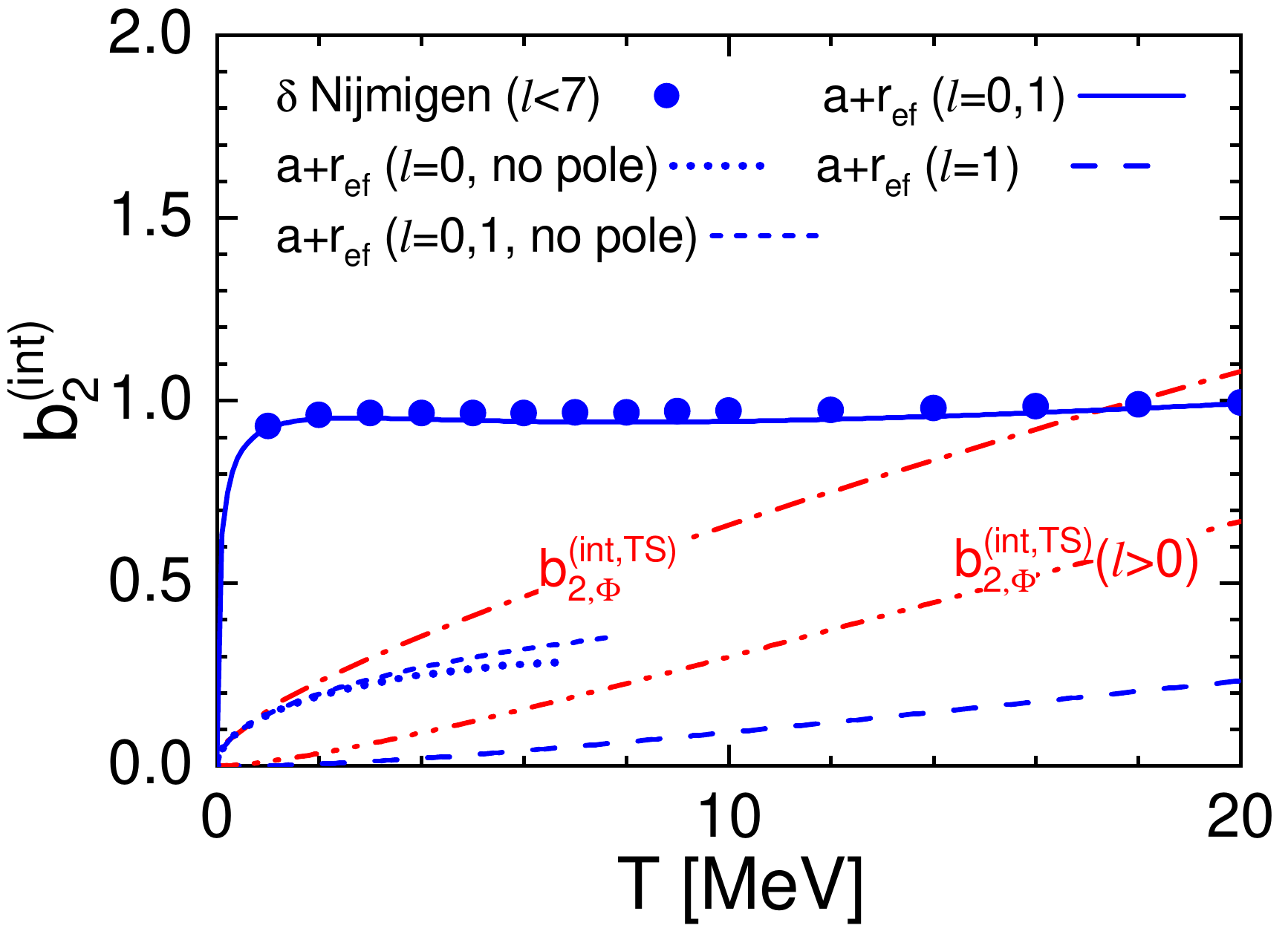}}
\caption{
The second virial coefficient as a function of temperature.
Bold dots show the results from Ref.~\cite{Horowitz2006_2} obtained with the Nijmegen phase shifts for all partial waves with $l<7$.
Solid line is calculated with  Eq.~(\ref{b-effr-l}), cf. Fig.~\ref{fig:b2-phase},
using the effective range representation (\ref{delta-effr}) of the s-wave and p-wave phase shifts with parameters (\ref{a-reff-param}). Long-dashed line is the same calculation with only p-waves taken into account, i.e., term $b_{2,l=1}^{\rm (int)}$. Dotted line represents the quantity $\widetilde{b}_{2,l=0}^{\rm (int)}$  calculated according to Eq.~(\ref{b2-no-pole}), where the quasi-bound state contribution from the s-wave is subtracted. Short-dashed line shows the sum  $\widetilde{b}_{2,l=0}^{\rm (int)}+b_{2,l=1}^{\rm (int)}$.
Dash-dotted line shows the coefficients $b_{2,\Phi}^{\rm (int,TS)}$ calculated  within the $\Phi$ derivable approach with Eq.~(\ref{b2-2}). Dash-double-dotted line stands for the coefficient $b_{2,\Phi,l>0}^{\rm (int,TS)}$  calculated according to Eq.~(\ref{b2-l=0}) with the s-wave excluded. In the $\Phi$ derivable approach we use the Yukawa potential (\ref{U-potR}) with parameters (\ref{Uomsig-par-W}).
}
\label{fig:F2rat-2}
\end{figure}

The contribution from the p-waves should be included into the second virial coefficient to proceed to higher temperatures. To analyze the p-wave contributions let us subtract the s-wave parts from $b_2^{\rm (int)}$ completely. For the coefficient $b_2^{\rm (int)}$ calculated with Eq.~(\ref{b-effr-l}) it can be done easily by dropping the terms for $l=0$ and keeping term $b_{2,l=1}^{\rm (int)}$.
Using the representation of the coefficient $b_{2,\Phi}^{\rm (int,TS)}$ through the potential Eq.~(\ref{b2-2}) and the phase shifts in the Born approximation Eq.~(\ref{b2-U-Born}) we can subtract the s-wave as follows,
\begin{align}
b_{2,\Phi,l>0}^{\rm (int,TS)} &=-\frac{2^{3/2}}{T}\!\!\! \intop\!\!\frac{2\rmd^3p}{(2\pi)^3}
e^{-\frac{p^2}{mT}}\Big(\Umom(0) -  \frac12\Umom(2\vec{p}\,)
\nonumber\\
& +\frac{2\pi}{m_f}\frac{\delta_{{\rm B},0}^{(0)}(p)}{p} \Big) \,,
\label{b2-l=0}
\end{align}
where the s-wave phase shift in the Born approximation is calculated using Eq.~(\ref{FBorn-delt}) for the Yukawa potential (\ref{U-potR}),
\begin{align}
\frac{\delta_{{\rm B},0}^{(0)}(p)}{p}=-\frac{m_f}{16\pi p^2}\Big[ g_\om^2\ln\frac{m_\om^2 + 4p^2}{m_\om^2} -
g_\sigma^2\ln\frac{m_\sigma^2+4 p^2}{m_\sigma^2}
\Big].
\end{align}
The coefficients $b_{2,\Phi}^{\rm (int,TS)}[l>0]$ and $b_2^{\rm (int)}[l=1]$   are shown in Fig.~\ref{fig:F2rat-2} by dash-double-dot and long dashed lines, respectively. The latter coefficient is calculated with parameters (\ref{a-reff-param}). The threshold behaviour of both coefficients is the same $\propto T^{3/2}$, that is in agreement with expansion (\ref{b2-l1-effr-large-a}) for the case $|a_1^{(1)}|\ll  \frac{\lambda^3}{\sqrt{2\pi}}$ and for the $b_{2,\Phi,l>0}^{\rm (int,TS)}$ coefficient it occurs since $\Umom(0) -  \Umom(2\vec{p}\,)/2  +2\pi \delta_{{\rm B},0}^{(0)}(p)/(m_f p) \to 3p^2\big( g_\om^2/m_\om^4 -  g_\sigma^2/m_\sigma^4\big)$ for $p\to 0$. However the slopes of dash-double-dot and long dashed lines are different, since the experimental nucleon-nucleon p-waves include also non-central interactions (tensor and spin-orbit ones). The sum  $\widetilde{b}_{2,l=0}^{\rm (int)}+b_{2,l=1}^{\rm (int)}$ is shown by short-dashed line, which moves closer to the result obtained in the $\Phi$-derivable approach with the TS diagrams.

A systematic approach, which allows to take into account the in-medium modification of particle interaction, is the generalized Beth-Uhlenbeck approach~\cite{ZS,RMS-NPA379,SRS1990,BBFR-18}. Within this approach the statistical sum is written as the combination of the contributions from the free quasiparticles with the in-medium spectrum (\ref{MfG}) and the correlated scattering states of these quasiparticles described by the in-medium scattering phase shifts, $\delta^{\rm med}$ following from the solution of the Lippmann-Schwinger equation including the in-medium spectra and Pauli-blocking for the intermediate two-particle states. The formal structure of the correlated state contribution is similar to the second virial coefficient (\ref{b2-phase}) but with the replacement $\frac{\rmd \delta}{\rmd p}\to 2\sin^2(\delta^{\rm med})\frac{\rmd \delta^{\rm med}}{\rmd p}$. The additional  $2\sin^2(\delta^{\rm med})$ factor reduces the two-body scattering contribution because a part of the two-body correlation effect is shifted to the self-energies of the quasiparticles. In Ref.~\cite{ZS} it is argued that in the low-density limit the generalized Beth-Uhlenbeck approach reproduces the standard Beth-Uhlenbeck result.

We can conclude that, if one assumes dissolution of the quasi-bound state in matter, the second virial coefficients, calculated in the Beth-Uhlenbeck approach employing the experimental scattering data and in the $\Phi$-derivable approach, agree qualitatively and even  quantitatively, see dash-dotted and dotted lines for low $T$ in Fig. \ref{fig:F2rat-2}. This dissolution of the quasi-bound state should occur due to the Pauli blocking effect in nuclear matter already at very low densities according to our estimates done above.

\subsection{Third virial coefficient}\label{ssec:3dvirial}

In contrast to the second virial coefficient, which is expressed in closed form in terms of the two-body scattering phase shifts, the third and higher virial coefficients are much more difficult to calculate especially for quantum systems, where the traces of statistical operators, see e.g. Eq.~(\ref{bN-def}),  involving the many-body Hamiltonians have to be evaluated. After Lee and Yang developed their binary-collision expansion method~\cite{Lee-Yang-I,Lee-Yang-II}, the attempts to calculate the third virial coefficient were undertaken by Pais and Uhelnbeck in~\cite{Pais-Uhlenbeck} and by Larsen~\cite{Larsen}. However, problems with the convergence of the series expansion and exceeding difficulties in calculations beyond the simplest terms diminish the practical usage of these calculations. In Ref.~\cite{Larsen-Mascheroni} Larsen and Mascheroni showed how the third virial coefficient can be expressed through the three-body phase shifts for the case of Boltzmann statistics and in the absence of bound states. The attempts to introduce the modern scattering theory by Faddeev~\cite{Faddeev61} in the calculation of the third virial coefficient were undertaken in Refs.~\cite{Reiner66,Baumgartl67} using the integral representation of the resolvent of the Hamiltonian by the Watson method~\cite{Watson56}, see Eq.~(\ref{integ-resolvent}). In Ref.~\cite{Baumgartl67} the third coefficient was expressed through the two-particle scattering amplitude, $t(\vec{p}-\vec{p\,}')=\frac14 \sum_{S=0,1}(2S+1)\bra{\vec{p\,}',S}\hat{\mathcal{T}}(E_p\pm i\eta) \ket{\vec{p\,},S}$, satisfying Eq.~(\ref{LSE-me}). The term in $b_3^{(\rm int)}$ involving two scattering amplitudes $t(\vec{p}-\vec{p\,}')$, see Eqs.~(82) and (13) in Ref.~\cite{Baumgartl67}, can be written with the help of relation (\ref{B-b-rel}) as
\begin{align}
b_3^{\rm (int)} &=-\frac{b_1^3}{2\lambda^6}\big( C-4B^2)
\nonumber\\
&=\frac{\lambda^3}{2 T^2}\intop \frac{2\rmd^3 p_1}{(2\pi)^3}\intop \frac{2\rmd^3 p_2}{(2\pi)^3}\intop
\frac{2\rmd^3 p_3}{(2\pi)^3} e^{-\frac{\epsilon_{\vec{p}_1}+\epsilon_{\vec{p}_2}+ \epsilon_{\vec{p}_3}}{T}}
\nonumber\\
&\times (2\pi)^6 \Re t(\vec{p}_2-\vec{p}_3) \Re t(\vec{p}_3-\vec{p}_1)\,.
\label{b3-tt}
\end{align}
The remaining terms are of the third and higher orders in $\Re t$ and depend on the energy dependence of the scattering amplitude beyond mass-shell, see Section 7 in~\cite{Baumgartl67}. In Ref.~\cite{Grossmann64} Eq. (\ref{b3-tt}) was introduced from the consideration of the Landau transport equation. In the Born approximation we can replace the full scattering amplitude $t(\vec{p}_2-\vec{p}_3)$ by the Fourier transform of the potential. Then, keeping only the first term in Eq.~(\ref{LSE-me}) and using Eq.~(\ref{U-spinaver}) we find $\Re t(\vec{p}-\vec{p\,}')\to\frac{1}{(2\pi)^3} \overline{\Umom^{(S)}}(\frac{\vec{p}-\vec{p\,}'}2,\frac{\vec{p}-\vec{p\,}'}{2}) $. After doing this replacement Eq.~(\ref{b3-tt}) becomes equivalent to the first term in Eq.~(\ref{b3-U}) for $b_{3,\Phi}^{\rm (int,TS)}$ obtained in the fugacity expansion of the pressure derived within the $\Phi$-derivable approach. The second (linear in $\Umom$) term in Eq.~(\ref{b3-U}) is a quantum mechanical correction of the order $\hbar^3$, see Eq.~(\ref{b3Phi-H}). Thus, we can  conclude that the $\Phi$ derivable approach allows to recover the classical part of the third virial coefficient evaluated with the scattering amplitude taken in the Born approximation.
Contributions to the third virial coefficient can be also recovered within the generalized Beth-Uhlenbeck approach.

Summarising, in this Section we  explicitly demonstrated the agreement between the tadpole-sandwich $\Phi$-derivable approach and the virial expansion within the region of their common applicability. Additionally, we provided explanation for the differences between these two methods outside of this region. We did not see a similar  analysis in the literature.

\section{Conclusion}\label{Conclusion}

Our study is focused on the conserving $\Phi$-derivable and the virial equation of state  approaches, applied to dilute strongly interacting systems in equilibrium. We investigated these approaches in the context of a system of non-relativistic fermions of one kind, e.g., the neutron matter, interacting via a pair potential, $\Ucor$, in the limit of low fermion number density and thermal wavelength such that $n\lambda^3\ll 1$. Our investigation encompassed a detailed analysis of the equation of state within the $\Phi$-derivable approach and the virial expansion scheme, particularly focusing on their agreement and the origins of any differences observed. We successfully demonstrated the consistency between the equation of state derived from the $\Phi$-derivable approach, specifically including in $\Phi$ the leading order diagrams $\propto U$, and the virial expansion scheme within the Born approximation. In addition, to numerically exemplify our findings and offer broader insights into our theoretical work, we considered some test pair potentials including the Yukawa central nucleon-nucleon potential within Walecka, CD Bonn, and Reid parameterizations, as well as classical Lennard-Jones potential.

Section~\ref{Phi} provides a comprehensive discussion of the general properties of the $\Phi$-derivable approach, exploring variational properties, the $\Phi$ derivable approximation scheme, and introducing key spectral functions $A_f$, $B_f$, and $A_{s, f}$.  These spectral functions obey the exact sum-rules and can be used to express the thermodynamic quantities such as particle density, energy, entropy, and pressure in a thermodynamically consistent way. Two of these functions, $A_f$ and $A_{s,f}$, enter  respectively the Noether and Botermans-Malfliet particle densities. We derived explicit expressions for the latter quantities. We  also demonstrated that the knowledge of only the spectral function $A_f$ and particle occupations does not allow, in general case, to calculate the entropy and the pressure without additional calculation  of the corresponding $\Phi$ functional diagrams.

In Section~\ref{ssec:test-pot} we considered examples of several test pair potentials such as the Yukawa-type central nucleon-nucleon potentials in Walecka, modified Walecka, Reid soft core  and CD Bonn fit parameterizations, and the corresponding classical Lennard-Jones potentials, cf. Fig.~\ref{fig:U-pot}. Also, we studied the model of the square barrier-well potential allowing for the analytical treatment. The variety of potentials of the quantum and classical origin considered in this Section were then used for the quantitative comparison of the results obtained within the $\Phi$-derivable and virial approaches, in particular, the classical and quantum limits of the second viral coefficients.

In Section~\ref{sec:Phi-U} we studied the approximation of the first two tadpole and sandwich (TS) diagrams in the $\Phi$ functional.
In this case, the self-energy is not frequency-dependent, however, remaining to be a complicated function of the pair potential. We,  first,  presented expressions for all thermodynamic quantities, verified the thermodynamic consistency of the TS approximation and obtained the compact relation for the entropy (\ref{TS-short}). We discussed equivalence of the Botermans-Malfliet and Noether particle densities  in case of the considered TS $\Phi$. Expanding the pressure in terms of the fugacity $z = e^{\mu/T}$, we obtained second $b_{2,\Phi}^{\rm (TS)}$ and third-order $b_{3,\Phi}^{\rm (TS)}$ virial coefficients expressed in the terms of the pair potential. The potential $\Ucor$ enters $b_{2,\Phi}^{\rm (TS)}$ only linearly, while other coefficients such as $b_{3,\Phi}^{\rm ( TS)}$, etc contain higher-order terms in $\Ucor$. This distinction indicates the difference between $b_{2,\Phi}^{\rm (TS)}$ derived from $\Phi^{\rm (TS)}$ and the coefficient $b_{2}$ derived of the virial expansion, which includes a full series in $\Ucor$. We showed that in the fugacity expansion of the pressure in the $\Phi$-derivable approach, each term corresponding to a specific order $k$ receives contributions only from those diagrams in  $\Phi$, which are of the order equal to or less then $k$ in $\Ucor$. Thus, by considering only the tadpole and sandwich diagrams, we can ensure accurate expressions up to the linear term in $\Ucor$ for each virial coefficient $b_k$. This approximation aligns with the Born approximation for fermion-fermion scattering amplitude. Although this approach maintains thermodynamic consistency and exact conservation of energy-momentum and current, it should be noted that it deviates from the full theory that incorporates an infinite number of diagrams, including terms like $U^2$ and higher-order terms in $U$. By conducting a numerical analysis, we have gained insights into the temperature-density limits of the virial expansion's applicability. The results, depicted in Figure ~\ref{fig:Phi-U-bs-P}, provide a clear overview of the range where the Born approximation and the second virial coefficient are suitable for describing the equation of state. Furthermore, our findings shed light on the relationships between Hartree and Fock contributions to $b_{2,\Phi}^{\rm (TS)}$  and $b_{3,\Phi}^{\rm (TS)}$  across different temperature regimes.

In Sections ~\ref{virial}-\ref{ssec:class-virial}, we introduced the classical virial expansion for the equation of state in terms of density and fugacity and its thermodynamically consistent extension.
We demonstrated that in the classical limit, one can separate contributions from the repulsive and attractive parts of the potential in the second virial coefficient. We showed that popular models, describing the effects of a first-order phase transition, e.g., the liquid--gas phase transition in nuclear matter, are based on the perturbative treatment of the attractive contribution, whereas some extrapolation scheme is applied to the repulsive contribution. As a result, the phenomenologically adjusted parameters of such models do not agree with those that follow from the typical nuclear potential. This is an important point, as there are studies that rely on purely repulsive excluded volume interaction to describe the high-temperature, low-density behavior in heavy-ion collisions. We found that for all considered Yukawa-based nuclear potentials the repulsive part exceeds the attractive part only for unreasonably high values of the temperature ($T>400$MeV). Our finding, which indicates the absence of a Boyle temperature except for classical LJ potentials, shows a limitation of the aforementioned excluded volume phenomenological models.
An interesting result derived in this Section is that in the limit $T\gg |U_{\rm typ}|$, where $U_{\rm typ}$ is the potential $\Ucor(r)$ at distances at which it changes significantly, the classical limits of the second and third virial coefficients, $b_2^{\rm (cl)}(T\gg |U_{\rm typ}|)$ and $b_3^{\rm (cl)}(T\gg |U_{\rm typ}|)$, coincide with the Hartree contributions obtained within the $\Phi$ derivable approach with the only tadpole diagram. Thus, the classical limit of the virial coefficients can be reproduced within the $\Phi$ derivable approach with account of the first Hartree diagram.

In Section~\ref{sssec:cluster}, we studied  quantum virial expansion, and the second virial coefficient $b_2$ was treated in a fully quantum approach. We presented the derivation of the Beth-Uhlenbeck expression for $b_2$, using the optical theorem and not the artificial zero boundary condition.
We derived parameters of the effective-range approximation for the p-wave scattering amplitude obtained by the Nijmegen group~\cite{Rijken-06}, see Fig.~\ref{fig:delta-NN}. The experimental scattering parameters are compared with the results of the Born approximation, which is demonstrated to be inapplicable at the hand of explicit calculation of the double-scattering contribution to the scattering amplitude, see Fig.~\ref{fig:F2rat}. Nevertheless, we showed that the Born approximation could be applied for the polarized fermion gas, where all collisions occur in the channel with total spin equal to one. We studied the temperature dependence of the second virial coefficient using the effective range approximation and derived general analytical expressions for the s-wave and p-wave contributions in terms of the scattering lengths and the range terms. We demonstrated in  Fig.~\ref{fig:b2-phase} that for $E<50$\,MeV the $b_2$ coefficient, calculated within the effective range approximation, closely agrees with the results obtained in Ref.~\cite{Horowitz2006_2} including experimental partial wave with $l<7$. The experimental $nn$ scattering length is large and negative in the $^1$S$_0$ channel owing to the presence of the quasi-bound state close to the scattering threshold, which parameters and the background scattering length are identified. Because of this state the $b_2$ coefficient is nearly constant (equal to $\sqrt{2}$ in the unitary limit) as a function of temperature for $2\,{\rm MeV} < T< 20$\,MeV and its magnitude is insensitive to the magnitude of the scattering length.

In Section~\ref{Virsemicl} we considered transition from the quantum expression for $b_2$ to the semiclassical and classical ones to clarify the appropriate scenarios for utilizing the full quantum framework for the virial expansion versus the classical and semiclassical frameworks.
The semiclassical expansion was conducted involving the small parameter $\lambda/r_{\rm typ}\ll 1$, where $r_{\rm typ}$ is the typical distance, at which the potential undergoes a substantial variation. By employing obtained semiclassical results, we performed an analytical evaluation of the second virial coefficient for a smoothed square barrier-well potential in order to demonstrate all relevant small parameters involved in the calculation. Additionally, we performed numerical calculations for the Yukawa-type Walecka central pair potential (cf. Fig~\ref{fig:Bquant}). The semiclassical approximation is shown to be valid for relatively high temperatures $T\gsim 500$ MeV, whereas it  breaks down completely at low temperatures $T\lsim 50$ MeV.

In the last Section~\ref{sec:comp} we compared the second virial coefficients found within the $\Phi$-derivable approach for the set of TS diagrams with the second virial coefficients calculated with the Beth-Uhlenbeck expression using the empirical phase shifts. We showed that keeping only the linear in $U$ terms in $b_2$ we recover exactly the term $b_{2,\Phi}^{\rm (TS)}$. In order to completely reproduce the second virial coefficient $b_2$ from the $\Phi$-derivable approximation, it is necessary to take into account an infinite subset of ring diagrams in  $\Phi$ (cf. Fig~\ref{fig:ring-diag}). The nucleon scattering amplitude entering $b_2^{}$ in the Beth-Uhlenbeck approach cannot be described by the leading $U$ terms only, since the Born approximation is not applicable for nuclear potentials. It results in the drastically different temperature dependence of $b_2^{}$ and  $b_{2,\Phi}^{\rm (TS)}$ coefficients even at low temperatures. The origin of this difference is the presence of a quasi-bound state slightly below the $NN$ threshold.
Some authors continue using the conventional (Beth-Uhlenbeck) second-order virial expansion employing experimental phase shifts up to relatively moderate densities, e.g. for $n\lsim (0.1-0.3) n_0$ at $T\lsim 10$ MeV. It disagrees with the results of \cite{RopkeSchnell}, which suggest that the influence of the Pauli blocking causes the quasi-bound state to be affected in nuclear matter even at much lower densities $n\ll 0.01 n_0$ at $T\lsim 10$MeV. Thus, we specially focused our attention on the problem of the anomalously large value of the nucleon-nucleon scattering length,  which arises due to the presence of the quasi-bound state in nucleon-nucleon scattering. We presented results for the second virial coefficient by subtracting this anomalous term, which can be interpreted as supporting the statement of \cite{RopkeSchnell}. We argued that for densities, at which $p_{\rm F}(n)\gsim 1/|a_0^{\rm nn}|$, i.e. already for $n\gsim 10^{-3} n_0$, the quasi-bound state can  be destroyed because of the action of the Pauli blocking. By subtracting the contribution of the quasi-bound state from the nucleon-nucleon scattering amplitude using expansion (\ref{LE-epxansion}) and evaluating the virial coefficient $b_2^{}$ with the modified effective scattering length, we achieved both qualitative and even quantitative agreements with the results obtained through the $\Phi$-derivable approach at low temperatures, see dash-dotted and dotted lines for low $T$ in Fig. \ref{fig:F2rat-2}. We also examined applicability of these approaches to calculations of the third virial coefficient and compared the part of the $b_{3,\Phi}^{\rm (TS)}$ that is proportional to the $U^2$ with the double scattering contribution to the $b_2^{}$ coefficient derived in the literature. Both results coincide within the framework of the Born approximation.

For very large temperatures, $T\gg |\Ucor_{\rm typ}|$, the second and third virial coefficients are dominated by the Hartree contributions, and the coefficient $b^{\rm (int,TS)}_{2,\Phi}$ calculated fully quantum-mechanically within the $\Phi$-derivable approach with the
first Hartree diagram agrees with the  classical high-temperature expression for $b^{\rm (cl)}_2$, see Eq.~(\ref{b2cl-H}).
In this limit the sign of $b^{\rm (cl)}_2$ for the typical nuclear potentials of the Yukawa type corresponds to the attractive interaction, since the hard-core radius is small $\sim 0.2$--0.3\,fm and the attractive pocket is sufficiently deep ($\sim 10^2$\,MeV) and broad $r_{\rm U}\sim 1.5$\,fm. Of course, one should bear in mind that  the description of the nucleon-nucleon interaction in terms of the non-relativistic pair potential generated by the static boson exchange is limited at the high temperature. One has to transit to the fully relativistic picture of the $NN$ interaction, and take into account the energy dependence (retardation effects) in propagators of intermediate mesons $G_{\rm ex}(\om,\vec{q}\,)=1/(\om^2-m_{\rm ex}^2-\vec{q\,}^2)$. Taking into account that in case $n\lambda^3\ll 1$, which we have studied in this paper, the typical exchange energies and momenta are $\om\sim T$ and $q\sim\sqrt{ m_N T}$, we can estimate the range of the validity for the pair potential description as  $T\ll m_{\rm ex}, m_N$. The existence of pion, the anomalously light Goldstone boson, limits this description to temperatures $T\ll m_{\pi}=140$\,MeV. Importance of retardation effects in description of the pion degrees of freedom in nuclear matter has been extensively studied in the literature, cf. \cite{Migdal:1978az,Migdal:1990vm} and references therein.

Concluding, we studied in detail conserving approximations to dilute equilibrium systems
in the virial limit, $n\lambda^3\ll 1$,  employing pair potential interaction and exploiting the $\Phi$ derivable and virial expansion approaches. For that we focused attention on the study of  one component fermion system (like the neutron gas). We traced the transition of quantum mechanical expressions to their classical limits with a temperature increase, and pointed out peculiarities specific for nuclear potentials originated by the boson exchanges. We demonstrated that the $\Phi$ derivable and virial expansion approaches coincide in the Born approximation for the $NN$ scattering amplitude and studied differences beyond this approximation.

\acknowledgments
We thank F.~Akhmetov,  D.~Blaschke, Yu.B.~Ivanov and K.A.~Maslov for the discussions. The work reported in Sections~I--V was supported by the grant VEGA~1/0353/22, Sections~VI-IX were supported by Russian Scientific Foundation with grant RSF 21-12-00061.

\appendix

\section{Thermodynamic potential for ideal non-relativistic Fermi gas}\label{ideal}

The free Matsubara Green's functions satisfy the integral representation,
\begin{align}
G_0(\epsilon_n,\vec{p}\,) &= \intop_{-\infty}^\infty\frac{\rmd {\epsilon}}{2\pi} \frac{A_{f,0}({\epsilon},\vec{p}\,)}{\epsilon_n -{\epsilon}}\,,\,\,
\label{G0}
\\
A_{f,0}(\epsilon,\vec{p}\,) &=2\pi\delta(\epsilon-\epsilon_{\vec{p}})\,,\,\,
\epsilon_{\vec{p}}=\frac{\vec{p\,}^2}{2m_f}
\nonumber
\end{align}
for the discrete Matsubara frequencies, $\epsilon_n=(2n+1)i\pi T+\mu_f$.
Here $m_f$ is the value of the physical mass of the free fermion, and the nonrelativistic chemical potential $\mu_f$ is counted from the fermion mass. If the system contains  several fermion species the fermion frequency and chemical potential are counted from the smallest mass $m_f$.

After the Matsubara frequency summation we obtain on the right-hand side  of  Eq.~(\ref{Msum-f}) the retarded Green's functions defined as the analytical continuation of (\ref{G0}) the upper complex semi-plane
 \begin{align}
G_0^R(\epsilon,\vec{p}\,)=G_0(\epsilon+i0,\vec{p}\,)\,.
\label{G0D0-R}
\end{align}
The spectral function is related to the retarded Green's functions as $$A_{f,0}=-2\Im G_0^R\,.$$

To bring the thermodynamic potential (\ref{free-Omega-f}) to a more conventional form we  use the identity
\begin{align}
f(\epsilon) &=-T{\partial}_\epsilon
%{\partial\epsilon}
\ln\big(1+e^{-\epsilon/T}\big)\,,
\label{nf-to-log}
\end{align}
and integrate by parts. Employing that
\begin{align}
&2{\partial}_\epsilon
%{\partial\epsilon}
\Im\ln( [G_0^{R}(\epsilon,\vec{p}\,)]^{-1})
%=2\Im(G_0^{R}(\epsilon,\vec{p}\,))
%\nonumber\\
%&\qquad
=-A_{f,0}(\epsilon,\vec{p}\,) \,,
\end{align}
we obtain
\begin{align}
\Omega_{f,0}(T,\mu_f) &= -d_f \intop
\frac{V\rmd^3 p}{(2\pi)^3}\intop_{-\infty}^{+\infty}\frac{\rmd \epsilon}{2\pi} A_{f,0}(\epsilon,\vec{p}\,)
\nonumber\\
&\qquad\times T\ln\big(1+e^{-(\epsilon-\mu_f)/T}\big)\,.
\end{align}

Finally, using $A_{f,0}$ from Eq.~(\ref{G0}), we bring the expression for the thermodynamic potential of the ideal gas of non-relativistic fermions in the standard form
\begin{align}
\Omega_{f,0}(T,\mu_f) &= -d_f\int
\frac{V\rmd^3 p}{(2\pi)^3}
T\ln\big(1+e^{-(\epsilon_{\vec{p}}-\mu_f)/T}\big)\,.
\label{free-Omega}
\end{align}

The chemical potential of the nonrelativistic fermion is found from the relation
\begin{align}
N_{f,0} =-\frac{\partial \Omega_{f,0}}{\partial \mu_f}\Big|_T= d_f\int\frac{V\rmd^3 p}{(2\pi)^3} f(\epsilon_p-\mu_f)\,,
\end{align}
where $N_{f,0}$ is the average number of the fermions.

\section{Diagrammatic rules in Matsubara formalism}\label{DiagramRules}
%\subsection{diagram rules}

\subsection{Pair potential:}
Consider now the system of one type of fermions interacting via a spin-independent pari-potential.

\begin{figure}[t]
\begin{align}
{}_\beta\parbox{2cm}{\includegraphics[width=2cm]{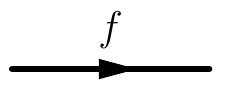}}{}_{\alpha} &= - [\hat{G}(\epsilon_n,\vec{p}\,)]_{\alpha\beta}
\nonumber\\
&=-\delta_{\alpha\beta}G(\epsilon_n,\vec{p}\,)\,
\end{align}
\caption{Fermion line assignment for diagrammatic rule 3 }
\label{fig:app-line}
\end{figure}

%%%%
\begin{figure}[t]
\begin{align}
\parbox{1.5cm}{\includegraphics[width=1.5cm]{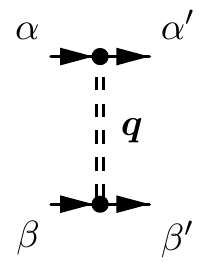}} &=
-\Umom(|\vec q\,|) \delta_{\alpha'\alpha}  \delta_{\beta'\beta}\,.
\end{align}
\caption{Pair potential assignment for diagrammatic rule 3 }
\label{fig:app-pot}
\end{figure}
%%%%

To compute diagrams shown in Figs.~\ref{fig:Phi-U-ser}, and  \ref{fig:Sig-U} one should:
\begin{enumerate}
\item With  series of diagrams shown in Fig. ~\ref{fig:Phi-U-ser} associate $-\Phi$. Each diagram of $\Phi$ assign symmetry factor $1/n_G$, where $n_G$ is the number of fermion Green's functions in the diagram.
\item With  series of diagrams shown in Fig. ~\ref{fig:Sig-U} associate $-\Sigma (\epsilon_n,  \vec{p})$.

\item  To each fermion line and the pair interaction assign a line, see Fig.~\ref{fig:app-line}.
For the case when fermions interact via a spin-independent pair potential, see Fig.~\ref{fig:app-pot} assign Greek indices $\alpha$, $\beta$ are indices in the spin-isospin space.

\item In each vertex require conservation of momentum and discrete fermion frequency $\epsilon_n= (2n+1)\pi i T +\mu_f$. One should integrate over remained not fixed 3-momenta $\int\frac{\rmd^3 p}{(2\pi)3}$ and sum over frequencies of fermions $T\sum_{n=-\infty}^{\infty}$.

\item Sum up over all internal spin-isospin indices of all intermediate lines. Multiply the expression by $(-1)^{L_f}$, where $L_f$ is number of fermion loops.

\item To each line closed on itself or joined by the same interaction line insert the factor  $e^{i \epsilon_n \eta}$, $\eta =+0$.

\item For a closed diagram multiply the result by the volume $V$.
\end{enumerate}
\begin{widetext}
Let us apply these rules to diagrams (a) and (b) in Fig.~\ref{fig:Phi-U-ser}:
\begin{align}
-\frac{\Phi}{V} &= \frac12\Bigg(T\sum_{n=-\infty}^{+\infty}\int\frac{\rmd^3 p}{(2\pi)^3}
e^{i\epsilon_n \eta}
[-\hat{G}(\epsilon_n,\vec{p})]_{\alpha\beta}\Bigg)
\big(-\Umom(0)\delta_{\beta\alpha}\delta_{\beta'\alpha'}\big)
\Bigg(T\sum_{n'=-\infty}^{+\infty}\int\frac{\rmd^3 p'}{(2\pi)^3}
e^{i\epsilon_{n'} \eta}
[-\hat{G}(\epsilon_{n'},\vec{p\,}')]_{\alpha'\beta'}\Bigg)
\nonumber\\
&+(-1) \frac12T^2\sum_{n,n'=-\infty}^{+\infty}
\int\frac{\rmd^3 p\rmd^3 p'}{(2\pi)^6}
[-\hat{G}(\epsilon_n,\vec{p})]_{\alpha\beta}
\big(-\Umom(\vec{p}-\vec{p\,}')\delta_{\beta\alpha'}\delta_{\beta'\alpha}\big)
[-\hat{G}(\epsilon_{n'},\vec{p\,}')]_{\alpha'\beta'}.
\label{Phi-deriv-exampl}
\end{align}
\end{widetext}
Herewith we recover expression (\ref{Phi-U-Mats}).

\section{Derivative of thermodynamic potential with respect to $T$}\label{app:deriv}

In this Section we calculate the derivative of $\Omega$ with respect to $T$, determining expression for the entropy. We begin with expression (\ref{Omega-int-1}), where due to variational  relations (\ref{Om-G-var}) we have to differentiate  in $T$ only the particle occupations and $\Phi$. Thus we arrive at   Eq. (\ref{dOmdT-0}), where the derivative $\frac{\partial \Phi}{\partial T}\Big|_{\mu,A_f}$ is taken at fixed spectral densities.
%%%%%%%%%%%%%%%%%%%%%%%%%%%%%%%%%%%%%%%%
 The derivatives of the particle occupations with respect to $T$ can be converted in derivatives with respect to frequencies
\begin{align}
\frac{\partial f(\epsilon-\mu_f)}{\partial T} &= -
\frac{\partial f(\epsilon-\mu_f)}{\partial \epsilon}\frac{\epsilon-\mu_f}{T}\,.
\label{dfdt-de}
\end{align}
Besides in Eq.~(\ref{dfdt-de})  the derivatives with respect to $T$ can be rewritten as
\begin{align}
\frac{\partial f(\epsilon-\mu_f)}{\partial T} =-\frac{\partial \sigma_f(\epsilon-\mu_f)}{\partial\epsilon}
\,,
\label{dfdt-sig}
\end{align}
 the entropy-function $\sigma_f$ is defined in (\ref{sig-def}).

Substituting Eqs.~(\ref{dfdt-sig}) in Eq.~(\ref{dOmdT-0}) and integrating by parts we present entropy as
\begin{align}
&S^{\rm (loc)}=-d_f\intop\frac{V\rmd^3 p}{(2\pi)^3}
\intop_{-\infty}^{+\infty}\frac{\rmd \epsilon}{2\pi}
\sigma_f(\epsilon-\mu_f)
\nonumber\end{align}\begin{align}
&\quad\times
\frac{\partial}{\partial \epsilon} \left[2\Im\ln(-[G^R(\epsilon,\vec{p}\,)]^{-1})
-\Gamma_f (\epsilon,\vec{p}\,)\Re G^R(\epsilon,\vec{p}\,)\right]\,,
 \label{dOmdT-1}
\end{align}
and
\begin{align}
S^{\rm (mem)} &=
d_f\!\!\intop\!\frac{V\rmd^3 p}{(2\pi)^3}\!\!
\intop_{-\infty}^{+\infty}\!\!\frac{\rmd \epsilon}{2\pi}
\frac{\partial f(\epsilon-\mu_f)}{\partial T}
\nonumber\\
&\times
 \Re\Sigma^R(\epsilon,\vec{p}\,)A_f (\epsilon,\vec{p}\,)
- \frac{\partial\Phi}{\partial T}\Big|_{\mu,A_f}  \,.
\label{DS-def-app}
\end{align}

Expressing Eq. (\ref{dOmdT-1}) through the spectral functions $A_{s,f}$ defined in Eq.~(\ref{Asf-def}) we arrive at Eq.~(\ref{DSp-1}), and substituting (\ref{dfdt-sig}) in Eq.~(\ref{DS-def-app}) after integrating by parts we obtain Eq.~(\ref{DSp-0}).

To proceed further we need to know $\frac{\partial \Phi}{\partial T}\big|_{\mu,A_f}$.
There are two sources for the $T$ dependence of $\Phi$, if the spectral functions are kept fixed. One is through the Matsubara frequencies in the Green's functions and other is through the $T$ factors in front of the Matsubara sums, see Eq.~(\ref{Tr-def}). Using the variational property (\ref{Sigma-Phi}) and assuming as before that Green's functions and self-energies are diagonal in the spin-isospin space we obtain
\begin{align}
\label{c7}
&\frac{\partial \Phi}{\partial T}\Big|_{\mu,A_f}
=\Big(\frac{\partial \Phi}{\partial T}\Big)_{\rm vert}
+\Tr_p\big[\Sigma_p\frac{\partial G_p}{\partial T}\big]\Big|_{A_f}
\nonumber\\
&=\Big(\frac{\partial \Phi}{\partial T}\Big)_{\rm vert}+\Tr_p\big[\Sigma_p\frac{\epsilon_n-\mu}{T}\frac{\partial  G_p}{\partial \epsilon_n}\big]\Big|_{A_f}\,.
\end{align}
\begin{widetext}
The first ``vertex'' term on r.h.s. of this equation shows that we should only differentiate the $T$ dependence of $\Phi$ through the $T$ factors in front of the Matsubara sums. After the Matsubara frequency summation using Eq. (\ref{Msum-f})  we obtain
\begin{align}
\frac{\partial \Phi}{\partial T}\Big|_{\mu,A_f}
&=\Big(\frac{\partial \Phi}{\partial T}\Big)_{\rm vert}-
d_f\intop\frac{V\rmd^3 p}{(2\pi)^3}
\intop_{-\infty}^{+\infty}\frac{\rmd \epsilon}{2\pi}
f(\epsilon-\mu_f)\frac{\epsilon-\mu_f}{T}
2\Im\Big(\Sigma^R(\epsilon,\vec{p\,})\frac{\partial G^R(\epsilon,\vec{p\,})}{\partial \epsilon}\Big)\,.
\label{dPhidT-fin}
\end{align}

Now we replace Eq.~(\ref{dfdt-de})  in Eq.~(\ref{DS-def-app}). After integration by parts and combining  the result with Eq.~(\ref{dPhidT-fin}) we obtain
\begin{align}
S^{\rm (mem)} &=
d_f\intop\frac{V\rmd^3 p}{(2\pi)^3}
\intop_{-\infty}^{+\infty}\frac{\rmd \epsilon}{2\pi}
 f(\epsilon-\mu_f)\frac{\Re\Sigma^R(\epsilon,\vec{p}\,)}{ T}
 A_f (\epsilon,\vec{p}\,)
 - \Big(\frac{\partial \Phi}{\partial T}\Big)_{\rm vert}
 \nonumber\\
&+ d_f\intop\frac{V\rmd^3 p}{(2\pi)^3}
\intop_{-\infty}^{+\infty}\frac{\rmd \epsilon}{2\pi}
f(\epsilon-\mu_f)\frac{\epsilon-\mu_f}{T}\Big[\frac{\partial }{\partial \epsilon}
\Big( \Re\Sigma^R(p)A_f (p) \Big)
+
2\Im\Big(\Sigma^R(\epsilon,\vec{p\,})\frac{\partial G^R(\epsilon,\vec{p\,})}{\partial \epsilon}\Big)\Big]\,.
\label{Smem-app-1}
\end{align}

We see that the memory entropy contains various contributions. First  term is from the potential energy of fermion  mode.  Explicit form of the next contribution,  $-\big(\frac{\partial \Phi}{\partial T}\big)_{\rm vert}$,   depends on the interaction model. The last term can be expressed through  functions $\mathcal{C}_{f}$ defined in Eq.~(\ref{C-def}), they characterize the deviation of the spectral functions $A_{s,f}$ from the spectral functions $A_{f}$. Thus we arrive at Eq.~(\ref{Smem-2}).

For the derivation of the expression for the energy with the help of the thermodynamic relation (\ref{thermoidentity}) it is convenient to rewrite  expression for the entropy in another form.
First, making use the relation (\ref{dfdt-de})  we rewrite equation (\ref{dOmdT-0})  as,
\begin{align}
S &= -\frac{\partial \Phi}{\partial T}\Big|_{\mu,A_f}
-
d_f\intop\frac{V\rmd^3 p}{(2\pi)^3}
\intop_{-\infty}^{+\infty}\frac{\rmd \epsilon}{2\pi}
(-1)
\frac{\partial f(\epsilon-\mu_f)}{\partial \epsilon}\frac{\epsilon-\mu_f}{T}
 2\Im\big(\ln(-[G^R(p)]^{-1}) +\Sigma^R(p)G^R(p)\big)\,.\label{EntrAnother}
\end{align}
Then, integrating by part we obtain
\begin{align}
S&=-\frac{\partial \Phi}{\partial T}\Big|_{\mu,A_f}
-
\frac{\Omega-\Phi}{T}
-
d_f\intop\frac{V\rmd^3 p}{(2\pi)^3}
\intop_{-\infty}^{+\infty}\frac{\rmd \epsilon}{2\pi}
f(\epsilon-\mu_f)\frac{\epsilon-\mu_f}{T}
 2\frac{\partial }{\partial \epsilon}\Im\big(\ln(-[G^R(p)]^{-1}) +\Sigma^R(p)G^R(p)\big)
\nonumber\\
&=
-\frac{\partial \Phi}{\partial T}\Big|_{\mu,A_f}
-
\frac{\Omega-\Phi}{T}
+
d_f\intop\frac{V\rmd^3 p}{(2\pi)^3}
\intop_{-\infty}^{+\infty}\frac{\rmd \epsilon}{2\pi}
f(\epsilon-\mu_f)\frac{\epsilon-\mu_f}{T}\Big\{
A_f(p)+2\Im\big(G^R(p)\frac{\partial \Sigma^R(p)}{\partial \epsilon}\big)
-
 2\frac{\partial }{\partial \epsilon}\Im\big(\Sigma^R(p)G^R(p)\big)
 \Big\}\,.
\label{S-app-1}
\end{align}

Substituting (\ref{dPhidT-fin}) in (\ref{S-app-1}) we finally obtain
\begin{align}
S &=-
\frac{\Omega}{T}
-\frac{\mu_f}{T} N_f +
\frac{\Phi}{T}-\Big(\frac{\partial \Phi}{\partial T}\Big)_{\rm vert}
+
\frac{d_f}{T}\intop\frac{V\rmd^3 p}{(2\pi)^3}
\intop_{-\infty}^{+\infty}\frac{\rmd \epsilon}{2\pi}
f(\epsilon-\mu)\epsilon
A_f(p)
\,,
\label{dOmdT-2}
\end{align}
where  $N_f$ is the number of fermions $N_f=n_fV$ with $n_f$ given by Eq.~(\ref{nf-Noether}).
Finally combining this relation for the entropy with (\ref{nf-dOmdmu}) and using the thermodynamic identity  Eq.~(\ref{thermoidentity}) we obtain expression for the energy given by Eq.~(\ref{E-result}).
 %and (\ref{DeltaE-def}).
\end{widetext}

\section{Calculation of $\Big(\frac{\partial \Phi}{\partial T}\Big)_{\rm vert}$ for a specific model}\label{app:ddPhi}

Consider a theory with one type of vertices, where enter 4 fermion  fields with $\Ucor$ as the pair potential, see Fig.~\ref{fig:Phi-U-ser}. The pair potential does not depend on frequencies  but may depend on 3-momenta of entering particles.

The $\Phi$ functional can be presented as a series of diagrams
\begin{align}
\Phi=V\sum_{N_v} S_v \big\{\Ucor^{N_v} G^{I_f}T^{L}\big\}\,,
\label{Sig-S-P}
\end{align}
where summation goes over the number of potential lines in each diagram. The curly brackets symbolize integration over momenta and summation over Mutsubara frequencies. Each diagram carries  a symmetry factor $S_v$, which includes also necessary spin-isopin coefficients and signs.
Each Matsubara summation introduces  explicit dependence of the diagram on $T$. Hence the full $T$-dependence of a diagram is
$T^L$, where $L$ is the number of integrations/summation over the unconstrained momenta/frequency in the diagram remaining after taking into account momentum conservations in each vertex (given by dashed line) explicitly. The product of the Green's functions is $G^I_f$, where the number of internal lines for fermions is $I_{f}=2 N_v$. The number $L$ is given by the Euler relation $L=I_f-N_v+1$~\cite{Wein90}.

For the given $\Phi$, the self-energies of fermions  can be obtained via Eq.~(\ref{Sigma-Phi}) as
\begin{align}
\Sigma_p &= \sum_{N_v}  S_v I_f\big\{ \Ucor^{N_v} G^{I_f-1}T^{L}\big\}_p\,.
\label{Phi-S-P}
\end{align}
Here $\{\dots\}_p$ mean that Matsubara summations and momenta integration leave one external momentum and frequency free.

Now we differentiate $\Phi$ keeping the Green's functions fixed, i.e. we differentiate only the explicit $T$ dependence of each diagram because of the Matsubara summations. Then $\frac{\partial \Phi}{\partial T}\big|_{G}=\big(\frac{\partial \Phi}{\partial T}\big)_{\rm vert}$ can be calculated as follows
\begin{align}
\Big(\frac{\partial \Phi}{\partial T}\Big)_{\rm vert}
&=V\sum_{N_v} \frac{L}{T} S_v \big\{ \Ucor^{N_v} G^{I_f}T^{L}\big\}
\nonumber\\
&=V\sum_{N_v} \frac{I_f-N_v+1}{T} S_v \big\{ \Ucor^{N_v} G^{I_f}T^{L}\big\}\,.
\end{align}
Comparing with Eqs.~(\ref{Sig-S-P}) and (\ref{Phi-S-P}) we can write now
\begin{align}
&\Big(\frac{\partial \Phi}{\partial T}\Big)_{\rm vert}
=\frac{\Phi}{T}+ \frac{1}{2T}Vd_f\Tr_p[\Sigma_pG_p]  \,.
\label{dDPhidT-vert}
\end{align}

Relation (\ref{dDPhidT-vert})  can be used to simplify expressions for  thermodynamic quantities. Doing so, we  obtain a closed expression for the non-Markovian part of the entropy, $S^{\rm (mem)}$, see Eq.~(\ref{Smem-fb}). Also the quantity $\Delta E$ appearing in Eq.~(\ref{DeltaE-def}) can be written with the help of Eq.~(\ref{dDPhidT-vert}) as
\begin{align}
\Delta E &=\Phi -T\Big(\frac{\partial \Phi}{\partial T}\Big)_{\rm vert}
\nonumber\\
&=- \frac{1}{2}V\Tr_p[\Sigma_pG_p] \,.
\label{DE-general}
\end{align}
Performing the Matsubara summation, (\ref{Msum-f}) we obtain Eq.~(\ref{DE-def}).

\section{Matrix elements of $NN$ potential}\label{app:ME-NN-pot}

In this Appendix we consider matrix elements of the potential operator $\hat{U}$ in various representations.

We use the eigen states of position and momentum operators denoted as  $\ket{\vec{r}\,}$ and $\ket{\vec{p}\,}$, respectively. These states are normalized as follows
\begin{align}
\braket{\vec{p}\,|\vec{p\,}'}=\delta^{(3)}(\vec{p}-\vec{p\,}')
\,,\quad
\braket{\vec{r}\,|\vec{r\,}'}=\delta^{(3)}(\vec{r}-\vec{r\,}')\,.
\label{norma-pp-rr}
\end{align}
The completeness relations are
\begin{align}
\intop \rmd^3\vec{p}\,\ket{\vec{p}\,}\bra{\vec{p}\,}=1\,,
\quad
\intop \rmd^3\vec{r}\,\ket{\vec{r}\,}\bra{\vec{r}\,}=1\,.
\label{complete-pp-rr}
\end{align}
The wave function of the state with momentum $\vec{p}$ is
\begin{align}
\varphi_{\vec{p}}(\vec{r}\,)=\braket{\vec{r}\,|\vec{p}\,}
=\frac{e^{i\vec{p}\vec{r}}}{(2\pi)^{3/2}}.
\label{rp-wf}
\end{align}
The same convention is used also in Section~\ref{sssec:cluster}.

For the coordinate matrix elements we have obviously
\begin{align}
\bra{\vec{r}\,'}\hat{U}\ket{\vec{r}\,}= \Ucor(|\vec{r}\,|)\delta^{(3)}(\vec{r}\,'-\vec{r}\,).
\label{U-matrix-r}
\end{align}
Matrix elements in the momentum space are
\begin{align}
\bra{\vec{p}\,'}\hat{U}\ket{\vec{p}\,} &= \intop\rmd^3r\rmd^3r' \braket{\vec{p}\,'|\vec{r}\,'} \bra{\vec{r}\,'}\hat{U}\ket{\vec{r}\,}
\braket{\vec{r}\,|\vec{p}\,}
\nonumber\\
&= \intop\frac{\rmd^3 r}{(2\pi)^3} e^{i(\vec{p}-\vec{p}\,')\vec{r}}\, \Ucor(|\vec{r}\,|)
=\frac{\Umom(|\vec{p}\,'-\vec{p}|\,)}{(2\pi)^3}
\,,
\label{U-matrix-p}
\end{align}
with the property
\begin{align}
\bra{-\vec{p}\,'}\hat{U}\ket{\pm\vec{p}\,} = \bra{\vec{p}\,'}\hat{U}\ket{\mp\vec{p}\,} \,.
\label{U-matrix-p-sym}
\end{align}
Using the expansion for the plane wave for $\vec{p}=p \vec{n}_p$ and $\vec{r}=r\vec{n}_r$
\begin{align}
e^{i\vec{p}\,\vec{r}} = \sum_l (2l+1) P_l(\vec{n}_p \vec{n}_r) i^l j_l(p\,r)\,,
\end{align}
where $P_l$ are the Legendre polynomials normalized as
\begin{align}
\intop \frac{\rmd \Omega_{\vec{n\,}''}}{4\pi}
P_l(\vec{n}\vec{n\,}'') P_{l'}(\vec{n\,}''\vec{n\,}')=\frac{\delta_{ll'}}{2l+1}
P_l(\vec{n}\vec{n\,}')\,,
\label{Pl-integ}
\end{align}
and $j_l(x)$ are the spherical Bessel functions, cf. \cite{LLIII}, we can write the matrix element (\ref{U-matrix-p}) as
\begin{align}
\bra{\vec{p}\,'}\hat{U}\ket{\vec{p}\,} &= \sum_l (2l+1) P_l(\vec{n}_p'\vec{n_p})  \bra{p'}\hat{U}_l\ket{p} \,,
\label{U-exp-l}\\
\bra{p'}\hat{U}_l\ket{p} &= \intop_0^\infty \frac{\rmd r r^2}{2\pi^2} \Ucor(r) j_l(p\,r) j_l(p'\,r) \,.
\label{Ul}
\end{align}
Here we separate the part of the potential responsible for the scattering with the specific value of the angular momentum, $\bra{p'}\hat{U}_l\ket{p}$.
Using the expansion of the spherical Bessel function
\begin{align}
j_l(x)=\frac{2^l\,l!}{(2l+1)!} x^l \Big(1-\frac{1}{2l+3}\frac{x^2}{2}\Big),
\label{j_l-exp}
\end{align}
we can write the asymptotics of the matrix element (\ref{Ul}) at small momenta
\begin{align}
\bra{p'}\hat{U}_l\ket{p} &\approx \Big[\frac{2^l\,l!}{(2l+1)!}\Big]^2 (p p')^l\intop_0^\infty \frac{\rmd r r^{2(l+1)}}{2\pi^2} \Ucor(r)\,,
\nonumber\\
 p^2 ,{p'}^{2} &\ll  \frac{\Big|\intop_0^\infty\rmd r r^{2(l+1)}\Ucor(r)\Big|}{\Big|\intop_0^\infty\rmd r r^{2(l+2)}\Ucor(r)\Big|}.
 \label{Ul-epx}
\end{align}

Consider also matrix elements between momentum states $\ket{\vec{p},S}$ defined in Eq.~(\ref{b2-full-p}),
\begin{align}
\bra{\vec{p}\,',S=0}\hat{U}\ket{\vec{p},S=0} &= \bra{\vec{p}\,'}\hat{U}\ket{\vec{p}\,}+\bra{\vec{p}\,'}\hat{U}\ket{-\vec{p}\,},
\nonumber\\
\bra{\vec{p}\,',S=1}\hat{U}\ket{\vec{p},S=1} &= \bra{\vec{p}\,'}\hat{U}\ket{\vec{p}\,}-\bra{\vec{p}\,'}\hat{U}\ket{-\vec{p}\,},
\nonumber\\
\bra{\vec{p}\,',S=0}\hat{U}\ket{\vec{p},S=1} &= 0,
\label{US-def}
\end{align}
where we use Eq.~(\ref{U-matrix-p-sym}).
Substituting Eq.~(\ref{U-exp-l}) and taking into account that $P_{l}(-x)=(-1)^lP_l(x)$
we find that the projection of the potential on the spin-0 state contains only partial wave with even $l$, whereas the projection on the spin-1 state contains only partial waves with the odd $l$,
\begin{align}
\bra{\vec{p}\,',S=0}\hat{U}\ket{\vec{p}\,,S=0} &=
\bra{\vec{p}\,'}\hat{U}^{(0)}\ket{\vec{p}\,}
\nonumber\\
&=2\sum_{l,{\rm even}} (2l+1) P_l(\vec{n}_p'\vec{n_p})  \bra{p'}\hat{U}_l\ket{p},
\nonumber\\
\bra{\vec{p}\,',S=1}\hat{U}\ket{\vec{p}\,,S=1} &=
\bra{\vec{p}\,'}\hat{U}^{(1)}\ket{\vec{p}\,}
\nonumber\\
&=2\sum_{l,{\rm odd}} (2l+1) P_l(\vec{n}_p'\vec{n_p})  \bra{p'}\hat{U}_l\ket{p}.
\label{U-S-exp}
\end{align}

The spin projections of the potential can be expressed through the Fourier transforms of the potential (\ref{U-matrix-p}) as follows,
\begin{align}
\bra{\vec{p}\,'}\hat{U}^{(0)}\ket{\vec{p}\,} &= \frac{1}{(2\pi)^3}
\big( \Umom(|\vec{p}-\vec{p}\,'|) + \Umom(|\vec{p}+\vec{p}\,'|) \big) \,,
\nonumber\\
\bra{\vec{p}\,'}\hat{U}^{(1)}\ket{\vec{p}\,} &= \frac{1}{(2\pi)^3}
\big( \Umom(|\vec{p}-\vec{p}\,'|) - \Umom(|\vec{p}+\vec{p}\,'|) \big) \,,
\label{U-S-Umom}
\end{align}
and the spin-averaged potential is
\begin{align}
\overline{\Umom^{(S)}}(\vec{p\,}',\vec{p}\,) &=(2\pi)^3
\sum_{S=0,1}\frac{(2S+1)}{4}\bra{\vec{p}\,'}\hat{U}^{(S)}\ket{\vec{p}\,}
\nonumber\\
&=\Umom(|\vec{p}-\vec{p}\,'|) - \frac12\Umom(|\vec{p}+\vec{p}\,'|)
\,.
\label{U-spinaver}
\end{align}
We see that the self-energy and other thermodynamic quantities derived in Section.~\ref{sec:Phi-U} in the self-consistent $\Phi$-derivable approximation with the tadpole and sandwich approximation can be expressed through the spin-averaged potential.

\section{Two-body scattering}\label{app:scattering}

In this Appendix we introduce formal definitions and relations for the scattering amplitude and scattering phases, which are used in Sections~\ref{ssec:fugacity-Phi} and \ref{ssec:fullquant-b2}, to express the coefficients of the fugacity expansion of the pressure in Eq.~(\ref{P_TS}) and the second virial coefficient in Eq.~(\ref{b2-full}) in terms of the scattering phase.

Consider two fermions, which interaction in the center-of-mass system is described by the Hamiltonian
$\hat{H}^{(2)}_{\rm rel}=\hat{T}_{\rm rel} + \hat{U},$ where in the coordinate representation the kinetic energy of the relative motion is $\hat{T}_{\rm rel}=-\frac{1}{m_f}\nabla_{\vec{r}}^2$ and the potential operator is the function $\hat{U}=\Ucor(|\vec{r}\,|)$ of the relative distance between the fermions $\vec{r}$.
Consider resolvent operators
\begin{align}
\hat{\mathcal{G}}_0(z) = \big(z - \hat{T}_{\rm rel}\big)^{-1}\,
\quad
\hat{\mathcal{G}}(z) = \big(z-\hat{H}^{(2)}_{\rm rel}\big)^{-1},
\label{resolvent}
\end{align} which are defined for some complex energy $z$.
Since  $\hat{H}_{\rm rel}^{(2)}=\hat{T}_{\rm rel} + \hat{U}$, the resolvents can be related as
$\hat{\mathcal{G}}^{-1}(z)=\hat{\mathcal{G}}^{-1}_0(z)-\hat{U}$ and, therefore,
\begin{align}
\hat{\mathcal{G}}(z) &= \hat{\mathcal{G}}_0(z) + \hat{\mathcal{G}}_0(z) \hat{U} \hat{\mathcal{G}}(z)=
\hat{\mathcal{G}}_0(z)[1 + \hat{U} \hat{\mathcal{G}}(z)]
\nonumber\\
& =\hat{\mathcal{G}}_0(z)\big[1-\hat{U}\hat{\mathcal{G}}_0\big]^{-1} =\big[1-\hat{\mathcal{G}}_0\hat{U}\big]^{-1} \hat{\mathcal{G}}_0(z)\,.
\label{G-G0}
\end{align}
From the definition of the resolvent  we have $[\hat{\mathcal{G}}_0(z)]^{-1} - [\hat{\mathcal{G}}_0(z')]^{-1}=z-z'$, and the same relation for  $\hat{\mathcal{G}}$, which can be rewritten as
\begin{align}
\hat{\mathcal{G}}_0(z')-\hat{\mathcal{G}}_0(z) &= (z-z')
\hat{\mathcal{G}}_0(z') \, \hat{\mathcal{G}}_0(z)\,,
\nonumber\\
\hat{\mathcal{G}}(z')-\hat{\mathcal{G}}(z) &= (z-z')
\hat{\mathcal{G}}(z') \, \hat{\mathcal{G}}(z)\,.
\label{G-G}
\end{align}
Two-particle scattering is described by the scattering operator $\hat{\mathcal{T}}$,
\begin{align}
\hat{\mathcal{T}}(z)=\hat{U}+\hat{U}\hat{\mathcal{G}}(z)\hat{U}.
\label{T-def}
\end{align}
Here the resolvent $\hat{\mathcal{G}}$ describes intermediate states of the scattering process.
Using Eq.~(\ref{G-G0}) we obtain
\begin{align}
\hat{\mathcal{G}}_0\hat{\mathcal{T}}
=\hat{\mathcal{G}}_0[1+\hat{U}\hat{\mathcal{G}}]\hat{U}
=\hat{\mathcal{G}}\hat{U}
\label{G0T=GU}
\end{align} and analogously
\begin{align}
\hat{\mathcal{T}}\hat{\mathcal{G}}_0=\hat{U}\hat{\mathcal{G}}\,.
\label{TG0=UG}
\end{align}
Using these relations we can derive the Lippmann-Schwinger equation
\begin{align}
\hat{\mathcal{T}}(z) &= \hat{U}\hat{\mathcal{G}}(z)\hat{\mathcal{G}}_0^{-1}(z)
=\hat{U} + \hat{U}\hat{\mathcal{G}}_0(z) \hat{U} \hat{\mathcal{G}}(z)\hat{\mathcal{G}}_0^{-1}(z)
\nonumber\\
&=\hat{U} +\hat{U}\hat{\mathcal{G}}_0(z)\hat{\mathcal{T}}(z).
\label{LSE}
\end{align}
The formal solution of the Lippmann-Schwinger equation
expresses the $\hat{\mathcal{T}}$ operator through bare resolvent $\hat{\mathcal{G}}_0$ and the potential $\hat{U}$,
\begin{align}
\hat{\mathcal{T}}(z)
=\hat{U}[1-\hat{\mathcal{G}}_0(z)\hat{U}]^{-1}
=[1-\hat{U}\hat{\mathcal{G}}_0(z)]^{-1}\hat{U}.
\label{LSE-sol}
\end{align}
The difference between  $\hat{\mathcal{G}}$ and $\hat{\mathcal{G}}_0$ can be expressed through the scattering operator, using Eqs.~(\ref{G-G0}) and (\ref{TG0=UG}),
\begin{align}
\hat{\mathcal{G}}(z)- \hat{\mathcal{G}}_0(z) = \hat{\mathcal{G}}_0(z) \hat{U} \hat{\mathcal{G}}(z)= \hat{\mathcal{G}}_0(z)\hat{\mathcal{T}}(z)
\hat{\mathcal{G}}_0(z)\,.
\label{G-G0-TGG}
\end{align}

From the definition of the scattering operator (\ref{T-def}) we can obtain two useful relations. First, we calculate the difference of the scattering operator for various arguments $\hat{\mathcal{T}}(z)-\hat{\mathcal{T}}(z')=\hat{U}\big[
\hat{\mathcal{G}}(z) - \hat{\mathcal{G}}(z')\big] \hat{U}$. With the help of Eqs.~(\ref{G-G}), (\ref{G0T=GU}), and (\ref{TG0=UG}) we find
\begin{align}
&\hat{\mathcal{T}}(z)-\hat{\mathcal{T}}(z')
%=\hat{U}\big(\hat{\mathcal{G}}(z) - \hat{\mathcal{G}}(z')\big) \hat{U}
%\nonumber\\
%&\,\,=\hat{U}\hat{\mathcal{G}}(z)\,(z-z') \, \hat{\mathcal{G}}(z')\hat{U}
%=\hat{\mathcal{T}}(z)\hat{\mathcal{G}}_0(z) \,(z-z') \,
%\hat{\mathcal{G}}_0(z')\hat{\mathcal{T}}(z')
%\nonumber\\ &\,\,
=
\hat{\mathcal{T}}(z)\big(
\hat{\mathcal{G}}_0(z)-\hat{\mathcal{G}}_0(z')\big)\hat{\mathcal{T}}(z').
\label{Tz-Tz}
\end{align}
Another relation we obtain by differentiating Eq.~(\ref{T-def}):
\begin{align}
\frac{\rmd \hat{\mathcal{T}}(z)}{\rmd z} = -\hat{U}\hat{\mathcal{G}}^2(z)\hat{U}
=-\hat{\mathcal{T}}(z) \hat{\mathcal{G}}_0^2(z)\hat{\mathcal{T}}(z)\,.
\label{T-deriv}
\end{align}

Consider now the matrix elements of Eqs.~(\ref{LSE}), (\ref{Tz-Tz}), and (\ref{T-deriv}) between
the scattering states of two particles with the relative momenta $\vec{p}$ ,  $\ket{\vec{p}\,}$, see Eq.~(\ref{b2-full-p}),
such that $\hat{T}_{\rm rel}\ket{\vec{p}\,}=E_{\vec{p}}\ket{\vec{p}\,}=\frac{\vec{p\,}^2}{m_f}\ket{\vec{p}\,}$.
We start with Eq.~(\ref{Tz-Tz}). We substitute in Eq.~(\ref{Tz-Tz}) $z=E\pm i\eta$ and $z'=E\mp i\eta$
with real $E$ and $\eta\to 0$ and use the relation
\begin{widetext}
\begin{align}
\bra{\vec{p\,}'}\hat{\mathcal{G}}_0(E\mp i\eta)-\hat{\mathcal{G}}_0(E\pm i\eta)
\ket{\vec{p\,}}=\pm 2\pi i \delta(E_{\vec{p}}-E) \delta^{3}(\vec{p}-\vec{p\,}') .
\end{align}
As the result Eq.~(\ref{Tz-Tz}) takes the following form
\begin{align}
\bra{\vec{p}\,'}\hat{\mathcal{T}}(E\pm i\eta) -
\hat{\mathcal{T}}(E\mp i\eta)\ket{\vec{p}\,}
= \mp 2\pi i \intop\rmd^3 p''
\bra{\vec{p}\,'}\hat{\mathcal{T}}(E\pm i\eta) \ket{\vec{p\,}''}
\delta(E_{\vec{p\,}''}-E)
\bra{\vec{p\,}''}\hat{\mathcal{T}}(E\mp i\eta) \ket{\vec{p\,}}\,.
\label{T-optic}
\end{align}
This is the off-shell unitarity relation for the T-matrix for the energy $E$, which is not constrained by the relation $E=p^2/m_f$. Particularly, for the forward scattering  with $\vec{p\,}'=\vec{p}$, this relation takes the familiar form of \emph{the optical theorem},
\begin{align}
\Im\bra{\vec{p}\,}\hat{\mathcal{T}}(E\pm i\eta) \ket{\vec{p\,}}
&= \mp\pi \intop\rmd^3 p''
\bra{\vec{p}\,}\hat{\mathcal{T}}(E\pm i\eta) \ket{\vec{p\,}''}
\delta(E_{\vec{p\,}''}-E)
\bra{\vec{p\,}''}\hat{\mathcal{T}}(E\mp i\eta) \ket{\vec{p}\,}
\nonumber\\
&= \mp\pi \intop\rmd^3 p''
\big|\bra{\vec{p}\,}\hat{\mathcal{T}}(E\pm i\eta) \ket{\vec{p\,}''}\big|^2
\delta(E_{\vec{p\,}''}-E)
\,.
\label{T-optic-2}
\end{align}
We notice that energies of intermediate states with momenta $\vec{p\,}''$ in this equation are reduced to the mass shell $E_{\vec{p\,}''}$.
The matrix elements of Eq.~(\ref{Tz-Tz}) between the momentum states corresponding to the scattering channel with specific spin $S$, $\ket{\vec{p},S}$, as in Eq.~(\ref{b2-full-p}), differ by additional factor 1/2 in the r.h.s, and instead of Eq.~(\ref{T-optic}) we have
\begin{align}
\bra{\vec{p\,}',S}\hat{\mathcal{T}}(E\pm i\eta) -
\hat{\mathcal{T}}(E\mp i\eta)\ket{\vec{p},S}
= \mp \pi i \intop\rmd^3 p''
\bra{\vec{p\,}',S}\hat{\mathcal{T}}(E\pm i\eta) \ket{\vec{p\,}'',S}
\delta(E_{\vec{p\,}''}-E)
\bra{\vec{p\,}'',S}\hat{\mathcal{T}}(E\mp i\eta) \ket{\vec{p},S}\,.
\label{T-optic-S}
\end{align}

For matrix elements of Eq.~(\ref{T-deriv}) we have relation
\begin{align}
\frac{\rmd }{\rmd z}\bra{\vec{p\,}',S}\hat{\mathcal{T}}(E + i\eta) \ket{\vec{p},S}
&=-\bra{\vec{p\,}',S}\hat{\mathcal{T}}(E+i\eta) \hat{\mathcal{G}}_0^2(E+i\eta)
\hat{\mathcal{T}}(E+i\eta) \ket{\vec{p},S}
\nonumber\\
&=-\frac12\intop\rmd^3\vec{p\,}''
\frac{
\bra{\vec{p\,}',S}\hat{\mathcal{T}}(E+i\eta)\ket{\vec{p\,}'',S}
\bra{\vec{p\,}'',S}\hat{\mathcal{T}}(E+i\eta) \ket{\vec{p},S}}
{(E-E_{\vec{p\,}''}+i\eta)^2}
\,.
\label{T-deriv-matrix}
\end{align}
\end{widetext}

We turn now to the Lippmann-Schwinger equation (\ref{LSE}). Taking the matrix element we obtain
\begin{align}
&\bra{\vec{p\,}',S} \hat{\mathcal{T}}(z) \ket{\vec{p}, S}=
\bra{\vec{p\,}',S} \hat{U} \ket{\vec{p}, S}
\nonumber\\
&\quad +\frac12
\intop \rmd^3 p''
\frac{\bra{\vec{p\,}',S} \hat{U} \ket{\vec{p\,}'', S} \bra{\vec{p\,}'',S}\hat{\mathcal{T}}(z) \ket{\vec{p}, S}}{z-E_{\vec{p\,}''}}
\,.
\label{LSE-me}
\end{align}
We can expand the scattering matrix element  in the partial waves, as we have done with the potential Eq.~(\ref{U-S-exp}),
\begin{align}
\bra{\vec{p\,}',S} \hat{\mathcal{T}}(z) \ket{\vec{p}, S} &=
\bra{\vec{p\,}'} \hat{\mathcal{T}}^{(S)}(z) \ket{\vec{p}\,}
\nonumber\\
&= 2\sum_{l}(2l+1) P_l(\vec{n}_p\vec{n\,}_p')
\bra{p'} \hat{\mathcal{T}}_l^{(S)}(z) \ket{p}
\,.
\label{T-S-exp}
\end{align}
Here $\bra{\vec{p\,}'} \hat{\mathcal{T}}^{(S)}(z) \ket{\vec{p}\,}$ is
%called as
the scattering matrix element
%for
in the spin-$S$ channel.
%According to the expansion of the potential (\ref{U-S-exp}) the scattering matrix for $S=0$ has contributions only from partial waves with even  $l$ and for $S=1$ only from waves with odd values of $l$.
For the reduced matrix elements we obtain the following equation
\begin{align}
&\bra{p'} \hat{\mathcal{T}}_l^{(S)}(z) \ket{p} = \bra{p'} \hat{U}_l^{(S)} \ket{p}
\nonumber\\
&\qquad + 4\pi
\intop \rmd p'' (p'')^2
\frac{\bra{p'} \hat{U}_l^{(S)} \ket{p''} \bra{p''}\hat{\mathcal{T}}_l^{(S)}(z) \ket{p}}{z-E_{\vec{p\,}''}}
\,.
\label{LSE-red}
\end{align}
The formal index $(S)$ escribed to the matrix element of the potential, $\bra{p'} \hat{U}_l^{(S)} \ket{p}$ indicates the selection of the partial waves, which follows from the definition of the potential with the specific spin (\ref{US-def}), only even $l$s contribute to the matrix elements $\bra{p'} \hat{U}_l^{(S=0)} \ket{p}$  and odd $l$s to $\bra{p'} \hat{U}_l^{(S=1)} \ket{p}$. In each case the values of the matrix elements are calculated with Eq.~(\ref{Ul}). For the zero matrix element of the potential, the solution of Eq.~(\ref{LSE-red}) vanishes, hence
 $\bra{p'}\hat{\mathcal{T}}_{2l+1}^{(S=0)}(z) \ket{p}=0=\bra{p'}\hat{\mathcal{T}}_{2l}^{(S=1)}(z) \ket{p}$.
%$F^{(S=0)}_{2l+1}(E_p)=0$ and $F^{(S=1)}_{2l}(E_p)=0$. {Ul}

The matrix element of the scattering operator between states with momenta $\vec{p}$ and $\vec{p\,}'$, such that $|\vec{p}\,|=|\vec{p\,}'|$, see Eq.~(\ref{LSE-me}), taken at the energy $E_p+i\eta$, defines the scattering amplitude
in the center-of-mass system,
\begin{align}
& F^{(S)}(E_{\vec{p}},(\vec{n}\vec{n\,}')) = -2\pi^2 m_f \bra{\vec{p\,}',S} \hat{\mathcal{T}}(E_{\vec{p}} + i\eta) \ket{\vec{p}, S},
%\nonumber\\
%&\quad = \frac12\Big(f^{(S)}(E_p,(\vec{n}\vec{n\,}'))+(-1)^{S}f^{(S)}(E_p,-(\vec{n}\vec{n\,}')) \Big)\,,
\label{f-def-app}
\end{align}
which is the function of the energy
%in the center of mass system
and the angle between momenta of incoming and outgoing particles, $\vec{p}= \vec{n} |\vec{p}\,|$ and $\vec{p\,}'= \vec{n\,}' |\vec{p}\,|$.
The cross section of the particle scattering in the spin $S$ channel is $\rmd \sigma_S=|F^{(S)}(E_{\vec{p}},(\vec{n}\vec{n\,}'))|^2\rmd \Omega_{n'}$,
and the total cross section is $\rmd \sigma_{\rm tot}= \frac34\rmd \sigma_{1}+ \frac14\rmd \sigma_0$.

Substituting expansion (\ref{T-S-exp}) in the definition (\ref{f-def-app}) we obtain the expansion for the scattering amplitude
\begin{align}
F^{(S)}(E_{\vec{p}},(\vec{n}\vec{n\,}')) &= 2\sum_{l}(2l+1) P_l(\vec{n}\vec{n\,}') f^{(S)}_l(E_{\vec{p}})\,
\label{F-pwa}
\end{align}
in partial waves defined as $$f^{(S)}_l(E_{\vec{p}})=-2\pi^2 m_f \bra{p} \hat{\mathcal{T}}_l^{(S)}(E_{\vec{p}}+i\eta) \ket{p}.$$
The partial waves $f^{(S)}_l(E_{\vec{p}})$ determine the scattering amplitude for non-identical particles by Eqs.~(\ref{f-pwa}) and (\ref{f-nonident}).

Substituting Eqs.~(\ref{f-def-app}) and (\ref{F-pwa}) in Eq.~(\ref{T-optic-S}) we obtain the optical theorem for the partial wave amplitudes
\begin{align}
\Im f_l^{(S)}(E_{\vec{p}})= p\big| f_l^{(S)}(E_{\vec{p}})\big|^2\,.
\label{T-optic-l}
\end{align}
The complex function satisfying this condition can be parameterized as
\begin{align}
f^{(S)}_l(E_{\vec{p}}) &= \frac{e^{2i\delta_l^{(S)}}-1}{2ip}= \frac{1}{p}e^{i\delta_l^{(S)}}\sin\delta_l^{(S)}
\nonumber\\
&=\frac{1}{p\cot \delta_l^{(S)}-ip}\,, \quad E_{\vec{p}}=p^2/m_f\,,
\label{F-delt-def}
\end{align}
through the real function $\delta_l^{(S)}(E_{\vec{p}})$, which is  the scattering phase shift.
From the expansions (\ref{Ul-epx}) and Eq.~(\ref{LSE-red}) it follows that for small scattering momenta the scattering amplitudes and the phase shifts behave like
\begin{align}
f^{(S)}_l(E_{\vec{p}})\propto p^{2l}\,\,\mbox{and}\,\, \delta_l^{(S)}(E_{\vec{p}})\propto p^{2l+1}\,.
\label{low-k-delt}
\end{align}

The iterative solution of  the Lippmann-Schwinger equation~(\ref{LSE-me}) is
\begin{align}
\bra{\vec{p\,}',S} \hat{\mathcal{T}}^{(S)}(E_{\vec{p}}) \ket{\vec{p}, S} &=
\bra{\vec{p\,}',S} \hat{\mathcal{T}}_1^{(S)}(E_{\vec{p}}) \ket{\vec{p}, S}
\nonumber\\
&+ \bra{\vec{p\,}',S} \hat{\mathcal{T}}_2^{(S)}(E_{\vec{p}}) \ket{\vec{p}, S}
+\dots,
\nonumber\\
\bra{\vec{p\,}',S} \hat{\mathcal{T}}_1^{(S)}(E_{\vec{p}}) \ket{\vec{p}, S} &= \bra{\vec{p\,}',S} \hat{U} \ket{\vec{p}, S},
\nonumber\\
\bra{\vec{p\,}',S} \hat{\mathcal{T}}_2^{(S)}(E_{\vec{p}}) \ket{\vec{p}, S} &=
\intop \frac{\rmd^3 p''}{2}
\frac{\bra{\vec{p\,}',S} \hat{U} \ket{\vec{p\,}'', S} \bra{\vec{p\,}'',S}\hat{U}\ket{\vec{p}, S}}{E_{\vec{p}}-E_{\vec{p\,}''}+i\eta}.
\label{LSE-iter}
\end{align}
If the scattering potential can be considered as a perturbation, so that
$|\bra{\vec{p\,}',S} \hat{\mathcal{T}}_2^{(S)}(E_{\vec{p}}) \ket{\vec{p}, S} \ll \bra{\vec{p\,}',S} \hat{\mathcal{T}}_1^{(S)}(E_{\vec{p}}) \ket{\vec{p}, S}|$,
 one can keep only the first-order term
\begin{align}
&\bra{\vec{p\,}',S} \hat{\mathcal{T}}^{(S)}(E_{\vec{p}}) \ket{\vec{p}, S}
\approx
\bra{\vec{p\,}',S} \hat{\mathcal{T}}_1^{(S)}(E_{\vec{p}}) \ket{\vec{p}, S}
\nonumber\\
&\,\,= \bra{\vec{p\,}',S} \hat{U} \ket{\vec{p}, S}
\nonumber\\
&\,\,=\frac{1}{(2\pi)^3}
\big( \Umom(|\vec{p}-\vec{p}\,'|) +(-1)^S \Umom(|\vec{p}+\vec{p}\,'|) \big)
%=\frac{1}{(2\pi)^3} \overline{\Umom^{(S)}}(\vec{p\,}',\vec{p}\,)
\,,
\label{T-Born}
\end{align}
that corresponds to \emph{the Born approximation}.

Applicability conditions of the Born approximation can be easily formulated in the coordinate representation~\cite{LLIII} (see \S\,126), it depends on the typical range of the potential action, $R$, the typical magnitude of the potential, $\Ucor_{\rm typ}$, and the particle momentum
\begin{align}
|\Ucor_{\rm typ}|\ll \frac{1}{(m_f R^2)}\,, \quad pR\lsim 1\,,
\label{slowB}
\end{align}
for  slow particles with  $pR\lsim 1$,
and
\begin{align}
|\Ucor_{\rm typ}|\ll \frac{pR}{m_f R^2}\,, \quad pR\gsim 1
\label{fastB}
\end{align}
for rapid particles, i.e.  $pR\gsim 1$.

In the Born approximation we have    for the scattering amplitude
\begin{align}
&F^{(S)}_{\rm Born}(E_p,(\vec{n}\vec{n\,}')) = -2\pi^2 m_f\bra{\vec{p\,}'} \hat{U}^{(S)} \ket{\vec{p}\,}
\nonumber\\
&\qquad =-\frac{m_f}{4\pi}\big( \Umom(p|\vec{n}_p-\vec{n}_p\,'|) + (-1)^S \Umom(p|\vec{n}_p+\vec{n\,}'_p|) \big)\,,
\label{f-Born}
\end{align}
where  we used Eq.~(\ref{U-S-Umom}).

The spin-averaged scattering amplitude is expressed through the spin averaged potential (\ref{U-spinaver}),
\begin{align}
\overline{F^{(S)}_{\rm Born}}(E_p,(\vec{n}\vec{n\,}')) & =
\sum_{S=0,1}\frac{(2S+1)}{4}F^{(S)}_{\rm Born}(E_p,(\vec{n}\vec{n\,}'))
\nonumber\\
&=-\frac{m_f}{4\pi} \overline{\Umom^{(S)}}(\vec{p\,}',\vec{p}\,)\,.
\label{f-Born-spinaver}
\end{align}

Using expansions (\ref{F-pwa}) and (\ref{U-exp-l}) on the left and right sides of the definition of the Born amplitude Eq.~(\ref{f-Born}) we can rewrite the Born partial amplitudes as
\begin{align}
f^{(S)}_{{\rm Born},l}(E_p) =-\frac{m_f}{4\pi}\frac{1+(-1)^{S+l}}{2}\intop \rmd^3 r \Ucor(r) j_l^2(rp)\,.
\end{align}
%{\bf We see that as in the Born approximation the scattering amplitude is real, and the optical theorem (\ref{T-optic-l}) can be satisfied only perturbatively under condition that in expansion of Eq.~(\ref{F-delt-def}), $p f_l^{(S)}\approx \delta_l^{(S)} +i\delta_l^{(S)2}$ the quadratic term can be neglected, i.e $|\delta_l^{(S)}|\ll 1$. Ne ponyal }
The corresponding phase shift is then given by
\begin{align}
\delta_{{\rm Born},l}^{(S)}=-\frac{1+(-1)^{S+l}}{2}\frac{m_fp}{4\pi}\intop \rmd^3 r \Ucor(r) j_{l}^2(rp),
\label{FBorn-delt}
\end{align}
Since $|\delta_{{\rm Born},l>0}^{(S)}|\le |\delta_{{\rm Born},l=0}^{(S)}|$, it sufficient that
\begin{align}
|\delta_{{\rm Born},l=0}^{(S)}|\ll 1,
\label{delt-Born-cond}
\end{align}
the later condition can be reduced to conditions~(\ref{slowB}) and (\ref{fastB}).

As follows from (\ref{f-Born-spinaver}), the forward scattering amplitude in the Born approximation takes the simplest form,
\begin{align}
F^{(S)}_{\rm Born}(E_p,1) = -\frac{m_f}{4\pi}\big( \Umom(0) + (-1)^S\Umom(2p) \big)\,.
\end{align}
Taking into account that in the Born approximation $f_l^{(S)}\approx \delta_l^{(S)}/p$, the spin averaged forward scattering amplitude
can be written as
\begin{align}
&\overline{F^{(S)}}_{\rm Born}(E_p,1)=-\frac{m_f}{4\pi} \overline{\Umom^{(S)}}(\vec{p\,},\vec{p}\,)
\nonumber\\
&\qquad =
\frac12\sum_S\sum_{l[S]}(2S+1)(2l+1)\frac{\delta_{{\rm Born},l}^{(S)}}{p}\,.
\label{F-delta-Born}
\end{align}
This relation was used in derivation of Eq.~(\ref{b2-U-Born}).

Applying Eq.~(\ref{j_l-exp}) in  Eq.~(\ref{FBorn-delt}) we can find the parameters of the low-momentum expansion ($pR\ll 1$) of the phase shift
\begin{align}
\delta_{{\rm Born},l}^{(S)}\approx -a_{{\rm B},l}^{(S)} p^{2l+1}\Big(1+\frac12 a_{{\rm B},l}^{(S)} r_{{\rm B, ef},l}^{(S)}\,p^2\Big)\,,
\end{align}
being expressed through the integrals of the potential,
\begin{align}
a_{{\rm B},l}^{(S)} &=\frac{1+(-1)^{S+l}}{2} \Big[\frac{2^l\,l!}{(2l+1)!}\Big]^2\frac{m_f}{4\pi}\intop \rmd^3 r \Ucor(r) r^{2l}\,,
 \nonumber\\
a_{{\rm B},l}^{(S)} r_{{\rm B,ef},l}^{(S)} &=-\frac{1+(-1)^{S+l}}{2}\frac{2}{2l+3}\frac{\intop \rmd^3 r \Ucor(r)r^{2l+2}} {\intop \rmd^3 r \Ucor(r) r^{2l}}\,.
\label{lowp-param-Born}
\end{align}
Particularly for s- and p-waves we have
\begin{align}
&a_{{\rm B,0}}^{(0)} = \frac{m_f}{4\pi}\intop \rmd^3 r \Ucor(r)
\,,\,\,
a_{{\rm B},1}^{(1)} =\frac{m_f}{36\pi}\intop \rmd^3 r \Ucor(r) r^2\,,
\nonumber\\
&a_{{\rm B},0}^{(0)} r_{{\rm B,ef}, 0}^{(0)} = -\frac{2\intop \rmd^3 r \Ucor(r)r^{2}} {3\intop \rmd^3 r \Ucor(r)}\,,\,\,
a_{{\rm B},1}^{(1)}  r_{{\rm B,ef}, 1}^{(1)} = -\frac{2\intop \rmd^3 r \Ucor(r)r^{4}} {5\intop \rmd^3 r \Ucor(r) r^2}.
\label{lowp-param-L01}
\end{align}

\section{Semiclassical expansion }\label{app:QPexp}

In this Appendix we find the first two functions $\chi_{k\le 2}$ in the expansion (\ref{I-exp-1}), which satisfy the differential equation (\ref{I-eq}). First we rewrite it in the explicit form,
\begin{align}
&\frac{\partial \chi_k}{\partial \beta} =
-i\frac{2}{m}p_a (\beta (\nabla_a U) - \nabla_a) \chi_{k-1}
\nonumber\\
&+\frac{1}{m}\Big(\beta^2(\nabla U)^2-\beta (\nabla^2 U)
-2\beta(\nabla_a U)\nabla_a +\nabla^2\Big)\chi_{k-2} \,.
\label{I-eq-app}
\end{align}
Here $\beta=1/T$ and each operator $\nabla$ standing outside of the bracket acts on everything to the right of it. Operator $\nabla$ within brackets, e.g. $(\nabla_a \Ucor)$, acts only on the nearest function. Boundary conditions are $\chi_0=1$ and $\chi_{k}(\beta=0)=0$ for $k > 0$.
For $k=1$, Eq.~(\ref{I-eq-app}) gives the equation for the function $\chi_1$,
\begin{align}
\frac{\partial\chi_1}{\partial \beta}=-i\frac{2}{m}\beta p_a(\nabla_a U),
\end{align}
which yields the solution
\begin{align}
\hbar\chi_1=-i\frac{\hbar p_a}{m}\beta^2 (\nabla_a U)\,.
\end{align}
Substituting this result in Eq.~(\ref{I-eq-app}) and $\chi_0=1$ for $k=2$ we get the equation for $\chi_2$,
\begin{align}
\frac{\partial \chi_2}{\partial \beta} =&
-\frac{2}{m^2} \big(\beta^3 (p_a(\nabla_a U))^2 - \beta^2p_a p_b(\nabla_a\nabla_b U)\big)
\nonumber\\
&+\frac{1}{m}\big(\beta^2(\nabla U)^2-\beta (\nabla^2 U)\big) \,,
\end{align}
with the solution
\begin{align}
\hbar^2\chi_2 = & -\frac{\hbar^2 p_a p_b}{2m^2}\beta^3
\Big(\beta (\nabla_a U) (\nabla_b U) - \frac{4}{3} (\nabla_a\nabla_b U) \Big)
\nonumber\\
&+\frac{\hbar^2\delta_{ab}}{2m}\beta^2
\Big(\beta\frac{2}{3}(\nabla_a U)(\nabla_b U) - (\nabla_a\nabla_b U)\Big) \,.
\end{align}

The functions $\chi_k$ have to be integrated over momentum in Eq.~(\ref{h-exp}). We introduce the notation
\begin{align}
\big\langle \dots \big\rangle_p = \intop \frac{\rmd^3 p}{(2\pi\hbar)^3} (\dots) e^{-\frac{p^2}{m T}+\frac{i}{\hbar}\vec{p}\vec{\xi}}
\end{align}
for the momentum average and  use the  integrals
\begin{align}
&\big\langle 1 \big\rangle_p = \frac{e^{-\pi \vec{\xi\,}^2/2\lambda^2}}{2^{3/2}\lambda^3}\,,
%\nonumber\\
\quad
\big\langle p_a \big\rangle_p
%\intop \frac{\rmd^3 p}{(2\pi)^3} p_a e^{-\frac{p^2}{m T}+\frac{i}{\hbar}\vec{p}\vec{r}}
= \big\langle 1 \big\rangle_p i\pi\,\frac{\hbar}{\lambda}\frac{\xi_a}{\lambda} \,,
\nonumber\\
&\big\langle p_ap_b \big\rangle_p
%\intop \frac{\rmd^3 p}{(2\pi)^3} p_ap_b e^{-\frac{p^2}{m T}+\frac{i}{\hbar}\vec{p}\vec{r}}
=\big\langle 1 \big\rangle_p\frac{\pi\hbar^2}{\lambda^2} \Big(\delta_{ab} -\pi\frac{\xi_a \xi_b}{\lambda^2}\Big)\,.
%\nonumber\\
%\big\langle p_ap_b p_c \big\rangle_p
%&=\big\langle 1 \big\rangle_p
%\frac{i\pi^2\hbar^3}{\lambda^3}
%\Big(\frac{r_a}{\lambda}\delta_{bc}+\frac{r_b}{\lambda}\delta_{ac}
%\nonumber\\
%&\quad \qquad
%+\frac{r_c}{\lambda}\delta_{ab} - \pi\frac{r_ar_br_c}{\lambda^3}
%\Big)
\label{p-aver}
\end{align}
For the direct terms the matrix element (\ref{h-exp}) is evaluated for $\xi=\vec{r}-\vec{r'}=0$ and for the exchange term for $\xi=\vec{r}-\vec{r'}=2\vec{r}$.
Using these relations for calculations of the momentum averages of functions $\chi_{1,2,3}$ we obtain
\begin{align}
\big\langle \hbar\chi_1(\vec{r},\vec{p}\,) \big\rangle_p %= \frac{\hbar^2\pi}{mT}\frac{(\vec{r}\,\nabla U)}{\lambda^2T}\big\langle 1 \big\rangle_p
&=\frac12 \frac{(\vec{\xi}\,\nabla U(r))}{ T}\big\langle 1 \big\rangle_p,
\nonumber\\
\big\langle \hbar^2\chi_2(\vec{r},\vec{p}\,) \big\rangle_p
%=
% -\frac{\lambda^2}{4\pi}\big\langle 1 \big\rangle_p
%  \nonumber\\
%  &\times \Big( \frac{(\nabla U)^2}{T^2}  - \frac{4}{3} \frac{\nabla^2 U}{T}
%  - \frac{\pi}{\lambda^2}\big( \frac{(\vec{r}\, \nabla U)^2}{T^2} - \frac{4}{3} \frac{r_a r_b\nabla_a\nabla_bU}{T} \big)\Big)
%  \nonumber\\
%&+\frac{\lambda^2}{4\pi}\big\langle 1 \big\rangle_p \Big(\frac{2}{3}\Big(\frac{\nabla U}{T}\Big)^2 - \frac{\nabla^2 U}{T}\Big)
%\nonumber\\
&= \frac{\lambda^2}{4\pi}\big\langle 1 \big\rangle_p\frac16\Big( \frac{\big(\nabla U(r)\big)^2}{T^2}  -2\frac{\nabla^2 U(r)}{T}\Big)
  \nonumber\\
  &+\frac{1}{8}\big\langle 1 \big\rangle_p
  \Big( \frac{\big(\vec{\xi}\, \nabla U(r)\big)^2}{T^2} - \frac{4}{3} \frac{\xi_a \xi_b\nabla_a\nabla_bU(r)}{T} \Big)\,.
  \label{chi12-paver}
\end{align}
Using this average in Eq.~(\ref{h-exp}) we can cast the semiclassical expansion of the matrix elements in (\ref{b2int-full}) in terms of direct terms (\ref{b2-dir}) with $\vec{\xi} = 0$ and the exchange terms (\ref{b2-exch}) with $\vec{\xi}=2\vec{r}$.

To lower the order of potential derivatives in Eqs.~(\ref{b2-dir}) and (\ref{b2-exch}) we can use the  relations
\begin{align}
&\intop\rmd^3 r e^{-\frac{U(|\vec{r}\,|)}{T}} \nabla^2 U=\intop\rmd^3 r e^{-U(|\vec{r}|)/T}\frac{(\nabla U)^2}{T}\,,
\label{byparts-1}\\
&\intop\rmd^3 r e^{-\frac{U(|\vec{r}\,|)}{T}-\frac{2\pi r^2}{\lambda^2}}\nabla^2 U =
\intop\rmd^3 r e^{-\frac{U(|\vec{r}\,|)}{T}-\frac{2\pi r^2}{\lambda^2}}
\nonumber\\
&\qquad
\times\Big[ \frac{(\nabla U)^2}{T} + \frac{4\pi}{\lambda^2}(\vec{r\,}\nabla U)\Big]\,,
\label{byparts-2}\\
&\intop\rmd^3 r e^{-\frac{U(|\vec{r}\,|)}{T}-\frac{2\pi r^2}{\lambda^2}}r_a r_b \nabla_a\nabla_b U =
\intop\rmd^3 r e^{-\frac{U(|\vec{r}\,|)}{T}-\frac{2\pi r^2}{\lambda^2}}
\nonumber\\
&\qquad
\times\Big[ \frac{(\vec{r\,}\nabla U)^2}{T} + \Big(\frac{4\pi r^2}{\lambda^2} - 4\Big)(\vec{r\,}\nabla U)\Big]\,,
\label{byparts-3}
\end{align}
obtained with the help of the integration by parts.

\section{On-shell reduction of scattering amplitude in Eq.~(\ref{b2-T-1}) }\label{app:onshell}

In Section~\ref{ssec:fullquant-b2}  we have derived Eq.~(\ref{b2-T-1}), which allows to express the second virial coefficient through the matrix element of the scattering amplitude $\mathcal{T}$, being determined by the inter-particle potential $U$ obeying the Lippmann-Schwinger equation (\ref{LSE}). Although the integration over energy $z$ runs formally for all $z>0$, the imaginary part reduces the integration to the mass shell $z=E_{\vec{p}}=\vec{p\,}^2/m$.

Now we turn to the integral in Eq.~(\ref{b2-T-1}), which we denote as
\begin{align}
I_b(\vec{p}\,)=\intop_0^\infty\frac{\rmd E}{\pi} e^{-E/T}  \Im \frac{\bra{\vec{p},S}\hat{\mathcal{T}}(E+i\eta) \ket{\vec{p},S} }{(E+i\eta-E_{\vec{p}})^2}\,.
\label{b2-T-1-app}
\end{align}
\begin{widetext}
As the first step, take imaginary part in Eq.~(\ref{b2-T-1-app}) for finite value of $\eta$, then
\begin{align}
I_b(\vec{p}\,)
&=\intop_0^\infty\frac{\rmd E}{\pi} e^{-E/T}
\Im \bra{\vec{p},S}\hat{\mathcal{T}}(E+i\eta) \ket{\vec{p},S}
\Re\frac{1}{(E_{\vec{p}}-E-i\eta)^2}
-\intop_0^\infty\frac{\rmd E}{\pi} e^{-E/T}
\Re\bra{\vec{p},S}\hat{\mathcal{T}}(E + i\eta) \ket{\vec{p},S}\frac{2\eta(E-E_p)}
{((E-E_p)^2+\eta^2)^2}
\,.
\end{align}
In the second integral, the limit $\eta\to 0_+$ gives
\begin{align}
\lim_{\eta\to 0_+}\frac{2\eta(E - E_{\vec{p\,}})}
{((E - E_{\vec{p\,}} )^2+\eta^2)^2}=-\pi\frac{\rmd}{\rmd E}\delta(E-E_{\vec{p\,}})\,,
\end{align}
and in the first integral we can use the optical theorem (\ref{T-optic-S}). After integration over $E$ we get
\begin{align}
I_b(\vec{p}\,) &= - \frac12
\intop\rmd^3 p'e^{-E_{\vec{p\,}'}/T}
\Re \frac{\bra{\vec{p},S}\hat{\mathcal{T}}(E_{\vec{p\,}'}+i\eta) \ket{\vec{p\,}',S}
\bra{\vec{p\,}',S}\hat{\mathcal{T}}(E_{\vec{p\,}'}-i\eta) \ket{\vec{p},S}
 }{(E_{\vec{p\,}}-E_{\vec{p\,}'}-i\eta)^2}
\nonumber\\
& + \frac{1}{T}e^{-E_{\vec{p\,}}/T}
\Re\bra{\vec{p},S}\hat{\mathcal{T}}(E_{\vec{p\,}}+i\eta) \ket{\vec{p},S}
- e^{-E_{\vec{p\,}}/T} \Re\Big[\frac{\rmd}{\rmd E}\bra{\vec{p},S}\hat{\mathcal{T}}(E+i\eta) \ket{\vec{p},S}\Big]_{E=E_{\vec{p\,}}}\,.
\label{Ib-onshell}
\end{align}
Thus, we obtain the relation, in which the integration over the energy is replaced by the integration over the momenta and the scattering amplitudes are evaluated on shell, i.e., for energies expressed through the particle momenta.

The momentum integral of the function $I_b(\vec{p}\,)$ can be brought in a more convenient form.
In the last term in (\ref{Ib-onshell}) we can use Eq.~(\ref{T-deriv-matrix}), then the momentum integration gives
\begin{align}
\int\rmd^3 p I_b(\vec{p}\,)
& = - \frac12\intop\rmd^3 p\rmd^3 p'
e^{-E_{\vec{p\,}'}/T}
\Re \frac{\bra{\vec{p},S}\hat{\mathcal{T}}(E_{\vec{p\,}'}+i\eta) \ket{\vec{p\,}',S}
\bra{\vec{p\,}',S}\hat{\mathcal{T}}(E_{\vec{p\,}'}-i\eta) \ket{\vec{p},S}
 }{(E_{\vec{p\,}}-E_{\vec{p\,}'}-i\eta)^2}
  \nonumber\\
&+\intop\rmd^3 p\frac{e^{-E_{\vec{p\,}}/T}}{T}
 \Re\bra{\vec{p},S}\hat{\mathcal{T}}(E_{\vec{p\,}} + i\eta) \ket{\vec{p},S}
\nonumber\\
&+ \frac12\intop\rmd^3 p\,\rmd^3p' e^{-E_{\vec{p\,}}/T}
\Re\frac{\bra{\vec{p},S}\hat{\mathcal{\mathcal{T}}}(E_{\vec{p\,}} + i\eta)\ket{\vec{p\,}',S}
\bra{\vec{p\,}',S}\hat{\mathcal{T}}(E_{\vec{p\,}} + i\eta)\ket{\vec{p},S}}
{(E_{\vec{p}}-E_{\vec{p\,}'}+i\eta)^2} \,.
\end{align}
Changing variables ($\vec{p\,}'\leftrightarrow \vec{p}$) in the first integral and combining it with the third one we get
\begin{align}
\int \rmd^3 p I_b(\vec{p}\,) & =
\intop\rmd^3 p\frac{e^{-E_{\vec{p\,}}/T}}{T} \Re\bra{\vec{p},S}\hat{T}(E_{\vec{p\,}} + i\eta) \ket{\vec{p},S}
\nonumber\\
&- \frac12\intop\rmd^3 p\rmd^3 p'
e^{-E_{\vec{p\,}}/T}
\Re\Bigg[
 \frac{\bra{\vec{p\,}',S}\hat{\mathcal{T}}(E_{\vec{p\,}}+i\eta) \ket{\vec{p\,},S} } {(E_{\vec{p\,}'}-E_{\vec{p\,}}-i\eta)^2}
 \bra{\vec{p\,},S}\hat{\mathcal{T}}(E_{\vec{p\,}}-i\eta) -\hat{\mathcal{\mathcal{T}}}(E_{\vec{p\,}}+i\eta) \ket{\vec{p\,}',S}
 \Bigg]\,.
\end{align}
The last matrix element can be expressed using the optical theorem (\ref{T-optic}). Then we obtain
\begin{align}
\int\rmd^3 p I_b(\vec{p}\,)
&=  \intop\rmd^3 p\frac{e^{-E_{\vec{p\,}}/T}}{T} \Re\bra{\vec{p},S}\hat{\mathcal{T}}(E_{\vec{p\,}} + i\eta) \ket{\vec{p},S}
 + \frac{\pi}{2}\intop\rmd^3 p\rmd^3 p'
e^{-E_{\vec{p\,}}/T}  \intop\rmd^3 p''\delta(E_{\vec{p\,}''}-E_{\vec{p\,}})
\nonumber\\
&\times
\Im\Bigg[\bra{\vec{p},S}\hat{\mathcal{T}}(E_{\vec{p\,}}-i\eta) \ket{\vec{p\,}'',S}
 \frac{\bra{\vec{p\,}'',S}\hat{\mathcal{T}}(E_{\vec{p\,}}+i\eta) \ket{\vec{p\,}',S}
 \bra{\vec{p\,}',S}\hat{\mathcal{T}}(E_{\vec{p\,}}+i\eta) \ket{\vec{p\,},S}
 } {(E_{\vec{p\,}}-E_{\vec{p\,}'}+i\eta)^2}
 \Bigg]
\end{align}
To transform the last term we use Eq.~(\ref{T-deriv-matrix}) again and finally obtain
\begin{align}
\int\rmd^3 p I_b(\vec{p}\,) &=   \intop\rmd^3 p \frac{e^{-E_{\vec{p\,}}/T}}{T} \Re\bra{\vec{p},S}\hat{\mathcal{T}}(E_{\vec{p\,}} + i\eta) \ket{\vec{p},S}
\nonumber\\
& - \pi \intop\rmd^3 p
e^{-E_{\vec{p\,}}/T}  \intop\rmd^3 p''\delta(E_{\vec{p\,}''}-E_{\vec{p\,}})
\Im\Bigg[
\bra{\vec{p\,},S}\hat{\mathcal{T}}(E-i\eta) \ket{\vec{p\,}'',S}
 \frac{\rmd }{\rmd E}\bra{\vec{p\,}'',S}\hat{\mathcal{T}}(E+i\eta) \ket{\vec{p},S}
 \Bigg]_{E=E_{\vec{p\,}}}\,.
\label{int-onshell}
\end{align}
Integrating by parts the first integral here
\begin{align}
\intop\rmd^3 p \frac{e^{-E_{\vec{p\,}}/T}}{T} \Re\bra{\vec{p},S}\hat{\mathcal{T}}(E_{\vec{p\,}} + i\eta) \ket{\vec{p},S}
%=\intop\frac{\rmd \Omega\rmd E_p  }{(2\pi)^3} \frac12 (m E_p)^{1/2}m\frac{e^{-E_{\vec{p\,}}/T}}{T} \Re\bra{\vec{p},S}\hat{\mathcal{T}}(E_{\vec{p\,}} + i\eta) \ket{\vec{p},S}
%\nonumber\\
%=\intop\frac{\rmd \Omega\rmd E_p  }{(2\pi)^3}e^{-E_{\vec{p\,}}/T}
%\frac{\rmd}{\rmd E_p} \frac12 (m E_p)^{1/2}m\Re\bra{\vec{p},S}\hat{\mathcal{T}}(E_{\vec{p\,}} + i\eta) \ket{\vec{p},S}
%\nonumber\\
%=\intop\frac{\rmd \Omega\rmd p p   }{(2\pi)^3}\frac{2}{m}e^{-E_{\vec{p\,}}/T}
%\Big( \frac14\frac{m^{3/2}}{E_p^{1/2}} \Re\bra{\vec{p},S}\hat{\mathcal{T}}(E_{\vec{p\,}} + i\eta) \ket{\vec{p},S}
%+\frac12 m^{3/2} E_p^{1/2}\frac{\rmd}{\rmd E_p}\Re\bra{\vec{p},S}\hat{\mathcal{T}}(E_{\vec{p\,}} + i\eta) \ket{\vec{p},S}\Big)
%\nonumber\\
=\intop\rmd^3 p e^{-E_{\vec{p\,}}/T}
\Big( \frac1{2E_p}+ \frac{\rmd}{\rmd E_p}
\Big)\Re\bra{\vec{p},S}\hat{\mathcal{T}}(E_{\vec{p\,}} + i\eta) \ket{\vec{p},S}
\end{align}
and
taking into account that $\bra{\vec{p\,},S}\hat{\mathcal{T}}(E-i\eta) \ket{\vec{p\,}'',S}=
\bra{\vec{p\,}'',S}\hat{\mathcal{T}}(E+i\eta) \ket{\vec{p\,},S}^*$
we obtain Eq.~(\ref{int-onshell-2}).
\end{widetext}

\section{Double scattering amplitude}\label{app:double-scatt}

In order to judge about the applicability of the Born approximation we have to consider the second-order term, $O(U^2)$, in Eq.~(\ref{LSE-iter}):
\begin{align}
&\bra{\vec{p\,}'} \hat{\mathcal{T}}^{(S)}_2(E_p+i\eta) \ket{\vec{p}\,}=\frac12
\intop \rmd^3 p''
\frac{\bra{\vec{p\,}'} \hat{U}^{(S)} \ket{\vec{p\,}''} \bra{\vec{p\,}''}\hat{U}^{(S)} \ket{\vec{p}\,}}{E_p-E_{\vec{p\,}''}+i\eta}\,,
\label{T-second}
\end{align}
where $|\vec{p\,}'|=|\vec{p\,}|$.

For the nucleon-nucleon interaction described by two (attractive and repulsive) Yukawa  potentials (\ref{U-pot}),
we have to introduce the following integral for the iteration of Yukawa potentials:
\begin{align}
A_{ij}(p,\vec{n\,}',\vec{n}\,)&=\intop \rmd^3 q \frac{\mathcal{Y}((p\vec{n\,}' - \vec{q\,})^2,m_i^2)\mathcal{Y}((\vec{q\,} - p\vec{n})^2,m_j^2)}
{p^2-q^2+i\eta},
\nonumber\\
\mathcal{Y}(\vec{q\,}^2,m^2)&=\intop \frac{\rmd^3r}{4\pi} e^{-i\vec{q}\,\vec{r}} Y (m r)=\frac{1}{\vec{q\,}^2+m^2}\,.
\label{Aij}
\end{align}
Then we can express the second-order scattering matrix element (\ref{T-second}) as
\begin{widetext}
\begin{align}
&\frac{(2\pi)^6}{m_N}\bra{\vec{p\,}'} \hat{\mathcal{T}}^{(S)}_2(E_p+i\eta) \ket{\vec{p}\,} =
g_\om^4 A_{\om\om}(p,\vec{n\,}',\vec{n}) -g_\sigma^2 g_\om^2 \big(A_{\om\sigma}(p,\vec{n\,}',\vec{n}) + A_{\sigma\om}(p,\vec{n\,}',\vec{n})\big)
+ g_\sigma^4A_{\sigma\sigma}(p,\vec{n\,}',\vec{n})
\nonumber\\
&\quad + (-1)^S\Big(
g_\om^4 A_{\om\om}(p,\vec{n\,}',-\vec{n}) -g_\sigma^2 g_\om^2 \big(A_{\om\sigma}(p,\vec{n\,}',-\vec{n}) + A_{\sigma\om}(p,\vec{n\,}',-\vec{n})\big)
+ g_\sigma^4A_{\sigma\sigma}(p,\vec{n\,}',-\vec{n})
\Big).
\label{T-2}
\end{align}
Using the method proposed in Section 10.3.5 of book~\cite{Newton-book}
we can write integral (\ref{Aij}) in the form
\begin{align}
A_{ij}(p,\vec{n\,}',\vec{n}) &=
-\pi^2\intop_{-1}^1\frac{\rmd \alpha}{\Delta_{ij}^4(\alpha)+4\Delta_{ij}^2(\alpha)p^2 +p^2k^2(1-\alpha^2)}
 \Bigg[ i p + \frac{\Delta_{ij}^2(\alpha)}{\sqrt{4\Delta_{ij}^2(\alpha)+ k^2(1 - \alpha^2)}}
\Bigg] \,,
\nonumber\\
& \Delta_{ij}^2(\alpha)= \frac{m_i^2 + m_j^2}{2} + \alpha\frac{m_i^2 - m_j^2}{2}\,,\quad k^2=(\vec{p}-\vec{p\,}')^2=2p^2(1-  (\vec{n}\,\vec{n}'\,) ) \,.
\label{Aij-def}
\end{align}
\end{widetext}
There are several limits, when this general expression can be simplified.
So, for the diagonal elements $A_{ii}$ (Yukawa potentials of the same range) the final expression can be written in the compact form
\begin{align}
A_{ii}(p,\vec{n\,}',\vec{n}) &= \frac{i\, \pi^2}{ k\, h_i(p,k) } \ln\Big|\frac{h_i(p,k) - p k}{h_i(p,k) + p k}\Big|
\nonumber\\
& - \frac{2\pi^2}{kh_i(p,k)}
{\rm arctan}\frac{m_i\,k}{2h_i(p,k)}\,,
\nonumber\\
&h_i^2(p,k)=m_i^4 +4m_i^2p^2+p^2k^2\,.
\end{align}
For the forward scattering, $\vec{n\,}'=\vec{n}$ ($k\to 0$), integration in Eq.~(\ref{Aij-def}) gives the result
\begin{align}
A_{ij}(p,\vec{n\,},\vec{n}) &=-i\pi^2  \frac{\ln \left(\frac{ m_i^2(m_j^2+4 p^2)}{m_j^2 (m_i^2+4 p^2)}\right)}
{2 p(m_i^2-m_j^2)}
\nonumber\\
& - \pi^2 \frac{{\arctan}\big(\frac{m_i}{2p}\big)-{\rm arctan}\big(\frac{m_j}{2p}\big)}
{p(m_i^2 -m_j^2)}\,,
\label{Aij-forward}
\end{align}
which for the equal ranges of Yukawa potentials ($m_j = m_i$) reduces to the result from Section 9-c in~\cite{Taylor-book},
\begin{align}
&A_{ii}(p,\vec{n\,},\vec{n}) =  -\frac{\pi^2}{m_i^2(m_i - i2p)}\,.
\end{align}

In Fig.~\ref{fig:A-fun} we illustrate the momentum dependence of the function $A_{ij}$, see Eq.~(\ref{Aij-def}) for the iteration of $\sigma$ and $\om$ Yukawa potentials and various scattering angles $x=(\vec{n}\vec{n\,}')$.
Real parts of $A_{ij}$ are negative, and in the limit $p\to 0$
from Eq.~(\ref{Aij-forward}) we obtain $A_{ij}(p\to 0)=-2\pi^2/m_im_j(m_i+m_j)$. With the increasing $p$ the real part decreases on the scale  $p\sim 1/\sqrt{m_{i}m_j}$
The dependence of $A_{ij}$ on the scattering angle is the weaker the heavier the exchange mesons are (compare the results for $A_{\om\om}$ and $A_{\sigma\sigma}$).
Imaginary parts of $A_{ij}$ are linear in $p$ for small $p$, and reach maximum at $p\sim 1.8\mbox{--}2.0 m_\pi$ for $A_{\sigma\sigma}$ and $A_{\om\sigma}$, and
at $p\sim 2.7\,m_\pi$  for $A_{\om\om}$.

\begin{figure}
\centering
\includegraphics[width=6cm]{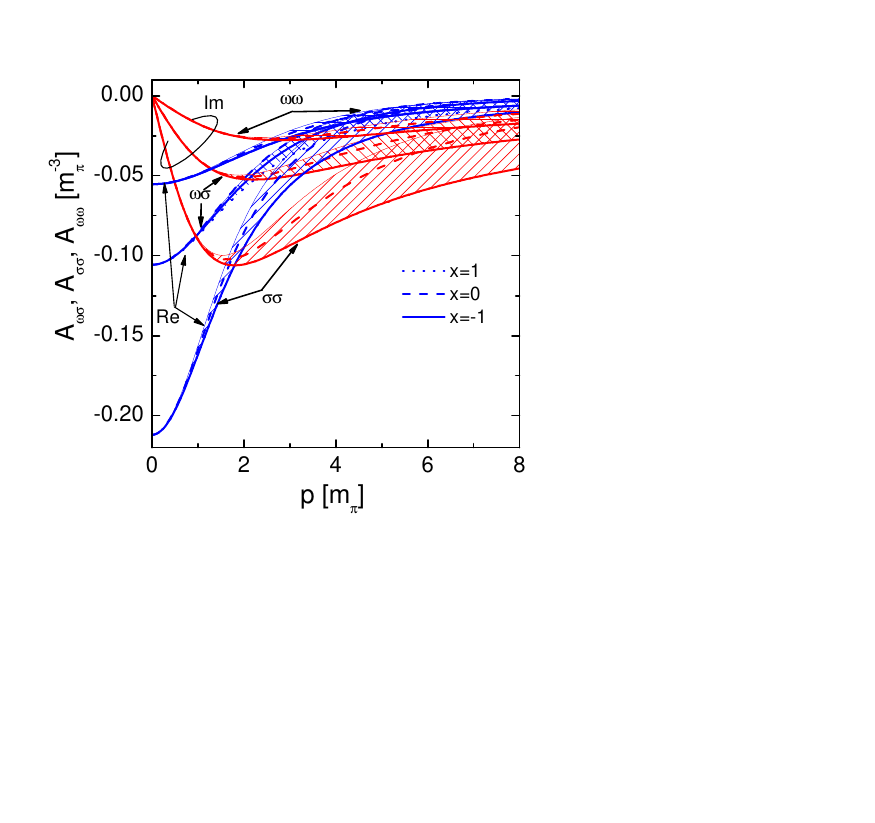}
\caption{Real and imaginary parts of functions (\ref{Aij-def}) for iteration of two Yukawa potentials with $\sigma$ and $\omega$ exchanges vs. momentum. Hatched regions correspond to variation of the scattering angle, $-1\le x=(\vec{n}\vec{n\,}')\le 1$. Dashed lines are drawn for $x=0$.}
\label{fig:A-fun}
\end{figure}

\begin{figure*}
\centering
\includegraphics[width=6cm]{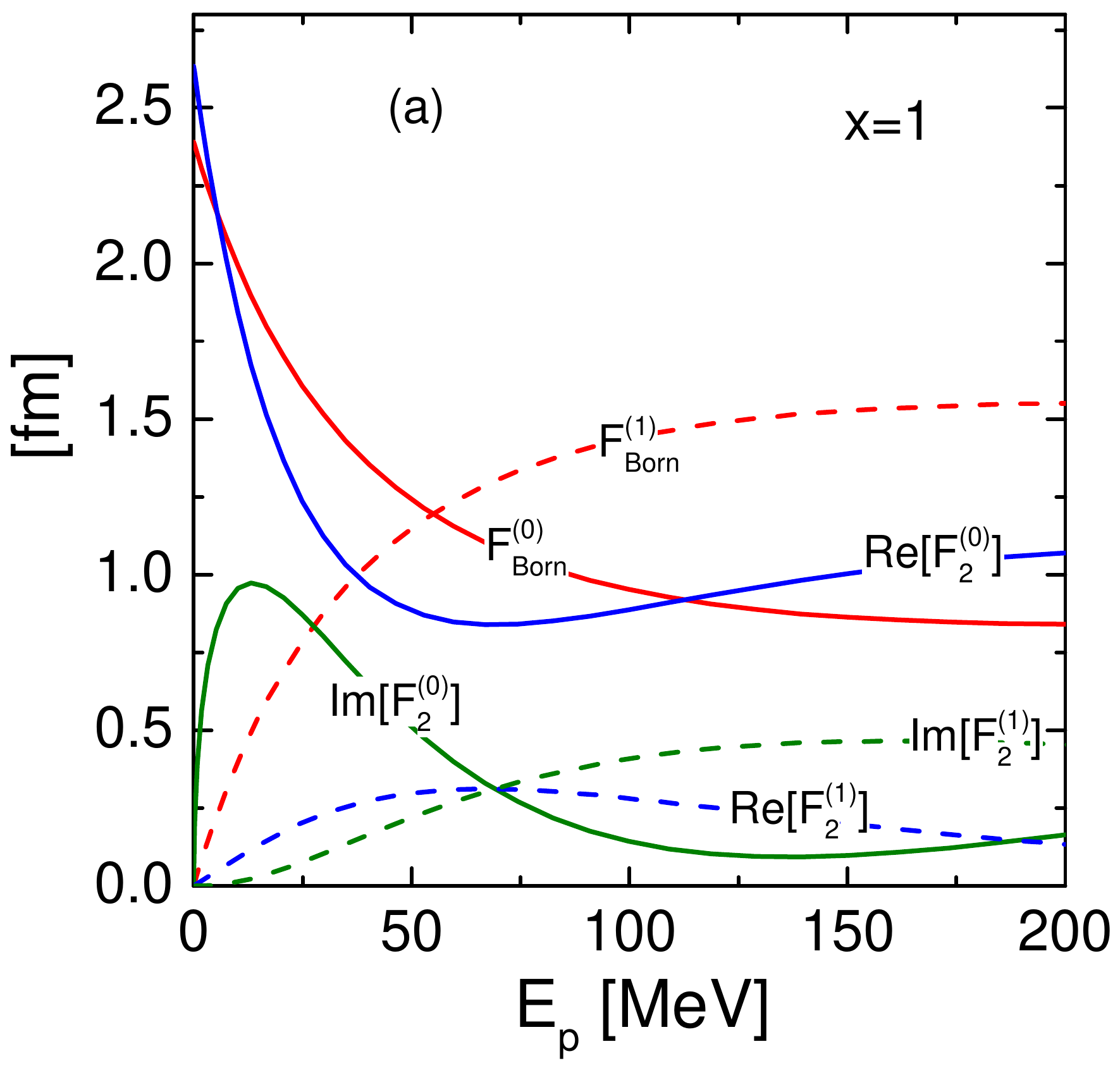}\quad
\includegraphics[width=6.6cm]{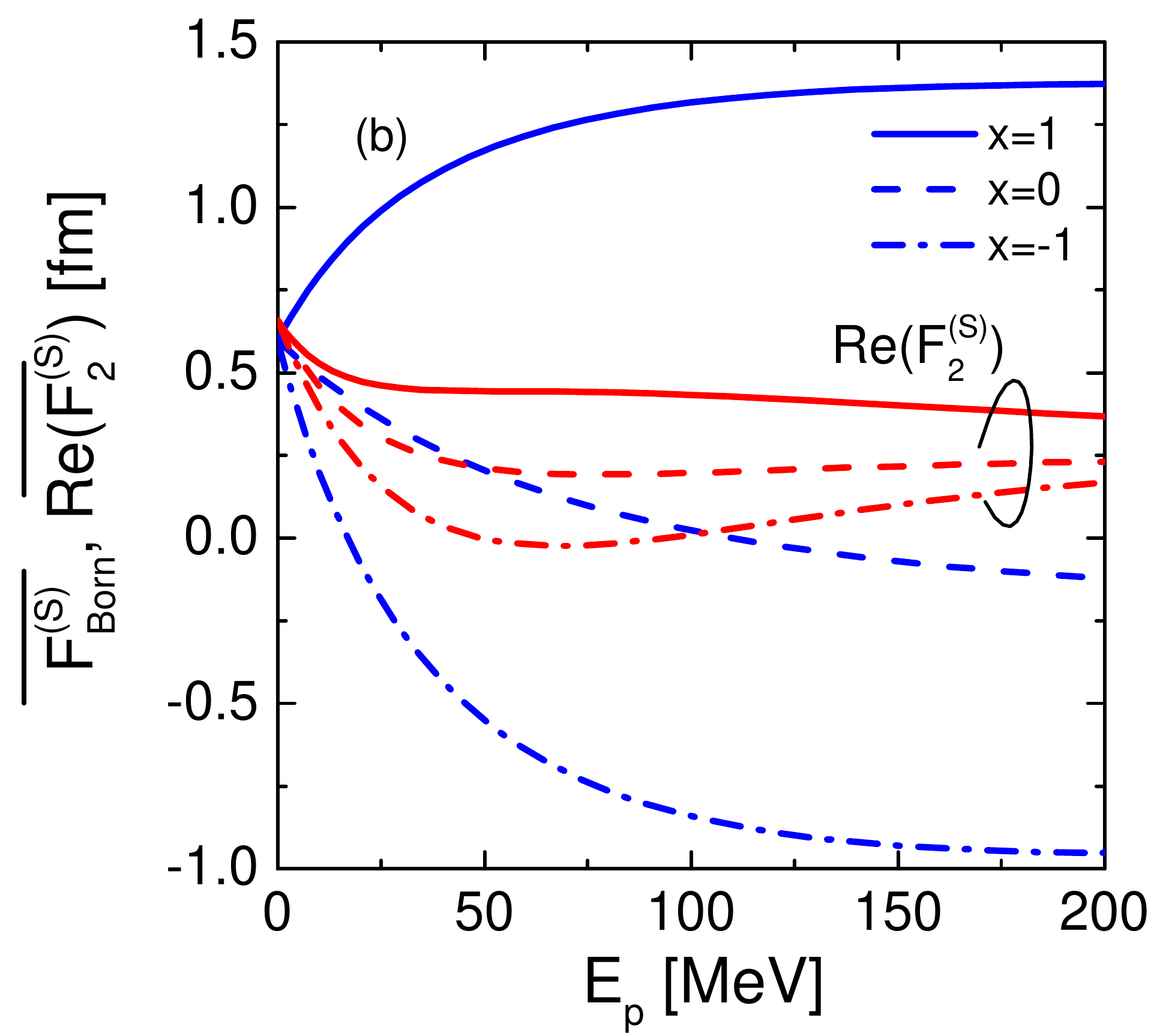}
\caption{Panel a): The scattering amplitudes in the Born approximation, Eq.~(\ref{f-Born}), and the doubly-iterated amplitude, Eq.~(\ref{FS2-def}) in spin 0 and spin 1 channels as functions of the energy of the relative motion and the forward kinematics $x=(\vec{n}\vec{n\,}')=1$. Panel b): The same as in panel a) but for the real parts of the isospin averaged amplitudes defined as in Eq.~(\ref{f-Born-spinaver}) for the Born amplitude and similarly for the $F^{(S)}_{2}$ amplitude. Solid, dashed and dash-dotted lines correspond to scattering angles $x=1$,0 and $-1$, correspondingly.}
\label{fig:F2-F1-comp}
\end{figure*}

Using the low-momentum expression for $A_{ij}$ we can estimate the ratio of the direct term in the second-order scattering amplitude $\bra{\vec{p\,}'} \hat{\mathcal{T}}^{(S)}_2(E_p+i\eta)\ket{\vec{p}\,}$ (first line in Eq.~(\ref{T-2})) and the first-order amplitude (Born approximation, Eq.~(\ref{T-Born}) and (\ref{U-pot})) for $p\to 0$,
%\begin{widetext}
\begin{align}
&\frac{|\bra{\vec{p\,}} \hat{\mathcal{T}}^{(S)}_2(E_p+i\eta)\ket{\vec{p}\,}|}{|\bra{\vec{p\,}} \hat{U}^{(S)}\ket{\vec{p}\,}|}\Bigg|_{p\to0 }
\simeq\frac{m_N \pi^2}{(2\pi)^3|g_\om^2/m_\om^2-g_\sigma^2/m_\sigma^2|}
\nonumber\\
&\qquad\times\Bigg|
\frac{g_\om^4}{m_\om^3} - \frac{4 g_\sigma^2\,g_\om^2}{m_\om m_\sigma(m_\om+m_\sigma)} +\frac{g_\sigma^4}{m_\sigma^3}
\Bigg|
\nonumber\\
&=\frac{1}{8\pi |C_\om^2 -C_\sigma^2|}\Bigg|
\frac{m_\om}{m_N}C_\om^4
- \frac{4 m_\om m_\sigma C_\om^2 C_\sigma^2}{m_N(m_\om+m_\sigma)}
+\frac{m_\sigma}{m_N}C_\sigma^4
\Bigg|
\nonumber\\
&\qquad \approx 1.1\,,
\label{F22F1-thresh}
\end{align}
%\end{widetext}
where we use parameters from Eq.~(\ref{Uomsig-par-W}). Thus, at small momenta the Born approximation is not applicable for the nucleon-nucleon scattering described by potential (\ref{U-S-Umom}). With increasing momentum the multiple-scattering contribution should become smaller, since the function $|A_{ij}|$ decreases, as seen in Fig.~\ref{fig:A-fun}.

In Fig.~\ref{fig:F2-F1-comp}a we compare scattering amplitudes in the Born approximation (\ref{f-Born}) and the double-scattering amplitudes,
\begin{align}
F^{(S)}_{2}(E_p,(\vec{n}\vec{n\,}')) = -2\pi^2 m\bra{\vec{p\,}'} \hat{\mathcal{T}}_2^{(S)} (E_p+i\eta)\ket{\vec{p}\,},
\label{FS2-def}
\end{align}
as functions of the energy $E_p$ for the forward kinematics $(\vec{n}\vec{n\,}')=1$. We see that in the spin 0 channel the real part of the double scattering amplitude is of the same order as the Born amplitude, $\Re(F^{(S=0)}_2)\sim F_{\rm Born}^{(S=0)}$. In the spin 1 channel the situation is different. Since the Pauli exclusion principle forbids two identical fermions to interact at the vanishing relative momentum, both
$F^{(1)}_{2}$ and $F^{(1)}_{\rm Born}$ are zero at $E_p=0$. With  increasing  $E_p$ the Born amplitude grows rapidly, while  $\Re(F^{(1)}_{2})$ increases more slowly, reaches maximum for $E_p\sim 60$\,MeV and then falls down. Two types of amplitudes possess also different dependence on the scattering angle, which is illustrated in Fig.~\ref{fig:F2-F1-comp}b, where we show spin averaged amplitudes, defined as in Eq.~(\ref{f-Born-spinaver}) for the Born amplitude and similarly for the $F^{(S)}_{2}$ amplitude. We see that the doubly iterated amplitude has much weaker angular dependence.
Ratios of the absolute values of the isospin-averaged and spin-1 amplitudes in the  Born approximation and the second order correction to it are depicted in Fig.~\ref{fig:F2rat} in Section~\ref{ssec:fugacity-Phi}, where it is used for the discussion of the applicability of the Born approximation.

\end{document}